\documentclass[twoside]{cernyrep}

\usepackage[utf8]{inputenc}
\DeclareUnicodeCharacter{2212}{-}
\counterwithin{table}{section}
\counterwithin{figure}{section}
\counterwithin{equation}{section}

\usepackage{comment} 
\usepackage{graphicx}
\usepackage{caption}
\usepackage{epstopdf}

\usepackage{booktabs}
\usepackage{subfloat}
\usepackage{acronym}
\usepackage{subfig}
\usepackage{amsmath}
\usepackage{color}
\usepackage{balance}
\usepackage{authblk}
\usepackage{fancyhdr}
\usepackage{multirow}
\usepackage{afterpage}
\usepackage{csvsimple}
\usepackage{pgfsys}
\usepackage{siunitx}

\usepackage[hang,flushmargin]{footmisc}
\fancypagestyle{plain}{}
\usepackage{emptypage}
\usepackage{lipsum}

\usepackage{texnames}
\usepackage{ctable}
\usepackage[T1]{fontenc}

\usepackage{blindtext}

\usepackage[bookmarks, colorlinks=true, linktoc=page, pdftex, linkcolor=blue, citecolor=blue, urlcolor=blue]{hyperref}
\usepackage{float}
\usepackage{xcolor}
\usepackage[toc,page]{appendix}
\usepackage[bookmarks, colorlinks=true, pdftex, linkcolor=blue, citecolor=blue, urlcolor=blue]{hyperref}
\newlength{\oddmarginwidth}
\newlength{\evenmarginwidth}
\fancyhfoffset[LO,RE]{\oddmarginwidth}
\fancyhfoffset[LE,RO]{\evenmarginwidth}
\usepackage{times,lipsum}
\usepackage[margin=1in]{geometry}
\usepackage[onehalfspacing]{setspace}

\begin{document}
\title{Interim report for the International Muon Collider Collaboration}
\pagenumbering{roman}
\setcounter{page}{1}

\thispagestyle{empty}
\setlength{\unitlength}{1mm}
\begin{picture}(0.001,0.001)
\put(-16,8){CERN Yellow Reports: Monographs}
\put(110,8){CERN-2024-002}

\put(-16,-80){\Huge \bfseries		
Interim report for the}
\put(-16,-90){\Huge \bfseries		
International Muon Collider Collaboration}

\put(-5,-110){\Large The International Muon Collider Collaboration }

\put(65,-250){\includegraphics{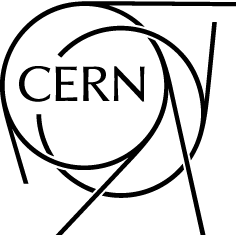}}

\end{picture}
\newpage

\thispagestyle{empty}
\mbox{}
\vfill

\begin{flushleft}
CERN Yellow Reports: Monographs\\
Published by CERN, CH-1211 Geneva 23, Switzerland\\[3mm]

\begin{tabular}{@{}l@{~}l}
 ISBN & 978-92-9083-667-4 (paperback) \\
 ISBN & 978-92-9083-668-1 (PDF) \\
 ISSN & 2519-8068 (Print)\\ 
 ISSN & 2519-8076 (Online)\\ 
 DOI & \url{https://doi.org/10.23731/CYRM-2024-002}\\
\end{tabular}\\[5mm]

Copyright \copyright{} CERN, 2024\\[1mm]
\raisebox{-1mm}{\includegraphics[height=12pt]{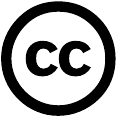}}
Creative Commons Attribution 4.0\\[5mm]

This volume should be cited as:\\[1mm]
Interim report for the International Muon Collider Collaboration,\\ The International Muon Collider Collaboration\\
CERN Yellow Reports: Monographs, CERN-2024-002 (CERN, Geneva, 2024)\\
\url{https://doi.org/10.23731/CYRM-2024-002}.\\[3mm]

Corresponding author: Christian Carli (\href{mailto:christian.carli@cern.ch}{christian.carli@cern.ch}).\\[1mm]
Accepted in Sep.\ 2024, by the \href{http://library.cern/about_us/editorial_board}{CERN Reports Editorial Board}
(contact \href{mailto:Carlos.Lourenco@cern.ch}{Carlos.Lourenco@cern.ch}).\\[1mm]
Published by the CERN Scientific Information Service (contact \href{mailto:Jens.Vigen@cern.ch}{Jens.Vigen@cern.ch}).\\[1mm]
Indexed in the \href{https://cds.cern.ch/collection/CERN\%20Yellow\%20Reports?ln=en}{CERN Document Server} and in \href{https://inspirehep.net/}{INSPIRE}.\\[1mm]
Published Open Access to permit its wide dissemination, as knowledge transfer is an integral part of the mission of CERN.

\end{flushleft}


\author {\it The International Muon Collider Collaboration}

\newpage
\thispagestyle{empty}

\vskip 10cm
\begin{center}
{\huge\bfseries
Interim report for the }
\vspace{1mm}

{\huge\bfseries\
International Muon Collider Collaboration }
\vspace{1mm}

{\huge\bfseries
(IMCC) }

\end{center}

\vspace{5mm}
{\small
\noindent
Carlotta~Accettura$^{1}$,
Simon~Adrian$^{2}$,
Rohit~Agarwal$^{3}$,
Claudia~Ahdida$^{1}$,
Chiara~Aim\'e$^{4}$,
Avni~Aksoy$^{1, 5}$,
Gian~Luigi~Alberghi$^{6}$,
Siobhan~Alden$^{7}$,
Luca~Alfonso$^{8}$,
Nicola~Amapane$^{9, 10}$,
David~Amorim$^{1}$
% ...
}

\vspace{3mm}

\begin{flushleft}

{\em\footnotesize
$^{1}$ {\href{https://ror.org/01ggx4157}{European Organization for Nuclear Research}, Geneva, Switzerland}  \\
$^{2}$ {\href{https://ror.org/03zdwsf69}{University of Rostock}, Rostock, Germany}  \\
$^{3}$ {\href{https://ror.org/02jbv0t02}{Lawrence Berkeley National Laboratory}, Berkeley, CA, United States}  \\
$^{4}$ {\href{https://ror.org/01st30669}{INFN Sezione di Pavia}, Pavia, Italy}  \\
$^{5}$ {\href{https://ror.org/01wntqw50}{Ankara University}, Ankara, T\"urkiye}  \\
$^{6}$ {\href{https://ror.org/04j0x0h93}{INFN Sezione di Bologna}, Bologna, Italy}  \\
$^{7}$ {\href{https://ror.org/04g2vpn86}{Royal Holloway University of London}, Egham, United Kingdom}  \\
$^{8}$ {\href{https://ror.org/02v89pq06}{INFN sezione di Genova}, Genova, Italy}  \\
$^{9}$ {\href{https://ror.org/048tbm396}{University of Turin}, Turin, Italy}  \\
$^{10}$ {\href{https://ror.org/01vj6ck58}{INFN Sezione di Torino}, Turin, Italy}  \\
% ...
}
\end{flushleft}



\begin{abstract}
This document summarises the International Muon Collider Collaboration (IMCC) progress and status of the Muon Collider R\&D programme.
\end{abstract}

\maketitle 
\clearpage
\tableofcontents

\cleardoublepage

\pagenumbering{arabic}
\setcounter{page}{1}
\section*{Executive summary 
}
\label{sec:Section01}
\addcontentsline{toc}{section}{\nameref{sec:Section01}}
\noindent

\long\def\nix#1{}

\def\myparagraph#1{\noindent{\bf #1}\\*}

\noindent
The recent European Strategy for Particle Physics Update (ESPPU) and the Snowmass processes have firmly established the need to get to the 10 TeV partonic center-of-mass (pCOM) energy collisions in order to make a major step forward in our understanding of particle physics.
The International Muon Collider Collaboration (IMCC) was formed in 2020 to secure a novel, sustainable path toward this goal by developing the muon collider concept and addressing the related challenges.
The collaboration obtained strong support by many partner institutes and funding agencies across the globe, the European Union and recently the Particle Physics Project Prioritization Panel (P5) in the US. It is currently hosted at CERN and has 50 full and several associated members. IMCC aims to develop a staged approach to a 10 TeV, high-luminosity muon collider.

The report summarizes progress made since the inception of IMCC and describes what is expected to be achieved by 2026 with currently secured resources. It also outlines important additional accelerator and detector studies that could be addressed and completed by 2026, if additional resources were made available.
In 2025, IMCC will provide the ESPPU with an evaluation report that assesses the muon collider potential and an R\&D path plan. Later it will submit the updated information to strategy process in the U.S.\,and potentially other regions and countries.

\subsection*{Motivation}
A muon collider with 10~TeV energy would enable an exceptionally broad physics programme. It could discover new particles with presently inaccessible mass, including weakly interacting massive particles (WIMP) dark matter candidates. It could discover cracks in the Standard Model by the precise study of the Higgs boson, including the direct observation of double-Higgs production and the precise measurement of triple Higgs coupling. It will uniquely pursue the quantum imprint of new phenomena in novel observables by combining precision with energy. It gives unique access to new physics coupled to muons and delivers beams of neutrinos with unprecedented properties from the muons' decay. 

This unmatched physics program can be realised because the muon collider offers major advantages in terms of the footprint, cost, and power consumption when compared to the next generation of high-energy lepton and hadron machines. In order to directly probe the existence of new heavy states, the luminosity of colliders has to increase with the square of the collision energy to compensate for the reduction in $s$-channel cross sections. The potential of muon colliders to improve the luminosity to beam power ratio at high energies is therefore a major benefit. Based on a combination of physics and technical considerations, the integrated luminosity target was set to 10\,ab$^{-1}$ at 10\,TeV.

\subsection*{Work programme and resources}
The IMCC, the muon beam panel of the European Large National Laboratories Directors Group (LDG) and the Snowmass process in the U.S.\,have all assessed the muon collider concept. They concluded that the concept is less mature than linear colliders but that there are no insurmountable obstacles. The concept promises a unique path toward high energy and high luminosity with limited cost and power consumption; this makes it important to address the challenges of the concept and open this path to the future.

A vigorous R\&D program is needed to develop an end-to-end design and to unequivocally demonstrate feasibility of the machine and the detector performance. With involvement of the global community, a concise set of work-packages has been developed for the European Strategy for Particle Physics---Accelerator R\&D Roadmap by the Laboratory Directors Group, hereinafter referred to as the European LDG roadmap. This process provided an excellent basis for the global collaboration and planning of future work.
The short-term goal is to assess initial technical feasibility of a 10 TeV collider, identify key R\&D items, develop an implementation timeline, and provide an initial cost estimate in time for the next ESPPU process. Further studies will address strategic needs in the U.S.\,and other regions.

IMCC is actively pursuing resources to complement the resources already foreseen by CERN and from other collaboration partners. In 2023 European Commission support was obtained for a design study of a muon collider. This project started in March 2023 and its work-packages are well aligned with the overall IMCC studies. The secured resources are used following the priorities established in the European LDG roadmap and will be able to cover about 40\% of the programme by the end of 2026.

The project continues to make important progress and the collaboration steadily grows as detailed in this report.
Examples of the ongoing efforts include studies and prototyping plans of the magnet and RF systems and their integration into a muon cooling cell. Progress is being made in designs of the muon production target, the muon cooling section, the fast-ramping accelerators, and the collider ring. Detailed studies of the machine-detector interface and the beam-induced background in the detector are also advancing.

A further increase in the design effort for the accelerator complex and the detector as well as the associated technologies is required to achieve the goals presented in the European LDG roadmap. In addition, the start of an experimental program is essential to develop the technologies, in particular the muon cooling cell magnets, RF systems and absorbers as well as their integration. Other key items are the muon production target and its solenoid, which is similar to what is required for fusion reactors. The fast-ramping magnets and power converter of the acceleration system. Several test infrastructures are instrumental to achieve progress, including one to test RF cavities in high magnetic field and one to test a muon cooling module prototype.

In December 2023, the U.S.\,Particle Physics Projects Prioritization Panel (P5) recommended that the U.S.\,should develop a collider with 10\,TeV parton collision energies, such as a muon collider, a proton collider, or possibly an electron-positron collider. The report states: ``In particular, a muon collider presents an attractive option both for technological innovation and bringing the energy frontier back to the U.S. The U.S.\,should participate to the IMCC and have the ambition to host a future collider.'' Several U.S.\,institutes have already officially joined the IMCC and members of the U.S.\,community are actively engaged in the collaboration's work. The amount of dedicated resources in the U.S.\,is not finalized and a considerable amount of time will be needed to ramp up the effort. IMCC plans to revise both the organization and the work plan together with colleagues from the U.S.\,and other regions. The ambition of the U.S.\,to host a muon collider facility further strengthens motivation for the R\&D and facilitates the demonstrator planning at a faster pace. Besides the U.S., collaborative ties are being established with other regions of the world.

\subsection*{Site and timeline}
The study is currently focusing on a site-agnostic design. The benefits of reusing the existing infrastructure, such as the existing proton complexes and collider tunnels, will be considered but detailed studies will only follow at a later time. Potential sites in Europe near CERN and in the United States at Fermilab are being explored and other regions may propose alternative sites as the study progresses.
Close to CERN, a potential location and orientation of the collider ring have been identified that can mitigate effects of the neutrino flux.
Muon collider implementation at Fermilab requires that the accelerator rings all fit within the laboratory's boundaries and first considerations indicate that this is possible.

There are many global uncertainties in planning of the next generation of collider facilities. These include budgetary and geopolitical considerations that are often outside of control for our field.
The muon collider is an important concept in the long run since it allows highest lepton collision energies; it could operate after or in parallel to a Higgs
factory.
In the U.S., the muon collider can be the next high-energy frontier project for particle physics. If no Higgs factory is being built Europe, the muon collider
could be the next project after the High-Luminosity Large Hadron Collider (HL-LHC).  
In light of these uncertainties, we focus on development of the fastest implementation timeline, while maintaining as much flexibility as possible.
This timeline is based on technical considerations, such as the typical amount of time needed to make one iteration on
the magnet technology, and hence we refer to it as ``technically limited''. We assume that no show-stoppers will occour
during the course of the R\&D programme and that the required funding and resources will be made available.

We consider a staged implementation of the collider, based on the anticipated maturity of the relevant technologies in about 15 years,
in particular the most critical ones: the muon cooling technology, the detector technologies and the high temperature superconductor (HTS) solenoid technology.
Only HTS-based collider ring magnet technology may not be mature at this timescale and is not planned for the first stage.
Depending on the physics needs and the funding situation, energy staging or luminosity staging can be considered.
One can start with a cheaper, lower energy stage (e.g.\,3 TeV) at full luminosity performance that is later extended to 10 TeV or higher with improved technology.
Alternatively, one can implement a 10 TeV collider directly, albeit with reduced luminosity, which can later be upgraded to the nominal
luminosity, similarly to HL-LHC.
Both approaches do not require HTS dipoles and allow a fast implementation of a muon collider with a start of operation before 2050, provided
the decision-making process is well prepared and development resources are made available. A muon collider can thus be
the next high-energy frontier flagship project in all regions, e.g.\,in Europe it could follow directly after the HL-LHC.

\subsection*{Synergies and outreach}
Many young scientists have developed their scientific and technical
skills in the field of particle physics and the associated accelerators;
they also learned to work in large international collaborations. The muon collider is a novel concept and opens opportunities for junior researches to make innovative contributions that are much harder to make in long-established design approaches.

The muon collider has synergies with other particle and nuclear physics projects, e.g.\,in the detector technologies and
concepts, the magnet technology for hadron colliders, the high-efficiency RF technology and high-power targets for neutron
spallation sources.
It also needs several technologies that differ from other colliders. High-field solenoids based on high-temperature superconductor are a prime example and are also of interest for fusion reactors and power generators for off-shore windmills as
well as life and material sciences, e.g.\,nuclear magnetic resonance and magnetic reasonance imaging.
 
The test facility and the collider itself require a high power proton source. This allows to share technology and potentially even facilities with neutrino facilities (e.g.\,NuSTORM, neutrino factory, DUNE), lepton flavour violation experiments (e.g.\,Mu2e, Mu2e-II, COMET, AMF) and the next generation of low-energy, highly polarized muon beams.

\subsection*{Conclusion}
The unparalleled physics case coupled with sustainability arguments 
created a very strong recent interest in the muon collider concept. This interest led to formation of the International Muon Collider Collaboration. The near term goal for the collaboration is to assess initial technical feasibility of the machine, identify key R\&D items, develop an implementation timeline, and provide an initial cost estimate in time for the next ESPPU process. There has been a significant R\&D progress made by the collaboration and its partner institutions, including those from the U.S.\,during the Snowmass process. The recent P5 recommendations further stressed the importance of the R\&D program. Realization of a staged muon collider appears promising but the effort is in strong need for additional resources, in particular to strengthen the hardware developments.

\nix{
The recent European Strategy for Particle Physics Update (ESPPU) and Snowmass processes have firmly established the need to get to the 10 TeV partonic center-of-mass (pCOM) energy collisions with high luminosity in order to make a major step forward in our understanding of particle physics. Global sustainability targets require that future particle physics facilities are designed in a way that minimizes their environmental impact.

\nix{
A 10 TeV muon collider would enable an unprecedentedly broad physics program with both energy reach for direct discoveries of new particles and precision necessary for to observe new physics indirectly, while maintaining a relatively small footprint and low power consumption. However, realization of such a collider on a timescale of approximately two decades requires a significant R\&D program aimed at developing the design and demonstrating critical accelerator and detector technologies. 
}

A muon collider with 10~TeV energy would enable an exceptionally broad physics programme. It could discover new particles with presently inaccessible mass, including WIMP dark matter candidates. It could discover cracks in the Standard Model by the precise study of the Higgs boson, including the direct observation of double-Higgs production and the precise measurement of triple Higgs coupling. It will uniquely pursue the quantum imprint of new phenomena in novel observables by combining precision with energy. It gives unique access to new physics coupled to muons and delivers beams of neutrinos with unprecedented properties from the muons decay. 

In addition to an unmatched physics program, muon colliders offer major advantages in terms of the footprint, cost, and power consumption when compared to the next generation of high-energy lepton and hadron machines. In order to directly probe the existence of new heavy states, the luminosity of colliders has to increase with the square of the collision energy to compensate for the reduction in $s$-channel cross sections. The potential of muon colliders to improve the luminosity to beam power ratio at high energies is therefore a major benefit. Based on a combination of physics and technical considerations, the initial integrated luminosity targets was set to 10\,ab$^{-1}$ at 10\,TeV. Various energy and luminosity staging strategies are possible and will be discussed later in this document.

The International Muon Collider Collaboration (IMCC) was formed in 2020 with the goal to secure a sustainable path towards the highest energy collisions via development of a novel high-energy and high-luminosity muon collider concept and addressing the associated technical challenges. 
IMCC is developing a baseline design of a 10 TeV muon collider and the associated R\&D program necessary for realization of such a machine. Various implementation and staging strategies are being studied along with exploration of potential sites for various technology demonstrators as well as the final collider facility.

The collaboration obtained strong support by many partner institutes across the globe, the European Union and recently the Particle Physics Project Prioritization Panel (P5) in the US. The near term goal for the collaboration is to assess initial technical feasibility of the machine, identify key R\&D items, develop an implementation timeline, and provide an initial cost estimate in time for the next ESPPU process in 2026. 

The report summarizes progress made since the inception of IMCC and describes what is expected to be achieved by 2026 with currently foreseen resources. It also outlines important additional accelerator and detector studies that would require additional resources to be addressed and completed by the next ESPPU, making a case for such additional resources.

\subsection*{Technical challenges and progress}
In the recent years the IMCC, the muon beam panel of the European Large National Laboratories Directors Group (LDG) and the Snowmass process in the U.S.\,all assessed muon collider challenges and concluded that the concept was less mature than linear colliders but that there were no insurmountable obstacles. Past work on muon colliders produced component designs and provided partial demonstration of the technologies that allow to cool the initially diffuse muon beam and accelerate it to multi-TeV energy on a time scale compatible with the muon lifetime. However, a vigorous R\&D program is needed to develop an end-to-end design and to unequivocally demonstrate feasibility of the machine. 

With involvement of the global community, a concise set of work-packages has been developed for the European LDG roadmap~\cite{c:roadmap}. The program provided an excellent basis for the global collaboration and planning of future work towards a multi-TeV muon collider. The project continues to make important progress and the collaboration steadily grows. A specific goal of the European LDG roadmap is to provide two deliverables for the next ESPPU:
\begin{itemize}
\item A project evaluation report that assesses the muon collider potential;
\item An R\&D plan that describes a path towards the collider.
\end{itemize}

IMCC is currently hosted by CERN and covers topics related to the accelerator complex, detectors and physics for a future multi-TeV muon collider. IMCC is actively pursuing resources to complement the resources already foreseen at CERN and from other collaboration partners. In 2023 European Commission support was obtained for a design study of a muon collider. This project started in March 2023 and its work-packages are well aligned with the overall IMCC studies. More than 70 partner institutes are currently involved in the studies. 

The collaboration is making steady progress towards the deliverables mentioned above. Examples of the ongoing efforts include studies and prototyping plans of the magnet and RF systems and their integration into a muon cooling cell. Progress is being made in designs of the muon production target, the muon cooling section, the fast-ramping accelerators, and the collider ring. Detailed studies of the machine-detector interface and the beam-induced background in the detector are also advancing. The study continues to be resource limited with respect to what is needed to addressing all the European LDG roadmap targets. Additional resources are being actively sought across the collaboration.

The IMCC leadership is actively engaged in growing the collaboration. In December 2023, the U.S.\,Particle Physics Projects Prioritization Panel (P5) recommended that the U.S.\,should develop a collider with 10\,TeV parton collision energies, such as a muon collider, a proton collider, or possibly an electron-positron. The report states: ``In particular, a muon collider presents an attractive option both for technological innovation and bringing the energy frontier back to the U.S. The U.S.\,should participate to the IMCC and have the ambition to host a future collider.'' Several U.S.\,institutes have already officially joined the IMCC and members of the U.S.\,community are actively engaged in collaborations work. The amount of dedicated resources in the U.S.\,is not finalized and a considerable amount of time will be needed to ramp up the effort. IMCC plans to revise both the organization and the work plan together with colleagues from the U.S. The ambition of the U.S.\,to host a muon collider facility further strengthens motivation for the R\&D and facilitates the demonstrator planning at a faster pace. Besides the U.S., collaborative ties are being established with other regions of the world.

\subsection*{Urgent needs}
A further increase in the design effort for the accelerator complex and the detector as well as the associated technologies is required to achieve the goals presented in the European LDG roadmap. In addition, the start of an experimental program is essential to develop the technologies, in particular the muon cooling cell magnets, RF systems and absorbers as well as their integration. Other key items are the muon production target and its solenoid, which is similar to what is required for fusion reactors. The fast-ramping magnets and power converter of the acceleration system. Several test infrastructure are instrumental to achieve progress, including one to test RF cavities in high magnetic field and one to test a muon cooling module prototype.

\subsection*{Timeline considerations}
There are many global uncertainties in planning of the next generation collider facilities. These include budgetary and geopolitical considerations that are often outside of control for our field. In light of these uncertainties, we focus on development of the fastest implementation timeline, while maintaining as much flexibility as possible. 

In development of the timeline for the muon collider implementation, we assume that all the required funding and resources will become available. The timeline is thus based primarily on technical considerations, such as the typical amount of time needed to make one iteration on the magnet technology; we refer to it as "technically limited". We also assume that no show-stoppers will be identified during the course of the R\&D program. 


Three R\&D areas have been identified to be on the critical path and therefore define the technically limited timeline:
\begin{itemize}
\item The muon production and cooling technologies and their demonstration in a dedicated test facility;
\item Development of the magnets and in particular the HTS technology;
\item Detector technologies necessary to operate in the muon collider environment.
\end{itemize}
There are other R\&D areas that require work, but they should not constrain the timeline assuming that adequate funding is allocated.
At this moment the studies indicate that a fast implementation of a muon collider is possible with a start of operation before 2050, provided
the decision-making process is well prepared and development resources are made available. A muon collider can thus be the next high-energy frontier flagship project in all regions, e.g.\,in Europe it could follow directly after the HL-LHC.

The timeline assumes a staged approach of the collider implementation. Depending on the physics needs and the funding situation, we can start with a lower energy stage (e.g.\,3 TeV) that is later upgraded and extended with improved technology to 10 TeV. Alternatively, we can implement a 10 TeV collider directly albeit with reduced luminosity, which can later be upgraded to the nominal luminosity, similarly to HL-LHC.

\subsection*{Site studies}
The study is currently focusing on a site-agnostic design and the  implications of specific sites will be investigated at a later time. The interesting benefits of reusing the existing infrastructure, such as the existing proton complexes and collider tunnels, will be studied. Potential sites in Europe near CERN and in the United States at Fermilab have been identified and other regions may propose alternative sites as the study progresses. Muon collider implementation at Fermilab requires that the accelerator rings all fit within the laboratory's boundaries and first considerations indicate that this is possible. For the CERN implementation, potential location and orientation of the collider ring,  that can mitigate effects of the neutrino flux, have been identified.

\subsection*{Synergies and outreach}
Many scientists have developed their scientific and technical
skills in the field of particle physics and the associated accelerators;
they also learned to work in large international collaborations. The muon collider is a novel concept and opens opportunities for junior researches to make innovative contributions that are much harder to make in long-established design approaches.

The muon collider needs in several areas technologies that differ
from other colliders. High-field solenoids are a prime example.
In the past low-temperature superconductors such as the very mature NbTi and still developing Nb$_3$Sn were the technologies of choice for accelerators and most other applications. Now high-temperature superconductors (HTS) are becoming an important technology.
In particular they are of interest for fusion reactors, that have similar requirements than the one for the muon collider target solenoid. Highly-efficient superconducting motors and power generators, e.g.\,for off-shore windmills, also have strong synergy. Other relevant areas are life and material sciences; in particular, applications for nuclear magnetic resonance (NMR) and magnetic resonance imaging (MRI). In addition synergy exists with magnets for axion searches, neutron spectroscopy, physics detectors, and magnets for other particle colliders, such as hadron colliders.

The muon production target is synergetic with neutron spallation sources targets, in particular the alternative liquid metal concept.

The muon collider RF power sources have synergy with other developments of high-efficiency klystrons and superconducting cavities. Some RF systems need to work in high magnetic fields, an issue that also exists in fusion reactor designs.
 
The test facility and the collider itself require a high power proton source. This allows to share technology and potentially even facilities. Examples are neutrino facilities (e.g.\,NuSTORM, neutrino factory), lepton flavour violation experiments (e.g.\,Mu2e, Mu2e-II, COMET, AMF) and the next generation of low-energy, highly polarized muon beams.

\subsection*{Conclusion}
The unparalleled physics case coupled with sustainability arguments 
created a very strong recent interest in the muon collider concept. This interest led to formation of the International Muon Collider Collaboration. The near term goal for the collaboration is to assess initial technical feasibility of the machine, identify key R\&D items, develop an implementation timeline, and provide an initial cost estimate in time for the next ESPPU process in 2026. There has been a significant R\&D progress made by the collaboration and its partner institutions, including those from the U.S.\,during the Snowmass process. The recent P5 recommendations further stressed the importance of the R\&D program. Realization of a staged muon collider appears promising but the effort is in strong need for additional resources, in particular to strengthen the hardware developments.
}
\nix{
\begin{flushleft}
\interlinepenalty=10000

\end{flushleft}
}
\clearpage

\section{Overview of collaboration goals, challenges and R\&D programme}
The International Muon Collider Collaboration (IMCC)~\cite{c:IMCC} was established in 2020
following the recommendations of the European Strategy for Particle Physics
(ESPP) 
and the implementation of the European Strategy for Particle Physics---Accelerator R\&D Roadmap by the Laboratory Directors Group~\cite{c:roadmap}, hereinafter referred to as the the European LDG roadmap. The Muon Collider Study (MuC) covers the accelerator complex, detectors
and physics for a future muon collider. In 2023, European Commission support was
obtained for a design study of a muon collider (MuCol)~\cite{c:mucol}. This project started on
1$^{\mathrm{st}}$ March 2023, with work-packages aligned with the overall muon collider studies. In preparation of and during the 2021--22 U.S.\,Snowmass process, the muon collider project parameters, technical studies and physics performance studies were performed and presented in great detail. 
Recently, the P5 panel~\cite{c:P5} in the U.S.\,recommended a muon collider R\&D, proposed
to join the IMCC and envisages that the U.S.\,should prepare to host a muon
collider, calling this their ``muon shot''. In the past the U.S.\,Muon Accelerator Programme (MAP)~\cite{map_overview} has been instrumental in studies of concepts and technologies for a muon collider.

\subsection{Motivation}
High-energy lepton colliders combine cutting edge discovery potential
with precision measurements. Because leptons are point-like particles in
contrast to protons, they can achieve comparable physics at lower centre-of-mass
energies~\cite{c:white1,Delahaye:2019omf,AlAli:2021let,Accettura:2023ked}. However, to efficiently reach the 10+ TeV scale recognized by ESPP and P5 as a necessary target requires a muon collider. A muon collider with 10~TeV energy or more could discover new particles with presently inaccessible mass, including WIMP dark matter candidates. It could discover cracks in the Standard Model (SM) by the precise study of the Higgs boson, including the direct observation of double-Higgs production and the precise measurement of triple Higgs coupling. It will uniquely pursue the quantum imprint of new phenomena in novel observables by combining precision with energy. It gives unique access to new physics coupled to muons and delivers beams of neutrinos with unprecedented properties from the muons' decay. Based on physics considerations, an integrated luminosity target of 10\,ab$^{-1}$ at 10\,TeV was chosen.  However, various staging options are possible that allow fast implementation of a muon collider with a reduced collision energy or the luminosity in the first stage and reaches the full performance in the second stage.

In terms of footprint, costs and power consumption a muon collider has potentially very favourable properties.
\begin{figure}[b]
  \centerline{\raisebox{20pt}{\includegraphics[width=7.5cm]{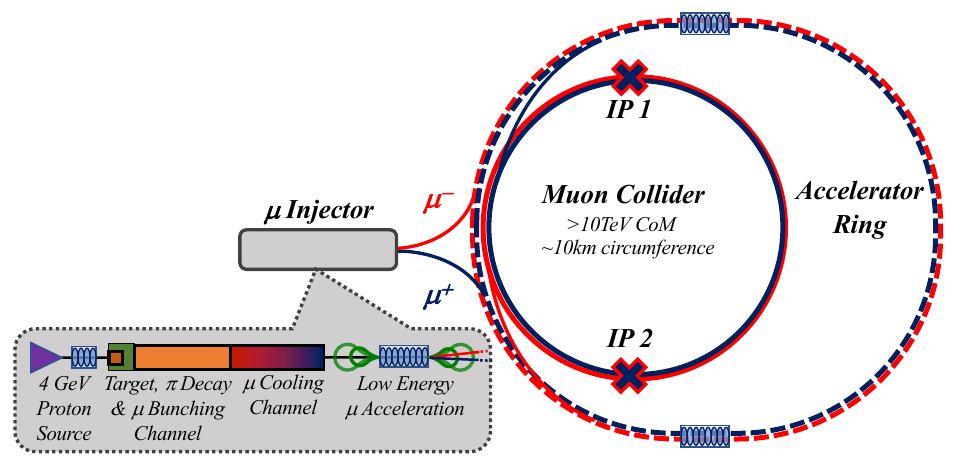}}\includegraphics[width=7.5cm]{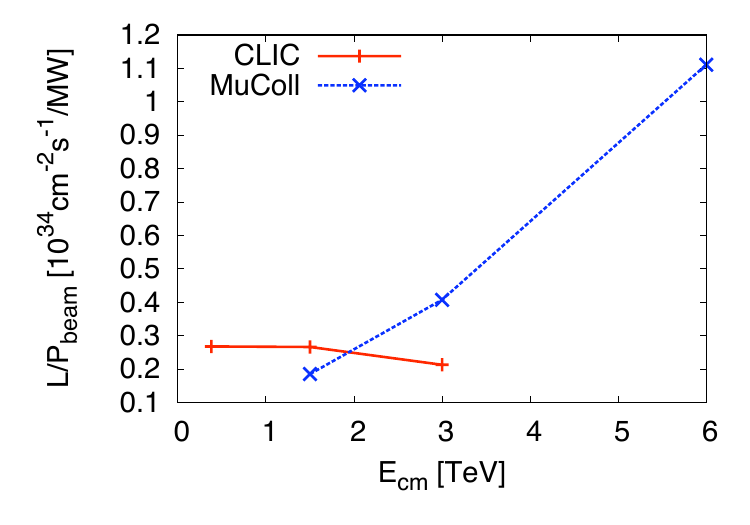}}
\caption{Left: Conceptual scheme of the muon collider. Right: Comparison of CLIC and a muon collider luminosities normalised
to the beam power and as a function of the centre-of-mass energy.}
\label{f:design}
\end{figure}
The luminosity of lepton colliders has to increase with the square of the
collision energy to compensate for the reduction in $s$-channel cross sections.
Figure~\ref{f:design} (right panel) compares the luminosities of the Compact Linear Collider (CLIC) and a muon collider, based on the U.S.\,Muon Accelerator Programme (MAP)
parameters~\cite{Delahaye:2019omf}, as a function of centre-of-mass energy.
The luminosities are normalised to the beam power.
The potential of muon colliders to improve the luminosity to beam power ratio
at high energies is one of the main advantages of the concept.

\subsection{The accelerator concept} 
IMCC studies a muon collider concept that has initially developed by MAP;
a schematic view is shown in Fig.~\ref{f:design} (left panel).

The proton complex produces a short, high-intensity proton pulse that hits the
target and produces pions. The decay channel guides the pions and collects the
produced muons into a buncher and phase rotator system to form a muon beam.
Several cooling stages then reduce the longitudinal and transverse emittance of
the beam using a sequence of absorbers and RF cavities in a high magnetic field.
A system of a linac and two recirculating linacs accelerate the beams to
63\,GeV followed by a sequence of high-energy accelerator rings; the optimum sequence needs to be determined based on the ongoing studies.
Finally the beams are injected at full energy into the collider ring. Here, they will circulate to
produce luminosity until they are decayed; alternatively they can be extracted
once the beam current is strongly reduced. 

A set of parameters has been defined for 10\,TeV and also 3\,TeV. These are target parameters to explore the limits of each technology and design. If they can be fully met, the integrated luminosity goal could be reached within five years (or $2.5$~years, with two detectors) of full luminosity operation. This provides margin for further design and technology studies and a realistic ramp-up of the luminosity. It also enables to consider initial stages that can be implemented faster but often with reduced luminosity performance in this stage.

\begin{table}[!hb]
\caption{Tentative target parameters for a muon collider at different
  energies. These values are only to give a first indication and correspond to the two staging scenarios discussed in Section~\ref{s:stagingSCD}. The estimated luminosity refers to the value that can be reached if all target specifications can be reached; it also includes the beam-beam effect.}
\label{t:facility_param}
\begin{center}
  \begin{tabular}{|*3{c|}|*2{c|}|*2{c|}|}
    \hline
    Parameter & Symbol & unit & \multicolumn{2}{c||}{Scenario 1} & \multicolumn{2}{c||}{Scenario 2}\\
    \hline
 & & & Stage 1 &Stage 2 & Stage 1 & Stage 2\\
\hline
    Centre-of-mass energy & $E_{\mathrm{cm}}$ & TeV & 3 & 10 & 10 &10\\
    Target integrated luminosity & $\int{\cal L}_{\mathrm{target}}$ & $\rm ab^{-1}$ & 1 & 10 & \multicolumn{2}{c||}{10}\\
    Estimated luminosity & ${\cal L}_{\mathrm{estimated}}$ & $10^{34}\rm cm^{-2}s^{-1}$ & 2.1 & 21 & tbc & 14\\
    Collider circumference& $C_{\mathrm{coll}}$ & $\rm km$ & 4.5 & 10 & 15 &15\\
    Collider arc peak field& $B_{\mathrm{arc}}$ & $\rm T$ & 11 & 16 & 11 & 11\\
    Luminosity lifetime & $N_{\mathrm{turn}}$ &turns& 1039 & 1558 & 1040 & 1040\\
    \hline
    Muons/bunch & $N$ & $10^{12}$ & 2.2 & 1.8 & 1.8 &1.8\\
    Repetition rate & $f_{\mathrm{r}}$ & $\rm Hz$ & 5 & 5 &5 &5\\
    Beam power  & $P_{\mathrm{coll}}$ & $\rm MW$ &5.3  & 14.4 & 14.4 &14.4\\
    RMS longitudinal emittance& $\varepsilon_\parallel$ & $\rm eVs$ & 0.025 & 0.025 & 0.025 &0.025\\
    Norm.\,RMS transverse emittance& $\varepsilon_\perp$ & \textmu m & 25 & 25 & 25 &25\\
    \hline
    IP bunch length& $\sigma_z $ & $\rm mm$ & 5 & 1.5 & tbc &1.5\\
    IP betafunction& $\beta $ & $\rm mm$ & 5 & 1.5 & tbc & 1.5\\
    IP beam size& $\sigma $ & \textmu m & 3 & 0.9 & tbc & 0.9\\
    \hline
    Protons on target/bunch & $N_{\mathrm{p}}$ & $10^{14}$ & 5 & 5 & 5 & 5\\
    Protons energy on target  & $E_{\mathrm{p}}$ & $\rm GeV$ & 5 & 5 & 5 & 5\\
    \hline
    BS photons &$ N_{\mathrm{BS},0}$ &per muon& 0.075 &0.2 & tbc & 0.2\\
    BS photon energy & $E_{\mathrm{BS,0}}$ & MeV & 0.016 & 1.6  &tbc & 1.6\\
    BS loss/lifetime (2 IP)&$E_{\mathrm{BS,tot}}$&GeV& 0.002 &1.0 &tbc &0.67\\
    \hline
  \end{tabular}
\end{center}
\end{table}

\begin{table}
  \caption{Tentative target beam parameters along the acceleration chain.
    A 10~\% emittance growth budget has been foreseen in the transverse and
    longitudinal planes, both for 3 and 10 TeV. This assumes that the
    technology and tuning procedures will have been improved between the two
    stages. The very first acceleration is assumed to be part of the final cooling. This choice allows to optimise the energy in the last absorber with no strong impact on the acceleration chain.}
  \label{t:beam_param}
\begin{center}
  \begin{tabular}{|*6{c|}}
    \hline
    Parameter & Symbol & Unit &  Final cooling & at 3 TeV & at 10 TeV \\
    \hline
    Beam total energy & $E_{\mathrm{beam}}$ & GeV & 0.255 & 1500 & 5000 \\
    \hline
    Muons/bunch & $N_{\mathrm{b}}$ & $10^{12}$ & 4 & 2.2 & 1.8 \\
    Longitudinal emittance& $\varepsilon_\parallel$ & $\rm eVs$ & 0.0225 & 0.025 & 0.025 \\
    RMS bunch length& $\sigma_z$ & $\rm mm$ & 375 & 5 & 1.5 \\
    RMS rel.\,momentum spread& $\sigma_P/P$ & $\rm \%$ & 9 & 0.1 & 0.1 \\
    Transverse norm.\,emittance& $\varepsilon_\perp$ & \textmu m & 22.5 & 25 & 25 \\
    \hline
    Aver.\,grad.\,$0.2$--$1500\rm\,GeV$&$G_{\mathrm{avg}}$&$\rm MV/m$& --- & 2.4 &\\
    Aver.\,grad.\,$1.5$--$5\rm\,TeV$&$G_{\mathrm{avg}}$&$\rm MV/m$& --- &  & 1.1\\
    \hline
  \end{tabular}
\end{center}
\end{table}

\subsection{Muon collider challenges}
The Muon Collider Collaboration, the muon beam panel of the Laboratory Directors Group (LDG) and the Snowmass process in the U.S.\,have all assessed the muon collider challenges with the support of the global community. Key conclusions are that although the muon collider concept is less mature than several linear collider concepts no insurmountable obstacles have been identified, and that important design and technical challenges have to be addressed with a coherent international effort. Furthermore, past work, in particular within the U.S.\,Muon Accelerator Programme (MAP)~\cite{map_overview}, has demonstrated several key MuC technologies and concepts, and gives confidence that the overall concept is viable. Since then further component designs and technologies have been developed that provide increased confidence that one can cool the initially diffuse beam and accelerate it to multi-TeV energy on a time scale compatible with the muon lifetime. However, a fully integrated design has yet to be developed and further development and demonstrations of technology are required. 

The current IMCC programme prepares the way towards a full conceptual design report (CDR) and a demonstration programme for a muon collider.
The IMCC efforts cover physics potential as well as detector and accelerator design and performance studies. The programme will provide the necessary information and input for the next European Strategy for Particle Physics Update and other international strategy processes, allowing stakeholders to make informed decisions about these next steps. 

Considering the facility design and technical challenges key performance drivers and goals have been identified, providing guidance for prioritising studies and efforts. They include the challenges that have previously identified by the
MAP study but represent a wider set. Particularly important are: 
\begin{itemize}
\item {\bf Environmental impact.}
The compact footprint, limited cost and power consumption are intrinsic
features that motivate the muon collider study in the first place. 
Radiation protection measures will ensure a negligible impact of the facility
on the environment, similar to the LHC. Based on the experience of LHC and other accelerators,
one can expect to mitigate the impact of losses in different parts of the accelerator complex. Particular attention will be paid to the
neutrino flux that is produced by the decays of the muons in the collider ring
and that exits the ground far from the collider. A mover system has been
proposed that will avoid significant localised neutrino flux from the arcs
by vertically changing the beam angle between $- 1$\,mradian to $1$\,mradian and
indeed can achieve a negligible level. The mover system is currently under
study and the impact it has one the beam will be studied soon.
The orientation of the
The straight experimental insertions of the collider ring will lead to a
localised higher neutrino flux. Civil engineering studies identified a position
and an orientation of the collider ring in which the neutrinos will exit into
the mediteranian sea and on the uninhabited side of a mountain, which could
be fenced.

\item {\bf Machine-detector interface and detector.}
The muons of the beams that circulate in the collider ring decay, each
producing two neutrinos and one electron or positron.
The latter will hit the aperture and create showers. Tungsten masks protect
the detector from this beam induced background. The detector will
reduce the impact of the residual background by using components with high time and space resolution.
Also the potential background from beam-beam effects is being explored but should be less severe.
The studies indicate that the radiation in the detector
is roughly similar to HL-LHC. A part of the physics measurements is not affected
by the background, some part shows some residual impact. Further optimisation is ongoing to also
remove this impact.

\item {\bf Proton complex.}
In the baseline, a proton beam power of around 2\,MW at 5\,Hz is used for
muon production. Designs for proton facilities with similar or larger power
exist. The main proton complex challenge arises from the combination
of the protons into short, high-charge bunches. The corresponding workpackage of the collaboration
is addressing this.

\item {\bf Muon production.}
The key challenge for the high-power target is the survival of the target
itself under the shock waves of the incoming beam pulses and the temperature
gradients to remove the deposited heat.
Three technologies are under consideration, a solid graphite target, a liquid metal target or
an fluidised tungsten target.
The target is immersed into a 20\,T solenoid field to efficiently collect the muons.
This solenoid requires a large aperture to allow for sufficient shielding against
the transverse showers from the target. Currently the studies indicate that a graphite target and its
solenoid are feasible and can provide a sufficient number of muons.

A liquid metal or fludised tungsten target would allow to sustain a higher proton beam power and
produce a larger number of muons. This can lead to improved performance and provides margin with the muon
bunch charge that could be used to reduce the cost of the downstream systems. The FCC-ee also plans to use
a liquid metal target to dump the beamstrahlung photons from the interaction region.

\item {\bf Muon cooling design.}
Muon ionisation cooling increases the muon beam brightness by repeatedly
slowing it in absorbers and re-accelerating it in RF cavities; both inside of
strong solenoid fields to keep the beam focused. This principle has been demonstrated
in MICE~\cite{mice}. The target parameters anticipate some improvement of the cooling complex
performance, e.g.\,a factor two reduction of the transverse emittance in the final cooling.
Studies of the physical limitations of the cooling indicate that the target can be reached in principle.
The current effort focuses on improving the lattice with the aim to reach this principle limit.

\item {\bf Muon cooling technology.}
  The muon cooling system requires close integration of absorbers and RF
  cavities in a strong magnetic field. The operation of RF cavities in a
  magnetic field can strongly limit the gradient that can be achieved with no
  breakdown. This has been addressed by the MAP study, which showed that
  gradients in excess of the design target can be achieved by using cavities
  with beryllium endplates or by filling the cavities with hydrogen~\cite{}.
  Unfortunately, the RF test stands in which these experiments were carried out
  has been dismanteled. The collaboration is developing a design for a new
  facility to test the RF in high magnetic field. IMCC is actively searching
  for the resources to implement such a facility and considers this a prime goals
  of the collaboration.

  The collaboration started to develop and engineering design of a cooling cell.
  This will address the challenges of its components and their integration into
  one unit. The findings may have important impact on the lattice design.

  The final cooling uses highest-field small aperture solenoids. Currently,
  fields of more than 30\,T can be achieved using HTS in the whole solenoid.
  The goal for the next generation is 40\,T and we use this value in the current
  design effort for the final cooling.

\item {\bf Muon acceleration.}
The lion share of the muon beam acceleration will be performed by a sequence of pulsed
synchrotrons (RCS); an alternative use of fixed field accelerators (FFAs) is also considered. In each RCS
the magnet field is ramped up in proportion to the energy gain of the beam.
Some synchrotrons are based on a hybrid design where the fast-ramping
magnets are interleaved with static superconducting ones. The pulsed
synchrotrons face challenges in terms of optics design, the fast-ramping magnet
systems and the RF systems. Field ramp rates between a few hundred T/s in the
largest final and several kT/s in the smallest first ring are currently
foreseen. The latter requires normal-conducting magnets
while for the former superferric or HTS magnets can also be considered.
Studies indicate that the ramp speeads can be obtained. A specific challenge
is the large stored energy in the magnets (in total in the range of O(100\,MJ)),
which requires demanding power converters with very efficient
recovery of the energy of each pulse for the subsequent one. 

Ongoing studies of the RCS considering lattice design, pulsed magnets and RF are promising.
An integrated cost model is being developed that allows to identify the ramp shape with the
best trade-off between RF and magnet demands.

\item {\bf Collider ring.}
The collider ring requires a small beta-function at the collision point,
resulting in significant chromaticity that needs to be compensated.
It also needs to maintain a short bunch.

High-energy electrons and positrons that arise from muon decay and strike the
collider ring magnets can cause radiation damage and unwanted heat load.
Studies show that this can be mitigated with sufficient tungsten shielding and
concepts of efficient cooling of these shields have been made.
The shielding requires a larger magnet aperture in the range which is taken
into account in their concept.

A solution for 3\,TeV collider ring lattice has been
developed by the MAP study and successfully addresses the challenges. The
design is based on Nb$_{3}$Sn magnets with performances similar to the ones for
HL-LHC. The magnet experts expect that this technology is fully mature in 15
years.

A design of 10\,TeV is more challenging because it requires an even smaller
beta-function. The current studies with higher field HTS and hybrid magnets
currently achieves the beta-function but does not yet achieve the target energy
acceptance. Further efforts aim to improve this. Otherwise one would need to
reduce the beam energy spread by about a factor two, potentially leading to a
30\% luminosity reduction.

\item {\bf Collective effects.} A very high muon bunch charge is required
  to achieve the luminosity goal. This can lead to important collective
  effects---such as space charge, beam loading, wakefields---that might
  limit the collider performance.
  
\item {\bf Cost and power consumption.} The cost and power consumtion of the
  faciliy is an important challenge since sustainability must be prime concern
  for future scientific projects. 
\end{itemize}

\subsection{Developing the muon collider study}
During and after the development of the European LDG roadmap, several
steps were important to generate the framework for the muon collider studies:
\begin{itemize}
\item The muon collider has been reviewed by the global community.
  Both, the muon beam panel of the European LDG roadmap as well as
  the Energy Frontier and Accelerator Frontier working groups of the U.S.\,Snowmass process
  used the world-wide expertise in assessing the challenges of the concept.
  This has been instrumental in identifying the required R\&D programme.
\item The International Muon Collider Collaboration has been established.
  The collaboration provides the organisational framework for all the efforts
  to develop the collider concept.
\item The inclusion of the muon collider in the medium term plan of CERN in
  2021, providing an important part of the resources.
\item The decision of the European Union to fund the proposal for a
  co-financed design study (MuCol) that started in 2023. This also served to
  initiate the additional contributions from many of the collaboration partners.
\item At the end of 2023, the U.S.\,P5 process recommended that the U.S.\,join the
  IMCC and to consider hosting such a facility. The collaboration organisation
  will adapt to fully include the U.S.\,as efforts will ramp up. The goals set out for the muon collider studies during the Snowmass process and in the P5 recommendations are being integrated into the IMCC future programme.  
\end{itemize}
The European LDG roadmap identifies a prioritised R\&D programme to make fully informed decisions at the next strategy process.
It estimated the required resources to complete this programme to be a total of 446 FTE-years and 11.9 MEUR.
With the currently secured funding the collaboration can provide slightly less than 200 FTE-years until and including
2025. Almost no material budget is available at this moment. However, the collaboration is activley seeking to
increase the funding.

\subsubsection{IMCC short-term goals}
The International Muon Collider Collaboration has formed with the short term goal to address the European LDG roadmap
in time for the next European Strategy for Particle Physics Update. More than 60 partner
institutions are currently involved in the muon collider studies. Three main deliverables are foreseen:
\begin{itemize}
\item a project evaluation report that assesses the muon collider potential;
\item a R\&D plan that describes a path towards the collider;
\item an interim report early in 2024 (the present document) that documents progress and allows the wider community to be updated on the concept and to give feedback to the collaboration.
\end{itemize}
IMCC envisages to study the 10\,TeV option, and also
explore lower and higher energy options, e.g., a 3\,TeV option as a step toward
10\,TeV. The details of the required work and the required resources are documented in the European LDG roadmap~\cite{c:roadmap}.

Overall facility parameters as cost, power/energy consumption and carbon footprint will be estimated
as the design matures, but are in many cases driving technical choices and optimisation. 

\subsubsection{IMCC organisation}
A memorandum of cooperation (MoC) for the IMCC has been drawn up and the collaboration is rapidly growing. 
The formal structure of the collaboration, as shown in Fig.~\ref{f:governance} consists of the following main bodies: 
\begin{itemize}
\item International collaboration board (ICB):
The ICB role is to provide a forum for participants to examine ongoing activities,
assure appropriate, well directed use of contributions and ensure
a balanced portfolio of engagements.
The chair of the ICB is elected by its members (simple majority).
The ICB consists of one representative of each participating institute that has signed the MoC. The ICB has been active since October 2022. 
\item Steering board (SB):
The SB oversees the execution of the muon collider study by assessing the progress, resource usage and providing guidance.
The SB reports to the LDG in Europe and to similar regional supervising organisations in other regions as required.
The SG is active and operational since Spring 2023. 
\item International advisory committee (IAC): The mandate of the IAC is to review the scientific and technical progress of the study. This body is active since Spring 2024.
\item Coordination committee (CC): The CC performs the overall coordination and execution of the muon collider study.
It is chaired by the study leader. This body consists of the workpackage leaders and some additional experts of the collaboration and is very active. It provides the daily guidance of the study.
\end{itemize}
The workpackages of the collaboration are shown in Fig.~\ref{f:governance}.

\begin{figure}
\centerline{\includegraphics[width=12cm]{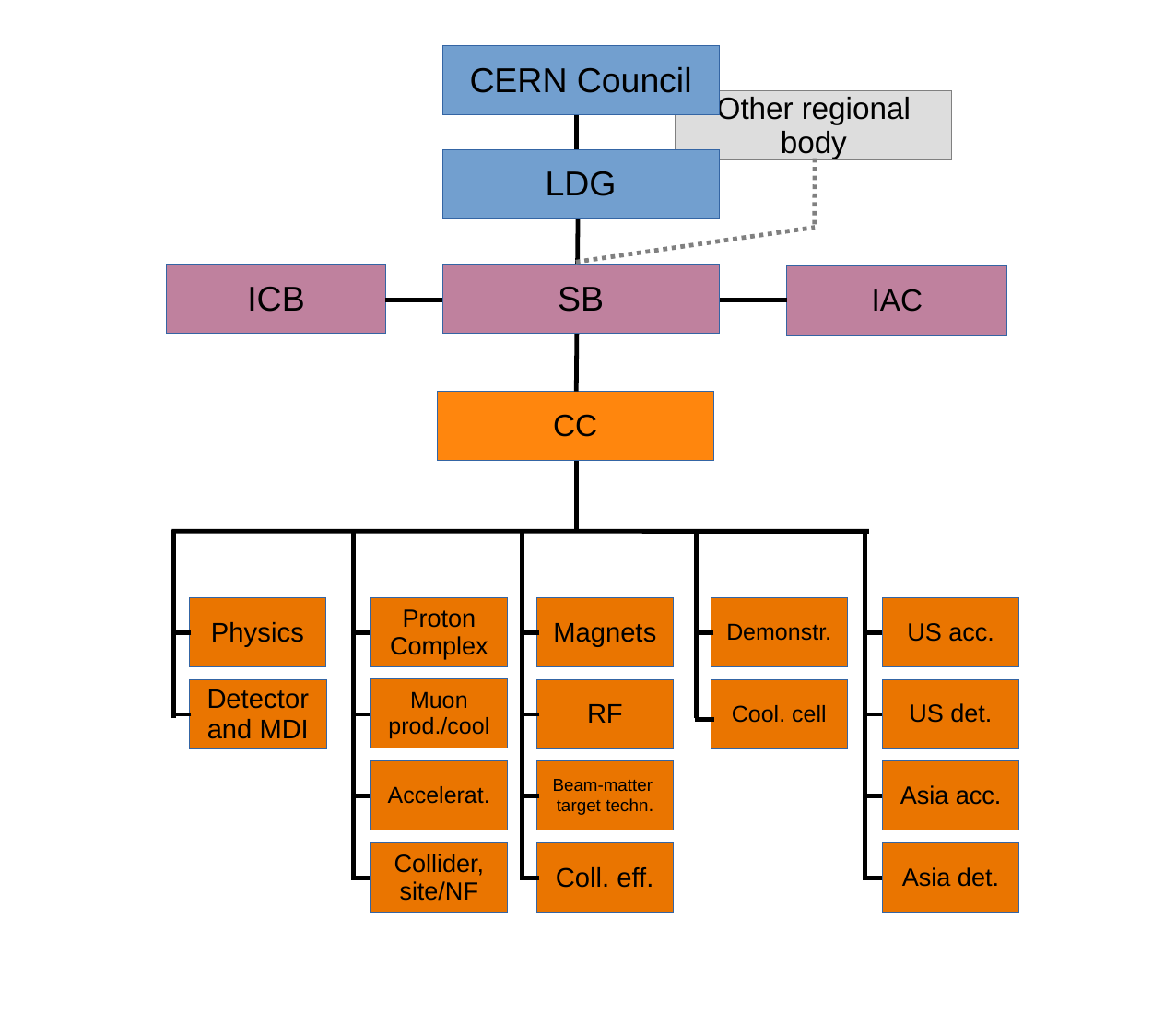}}
\caption{Organigram of the muon collider study governance structure}
\label{f:governance}
\end{figure}

The study leader is proposed by the host organization, and endorsed by the ICB. The study leader was appointed in 2022 and has led the work to build up the collaboration and activities since then.

\subsubsection{The EU co-funded design study MuCol}
IMCC successfully applied for an EU cofunded design study (named MuCol)~\cite{c:mucol}. The project started on $1^{\mathrm{st}}$~March 2023. The total co-funding of MuCol amounts to 3\,MEUR, provided by
the European Commission, the U.K.\,and Switzerland. In the design study, CERN only receives limited contributions for
administrative support and travel, but has, in support of the successful design study bid, increased its contribution to the muon collider study.

The design study is fully integrated in the overall muon collaboration. The technical meetings and leaders are in common and governance and management are also synchronised. The organisation is shown in Fig.~\ref{f:MuColwp}.
Important activities in the collaboration are directly supported by the MuCol work-packages. In particular MuCol contains the workpackges Physics and Detector Requirements, Proton Complex, Muon Production and Cooling, High-energy Complex, Radio Frequency Systems, Magnet Systems and Cooling Cell Integration.

\begin{figure}
\centerline{\includegraphics[width=10cm]{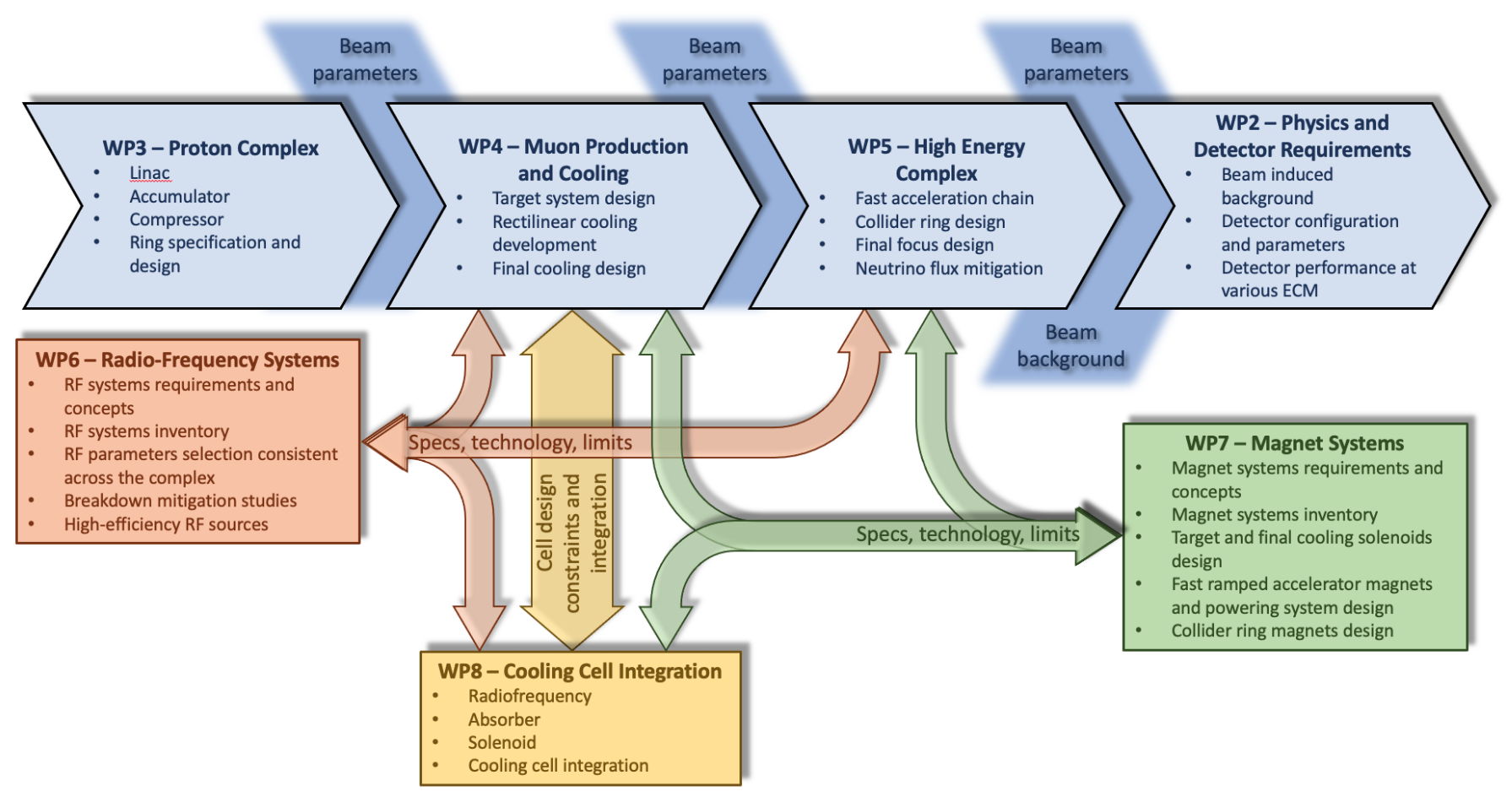}}
\caption{The MuCol work-packages and their interactions.}
\label{f:MuColwp}
\end{figure}

\subsubsection{Extending the collaboration---US plans after P5}
In the US, the Particle Physics Projects Prioritization Panel (P5) recommended in December 2023
that the U.S.\,should develop a collider with 10\,TeV parton collision energies,
such as a muon collider, a proton collider, or maybe an electron-positron collider.
The report states: ``In particular, a muon collider presents an attractive
option both for technological innovation and bringing the energy frontier back
to the US. The U.S.\,should participate to the IMCC and have the ambition to host
a future collider.''

U.S.\,groups are very active and in leading roles in the IMCC, e.g., in the Coordination Committee and the Publications and Speakers Committee, even though in several cases the formal MoC signatures and effort ramp-up the will need to wait for the forthcoming implementation planning of the P5 recommendations.

The collaboration plans to revise both the organisation and the workplan together with  U.S.\,colleagues. This will also impact the work programme laid out in this report. In particular, the ambition of the U.S.\,to host such a facility strengthens the motivation for R\&D and demonstrator planning at a fast pace.

The collaboration also plans to attempt further strengthen other regions' participation during this process. There is at this moment an important potential for substantial expansion of the collaboration, resources and efforts in Asia, and several European countries, Canada and South America.  

\subsection{R\&D programme}
The novelty of the muon collider concept implies that the R\&D programme
contains more uncertainty and challenges than for more conservative collider
approaches.

A concise set of work-packages has been developed for the European LDG roadmap~\cite{c:roadmap}. The programme aims to provide a broad basis for the global collaboration and planning of future work towards a
multi-TeV muon collider. The programme is estimated to require about
450~full-time equivalent (FTE) years of labour and 12~MEUR material budget for the accelerator studies.

Substantial progress has been made since the definition of the European LDG roadmap. The approval of the EU co-funded design study MuCol, contributions from the collaboration members as well from U.S.\,partners during the Snowmass process and an increase of the budget at CERN were instrumental in supporting the muon collider study.

With the existing funding until 2026, the R\&D cannot cover the full programme as described in the European LDG roadmap. The focus is on the most critical challenges and is guided by the priorities established by the roadmap. The planned effort is roughly consistent with the minimal scenario in the roadmap.

\nix{
Among the work needed to prepare the ground for a CDR and demonstrator programme for a muon
collider, some studies were high-lighted in Ref.~\cite{c:roadmap} as particularly important. This priority list is reported below, with links to the sections of the present document that describe the progress done in each area:
\begin{itemize}
\item 
{\emph{
The {\textbf{physics potential}} has to be further explored; 10~TeV is uncharted territory.}} The recent explosion of physics studies developed a robust physics case and identified directions for future advances, see Section~\ref{sec:Section03}. These targets are being translated into increasingly accurate specification requirements for the detector, in Section~\ref{sec:Section04}.
\item \emph{The {\textbf{environmental impact}} must be minimised and at least one {\bf potential site} for the collider identified.} A potential site close to CERN has been identified and will be explored; a set of parameters that would allow to remain completely on the Fermilab site has been developed. The mechanical system to mitigate the neutrino flux are being studied. See Sections~ref{sec:Section03} and~\ref{sec:Section07} for details.
\item {\emph{The impact of {\textbf{beam induced background}} in the detector might limit the physics reach and has to be minimised.}} Promising physics performances, based on novel accurate simulations of the beam induced background, are described in Section~\ref{sec:Section05}.
\item {\emph{The muon {\textbf{ acceleration and collision}} systems become more demanding at higher energies and are the most important cost and power consumption drivers.}} The concept and technologies for the accelerating pulsed synchrotrons and the collider ring have progressed and are being further developed, see Sections~\ref{sec:Section06} and \ref{sec:Section07}. 
\item \emph{The muon {\textbf{production and cooling system}} are challenging novel systems and call for development and optimisation beyond the MAP designs.} The target studies are very promising and the muon cooling system design is progressing in particular also addressing engineering issues, see Sections~\ref{sec:Section06} and \ref{sec:Section07}.
\end{itemize}
The details of the workpackage scope and the estimates of required resources can be found in the European LDG roadmap~\cite{c:roadmap}.
}

The focus is on the most critical challenges:
\begin{itemize}
\item The development of the physics case.
\item The possibility to find a site and mitigate the neutrino flux.
\item The detector concept, in particular, the machine-detector interface which drives the physics capabilities of the detector.
\item The development of the high energy accelerator design concepts, in particular the pulsed synchrotrons and the collider ring.
  These are vital to reach the high energy beyond the scale studied by MAP.
\item The further development of the muon production and cooling complex. The focus is on an engineering design of the cooling cell
  and the evaluation of the target concept. An overall optimisation of the complex is desirable but very resource demanding.
\item The magnet technologies that drive the performance, cost and power consumption.
\item The RF technologies that drive the performance, cost and power consumption.
\item Other key technologies, for example shielding and cryogenics.
\end{itemize}

\subsubsection{Work organisation}
The distribution of the work onto workpackages has been modified since the publication of the European LDG roadmap to adjust to the availability of effort and experts. The document reflects this change. The areas represented in the coordination committee are:
\begin{itemize}
\item {\bf The physics potential} links to all studies of the muon collider
  physics potential.
\item {\bf The detector and MDI} Links to the detector design studies and to the
  machine detector interface.
\item {\bf The proton complex} addresses the key challenge of the
accumulation of the protons in very high-charge bunches, by addressing in
detail the proton complex design, and provides the basic parameters of
the complex and the characteristics of the beam impacting on the production
target.
\item {\bf The muon production and cooling} addresses the production of the
muons by the proton beam hitting a target and the subsequent cooling,
including some of the specific technologies, such as for the production target
and the absorbers that reduce the beam phase space volume.
\item {\bf The muon acceleration complex} (in MuCol part of
  {\it the high-energy complex}) studies the acceleration complex of the muons.
\item {\bf The muon collider ring and siting} studies the design of the collider
  ring. Because the muon collider siting is foremost driven by considerations on
  the neutrino flux from the collider ring the package also contains these
  activities.
\item {\bf Magnet systems} will establish a complete inventory of the necessary
magnets to optimize and standardise the design, and address the most critical
ones. In particular it focuses on the solenoids of the muon production and
cooling, which are specific to the muon collider, and the fast-ramping magnet
system, which have ambitious requirements on power flow and power efficiency
and limits the energy reach of the collider.
\item {\bf Radio frequency (RF) systems} addresses the muon cooling ensuring coherence in frequency choice and
synchronization among the various stages. It contributes to sustainability studies by its work on high efficiency RF power sources.
\item
  {\bf The beam-matter interaction and target design} links to the different
  shielding studies, in particular for the muon production target and the collider ring.
\item {\bf The collective effects} area includes all beam dynamics studies across
  the complex.
\item {\bf The demonstrator} area connects to all the activities regarding the
  muon cooling demonstration.
\item {\bf Cooling cell integration} addresses the design of the muon
cooling cell, which is a unique and novel design and that faces integration
challenges.
\end{itemize}

\subsubsection{Recent progress and studies until 2026}
The studies are summarised below, largely following the European LDG roadmap workpackages, which are indicated in brackets.
\begin{itemize}
\item
For the {\bf site and neutrino flux mitigation} ({\it MC.SITE, MC.NF}) the following studies
are ongoing and planned by 2026:
\begin{itemize}
\item Verification of the neutrino flux impact using FLUKA and radiation protection knowledge.
\item Exploration of a site and orientation of the collider ring to mitigate the neutrino flux, in particular from the experimental insertions. This is the most difficult challenge.
\item Assessment of the mover concept to mitigate neutrino flux from the arcs.
\item An assessment if the neutrino flux mitigation with movers impacts beam operation.
\end{itemize}
A number of important points cannot be covered with the current resources.
This includes an assessment of the neutrino flux mitigation for the other
parts of the complex, the development of a concept to measure and verify
the neutrino flux in operation and the alignment concept that allows to link
the machine reliably to the surface. 

\item
The {\bf MDI studies} ({\it MC.MDI}) are progressing:
\begin{itemize}
\item The simulation tools for the background prediction have been developed.
\item A masking system has been studied and optimisation is planned.
\item A programme to simulate beam-beam induced background has been developed.
\item Different lattice designs for the interaction region have been compared
for the beam-induced background and limited differences were found.
\end{itemize}
The main effort to mitigate the background will have to be carried by the
detector design.
 
\item
The {\bf accelerator concept and beam dynamics} ({\it MC.ACC}) studies cover the key
activities:
\begin{itemize}
\item The combination of the high-power proton beam into short intense pulses.
\item A design of the experimental insertion and the arc lattices of the
collider ring for 10 TeV.
\item A design of the arc lattice of the pulses synchrotrons of the muon beam acceleration system.
\item An improved concept of the muon cooling system.
\item Assessment of key collective effects and their impact on the performance and specifications.
\end{itemize}
Studies that are not covered are the concepts of other parts of the collider
complex, e.g., the low energy muon acceleration or the proton complex as well
as full start-to-end simulations. Also alternatives for the muon cooling and
acceleration cannot be covered. Further the studies do not cover the 3~TeV
design. We anticipate that a solution for 10~TeV is also valid at 3~TeV.

\item
The {\bf high-field magnet} ({\it MC.HFM}) studies are ongoing
\begin{itemize}
\item The muon production target solenoid concept is being developed, based on HTS technology.
\item A database with the magnet performance specifications from the previous MAP design has been developed and the resulting challenges identified. A tool to systematically derive optimised magnet configurations that create similar fields for the beam is under development and will be used on subsequent lattice designs.
\item 40\,T has been identified as a realistic goal for the final cooling magnets and is used in the lattice design effort. The magnet study addresses the technology challenges that stem from this field.
\item Initial limits for the collider ring magnets have been identified and show the need for 
improved magnet protection and stress management, which will be addressed in further studies.
Additional resources are required to perform more cable tests and to
build magnet models to verify the simulations and improve the technology.
\end{itemize}
An experimental programme would be important to verify magnet performances and to develop the technology, but the material budget for this is currently very limited. Also measurement of the radiation hardness of the HTS magnets
would be instrumental in establishing the performance limits.

\item
The {\bf design of the power converters and fast-ramping normal magnets} ({\it MC.FR}) for
the RCSs is ongoing in close interaction with the RF design and the lattice
design. An overall optimisation will be performed of magnets, power converter and RF systems.
These systems have a very important impact on cost and additional funding would allow experimental verification of the performances. The development of HTS-based alternatives is not covered.

\item
The {\bf RF system} ({\it MC.RF}) studies are ongoing:
\begin{itemize}
\item  The acceleration of high-charge muon bunches to high energies is being developed using superconducting
cavities. The longitudinal beam dynamics studies along the high-energy complex is progressing
and shows that cavities developed for the International Linear Collider (ILC) might be applicable.
\item The synchronisation is being studied between the RF and the ramping of the magnets in the RCS together with its impact on the cost. 
\item The design is progressing of normal conducting cavities for the muon cooling cell.
\end{itemize}

\item
The {\bf muon production target} ({\it MC.TAR}) studies focus on the ability of the target to
withstand the beam power and the required shielding of the surrounding solenoid.
Indications are that a graphite target can sustain about 2\,MW. Higher power
liquid metal target alternatives will be explored. The current resources do not
allow a complete design of the target and study of the muon capture systems.

\item
The {\bf engineering design of one muon cooling cell} ({\it MC.MOD}) is ongoing and is
instrumental to maintain the option of a timely implementation of a
demonstrator facility.
Currently the resources to actually build models of the components and
ultimately a prototype of the cell are not secured.

Concepts of an {\bf RF test stand} are being developed and allow to apply for funding to overcome the current lack of infrastructure of this type.

\item
The initial scope for the {\bf muon cooling demonstrator} ({\it MC.DEM}) has been determined
and high-level concepts of the facility are being developed to 
understand the footprint requirements. Initially two potential sites at CERN
have been identified, further options exist at Fermilab and we will invite more options. More resources are needed to develop a full conceptual design of the facility. 

The estimate of the {\bf cost scale} ({\it MC.INT}) has started and will use a mixture of
high-level estimates with detailed estimates of key components.
\end{itemize}

The physics potential and detector R\&D goals are not part of the European LDG roadmap.

The {\bf physics potential} requires sustained study to fully develop the understanding of the different measurements that can be performed at the muon collider, in particular also in view of synergies with other physics instruments.

A set of {\bf physics benchmark processes} is being developed that allows to assess the
key detector concept capabilities with a limited number of studies.
The aim is to use these benchmark processes to demonstrate that the detector performance is adequate and that the background is well controlled.

Designs of the {\bf detector at 3 and 10 TeV detectors} is being developed with clear
specifications for the technology performance. A systematic review is important of these
performance requirements and an assessment which current technology developments will meet
them on which timescale. Also an exploration of alternative technologies is essential and the
derivation of a prioritised list of alternatives that should be further developed.
Of particular concern is the forward region, which suffers most from the background,
as well as the detection of small angle muons. Also some concept for the luminosity
monitoring is essential.

A 3~TeV detector concept exists based on a slightly modified version of the CLIC detector concept. Performance studies
with this concept show that the measurement resolution for some processes is not strongly
impacted by the beam-induced background while important impact exists for other channels.
A 10~TeV detector concept is under development.

A systematic performance study of the 10 TeV detector concept is required for the benchmark
processes to establish that the physics potential can be realised.
Therefore simulation of the detector is required without and with beam-induced background
in order to establish the latter has no significant impact on the physics performance.
The development of artificial intelligence-based methods will be important to reduce the background at the reconstruction and potentially the trigger level.

\subsection{Timeline and staging}
\label{s:stagingSCD}
The potential of a muon collider to reach 10 TeV parton-parton collisions with high luminosity makes it a very exciting future opportunity both on a longer and shorter timescale.

A roadmap towards a muon collider must remain flexible, and is an important part of a diverse international research programme in particle physics. New results, from the HL-LHC or non-accelerator based experiments, and potentially also lower energy experiments, might change the desired parameters for a future muon collider. Societal changes and priorities might also influence resource  availability and hence timelines for investments in basic science as needed for future colliders.

We already know that sustainability and environmental consideration needs to guide the design process much more than projects decades back. Staging scenarios, a clear focus on working with developing technologies (from HTS magnets to machine learning (ML) and artificial intelligence (AI) methods) with a wide impact and interest outside our field, synergetic studies and projects as described in next subsection, and a broad international collaboration, are all important to develop the muon collider as a scientifically and technically novel but robust future facility when it comes to implementation.

\nix{All the possible scenarios cannot be addressed in parallel so we are currently developing a timeline that allows to operate a muon collider by 2050. This demands that all needed technologies be mature in approximately 15 years.  This timeline is essentially driven by technical considerations and assumes that sufficient funding is available and that the R\&D is successful. This technically-limited, success-oriented timeline can serve the community and the decision makers as a basis to define the strategy and the budget for the future of the study.
}

The current timeline considerations identified that three technical developments are defining the
minimum time to implement a muon collider:
\begin{itemize}
\item The muon production and cooling technology and its demonstration in a test facility. At this moment we expect that this development can be mature enough for a decision in 15 years provided sufficient funding is made available.
\item The magnet technologies, in particular HTS superconductor technology. At this moment, we expect
  that the different HTS solenoids for the muon production and cooling are available within 15 years;
  the same is expected for the fast-ramping magnets. For the collider ring, one can expect 11~T Nb$_{3}$Sn
  magnets with an aperture of 16 cm to be mature. Higher performant HTS or hybrid collider ring
  magnets may take longer.
\item The detector and its technologies that impact the efficiency of background suppression and the quality of the measurements. At this moment, we expect this technology to be mature in 15 years.
\end{itemize}
Sufficient funding should enable to accelerate the R\&D of the other technologies and designs as to not constrain the timeline..

A staged implementation that anticipates use of Nb$_{3}$Sn collider ring magnets can provide a muon collider
by 2050. The scenarios are possible, see also Table~\ref{t:facility_param}:
\begin{itemize}
\item
  In the {\bf energy-staging} approach, the inital stage is at lower energy, for example 3\,TeV. There is an important physics case already at this energy. In this strategy, the cost of the initial stage is substantially lower than for the full project. This could accelerate the decision-making processes. The 3\,TeV design is consistent with Nb$_3$Sn magnets at 11\,T. In the second stage the whole complex will be reused with the exception of the collider ring. An RCS to accelerate to full energy and a new collider ring will be added.
\item
In the {\bf luminosity-staging} approach, the initial stage is at the full energy but using less performant collider ring magnets. This leads to an important reduction of the luminosity. If 11\,T Nb$_3$Sn dipoles are used instead of 16\,T HTS dipoles, the resulting increase of collider ring circumference reduces the luminosity by one third. In addition a further reduction arises from the interaction regions. A detailed study is required to quantify the loss, but a factor three reduction might be a good guess.
In the luminosity upgrade, the interaction region magnets will be replaced with more performant ones, similar to the HL-LHC. However one most likely will not replace the other collider ring magnets. In this scenario almost the complete project cost is required in the first stage, which could have important implications on the timeline.
\end{itemize}
The choice between the staging options will depend on physics needs and funding availability. Also the progress in magnet development will play an important role. Faster progress, in particular of HTS for the collider ring magnets, combined with strong funding support will make an earlier start of the 10\,TeV
option more attractive.

\subsection{Site considerations}

Different sites for the muon collider will be explored in the longer run. However, at this moment the international study to a large extent focuses on the technical and design challenges with no specific site in consideration. Specific sites however offer the benefit of existing infrastructure, e.g., existing proton complexes and collider tunnels. We will consider these benefits later for the different potential sites. Specific sites also can impose important constraints. For example, for Fermilab an option is preferred where the accelerator rings all fit on the existing site. First considerations indicate that this is possible. For CERN the most important consideration is to ensure that the neutrino flux can be mitigated to a negligible level. A first site and orientation of the collider ring that can achieve this has been identified and requires further study. In this case the extension of the experimental insertions point towards a steep uninhabited slope of the Jura on the one side and to the Mediterranean sea on the other side.
        
\subsection{Synergies and outreach}
Particle physics and the associated accelerator development have made important
contributions to society; both in training of young people and in developing
technologies.

Many young people have developed their scientific and technical
skills in the field; they also learned to work in fully international
collaborations. Because the muon collider is a novel concept it opens opportunities for young
researchers to make original contributions to the development that are much
harder to make in long-established design approaches.

The muon collider needs technologies in several areas that differ
from other colliders. High-field solenoids are a prime example.
In the past low-temperature superconductors such as the very mature NbTi and
still developing Nb$_3$Sn were the technologies of choice for accelerators and
most other applications.
Now high-temperature superconductors (HTS) are becoming an important technology.
In particular they are of interest for fusion reactors, that have similar requirements than the one for the muon collider target solenoid. Highly-efficient superconducting motors and power generators, e.g., for off-shore windmills, also have strong synergy. Other relevant areas are life and material sciences; in particular, applications for nuclear magnetic resonance (NMR) and magnetic resonance imaging (MRI). In addition synergy exists with magnets for neutron spectroscopy, physics detectors and magnets for other particle colliders, such as hadron colliders.

The muon production target is synergetic with neutron spallation sources targets, in particular the alternative liquid metal concept.

The muon collider RF power sources have synergy with other developments of high-efficiency klystrons and superconducting cavities. Some RF systems need to work in high magnetic fields, an issue that also exists in some fusion reactor designs.
 
The test facility and the collider itself require a high power proton source. This allows sharing technology and potentially even facilities. Neutron spallation sources such as SNS and ESS are major examples; other examples are neutrino facilities, such a NuSTORM, lepton flavour violation experiments, such as mu2e and COMET and the generation of low-energy, highly polarised muon beams.

\begin{flushleft}
\interlinepenalty=10000

\end{flushleft}

\nix{
\section{Overview of collaboration goals, challenges and R\&D programme}
The International Muon Collider Collaboration (IMCC)~\cite{c:IMCC} was established in 2020
following the recommendations of the European Strategy for Particle Physics
(ESPP)~\cite{c:ESPP} and the subsequent definition of the Accelerator R\&D Roadmap~\cite{c:roadmap}. The Muon Collider Study (MuC) covers the accelerator complex, detectors
and physics for a future muon collider. In 2023, European Commission support was
obtained for a design study of a muon collider (MuCol)~\cite{c:mucol}. This project started
1.3.2023, with work-packages aligned with the overall Muon Collider Studies.
Recently, the P5 panel~\cite{c:P5} in the U.S.\,recommended a muon collider R\&D, proposed
to join the IMCC and envisages that the U.S.\,should prepare to host a muon
collider, calling this their ``muon shot''. In the past the U.S.\,Muon Accelerator Programme (MAP)~\cite{map_overview} has been instrumental in studies of concepts and technologies for a muon collider.

\subsection{Motivation}
High-energy lepton colliders combine cutting edge discovery potential
with precision measurements. Because leptons are point-like particles in
contrast to protons, they can achieve comparable physics at lower centre-of-mass
energies~\cite{Delahaye:2019omf,AlAli:2021let,Accettura:2023ked}. However, to efficiently reach the 10+ TeV scale recognized by ESPP and P5 as a necessary target requires a muon collider. A muon collider with 10~TeV energy or more could discover new particles with presently inaccessible mass, including WIMP dark matter candidates. It could discover cracks in the SM by the precise study of the Higgs boson, including the direct observation of double-Higgs production and the precise measurement of triple Higgs coupling. It will uniquely pursue the quantum imprint of new phenomena in novel observables by combining precision with energy. It gives unique access to new physics coupled to muons and delivers beams of neutrinos with unprecedented properties from the muons decay. Based on physics considerations, an initial integrated luminosity target of 10\,ab$^{-1}$ at 10\,TeV was chosen.  However, various staging options are possible as well.

In terms of footprint, costs and power consumption a muon collider has potentially very favourable properties.
\begin{figure}[b]
  \centerline{\raisebox{20pt}{\includegraphics[width=7.5cm]{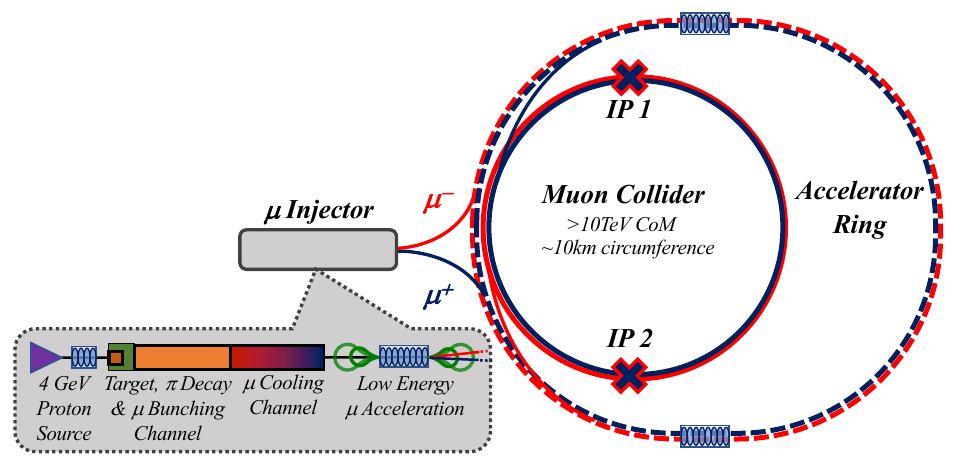}}\includegraphics[width=7.5cm]{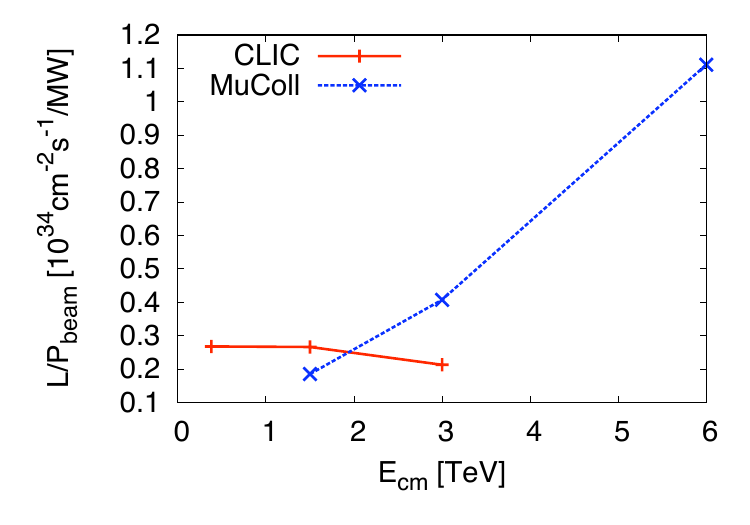}}
\caption{Left: Conceptual scheme of the muon collider. Right: Comparison of CLIC and a muon collider luminosities normalised
to the beam power and as a function of the centre-of-mass energy.}
\label{f:design}
\end{figure}
The luminosity of the lepton colliders has to increase with the square of the
collision energy to compensate for the reduction in $s$-channel cross sections.
Figure~\ref{f:design}-right compares the luminosities of CLIC and a muon collider, based on the U.S.\,Muon Accelerator Programme (MAP)
parameters~\cite{Delahaye:2019omf}, as a function of centre-of-mass energy.
The luminosities are normalised to the beam power.
The potential of muon colliders to improve the luminosity to beam power ratio
at high energies is one of the main advantages of the concept.

\subsection{The accelerator concept} 

The muon collider concept being studies by the IMCC, initially developed by MAP, is shown on the left panel of figure~\ref{f:design}.
The proton complex produces a short, high-intensity proton pulse that hits the
target and produces pions. The decay channel guides the pions and collects the
produced muons into a bunching and phase rotator system to form a muon beam.
Several cooling stages then reduce the longitudinal and transverse emittance of
the beam using a sequence of absorbers and RF cavities in a high magnetic field.
A system of a linac and two recirculating linacs accelerate the beams to
63\,GeV followed by a sequence of high-energy accelerator rings; the optimum sequence needs to be determined based on the ongoing studies.
Finally the beams are injected at full energy into the collider ring. Here, they will circulate to
produce luminosity until they are decayed; alternatively they can be extracted
once the beam current is strongly reduced. 

A set of parameters has been defined for 10\,TeV and also 3\,TeV. These are target parameters to explore the limits of each technology and design. If they can be fully met, the integrated luminosity goal could be reached within five years (or $2.5$~years, with two detectors) of full luminosity operation.  In practice, the operation time will be substantial longer for several reasons. For a new collider concept like a muon collider several years of luminosity ramp up time will be required. The peak luminosity might be somewhat lower due to technical limitations. The luminosity goals are also likely to change with further technical studies, and will also be driven by results obtained during operation.

\begin{table}[!hb]
\caption{Tentative target parameters for a muon collider at different
  energies. These values are only to give a first indication. The estimated luminosity refers to the value that can be reached if all target specifications can be reached; it also includes the beam-beam effect.}
\label{t:facility_param}
\begin{center}
  \begin{tabular}{|*3{c|}|*2{c|}|*2{c|}|}
    \hline
    Parameter & Symbol & unit & \multicolumn{2}{c||}{Scenario 1} & \multicolumn{2}{c||}{Scenario 2}\\
    \hline
 & & & stage 1 &stage 2 & stage 1 & stage 2\\
\hline
    Centre-of-mass energy & $E_{\mathrm{cm}}$ & TeV & 3 & 10 & 10 &10\\
    Target integrated luminosity & $\int{\cal L}_{\mathrm{target}}$ & $\rm ab^{-1}$ & 1 & 10 & \multicolumn{2}{c||}{10}\\
    Estimated luminosity & ${\cal L}_{\mathrm{estimated}}$ & $10^{34}\rm cm^{-2}s^{-1}$ & 2.1 & 21 & tbc & 14\\
    Collider circumference& $C_{\mathrm{coll}}$ & $\rm km$ & 4.5 & 10 & 15 &15\\
    Collider arc peak field& $B_{\mathrm{arc}}$ & $\rm T$ & 11 & 16 & 11 & 11\\
    Luminosity lifetime & $N_{\mathrm{turn}}$ &turns& 1039 & 1558 & 1040 & 1040\\
    \hline
    Muons/bunch & $N$ & $10^{12}$ & 2.2 & 1.8 & 1.8 &1.8\\
    Repetition rate & $f_{\mathrm{r}}$ & $\rm Hz$ & 5 & 5 &5 &5\\
    Beam power  & $P_{\mathrm{coll}}$ & $\rm MW$ &5.3  & 14.4 & 14.4 &14.4\\
    RMS longitudinal emittance& $\varepsilon_\parallel$ & $\rm eVs$ & 0.025 & 0.025 & 0.025 &0.025\\
    Norm. RMS transverse emittance& $\varepsilon_\perp$ & \textmu m & 25 & 25 & 25 &25\\
    \hline
    IP bunch length& $\sigma_z $ & $\rm mm$ & 5 & 1.5 & tbc &1.5\\
    IP betafunction& $\beta $ & $\rm mm$ & 5 & 1.5 & tbc & 1.5\\
    IP beam size& $\sigma $ & \textmu m & 3 & 0.9 & tbc & 0.9\\
    \hline
    Protons on target/bunch & $N_p$ & $10^{14}$ & 5 & 5 & 5 & 5\\
    Protons energy on target  & $E_p$ & $\rm GeV$ & 5 & 5 & 5 & 5\\
    \hline
    BS photons &$ N_{BS,0}$ &per muon& 0.075 &0.2 & tbc & 0.2\\
    BS photon energy & $E_{BS,0}$ & MeV & 0.016 & 1.6  &tbc & 1.6\\
    BS loss/lifetime (2 IP)&$E_{BS,tot}$&GeV& 0.002 &1.0 &tbc &0.67\\
    \hline
  \end{tabular}
\end{center}
\end{table}

\begin{table}
  \caption{Tentative target beam parameters along the acceleration chain.
    A 10~\% emittance growth budget has been foreseen in the transverse and
    longitudinal planes, both for 3 and 10 TeV. This assumes that the
    technology and tuning procedures will have been improved between the two
    stages. The very first acceleration is assumed to be part of the final cooling. This choice allows to optimise the energy in the last absorber with no strong impact on the acceleration chain.}
  \label{t:beam_param}
\begin{center}
  \begin{tabular}{|*6{c|}}
    \hline
    Parameter & Symbol & Unit &  Final Cooling & at 3 TeV & at 10 TeV \\
    \hline
    Beam total energy & $E_{beam}$ & GeV & 0.255 & 1500 & 5000 \\
    \hline
    Muons/bunch & $N_b$ & $10^{12}$ & 4 & 2.2 & 1.8 \\
    Longitudinal emittance& $\varepsilon_\parallel$ & $\rm eVs$ & 0.0225 & 0.025 & 0.025 \\
    RMS bunch length& $\sigma_z$ & $\rm mm$ & 375 & 5 & 1.5 \\
    RMS rel. momentum spread& $\sigma_P/P$ & $\rm \%$ & 9 & 0.1 & 0.1 \\
    Transverse norm. emittance& $\varepsilon_\perp$ & \textmu m & 22.5 & 25 & 25 \\
    \hline
    Aver. grad. $0.2$--$1500\rm GeV$&$G_{avg}$&$\rm MV/m$& --- & 2.4 &\\
    Aver. grad.  $1.5$--$5\rm TeV$&$G_{avg}$&$\rm MV/m$& --- &  & 1.1\\
    \hline
  \end{tabular}
\end{center}
\end{table}

\subsection{Maturity and R\&D challenges}
The muon collider collaboration, the muon beam panel of the Laboratory Directors Group (LDG) and the Snowmass process in the U.S.\,have all assessed the muon collider challenges. Key conclusions are that although the muon collider concept is less mature than several linear collider concepts no insurmountable obstacles have been identified, and that important design and technical challenges have to be addressed with a coherent international effort. Furthermore, past work, in particular within the U.S.\,Muon Accelerator Programme (MAP)~\cite{map_overview}, has demonstrated several key MuC technologies and concepts, and gives confidence that the overall concept is viable. Since then further component designs and technologies have been developed that provide increased confidence that one can cool the initially diffuse beam and accelerate it to multi-TeV energy on a time scale compatible with the muon lifetime. However, a fully integrated
design has yet to be developed and further development and demonstrations of
technology are required. 

The goal of the current IMCC effort is physics, detector and accelerator design and performance studies of a muon collider in preparation of the next European Strategy for Particle Physics Update and other international strategy processes. The programme prepares the way towards a full conceptual design report (CDR) and a demonstration programme, and allow stakeholder to make informed decisions about these next steps. The design and potential performance of such facility must be developed and detailed with some urgency to achieve this.

\subsubsection{The LDG R\&D roadmap}
A concise set of work-packages has been developed for the European LDG roadmap~\cite{c:roadmap}. The programme aims to provide a broad basis for the global collaboration and planning of future work towards a
multi-TeV muon collider. Among the work needed to prepare the ground for a CDR and demonstrator programme for a muon
collider, some studies were high-lighted in Ref.~\cite{c:roadmap} as particularly important. This priority list is reported below, with links to the sections of the present document that describe the progress done in each area:
\begin{itemize}
\item 
{\emph{
The {\textbf{physics potential}} has to be further explored; 10~TeV is uncharted territory.}} The recent explosion of physics studies developed a robust physics case and identified directions for future advances, see Section~\ref{sec:Section03}. These targets are being translated into increasingly accurate specification requirements for the detector, in Section~\ref{sec:Section04}.
\item \emph{The {\textbf{environmental impact}} must be minimised and at least one {\bf potential site} for the collider identified.} A potential site close to CERN has been identified and will be explored; a set of parameters that would allow to remain completely on the Fermilab site has been developed. The mechanical system to mitigate the neutrino flux are being studied. See Sections~ref{sec:Section03} and~\ref{sec:Section07} for details.
\item {\emph{The impact of {\textbf{beam induced background}} in the detector might limit the physics reach and has to be minimised.}} Promising physics performances, based on novel accurate simulations of the beam induced background, are described in Section~\ref{sec:Section05}.
\item {\emph{The muon {\textbf{ acceleration and collision}} systems become more demanding at higher energies and are the most important cost and power consumption drivers.}} The concept and technologies for the accelerating pulsed synchrotrons and the collider ring have progressed and are being further developed, see Sections~\ref{sec:Section06} and \ref{sec:Section07}. 
\item \emph{The muon {\textbf{production and cooling system}} are challenging novel systems and call for development and optimisation beyond the MAP designs.} The target studies are very promising and the muon cooling system design is progressing in particular also addressing engineering issues, see Sections~\ref{sec:Section06} and \ref{sec:Section07}.
\end{itemize}
The details of the workpackage scope and the estimates of required resources can be found in the European LDG roadmap~\cite{c:roadmap}.

\subsubsection{Facility challenges}
Considering the facility design and technical challenges key performance drivers and goals have been identified, providing guidance for prioritising studies and efforts. Particularly important are: 

\begin{itemize}
\item {\bf Environmental impact.}
The compact footprint, limited cost and power consumption are intrinsic
features that motivate the muon collider study in the first place. 
Radiation protection measures will ensure a negligible impact of the facility
on the environment, similar to the LHC. Particular attention will be paid to the
neutrino flux that is produced by the decays of the muons in the collider and
that exits the ground far from the collider.
\item {\bf Machine-detector interface and detector.}
The muons of the beams that circulate in the collider ring decay, each producing two neutrinos and one electron or positron. The latter will hit the aperture and create showers. Tungsten masks protect the detector
from this beam induced background in combination with detector components with high time and space resolution.

\item {\bf Muon production.}
A proton beam power of around 2\,MW at 5\,Hz is used for
muon production. Designs for proton facilities with similar or larger power
exist. The main proton complex challenge arises from the combination
of the protons into short, high-charge bunches.
The key challenge for the high-power target is the survival of the target
itself under the shock waves of the incoming beam pulses and the temperature
gradients to remove the deposited heat. 
The target is immersed into a 20\,T solenoid field.

\item {\bf Muon cooling.}
Muon ionisation cooling increases the muon beam brightness by repeatedly
slowing it in absorbers and re-accelerating it in RF cavities; both inside of
strong solenoid fields to keep the beam focused. This principle has been demonstrated
in MICE~\cite{mice}. A factor two improvement of
the transverse emittance in the final cooling will allow to reach the emittance goal. The cooling concept is based on
close integration of high-gradient normal conducting RF with the high-field
solenoids.
A facility to test the cavities in high magnetic fields is mandatory to
validate the muon collider performance predictions. Since the previous setups
in the U.S.\,do no longer exist, the design and construction of a new test stand
is a key goal. 

\item {\bf Muon acceleration.}
Most of the acceleration will be performed by a sequence of pulsed
synchrotrons; an alternative use of FFAs is also considered. The synchrotrons
can be based on a hybrid design where the fast-ramping magnets are interleaved
with static superconducting ones. The pulsed synchrotrons face challenges in terms of optics design, the magnet systems and the RF system.
Field ramp rates between a few hundred T/s in the largest and several \,kT/s in the smallest
ring are currently foreseen. The latter requires normal-conducting magnets
while for the former also superferric or HTS magnets can also be considered.  Finally, the large stored energy in the magnets (in total in
the range of O(100\,MJ)) requires demanding power converters with very efficient
recovery of the energy of each pulse for the subsequent one. 

\item {\bf Collider ring.}
The collider ring requires a small beta-function at the collision point,
resulting in significant chromaticity
that needs to be compensated. It also needs to maintain a short bunch. A
solution for 3\,TeV has been developed by the MAP study and successfully addresses the
challenges. A design of 10\,TeV is
more challenging and one of the key ongoing efforts.
High-energy electrons and positrons that arise from muon decay and strike the collider ring magnets can cause radiation damage and unwanted heat load. This can be mitigated with sufficient tungsten shielding. 
\end{itemize}

\subsection{The International Muon Collider Collaboration}
The International Muon Collider Collaboration has formed with the short term goal to address the R\&D Roadmap
in time for the next European Strategy for Particle Physics Update. More than 70 groups are currently involved in the muon collider studies. Three main deliverables are foreseen:
\begin{itemize}
\item a Project Evaluation Report that assesses the muon collider potential;
\item an R\&D Plan that describes a path towards the collider;
\item an Interim Report early 2024 (the present document) that documents progress and allows the wider community to be updated on the concept and to give feedback to the collaboration.
\end{itemize}
IMCC envisages to study the 10\,TeV option, and also
explore lower and higher energy options, e.g., a 3\,TeV option as a step toward
10\,TeV. The details of the required work and the required resources are documented in the LDG roadmap report~\cite{c:roadmap_overview}.

Overall facility parameters as cost, power/energy, carbon will be estimated as the design matures, but are in many cases driving technical choices and optimisation. 

\subsubsection{Collaboration organisation}
A memorandum of cooperation (MoC) for the IMCC has been drawn up and the collaboration is rapidly growing. 
The formal structure of the collaboration, as shown in figure~\ref{f:governance} consists of the following main bodies: 
\begin{itemize}
\item International collaboration board (ICB):
The ICB role is to provide a forum for participants to examine ongoing activities,
assure appropriate, well directed use of contributions and ensure
a balanced portfolio of engagements.
The chair of the ICB is elected by its members (simple majority).
The ICB consists of one representative of each participating institute that has signed the MoC. The ICB has been active since October 2022. 
\item Steering board (SB):
The SB oversees the execution of the Muon Collider Study by assessing the progress, resource usage and providing guidance.
The SB reports to the LDG in Europe and to similar regional supervising organisations in other regions as required.
The SG is active and operational since Spring 2023. 
\item International Advisory Committee (IAC): The mandate of the International Advisory Committee is to review the
scientific and technical progress of the Study. This body is in the process of being set up.
\item Coordination committee (CC): The CC performs the overall coordination and execution of the Muon Collider study.
It is chaired by the Study Leader. This body consists of the workpackage leaders and some additional experts of the collaboration and is very active. It provides the daily guidance of the study.
\end{itemize}
The workpackages of the collaboration are shown in figure~\ref{f:governance}.

\begin{figure}
\centerline{\includegraphics[width=12cm]{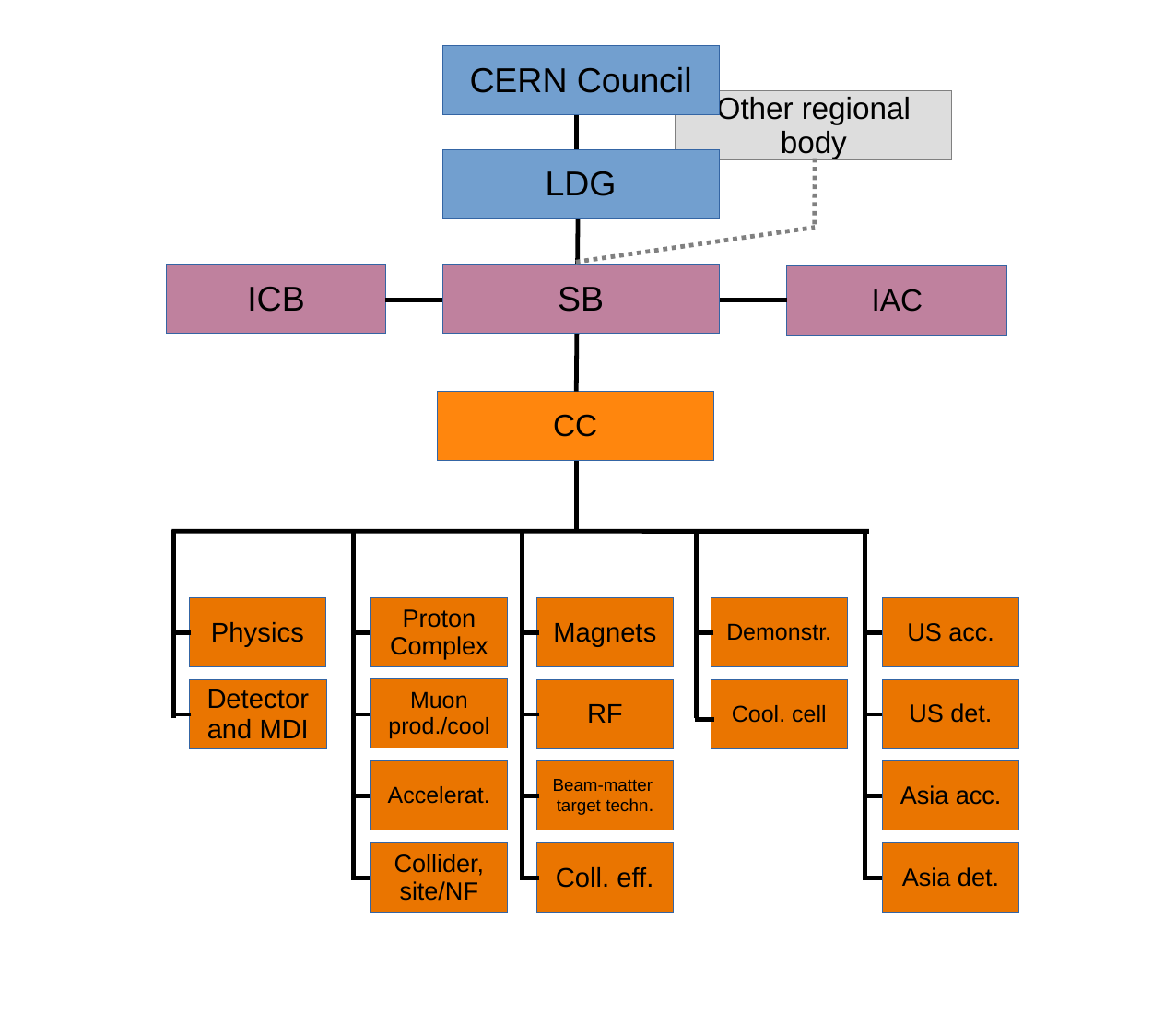}}
\caption{Organigram of the Muon Collider study governance structure}
\label{f:governance}
\end{figure}

The study leader is proposed by the host organization, and endorsed by the ICB. The study leader was appointed in 2022 and has led the work to build up the collaboration and activities since then.

\subsubsection{MuCol}
IMCC successfully applied for an EU cofunded design study (named MuCol)~\cite{c:mucol}. The project started 1.3.2023. The total co-funding of MuCol amounts to 3\,MEUR, provided by
the European Commission, the UK and Switzerland. In the design study, CERN only receives limited contributions for
administrative support and travel, but has, in support of the successful Design Study bid, increased its contribution to the muon collider study.

The design study is fully integrated in the overall muon collaboration. The technical meetings and leaders are in common and governance and management will also be synchronised. The organisation is shown in figure~\ref{f:MuColwp}.
Important activities in the collaboration, directly supported by the MuCol work-packages listed below, are:
\begin{itemize}
\item {\bf Physics and detector requirements} provides the link to the
physics and detector studies. Based on feedback from
the physics community, it will provide feedback and guidance to the
accelerator design. 
\item {\bf The Proton Complex} addresses the key challenge of the
accumulation of the protons in very high-charge bunches, by addressing in
details the proton complex design, and provides the basic parameters of
the complex and the characteristics of the beam impacting on the production
target.
\item {\bf The Muon Production and Cooling} addresses the production of the
muons by the proton beam hitting a target and the subsequent cooling,
including some of the specific technologies, such as for the production target
and the absorbers that reduce the beam phase space volume.
\item {\bf The High-energy Complex} studies the acceleration and collision complex of the muons.
\item {\bf Radio Frequency Systems} addresses the Radio Frequency (RF)
systems of the muon cooling ensuring coherence in frequency choice and
synchronization among the various stages. It contributes to sustainability studies by its work on high efficiency RF Power sources.
\item {\bf Magnet Systems} will establish a complete inventory of the necessary
magnets to optimize and standardise the design, and address the most critical
ones. In particular it focuses on the solenoids of the muon production and
cooling, which are specific to the muon collider, and the fast-ramping magnet
system, which have ambitious requirements on power flow and power efficiency
and limits the energy reach of the collider.
\item {\bf Cooling Cell Integration} addresses the design of the muon
cooling cell, which is a unique and novel design and that faces integration
challenges.
\end{itemize}

\begin{figure}
\centerline{\includegraphics[width=10cm]{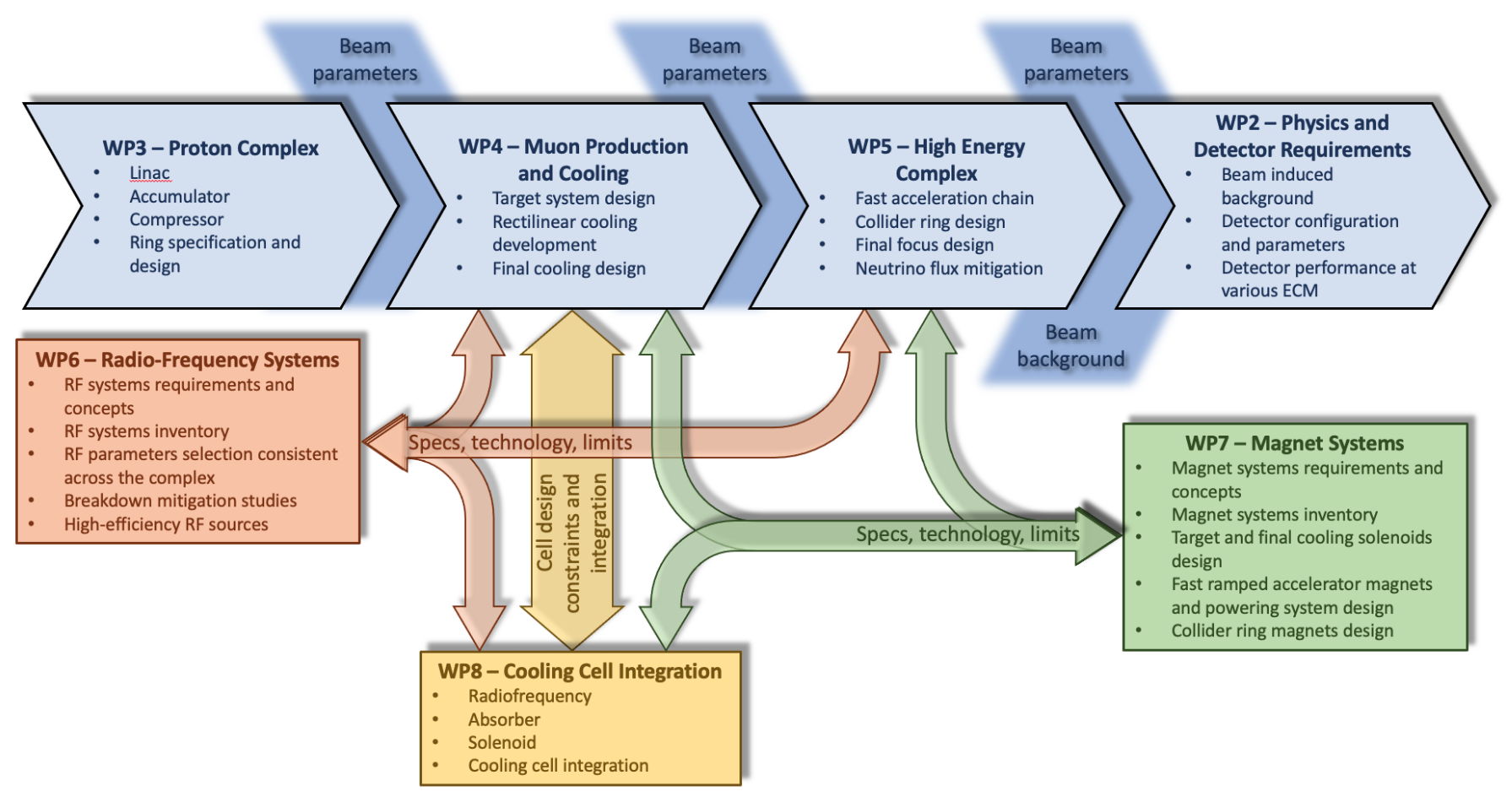}}
\caption{The MuCol work-packages and their interactions.}
\label{f:MuColwp}
\end{figure}

\subsubsection{Extending the collaboration--US Plans after P5}
In the US, the Particle Physics Projects Prioritization Panel (P5) December 2023 recommended
that the U.S.\,should develop a collider with 10\,TeV parton collision energies,
such as a muon collider, a proton collider, or possibly an electron-positron.
The report states: ``In particular, a muon collider presents an attractive
option both for technological innovation and bringing the energy frontier back
to the US. The U.S.\,should participate to the IMCC and have the ambition to host
a future collider.''

US groups are very active and in leading roles in the IMCC, e.g.\,in the Coordination Committee and the Publications and Speakers Committee, even though in several cases the formal MoC signatures and ramp up the efforts will need to wait for the forthcoming implementation planning of the P5 recommendations.

The collaboration plans to revise both the organisation and the workplan together with  U.S.\,colleagues. This will also impact the work programme laid out in this report. In particular, the ambition of the U.S.\,to host such a facility strengthens the motivation for R\&D and demonstrator planning at a fast pace.

The collaboration also plans to attempt further strengthening other regions' participation during this process. There is at this moment an important potential for substantial expansion of the collaboration, resources and efforts in Asia, and several European countries, Canada and South America.  

\subsection{Description of R\&D Programme until 2026}
The novelty of the muon collider concept implies that the R\&D programme
contains more uncertainty and challenges than for more conservative collider approaches.
With the existing funding until 2026, the R\&D cannot cover the full programme as described in the accelerator roadmap. The focus is on the most critical challenges and is guided by the priorities established by the roadmap. The planned effort is roughly consistent with the minimal scenario in the roadmap.

Substantial progress has been made since the definition of the Accelerator R\&D Roadmap. The approval of the EU co-funded Design Study MuCol, contributions from the collaboration members as well from U.S.\,partners during the Snowmass process and an increase of the budget at CERN were instrumental in supporting the muon collider study.

The focus is on the most critical challenges:
\begin{itemize}
\item The possibility to find a site and address the neutrino flux mitigation.
\item The magnet and RF technologies which drive the performance, cost and power consumption.
\item The integrated muon cooling cell design, which combines superconducting dipoles with normal-conducting RF, absorbers, vacuum and other components in a tight space.
\item Other key technologies, in particular the muon production target and the shielding of the accelerator components that drive the specifications.
\item The detector concept, in particular, the machine-detector interface which drives the physics capabilities of the detector.
\item The development of design concepts to address key challenges of the accelerator complex.
\end{itemize}
The distribution of the work onto workpackages has been modified since the roadmap to adjust to the availability of effort and experts. The document reflects this change. Below however, the studies are summarised largely following the roadmap workpackages, which are indicted in brackets.
\begin{itemize}
\item
For the {\bf site and neutrino flux mitigation} ({\it MC.SITE, MC.NF}) the following studies
are ongoing and planned by 2026:
\begin{itemize}
\item Verification of the neutrino flux impact using FLUKA and radiation protection knowledge.
\item Exploration of a site and orientation of the collider ring to mitigate the neutrino flux, in particular from the experimental insertions. This is the most difficult challenge.
\item Assessment of the mover concept to mitigate neutrino flux from the arcs.
\item An assessment if the neutrino flux mitigation with movers impacts beam operation.
\end{itemize}
A number of important points cannot be covered with the current resources.
This includes an assessment of the neutrino flux mitigation for the other
parts of the complex, the development of a concept to measure and verify
the neutrino flux in operation and the alignment concept that allows to link
the machine reliably to the surface. 

\item
The {\bf MDI studies} ({\it MC.MDI}) are progressing:
\begin{itemize}
\item The simulation tools for the background prediction have been developed.
\item A masking system has been studied and optimisation is planned.
\item A programme to simulate beam-beam induced background has been developed.
\item Different lattice design for the interaction region have been compared
for the beam-induced background and found limited differences.
\end{itemize}
The main effort to mitigate the background will have to be carried by the
detector design.
 
\item
The {\bf accelerator concept and beam dynamics} ({\it MC.ACC}) studies cover the key
activities:
\begin{itemize}
\item The combination of the high-power proton beam into short intense pulses.
\item A design of the experimental insertion and the arc lattices of the
collider ring for 10 TeV.
\item A design of the arc lattice of the pulses synchrotrons of the muon beam acceleration system.
\item An improved concept of the muon cooling system.
\item Assessment of key collective effects and their impact on the performance and specifications.
\end{itemize}
Studies that are not covered are the concepts of other parts of the collider
complex, e.g., the low energy muon acceleration or the proton complex as well
as full start-to-end simulations. Also alternatives for the muon cooling and
acceleration cannot be covered, Further the studies do not cover the 3 TeV
design. We anticipate that a solution for 10 TeV is also valid at 3 TeV.

\item
The {\bf high-field magnet} ({\it MC.HFM}) studies include
\begin{itemize}
\item The muon production target solenoid concept is being developed, based on HTS technology.
\item A database with the magnet performance specifications from the previous MAP design has been developed and the resulting challenges identified. A tool to systematically derive optimised magnet configurations that create similar fields for the beam is under development and will be used on subsequent lattice designs.
\item 40\,T has been identified as a realistic goal for the final cooling magnets and is used in the lattice design effort. The magnet study addresses the technology challenges that stem from this field.
\item Initial limits for the collider ring magnets have been identified and show the need for 
improved magnet protection and stress management, which will be addressed in further studies.
Additional resources are required to perform more cable tests and to
build magnet models to verify the simulations and improve the technology.
\end{itemize}
An experimental programme would be important to verify magnet performances and to develop the technology, but the material budget for this is currently very limited. Also measurement of the radiation hardness of the HTS magnets
would be instrumental in establishing the performance limits.

\item
The {\bf design of the power converters and fast-ramping normal magnets} ({\it MC.FR}) for
the RCSs is ongoing in close interaction with the RF design and the lattice
design. An overall optimisation will be performed of magnets, power converter and RF systems.
These systems have a very important impact on cost and additional funding would allow experimental verification of the performances. The development of HTS-based alternatives is not covered.

\item
The {\bf RF system} ({\it MC.RF}) studies are covering:
\begin{itemize}
\item  The acceleration of high-charge muon bunches to high energies using superconducting
cavities. The longitudinal beam dynamics along the high-energy complex is progressing
and shows that ILC-type cavities might be applicable.
\item The synchronisation is being studied between the RF and the ramping of the magnets in the RCS together with its impact on the cost. 
\item The design is progressing of normal conducting cavities for the muon cooling cell.
\end{itemize}

\item
The {\bf muon production target} ({\it MC.TAR}) studies focus on the ability of the target to
withstand the beam power and the required shielding of the surrounding solenoid.
Indications are that a graphite target can sustain about 2\,MW. Higher power
liquid metal target alternatives will be explored. The current resources do not
allow a complete design of the target and study of the muon capture systems.

\item
The {\bf engineering design of one muon cooling cell} ({\it MC.MOD}) is ongoing and is
instrumental to maintain the option of a timely implementation of a
demonstrator facility.
Currently the resources to actually build models of the components and
ultimately a prototype of the cell are not secured.

Concepts of an {\bf RF test stand} are being developed and allow to apply for funding to overcome the current lack of infrastructure of this type.

\item
The initial scope for the {\bf muon cooling demonstrator} ({\it MC.DEM}) has been determined
and high-level concepts of the facility are being developed to 
understand the footprint requirements. Initially two potential sites at CERN
have been identified, further options exist at Fermilab and we will invite more options. More resources are needed to develop a full conceptual design of the facility. 

The estimate of the {\bf cost scale} ({\it MC.INT}) has started and will use a mixture of
high-level estimates with detailed estimates of key components.
\end{itemize}

The physics potential and detector R\&D goals are not part of the Accelerator R\&D Roadmap.

The {\bf physics potential} requires sustained study to fully develop the understanding of the different measurements that can be performed at the muon collider, in particular also in view of synergies with other physics instruments.

A set of {\bf physics benchmark processes} is being developed that allows to assess the
key detector concept capabilities with a limited number of studies.
The aim is to use these benchmark processes to demonstrate that the detector performance is adequate and that the background is well controlled.

Designs of the {\bf detector at 3 and 10 TeV detectors} will be developed with clear
specifications for the technology performance. A systematic review is important of these
performance requirements and an assessment which current technology developments will meet
them on which timescale. Also an exploration of alternative technologies is essential and the
derivation of a prioritised list of alternatives that should be further developed.
Of particular concern is the forward region, which suffers most from the background,
as well as the detection of small angle muons. Also some concept for the luminosity
monitoring is essential.

A 3 TeV detector concept exists based a slightly modified version of the CLIC detector concept. Performance studies
with this concept show that the measurement resolution for some processes is not strongly
impacted by the beam-induced background while important impact exists for other channels.
A 10 TeV detector concept is under development.

A systematic performance study of the 10 TeV detector concept is required for the benchmark
processes to establish that the physics potential can be realised,
Therefore simulation of the detector is required without and with beam-induced background
in order to establish that latter has no significant impact on the pyhsics performance.
The development of artificial intelligence-based methods will be important to reduce the background at the reconstruction and potentially the trigger level.

\subsection{Timeline and staging}
The potential of a muon collider to reach 10 TeV parton-parton collisions with high luminosity makes it a very exciting future opportunity.
A roadmap towards a muon collider must remain flexible, and ideally be an important part of a wider diverse international research programme in particle physics. New results, from the HL-LHC or non-accelerator based experiments, and potentially also lower energy experiments, might change the desired parameters for a future muon collider. Societal changes and priorities might also influence resource  availability and hence timelines for investments in basic science as needed for future colliders. We already know that sustainability and environmental consideration needs to guide the design process much more than projects decades back. Staging scenarios, a clear focus on working with developing technologies (from HTS magnets to ML/AI methods) with a wide impact and interest outside our field, synergetic studies and projects as described in next subsection, and a broad international collaboration, are all important to develop the muon collider as a scientifically and technically novel but yet robust future facility when it comes to implementation.

All the possible scenarios cannot be addressed in parallel so we are currently developing a timeline that allows to operate a muon collider by 2050. This demands that all needed technologies be mature in approximately 15 years.  This timeline is essentially driven by technical considerations and assumes that sufficient funding is available and that the R\&D is successful. This technically-limited, success-oriented timeline can serve the community and the decision makers as a basis to define the strategy and the budget for the future of the study.

The current timeline considerations identified that three technical developments are defining the minimum time to implement a muon collider:
\begin{itemize}
\item The muon production and cooling technology and its demonstration in a test facility;
\item the magnet technologies, in particular HTS superconductor technology;
\item the detector and its technologies that impact the efficiency of background suppression and the quality of the measurements.
\end{itemize}
Sufficient funding should enable to accelerate the R\&D of the other technologies and designs sufficiently to not constrain the timeline.

At this moment the indications a staged implementation of a muon collider is possible with the first stage starting luminosity operation before 2050. It can thus be the next high-energy frontier flagship project in all regions, e.g., in Europe it could follow directly after the HL-LHC.

This scenario is based on a staged approach of the collider implementation. 
In the {\bf energy-staging} approach, the inital stage is at lower energy, for example 3\,TeV. There is an important physics case already at this energy. In this strategy, the cost of the initial stage is substantially lower than for the full project. This could accelerate the decision-making processes. The 3\,TeV design is consistent with Nb$_3$Sn magnets at 11\,T. In the second stage the whole complex will be reused with the exception of the collider ring. An RCS to accelerate to full energy and a new collider ring will be added.

In the {\bf luminosity-staging} approach, the initial stage is at the full energy but using less performant collider ring magnets. This leads to an important reduction of the luminosity. If 11\,T Nb$_3$Sn dipoles are used instead of 16\,T HTS dipoles, the resulting increase of collider ring circumference reduces the luminosity by one third. In addition a further reduction arises from the interaction regions. A detailed study is required to quantify the loss, but a factor three reduction might be a good guess.
In the luminosity upgrade, the interaction region magnets will be replaced with more performant ones, similar to the HL-LHC. However one most likely will not replace the other collider ring magnets. In this scenario almost the complete project cost is required in the first stage, which could have important implications on the timeline.

The choice between the staging options will depend on physics needs and funding availability. Also the progress in magnet development will play an important role. Faster progress, in particular of HTS for the collider ring magnets, combined with strong funding support will make an earlier start of the 10\,TeV
option more attractive.

\subsection{Site considerations}

Different sites for the muon collider will be explored in the longer run. However, at this moment the international study to a large extent focuses on the technical and design challenges with no specific site in consideration. Specific sites however offer the benefit of existing infrastructure, e.g., existing proton complexes and collider tunnels. We will consider these benefits later for the different potential sites. Specific sites also can impose important constraints. For example, for Fermilab an option is preferred where the accelerator rings all fit on the existing site. First considerations indicate that this is possible. For CERN the most important consideration is to ensure that the neutrino flux can be mitigated to a negligible level. A first site and orientation of the collider ring that can achieve this has been identified and requires further study. In this case the extension of the experimental insertions point towards a steep uninhabited slope of the Jura on the one side and to the Mediterranean sea on the other side.
        
\subsection{Synergies and outreach}
Particle physics and the associated accelerator development have made important
contributions to society; both in training of young people and in developing
technologies.

Many young people have developed their scientific and technical
skills in the field; they also learned to work in fully international
collaborations. Because the muon collider is a novel concept it opens opportunities for young
researches to make original contributions to the development that are much
harder to make in long-established design approaches.

The muon collider needs in several areas technologies that differ
from other colliders. High-field solenoids are a prime example.
In the past low-temperature superconductors such as the very mature NbTi and
still developing Nb$_3$Sn were the technologies of choice for accelerators and
most other applications.
Now high-temperature superconductors (HTS) are becoming an important technology.
In particular they are of interest for fusion reactors, that have similar requirements than the one for the muon collider target solenoid. Highly-efficient superconducting motors and power generators, e.g.\,for off-shore windmills, also have strong synergy. Other relevant areas are life and material sciences; in particular, applications for Nuclear Magnetic Resonance (NMR) and Magnetic Resonance Imaging (MRI). In addition synergy exists with magnets for neutron spectroscopy, physics detectors and magnets for other particle colliders, such as hadron colliders.

The muon production target is synergetic with neutron spallation sources targets, in particular the alternative liquid metal concept.

The muon collider RF power sources have synergy with other developments of high-efficiency klystrons and superconducting cavities. Some RF systems need to work in high magnetic fields, an issue that also exists in some fusion reactor designs.
 
The test facility and the collider itself require a high power proton source. This allows to share technology and potentially even facilities. Examples are neutrino facilities, such a NuSTORM, lepton flavour violation experiments, such as mu2e and COMET and the generation of low-energy, highly polarised muon beams.

\begin{flushleft}
\interlinepenalty=10000

\end{flushleft}}

\clearpage
\section{Physics opportunities 
}
\label{sec:Section03}

A muon collider with 10~TeV energy or more could discover new particles with presently inaccessible mass, including WIMP dark matter candidates. It could discover cracks in the SM by the precise study of the Higgs boson, including the direct observation of double-Higgs production and the precise measurement of the triple Higgs coupling. It will uniquely pursue the quantum imprint of new phenomena in novel observables by combining precision with energy. It gives unique access to new physics coupled to muons and delivers beams of neutrinos with unprecendeted properties from the muons decay.

\begin{figure}[t]
   \begin{center}
        \includegraphics[width=0.45\textwidth]{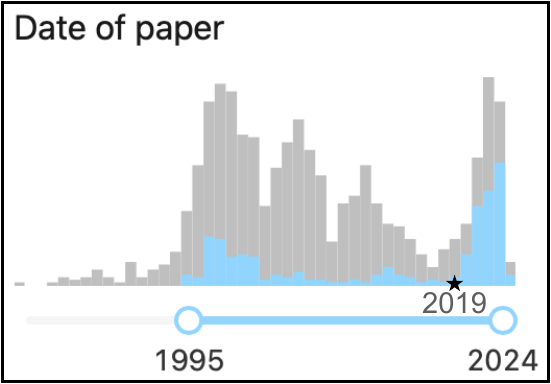}   \hspace{40pt}
        \includegraphics[width=0.38\textwidth]{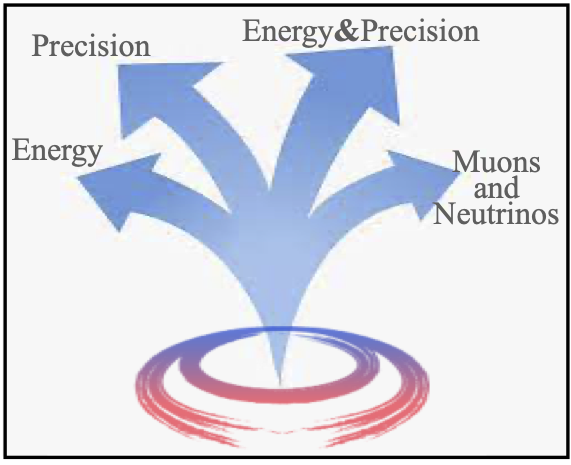}  
    \end{center}
 \caption{Left: Time distribution of the papers on muon colliders (grey) compared with those (blue) with subject ``Phenomenology-HEP''~\cite{inspire}. Right: Pictorial view of the muon colliders exploration opportunities.}
 \label{fig:papdist}
\end{figure}

For these reasons, the new European interest on muon colliders that emerged during the process and with the deliberation of the 2020 Strategy Update---and the creation of the IMCC---triggered an enthusiastic reaction of the theory and phenomenology community. Since 2019, INSPIRE-HEP has recorded more than 150 papers about muon colliders in the ``Phenomenology-HEP'' subject category~\cite{inspire}. This is about half of the papers ever written on this topic. The left panel of Fig.~\ref{fig:papdist} compares the time distribution of the phenomenology papers (in blue), with the papers about muon colliders in all subjects (in grey). In the past, muon collider phenomenology papers used to be a small fraction of the total and the development of the field was almost entirely driven by the advances in accelerator physics. Physics studies are instead a major component and a driver of the activity in the last few years, indicating  an unprecedented enthusiasm on muon colliders physics opportunities. A small fraction of these recent works is described below to exemplify some of the achieved accomplishments. For an extensive overview, see Refs.~\cite{PhysOpp_AlAli:2021let,PhysicsPotential_Black:2022cth} and, most recently, the review produced by the  International Muon Collider Collaboration~\cite{PhysicsPotential_Accettura:2023ked}.

Several workshops and seminars on muon colliders physics were held in the last few years, including a very successful series of events organised by the ``Muon Forum''~\cite{muonforum} in the context of the Snowmass 2021 Community Planning Exercise. The activities and the work triggered by the Forum~\cite{PhysicsPotential_Black:2022cth} strongly impacted the Snowmass Energy Frontier outcome~\cite{Narain:2022qud}, which recognised the muon colliders potential for the exploration of the energy frontier and advocated R\&{D} investments with the perspective of hosting a muon collider in the US. The P5 panel report confirmed and strengthened this view~\cite{P5}.

The rest of this section describes the work done, the open questions and the future directions for the development of muon colliders physics. In Section~\ref{subsec:EF} we outline the potential of a 10~TeV muon collider to explore the energy frontier conclusively and systematically by a number of different strategies, depicted on the right panel of Fig.~\ref{fig:papdist}. Section~\ref{subsec:TF} describes the future challenges for theory and phenomenology in the novel environment of multi-TeV muon collisions. These challenges are in fact opportunities for the theory frontier exploration of a new regime of the electroweak interactions. Section~\ref{subsec:SaS} is devoted to physics opportunities in addition to those entailed by the 10~TeV muons collisions. In particular, we 
illustrate the unprecendented properties of the collimated beam of neutrinos that emerges from muon decays close to the interaction point. 
Finally, we present in Section~\ref{subse:stage} a physics-driven assessment of possible muon colliders energy or luminosity staging plans.

\subsection{Exploring the energy frontier}\label{subsec:EF}

\subsubsection*{Motivations}

The LHC confirmed an unprecedentedly complete framework---the Standard Model (SM)---that enables a theoretical description of Nature up to extremely short distance scales, which are still experimentally unexplored. Testing the SM predictions at shorter and shorter scales is a prime motivation for studying higher and higher energy collisions. On the other hand, the SM success should not be overrated. The theoretical machinery of local field interactions the SM formulation builds upon is compatible with the profound principles of relativistic quantum field theory, but it is not uniquely selected by these principles, nor it is sufficient for a truly complete description of Nature that includes gravity. Even within this partial framework of local Lagrangian field theory, the SM is highly non-unique and its field content is merely engineered to accommodate the observed particles, and in fact not all of them because the SM lacks a dark matter candidate. Since our present knowledge of fundamental particles emerges from past observations, the existence or non-existence of other particles can be only established by future observations. If the new particles are heavy, this requires a high-energy collider.

The most mysterious sector of the SM is the one associated with the breaking of the electroweak~(EW) symmetry: the Higgs sector. The Higgs boson is not only the most recently discovered SM particle. It is the first manifestation of a description of the massive spin one EW force carriers---by the formalism of spontaneously-broken gauge theories---that can be extrapolated up to energies much above the EW bosons mass. Testing the SM predictions for the properties and the couplings of the Higgs particle is an important but indirect way to probe the validity of this novel description. The key innovation of the Higgs theory is to offer a modelling of EW interactions in the multi-TeV energy regime. Its most striking manifestations are thus the interplay between the Higgs and the EW bosons---and also the fermions, as also their mass requires the Higgs---in high energy collisions. 

At variance with other theoretical possibilities, the SM Higgs achieves the breaking of the EW symmetry by a fundamental scalar field, which results in an elementary (point-like) Higgs boson. The naturalness paradox famously displays the contradiction between the existence of elementary spin zero particles and the Wilsonian interpretation of quantum field theory. We will thus learn profound lessons by probing if the Higgs particle is elementary---namely, if its size is much smaller than its Compton wavelenght---or instead composite. Less radical possibilities to circumvent the naturalness paradox such as supersymmetry should be also investigated for the same reason.

In this context, the muon colliders' potential to explore Higgs and EW physics at high energy and without large backgrounds from the QCD interaction appears particularly appealing. A high-energy muon collider is arguably the best response to the contemporary challenges of fundamental particles and interactions physics.

\subsubsection*{Muon collider physics highlights}

The motivations for studying short-distance physics are strong, but broad. Therefore, a radical advance requires an equally broad and comprehensive program of energy frontier exploration that leverages many diverse strategies of investigation. The multiple exploration strategies available at a muon collider---see Fig.~\ref{fig:papdist}---make it ideally suited for this task. 

\begin{figure}[t]
   \begin{center}
        \includegraphics[width=0.4\textwidth]{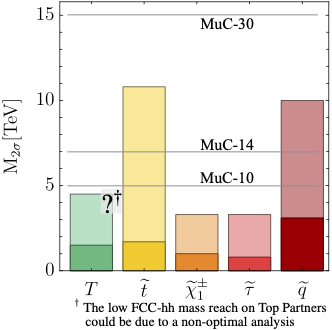} \;
        \hspace{20pt}
        \raisebox{16pt}{\includegraphics[width=0.4\textwidth]{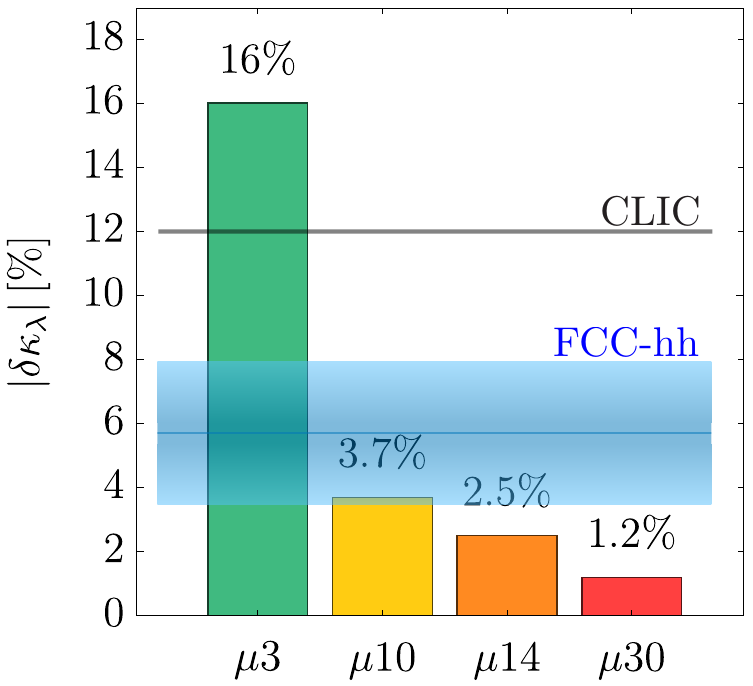}}  
    \end{center}
 \caption{Left: The HL-LHC and FCC-hh exclusion mass-reach for several beyond the Standard Model (BSM) particles. The exclusion (and discovery) reach of 10, 14 and 30~TeV muon colliders are reported as horizontal lines. Right: Sensitivity to the Higgs trilinear coupling modifier $\delta\kappa_\lambda$ of different future colliders proposals. The sensitivity of the 10~TeV muon collider ($\mu10$) is compared with the one of higher energy muon colliders ($\mu14$ and $\mu30$) and with the one of the low-energy stage at 3~TeV ($\mu3$).
 Plots from Ref.~\cite{PhysicsPotential_Accettura:2023ked}.}
 \label{fig:dir-3h-ch}
\end{figure}

The high available energy allows to search for new heavy particles, with a reach in mass that strongly extends the one of the LHC. In fact, the muons are elementary and their collision energy is entirely available to produce new particles. The protons instead are composite and their effective energy reach is limited to a fraction of the collider energy by the steep fall of the parton distribution functions. This is the reason why a muon collider with 10~TeV energy can access heavier particles than the 14~TeV LHC, as illustrated on the left panel of Fig.~\ref{fig:dir-3h-ch}. The figure shows the projected exclusion mass reach\footnote{We use the HL-LHC and FCC-hh sensitivity projections that have been presented for the 2020 Update of the European Strategy for Particle Physics, summarised in Ref.~\cite{EuropeanStrategyforParticlePhysicsPreparatoryGroup:2019qin}. The particles are labeled with a standard BSM notation; for instance $\widetilde{t}$ is the stop quark. See Ref.~\cite{PhysicsPotential_Accettura:2023ked} for additional details.} for a number of BSM particles at the HL-LHC (solid bars) and at the 100~TeV proton-proton collider FCC-hh (shaded bars). At a muon collider, these particles are produced in pairs by EW interactions and their production cross section does not depend on any BSM model detail other than the EW and spin quantum numbers of the states. The EW cross-sections for the particles considered in the figure range from $0.1$ to 10~fb~\cite{Buttazzo:2020uzc,Costantini:2020stv}  at the 10~TeV MuC, for masses almost up to the kinematic threshold of 5~TeV. With the target integrated luminosity of  10~ab$^{-1}$, enough (more than 1000) events will be available for discovery up to the threshold. The MuC mass-reach at 5~TeV is above the HL-LHC exclusion limit for all the BSM candidates. The 10~TeV muon collider has an even higher reach than a 100~TeV proton-proton collider on QCD-neutral particles such as charginos ${\widetilde{\chi}}_1^\pm$ and tau sleptons $\widetilde\tau$.

Along these lines, the ``Energy'' arrow in Fig.~\ref{fig:papdist} represents the possibility of searching for new heavy particles of very generic nature, or specific well-motivated candidates. Past works investigated the MuC sensitivity to a number of BSM scenarios ranging from WIMP dark matter, extended Higgs sectors, heavy neutral leptons, composite resonances, solutions to the $\mathrm{g}-2$ anomaly and more. An incomplete list of references include~\cite{Buttazzo:2018qqp,Ruhdorfer:2019utl,Liu:2021jyc,Asadi:2021gah,Huang:2021nkl,Liu:2021akf,Qian:2021ihf,Bandyopadhyay:2021pld,Capdevilla:2020qel,Capdevilla:2021rwo,Dermisek:2021ajd,Capdevilla:2021kcf,Homiller:2022iax,Mekala:2023diu,Li:2023tbx,Kwok:2023dck,Jueid:2023zxx,Liu:2023jta,DiLuzio:2018jwd,Franceschini:2022sxc,Han:2020uak,Bottaro:2021srh,Bottaro:2021snn,Bottaro:2022one,Capdevilla:2021fmj}. Few specific results are outlined below.

Reference~\cite{Buttazzo:2018qqp} (see also Refs.~\cite{PhysOpp_AlAli:2021let,Ruhdorfer:2019utl,Liu:2021jyc}) studied one extra EW-singlet extra Higgs scalar which is potentially responsible for the generation of a strong first-order EW phase transition in the Early Universe, and is present in other BSM scenarios as well. Such ``scalar singlet'' is a standard benchmark for future colliders, also in light of its peculiar coupling to the SM, which occurs only through a Higgs-portal interaction. The 10~TeV MuC mass-reach on this BSM scenario has been found superior to the one of the FCC-hh in the most motivated region of the parameter space of the model. In fact, the sensitivity is superior in the whole parameter space, if including~\cite{PhysicsPotential_Accettura:2023ked} the indirect MuC reach from Higgs coupling measurements. The MuC advantage over FCC-hh stems from the large MuC cross-section for the production of Higgs-portal coupled new physics in vector boson fusion. Similar findings have been reported in other Higgs-portal coupled BSM scenarios, making the muon collider an ideal option to cover this class of models at the multi-TeV scale.

It should be emphasised that the results described above---as well as the other findings reported in this section and the majority of the muon collider studies in the literature---are based on detailed phenomenological analyses that consider the relevant backgrounds as well as a parametric modeling of the detector effects. The assumed detector performances are those of the IMCC muon collider DELPHES card~\cite{deFavereau:2013fsa,delphes_card_mucol}, which match the performances of the CLIC detector and lie in between (see Section~\ref{sec:Section04_1}) the ``Baseline'' and ``Aspirational'' performances.

Several papers~\cite{DiLuzio:2018jwd,Franceschini:2022sxc,Han:2020uak,Bottaro:2021snn,Bottaro:2022one,Capdevilla:2021fmj} studied the observability of a variety of WIMP DM candidates at the muon collider (see Ref.~\cite{PhysicsPotential_Accettura:2023ked} for a summary). Detection strategies include mono-photon (or, more generally mono-$X$) searches, indirect searches from loop effects, and direct searches of the disappearing tracks produced by the charged particle in the dark matter multiplet. The beam-induced-background (BIB) from the decay of the muon beams has a potentially large impact on the disappearing track search, which is difficult to estimate and to parametrize faithfully in a fast detector simulation. For this reason, a complete study of disappearing tracks was performed in Ref.~\cite{Capdevilla:2021fmj}, based on realistic BIB Monte Carlo simulations. The observability of long-sought dark matter candidates such as the Higgsino and the Wino has been established up to the thermal mass.

The ``Precision'' arrow in Fig.~\ref{fig:papdist} represents accurate measurements in the EW and Higgs sector. These are possible at the muon collider exploiting the large rate for EW-scale scattering processes initiated by effective vector bosons, and the small QCD background that is typical of leptonic collisions. The perspectives of exploiting the 10 million Higgs boson produced by the collider in the vector boson fusion channel have been investigated in details~\cite{Han:2020pif,PhysOpp_AlAli:2021let,Forslund:2022xjq,MuonCollider:2022ded} based on fast simulation but also validated against the available full-simulation studies~\cite{MuonCollider:2022ded}. These sensitivity projections for Higgs signal-strength measurements were used in Ref.~\cite{PhysicsPotential_Accettura:2023ked} for a Higgs couplings fit in the same setup that was employed in Ref.~\cite{deBlas:2019rxi} for a global comparative assessment of future colliders sensitivity in preparation for the 2020 European Strategy Update process\footnote{Reference~\cite{PhysicsPotential_Accettura:2023ked} also presents an EFT fit, which includes several additional muon collider observables like high-energy measurements.}. The result is that a per mille level determination of the single Higgs boson couplings will be possible at the 10~TeV MuC, similarly to future low-energy $\mathrm{e}^+\mathrm{e}^-$ Higgs factories. A detailed inspection of the sensitivity to different couplings reveals the complementarity between MuC and low-energy Higgs factory couplings determinations.

The findings above refer to the ``kappa-0'' Higgs couplings fit~\cite{deBlas:2019rxi}, where no BSM Higgs decay channel is allowed. The closure of the ``kappa-1'' fit, where this assumption is relaxed, requires a measurement that is sensitive to the absolute value of at least one of the couplings of the Higgs, without dependence on the Higgs total width. The determination of the total (inclusive) Higgs cross-section in the Z-boson fusion production channel could provide such measurement at the MuC, enabling a very precise coupling determination also in the kappa-1 scheme~\cite{Li:2024joa}. However, this measurement relies on the observability of muons in the forward region by a dedicated detector, whose feasibility is still to be assessed (see Section~\ref{sec:Section04_1}). If such measurement turned out to be impossible, the closure of the kappa-1 fit will have to rely on the inclusive cross-section measurement at a low-energy $\mathrm{e}^+\mathrm{e}^-$ Higgs factory, providing one further element of complementarity with such machine. The flat direction in the kappa-1 fit could be also lifted by high-energy measurements at the MuC, by relying on extra assumptions~\cite{Forslund:2023reu}.

Unlike low-energy Higgs factories, the MuC can also measure the double-Higgs production cross-section and determine the Higgs trilinear coupling at the percent level, as shown on the right panel of Fig.~\ref{fig:dir-3h-ch}. This result~\cite{Han:2020pif,Buttazzo:2020uzc} is based on a parametric modeling of the detector effects---assuming CLIC-like performances---validated against the CLIC full simulation sensitivity projection~\cite{deBlas:2018mhx}. Figure~\ref{fig:dir-3h-ch} also reports the expected sensitivity of FCC-hh~\cite{Mangano:2020sao}, which ranges from $3\%$ to $8\%$ depending on the assumed detector performances. The MuC result has been found less sensitive to detector performances because the background is lower.

A unique feature of the muon collider is the simultaneous availability of energy and precision. This allows, very simply, to measure high-energy scattering cross-sections precisely. The possibility of exploiting energy and precision at once is represented by the ``Energy{\& Precision'' arrow in Fig.~\ref{fig:papdist} and it corresponds to an exploration strategy that is typical of high energy physics: higher energy observables are more sensitive to short-distance physical laws and give access to novel effects that were too tiny to be detected in previous experiments performed at lower energy. This mechanism produced some of the most striking past discoveries, including the discovery of neutral currents and the one of the finite radius---i.e., the composite nature---of the nucleons. Sufficient statistics will be available at the muon collider to measure 10~TeV $2\to2$ scattering cross sections with percent-level accuracy\footnote{Interestingly enough, the sensitivity is not limited to neutral $\mu^+\mu^-$ scattering processes. The copious emission of charged EW bosons at the 10~TeV scale gives effective access to charged $\mu\nu$ scattering, and to $\nu\bar{\nu}$ scattering.}, which roughly translates into the sensitivity to new physics at the 100~TeV scale. Notice for comparison that 100~TeV scale new physics produces unobservably small one part-per-millon effects on EW-scale observables. This explains the superior sensitivity~\cite{PhysicsPotential_Accettura:2023ked} to EFT interactions of the muon collider in comparison with Z-pole measurements at the FCC-ee.

Concrete illustrations of the physics potential from high-energy measurements include (see~\cite{PhysicsPotential_Accettura:2023ked} for an overview) the sensitivity to new neutral currents mediated by a very heavy ``$Z^\prime$'' boson of order 100~TeV mass~\cite{Chen:2022msz,Buttazzo:2020uzc} and the possibility of observing new quark- or lepton-flavour breaking transitions more effectively than low-energy experiments~\cite{PhysOpp_AlAli:2021let}. Furthermore, high-energy measurements give direct access to the possible finite size of the Higgs particle~\cite{Chen:2022msz,Buttazzo:2020uzc} testing Higgs compositeness more effectively than any other future collider project. This is shown on the left panel of Fig.~\ref{fig:fl} in the plane formed by the Higgs compositeness scale $m_*$---i.e., the inverse of the Higgs radius---and the effective coupling $g_*$ of the new strongly-interacting sector the Higgs emerges from. The ``Others'' line in the plot is the envelope of the sensitivity attained by other future collider projects. It assumes the entire FCC-ee physics program including EW precision tests at the Z pole, and direct searches at FCC-hh for new states with order $m_*$ mass that emerge in the Composite Higgs scenario under examination.

The ``Muons and neutrinos'' arrow in Fig.~\ref{fig:papdist} reminds us that energetic muon beams will be made available for the first time. Studying their collisions offers self-evident opportunities to discover new physics coupled primarily to muons and not for instance to electrons, whose collisions are extensively studied. Specific illustrations~\cite{PhysOpp_AlAli:2021let,PhysicsPotential_Accettura:2023ked} of the potential in this direction stem from the larger Yukawa coupling of the muon than of the electron, which suggests a stronger coupling to muons of extended Higgs sectors~\cite{Chakrabarty:2014pja}, from the connection with possible new physics in the muon anomalous magnetic moment~\cite{Capdevilla:2020qel,Buttazzo:2020ibd,Yin:2020afe,Capdevilla:2021rwo,Dermisek:2021ajd,Dermisek:2021mhi,Capdevilla:2021kcf,Bandyopadhyay:2021pld,Dermisek:2022aec,Paradisi:2022vqp}, as well as opportunities to test muon-philic scenarios~\cite{Huang:2021biu,Asadi:2021gah,Huang:2021nkl,Azatov:2022itm,Altmannshofer:2022xri,Bause:2021prv,Qian:2021ihf,Allanach:2022blr}. However, the agnostic exploration of new physics coupled to muons is arguably the strongest motivation. 

Beyond collisions, energetic muon beams offer additional unique opportunities. In particular, they produce a collimated beam of neutrinos with unprecedented properties described in Section~\ref{subsec:SaS}.

\subsubsection*{Future work and physics-detector integration}

The future development of muon collider physics should proceed along three complementary directions. First, by expanding the physics case with new targets and sensitivity projections. We saw that previous work already identified many promising avenues, but we are still far from a complete assessment of the muon collider physics opportunities. 

Second, by consolidating the physics case with increasingly detailed and optimised sensitivity projections in close connection with the evolution of the muon collider experiment and detector design. Current results assume the CLIC-like detector performances implemented in the IMCC muon collider DELPHES card~\cite{deFavereau:2013fsa,delphes_card_mucol}, as previously described. Novel and more specific (``Baseline'' and ``Aspirational'') target performances are defined in Section~\ref{sec:Section04_1}.

The third direction in which progress is needed is the development of the theoretical tools for making predictions at the muon collider. The next section summarises the main open questions in this area.

\subsubsection{Challenges and opportunities at the theory frontier}\label{subsec:TF}

EW interactions at 10~TeV display novel phenomena associated with the large separation between the collision energy and the mass of the EW particles, of order $100$~GeV. The EW bosons start behaving as nearly massless producing IR-enhanced radiative effects that are parametrically of order one at 10~TeV. The corresponding ``EW radiation'' phenomena share some similarity with electromagnetic and QCD radiation, but also remarkable differences that pose profound theoretical challenges. Being the EW symmetry broken at low energy, different particles in the same EW multiplet---i.e., with different EW gauge color---are distinguishable. This invalidates the cancellation between real and virtual radiation, obliging us to include the resummed effects of EW radiation in all observables, also in inclusive cross sections. Moreover, the resummation is needed for non color singlet observables that do not exist in QCD. In analogy with QCD and unlike QED, the IR scale for EW radiation is physical. However, at variance with QCD, the theory is weakly-coupled at the IR scale. A first-principle and accurate modelling of EW radiation is thus possible in principle, though based on all-orders resummation techniques that are still to be developed. On top of being functional to the exploitation of the MuC physics potential, this poses an interesting challenge to theory~\cite{Dawson:1984gx,Kane:1984bb,Kunszt:1987tk,Chanowitz:1985hj,Gounaris:1986cr,Kilgore:1992re,Fadin:1999bq,Ciafaloni:2000rp,Ciafaloni:2000df,Melles:2000gw,Manohar:2014vxa,Bauer:2017isx,Bauer:2018xag,Manohar:2018kfx,Fornal:2018znf,Chiu:2007dg,Chiu:2007yn,Chiu:2009ft,Denner:2000jv,Denner:2001gw,Borel:2012by,Wulzer:2013mza,Cuomo:2019siu,Bothmann:2020sxm,Chen:2016wkt,Han:2021kes,Ruiz:2021tdt,Chen:2022msz,Garosi:2023bvq}. 

Muon colliders will require at least as much theory progress as it was needed and achieved in the past decades for the exploitation of the LHC data. LHC studies greatly advanced fixed-order calculations, Monte Carlo showering and resummation techniques and produced radical progress in the theoretical (and later experimental) knowledge of the QCD interaction. Muon collider physics requires similar tools but in the EW sector. Its development will entail an equally radical advance of EW physics knowledge.

\subsection{Synergies and staging}\label{subsec:SaS}

Most of the recent work focuses on the opportunities offered by muon collisions at 10~TeV (or possibly higher) energy and with the target design luminosity of $10~{\textrm{ab}}^{-1}$ (or higher, at higher energy). Much less explored are instead alternative ways---other than collisions---to exploit the muon beams (see however Ref.~\cite{Cesarotti:2023sje} for a first study of physics at muon beam-dump experiments), or the opportunities that emerge during the facility construction in a staged approach. This is the subject Sections~\ref{subse:nu} and~\ref{subse:stage}, respectively.

\begin{figure}[t]
   \begin{center}       
   \includegraphics[width=0.45\textwidth, angle=0]{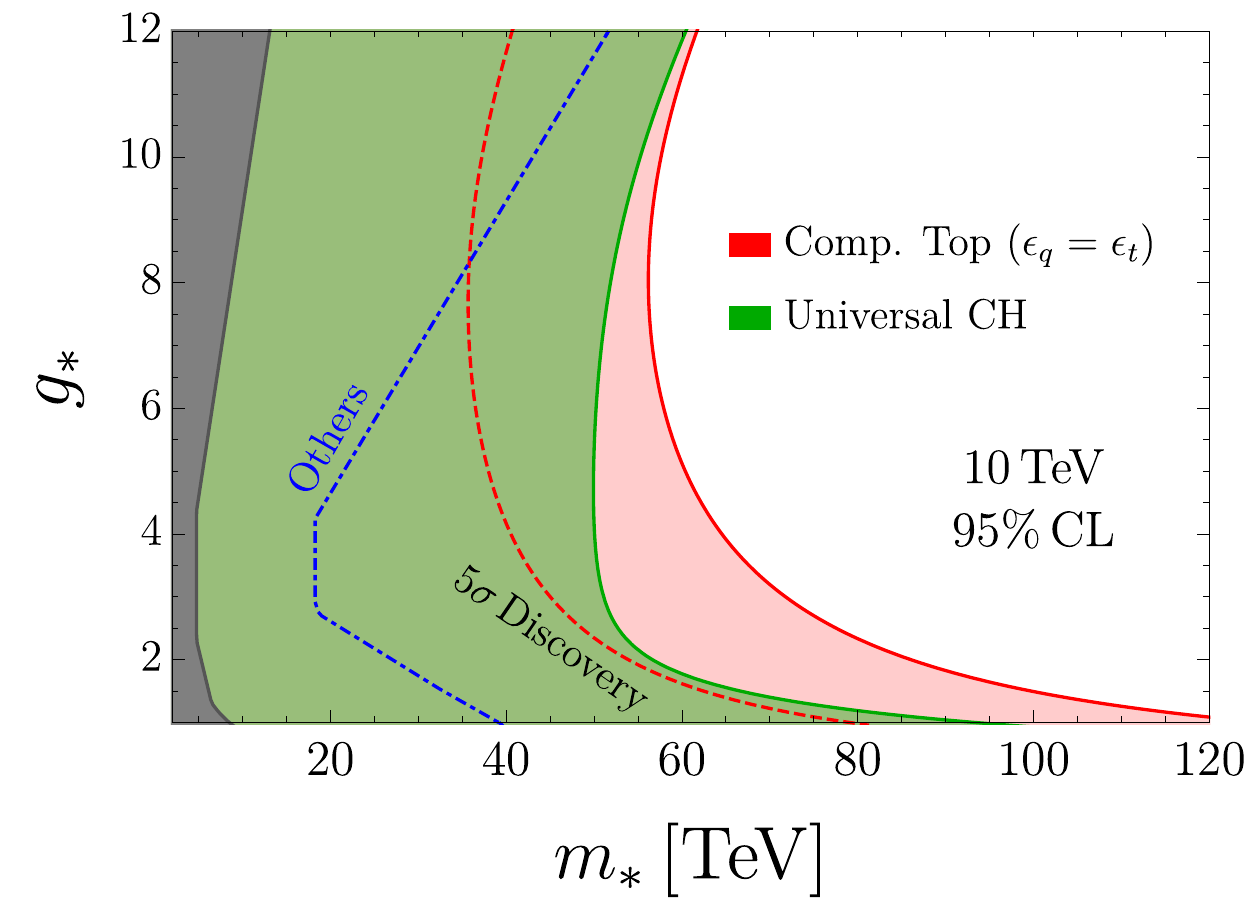} \hfill
   \raisebox{20pt}{\includegraphics[width=0.5\textwidth, angle=0]{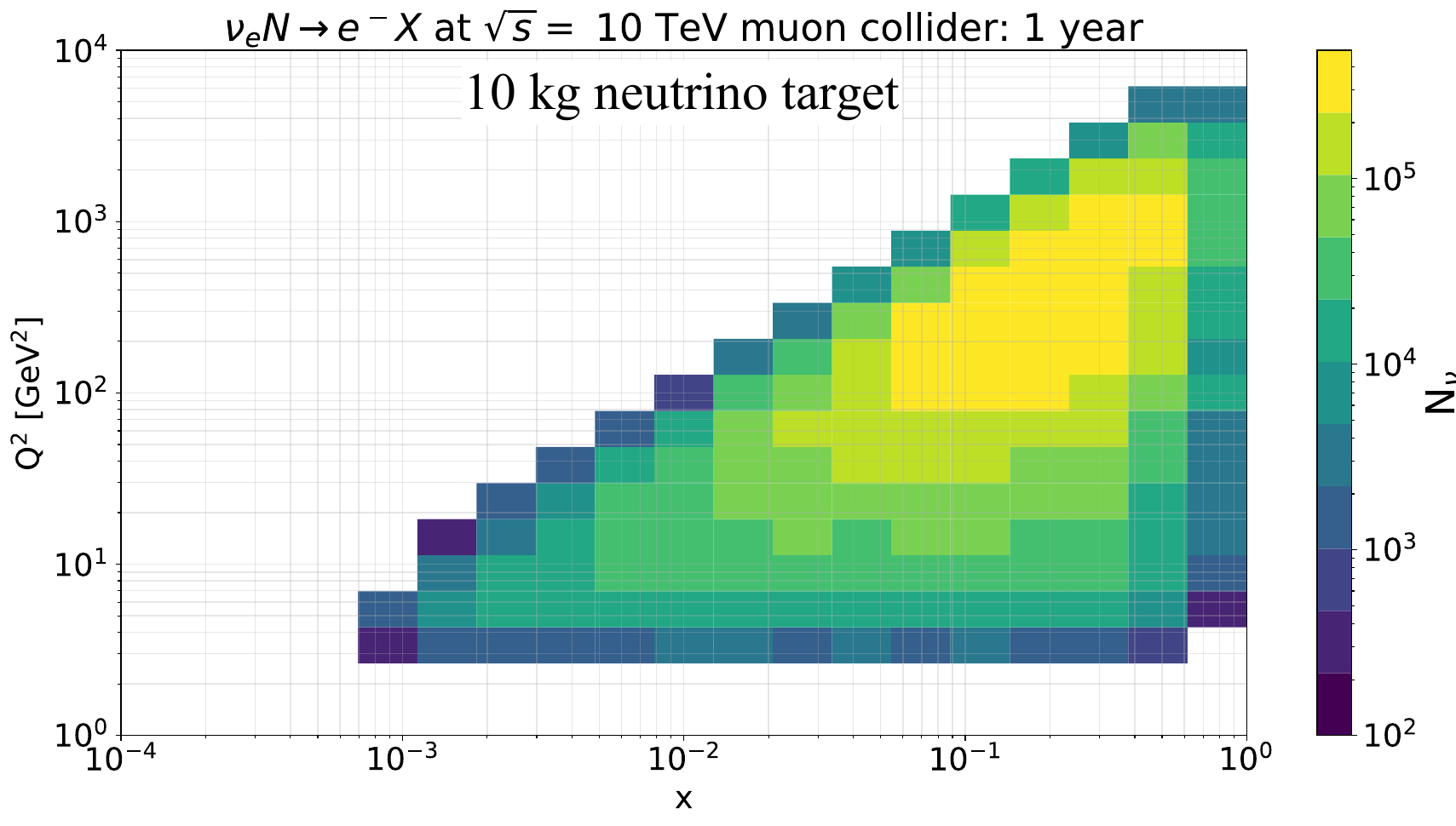}}    
   \end{center}
 \caption{
Left: The sensitivity (from~\cite{PhysicsPotential_Accettura:2023ked}) to Higgs compositeness. Right: the number of neutrino DIS events as a function of $x$ and $Q^2$ for a 10 TeV MuC after one year of data taking with a small 10~kg detector.}
 \label{fig:fl}
\end{figure}

\subsubsection{Neutrinos}\label{subse:nu}

When the muons in the beam decay in the straight section close to the Interaction Point (IP), they produce a collimated beam of very energetic neutrinos. Based on the conservative estimate of the neutrino flux presented in Section~\ref{sec:neutrino_interface}, we expect at least $9\times10^{16}$ neutrinos of each species emitted in a narrow cone of $0.6$~mrad average angle during one year of run at the 10~TeV MuC, and $2.4\times10^{17}$ neutrinos at the 3~TeV collider. A far-forward detector with realistic geometry (e.g., a 10~cm radius cylindrical target placed 100~m away from the IP, or a 1~m target at 1~km distance) would intercept all the emitted neutrinos. Each neutrino would thus have the chance to produce a detectable interaction, with probability
\begin{equation*}
    p_{\rm{int}}\simeq 6 \times10^{-8}\,
    \frac{\rho\cdot{L}}{{\rm{g\,cm}}^{-2}}\,\frac{E}{\rm{TeV}}\,,
\end{equation*}
where we assumed a typical cross-section on nucleons of $10^{-35} {\rm{cm}}^2\cdot{E}/{\rm{TeV}}$ for all neutrino species, with $E$ the neutrino energy. In the equation, $\rho$ is the density of the neutrino target and $L$ is the longitudinal extension of the target along the neutrino beam.

No design is currently available for the neutrino detector. One concept was developed long ago~\cite{King:1997dx}, with the  purpose of studying the neutrinos produced by a dedicated 250~GeV muons ring. The concept foresees a cylindrical target with 1~m length and 10~cm radius composed of 750 silicon CCD tracking planes, which would also act as a vertex detector. This would be followed by a magnetized spectrometer and calorimeter. The design of Ref.~\cite{King:1997dx} ensures excellent tracking and reconstruction performances for all the particles that are possibly produced in the interaction, including electrons, muons (and perhaps $\tau$'s) as well as bottom and charm hadrons tagging and discrimination. On the other hand, the target mass per unit area is relatively small: $\rho\cdot{L}=50$~g$\cdot{\textrm{cm}}^{-2}$, for a total target mass of only 10~kg. Contemporary neutrino detectors operating at similar energy and with similar physics needs employ denser and bigger targets and attain masses ranging from one ton (FASERv~\cite{FASER:2022hcn}, SND@LHC~\cite{SNDLHC:2022ihg}, CHORUS~\cite{CHORUS:2005cpn}) to several hundred tons (CHARM~\cite{CHARM:1987pwr}, NuTeV~\cite{NuTeV:2005wsg}). Waiting for dedicated studies, it seems reasonable to assume a target mass of at least one ton for the neutrino detector.

Figure~\ref{fig:nfl} reports the energy distribution and the total number of neutrino interactions that could be recorded by a 10~kg target (corresponding to Ref.~\cite{King:1997dx}) and by a 1~ton target during one year of run at the 10~TeV and 3~TeV muon colliders. The neutrinos available for a number of past and planned experiments (from Ref.~\cite{Ahdida:2023okr}) are also reported for comparison. Notice that only the anti-neutrinos from the decay of the $\mu^+$ are included in the figure. A second detector would be needed to collect the neutrinos from the $\mu^-$ decay, since they fly away from the IP in the opposite direction. Electron (anti-)neutrinos with a similar spectrum are also present.

The figure shows that one single year of muon collider operations would enable to collect orders of magnitude more interactions of TeV-energy neutrinos than current facilities such as FASER$\nu$ and SND@LHC. Even with the very small 10~kg detector, more neutrinos will be collected than at the proposed Forward Physics Facility (FPF)~\cite{Feng:2022inv}. The energy spectrum peaks at 1~TeV for the 3~TeV MuC, and at 4~TeV in the case of the 10~TeV collider, and more than 100 million interactions would be recorded at these energies by a 1~ton detector. There are many orders of magnitude more neutrino interactions in this energy range than any other past or planned neutrino experiment.  Furthermore, extremely small uncertainties are expected in the predictions for the energy spectrum of  the muon collider neutrinos,  to be contrasted with the large uncertainties in the spectrum of the neutrinos produced by the LHC arising from forward hadron production. On top of the superior statistics, neutrino physics measurements at a muon collider are thus expected to benefit from reduced systematic uncertainties in comparison with LHC-based experiments such as FASER or FPF.

\begin{figure}[t]
   \begin{center}        \includegraphics[width=\textwidth]{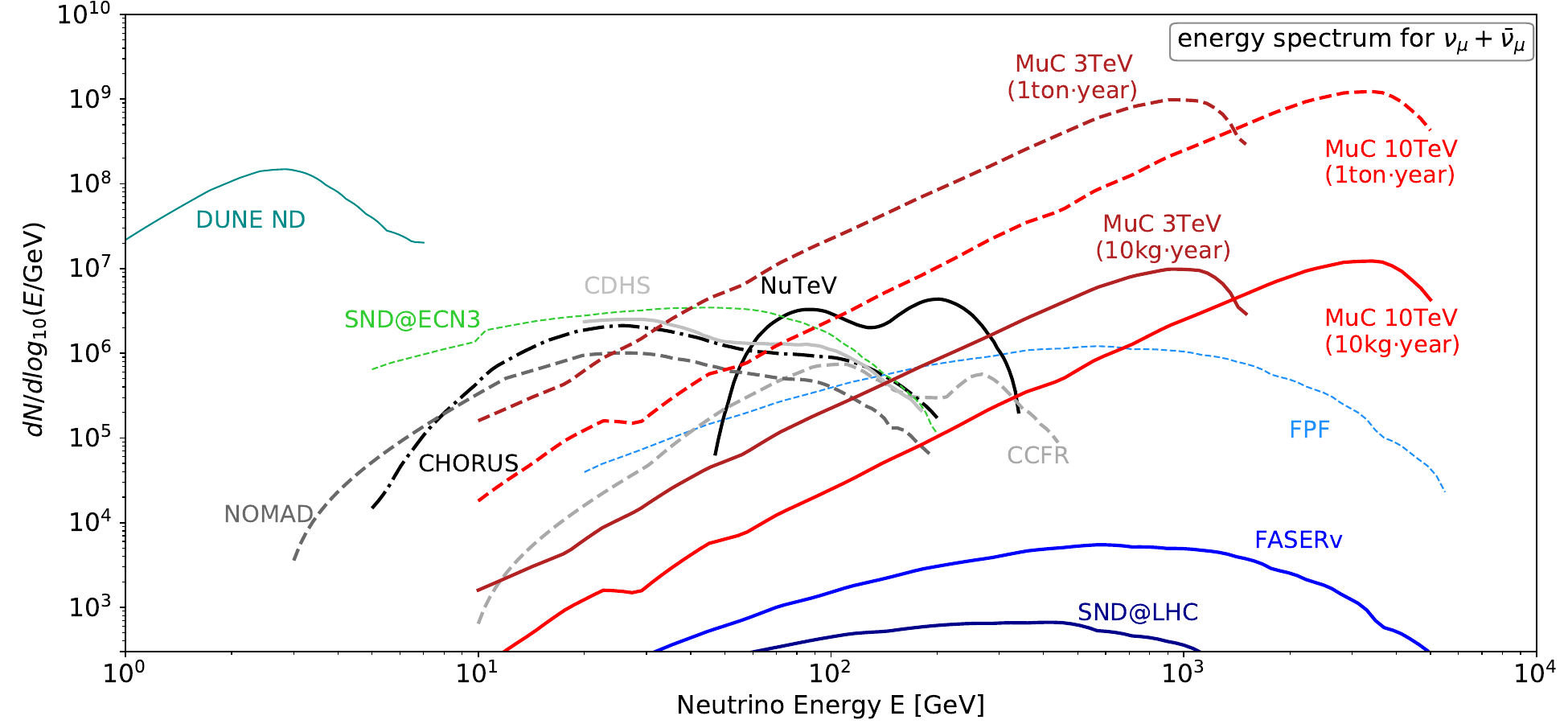}   
   \end{center}
 \caption{The energy spectrum of neutrino interactions produced by the 3~TeV and 10~TeV MuC in one year, overlaid with the summary plot in Ref.~\cite{Ahdida:2023okr} for past and planned neutrino experiments. The solid and dashed lines assume, respectively, a small 10~kg and a realistic 1~ton target mass.}
 \label{fig:nfl}
\end{figure}

The physics opportunities offered by the neutrino beams are still to be explored. Ideas discussed long ago~\cite{King:1997dx} include CKM quark mixing matrix, nucleon structure, EW precision and charm quark physics measurements. The contemporary relevance of these measurements should be assessed, and the sensitivity projections adapted to the higher-energy neutrino beams that would be available at the 3 and 10~TeV MuC. 

Progress can also come from the extrapolation of the physics reach of modern LHC-based neutrino experiments. For example, Ref.~\cite{Cruz-Martinez:2023sdv} studied the opportunities for QCD and hadronic physics studies, and Parton Distribution Functions (PDFs)~\cite{Gao:2017yyd} determination, offered by neutrino Deep Inelastic Scattering (DIS) at these facilities. The right panel of Fig.~\ref{fig:fl} shows the expected DIS event yields in the $x$--$Q^2$ plane at the 10~TeV muon collider. The figure conservatively assumes the 10~kg target. The unprecedented rate of events will make possible a very fine binning with permille-level statistical uncertainties in the whole kinematic region covered and enables multi-differential measurements, such as those required to access the 3D structure of the proton in terms of non-perturbative quantities such as transverse-momentum dependent PDFs or generalised PDFs. The large statistics can be positively compared with the predicted event yields at the Electron-Ion Collider (EIC)~\cite{AbdulKhalek:2021gbh}. A far-forward neutrino detector at the muon collider would therefore provide a charged-current analog of the EIC.

\subsubsection{Stages}\label{subse:stage}

We conclude this section with a physics-driven assessment of the  opportunities of muon colliders of lower energy or lower luminosity, to be possibly operated during the construction of the 10~TeV MuC facility in a staged approach. 

At order 100~GeV energy the luminosity is not competitive with the one of electron-positron colliders, however (see Ref.~\cite{PhysicsPotential_Accettura:2023ked} and references therein) muon colliders feature specific opportunities in this range of energy. One is to operate at 125~GeV energy in the centre of mass, exploiting the large muon Yukawa coupling to produce the Higgs boson in the $s$-channel. This would enable a percent-level direct determination of the Higgs boson width, an astonishingly precise Higgs mass measurement at one part per million and few permille measurement of the muon Yukawa coupling. Alternatively, or in addition, a muon collider operating at the threshold for the production of a pair of tops (which muon colliders could reach with a very compact design) would enable a conclusive assessment of the SM vacuum stability by a measurement of the top mass with 50~MeV uncertainty. Focused projects with specific and unique targets like the ones above would be definitely of interest.

The current IMCC staging plan~\cite{PhysicsPotential_Accettura:2023ked} foresees a first muon collider with 3~TeV energy in the centre of mass and a target integrated luminosity of 1~ab$^{-1}$ collected in~five years, or of 2~ab$^{-1}$ in 10~years if the 10~TeV collider construction is delayed. Such 3~TeV muon collider stage features striking physics opportunities, described in Ref.~\cite{MuonCollider:2022xlm} in details. These include probing Higgs compositeness at the 20~TeV scale (still competitive of better than other future collider project) by high-energy measurements and determining the Higgs trilinear coupling (see Fig.~\ref{fig:dir-3h-ch}). It will be also possible to improve HL-LHC single-Higgs couplings precision, but not to surpass it radically. Studying muon collisions for the first time entails strong opportunities, including those that stem from the neutrino beams previously described. 

The 3~TeV MuC can also search for new heavy particles. However, the window of opportunities beyond the HL-LHC sensitivity is limited by the pair-production mass reach of $1.5$~TeV. Progress could thus be possible only on relatively light motivated targets like extended Higgs sectors and Higgsino dark matter. Further studies and advances on the search of these specific scenarios would be an important addition to the 3~TeV MuC physics case. Recently, R.~Capdevilla \textit{et al.},studied the possibility of observing the soft tracks from the decay of the charged state in fully realistic experimental conditions including the BIB, and demonstrated that the Higgsino with thermal mass could be discovered by this strategy at the 3~TeV MuC~\cite{Capdevilla:2024bwt}.

An alternative plan is to operate the first muon collider directly at 10~TeV energy, but employing less performant technologies---available on a shorter timescale---which entail a luminosity reduction by around a factor of 10 or less. The physics reach of such low-luminosity 10~TeV MuC stage can be robustly extrapolated from the full-luminosity results. 

The candidate BSM particles considered in Fig.~\ref{fig:dir-3h-ch} would still be discovered or excluded up to the 5~TeV mass threshold. In fact, the particles considered in the figure are produced at a very high rate with the full design luminosity. More than $10^{3}$ or $10^4$ production events are expected with the baseline integrated luminosity of 10~ab$^{-1}$, which is collected in 5~years with the full luminosity. Around one month of run would be enable a discovery. With one tenth of the luminosity, discovery would merely take one year. The reduced luminosity would of course affect the excellent perspectives of  characterising the properties of the newly discovered particles with precision, but not the mass-reach.

The indirect physics reach described in Section~\ref{subsec:EF} will be impacted by the reduced luminosity because the corresponding measurements are dominated by statistical uncertainties and not by systematics or theory errors. However, the factor few increase of the uncertainties---by the square root of the luminosity---will not have a dramatic impact on the physics reach. While a detailed and comprehensive assessment is not yet available, it is expected that the reduced performances would still allow for very competitive studies of the Higgs single and trilinear couplings, and for probes of very heavy new physics by high-energy measurements including Higgs compositeness at tens of TeV. On the other hand, the total 10~ab$^{-1}$ target luminosity is needed in order to deliver the physics program described in Section~\ref{subsec:EF} in its full breadth and power.

 \begin{flushleft}
 \interlinepenalty=10000
 
\end{flushleft}
\clearpage
\section{Physics, detector and accelerator interface}
\label{sec:Section04}


\subsection{Physics and detector needs 
}
\label{sec:Section04_1}

\subsubsection{Overview}
The vast physics potential described in Section~\ref{sec:Section03} can be translated into requirements for the machine and the experiments that will harvest data from it.

The most notable requirements on the machine are: the mitigation of the beam-induced backgrounds (discussed in Section~\ref{sec:Section04_2}), the amount of delivered integrated luminosity and the ability to measure it with permille-level precision, and a precise determination of the beam energy.

The requirements on the detectors situated at the main interaction points are set in terms of ``baseline'' and ``aspirational'' targets. The latter are aimed at a muon collider operating at $\sqrt{s}=10$~TeV. They include requirements on the detector acceptance, particle detection and identification efficiency, as well as resolutions on the various particle properties inferred by the instrumental measurements.
The baseline requirements are set by the need to separate collision products from energy depositions from beam-induced backgrounds and the need to identify and measure particles over a wide range of energies with a precision at least comparable with the experiments taking data at the LHC.
The aspirational targets are set with the goal of fully exploiting the physics potential of the machine. They aim at a reconstruction performances comparable to those targeted by Higgs factories while extending to significantly higher energies. Furthermore, these requirements are set with the goal of  maintaining sensitivity to unconventional signatures as well as profiting from the unique potential of muon colliders to use forward muons to study vector boson fusion processes.
The preliminary targets for several key metrics discussed more in detail in the next Section are reported in Table~\ref{tab:detector_req}. 

\begin{table}[h]
\begin{center}
\caption{Preliminary summary of the ``baseline'' and ``aspirational'' targets for selected key metrics, reported separately for machines taking data at $\sqrt{s}=3$ and $10$~TeV. The reported performance targets refer to the measurement of the reconstructed objects in physics events after, for example, background subtraction and not to the bare detector performance.}
\label{tab:detector_req}
\begin{tabular}{lccc}
\hline\hline
 \textbf{Requirement} & \multicolumn{2}{c}{\textbf{Baseline}} & \textbf{Aspirational}\\
  & \textbf{$\sqrt{s}=3$~TeV} & \textbf{$\sqrt{s}=10$~TeV} & \\
\hline
Angular acceptance & $|\eta|<2.5$ & $|\eta|<2.5$ & $|\eta|<4$ \\
Minimum tracking distance [cm] & $\sim 3$ & $\sim 3$ & $< 3$\\
Forward muons ($\eta> 5$) & -- & tag & $\sigma_{p}/p \sim 10$\%\\
Track $\sigma_{p_T}/p^{2}_{T}$ [GeV$^{-1}$] & $4 \times 10^{-5}$ & $4 \times 10^{-5}$ & $1 \times 10^{-5}$ \\
Photon energy resolution & $0.2/\sqrt{E}$ & $0.2/\sqrt{E}$ & $0.1/\sqrt{E}$\\
Neutral hadron energy resolution & $0.5/\sqrt{E}$ & $0.4/\sqrt{E}$ & $0.2/\sqrt{E}$ \\
Timing resolution (tracker) [ps] & $\sim 30-60$ & $\sim 30-60$ & $\sim 10-30$ \\
Timing resolution (calorimeters) [ps] & 100 & 100 & 10 \\
Timing resolution (muon system) [ps] & $\sim 50$ for $|\eta|>2.5$ & $\sim 50$ for $|\eta|>2.5$ & $< 50$ for $|\eta|>2.5$ \\
Flavour tagging & $b$ vs $c$ & $b$ vs $c$ & $b$ vs $c$, $s$-tagging \\
Boosted hadronic resonance ID & $h$ vs W/Z & $h$ vs W/Z & W vs Z\\
\hline\hline
\end{tabular}
\end{center}
\end{table}

Initial considerations on the interplay between the accelerator complex and detectors dedicated to the study of neutrino physics are reported in Section~\ref{sec:neutrino_interface}.

\subsubsection{Key challenges}
\label{subsec:detector_challenges}

The requirements on a detector that can successfully extract physics from collision data at a muon collider are set by two categories of constraints: the rejection of the reducible beam-induced backgrounds (BIB) and the need for (precision) physics observables. These two categories can lead to conflicting choices that require an overall optimisation to be performed. A quantitative assessment of these requirements is being carried out based on a preliminary set of physics benchmark channels, discussed in the following paragraphs, which are used to study the overall performance of a detector design on selected key metrics.

The requirements on the detector acceptance, summarised in terms \textit{angular acceptance} and \textit{minimum tracking distance},  have to balance the need to maximise the sensitive volume for a high rate SM measurements, retain sensitivity to a broad range of non-prompt physics, while suppressing the BIB to a manageable level. 
For example, the presence of large absorber nozzles located in the forward region of the detector is key to reducing the high-energy showers from the beam decays to a soft, diffused, energy contribution in the detector volume. At the same time, especially when considering a machine operating at a centre of mass energy of 10~TeV, a sizeable fraction of the interesting events will include outgoing particles within the uninstrumented volume of the absorbers. 
The precision determination of the Higgs boson couplings in single and multi-Higgs boson production events are used as main benchmark to set the angular acceptance requirements. The requirements on the minimum tracking distance will be studied on high-rate Higgs boson studies exploiting flavour-tagging as well as searches for long-lived particles~\cite{disapp_tracks}. 

Thanks to their highly penetrating nature, muons offer a unique possibility to discriminate charged and neutral vector boson fusion processes. Dedicated \textit{forward muon} ($\eta>5$) detectors that could be placed beyond the detector shielding and the accelerator components to detect outgoing muons from neutral vector boson fusion processes. 
The needs of a comprehensive vector boson scattering measurement and BSM searches (e.g.\,extended Higgs sectors) programme set the goals of measuring not only the presence of the outgoing muons, but also their momenta with good resolution. A preliminary study of the placement of the detectors, their technology and the effects of BIB still needs to be performed.

It is assumed that excellent (>90\%) detection efficiencies must be achieved for energies between $\mathcal{O}(1)$~GeV to $\mathcal{O}(1)$~TeV for all measured particle species. The requirements on the resolution of the measurements of \textit{transverse momenta} and \textit{energies} are driven by the high rate Higgs and SM measurements. In particular, these requirements are set by the need of rejecting irreducible SM backgrounds that share the same final state signature. Searches for high-mass objects, where searches for heavy vector triplets (HVT) are being used as a benchmark, contribute with looser requirements since these are naturally less affected by both BIB and irreducible SM backgrounds. The same physics benchmarks are currently used for the determination of the requirements in terms of flavour tagging of hadronic jets.

The identification of boosted hadronic resonances is an important feature, required for the full exploration of the SM electroweak sector and several searches for new heavy particles. In particular, the ability to distinguish hadronically-decaying Higgs bosons from Z and W bosons in the whole detector acceptance is crucial. High-mass searches, e.g.\,searches for HVTs, would benefit from the ability to distinguish W and Z bosons for centrally-produced resonances. 

The availability of timing information from each of the sub-detectors of a next-generation collider experiment is taken to be the norm. Requirements on \textit{timing} resolutions are set by background suppression considerations. The time of arrival of BIB particles on the sensitive elements of the detector spreads over a few ns after the bunch crossing, providing a powerful handle for rejection. In the future, these requirements could be relaxed when improved reconstruction algorithms with alternative mitigation strategies (e.g.\,based on the event topology) become available.
The search for heavy (meta-)stable massive particles was identified as a benchmark channel that would profit directly from a precise determination of the time of flight by each of the detector systems. A minimum goal of retaining sensitivity to slow-moving particles with a relativistic velocity $\beta \geq 0.5$ was taken as baseline model-independent benchmark. The use of timing and time-of-flight information for particle identification is a potentially promising tool that has yet to be studied.

Unconventional signatures and the search for long-lived states still need to be explored in detail at a muon collider. However, it is reasonable to assume that they would impose requirements on the ability to reconstruct displaced decays using only a subset of the available detector systems. Foreseen benchmarks are for example a search for anomalous showers in the calorimeter systems, or searches for displaced decay vertices in the tracking detectors or the muon system.

The requirements on the detector acceptance and performance are less stringent when considering an energy stage at $\sqrt{s}=3$~TeV. The differences are driven by the smaller contribution of vector boson fusion processes to the total cross-section and by the lower maximum energy of the collision products.

A key part of the physics programme at a muon collider is the measurement of several cross-sections and branching ratios with percent-level precision. This immediately sets a stringent requirement on the determination of the delivered integrated luminosity. The current assumed target is a measurement with an uncertainty at the permille-level.

\subsubsection{Recent achievements}

Significant progress has been recently made on on the physics requirements of a muon collider detector.  Much of this was documented in the Muon Forum report~\cite{Black:2022cth} and a recent review~\cite{PhysDect_Accettura:2023ked}.  There have been two main directions of research undertaken in the pursuit of better understanding detector needs: those focused on phenomenological studies and those based on full detector simulation.  Phenomenological studies (discussed in Section~\ref{sec:Section03}), have been performed at a variety of levels of sophistication, from background free studies on truth level events, to fast detector simulation in DELPHES with all relevant backgrounds.  The utility of these studies is that they are able to cover a large array of physics possibilities and to focus on what is needed for detectors to disentangle the hard scattering physics backgrounds.  The studies based on full simulation allow one to also quantify the potential in the presence of beam-induced backgrounds.  As focused on in Refs.~\cite{Black:2022cth,PhysDect_Accettura:2023ked}, there have also been a small amount of studies of the same physics cases, where both fast and full detector simulations were performed.  While ultimately it would be ideal to have all studies that impacted detector performance requirements done at a full simulation level, in cases where there is overlap between fast and full simulation studies good agreement has been found. This in turn gives confidence in the use of fast simulations to help better delineate the physics performance goals that have been reported.

Notable highlights were the studies performed on Higgs measurements and dark matter searches with disappearing tracks. 

The Higgs measurements~\cite{PhysDect_Forslund:2022xjq,PhysDect_Forslund:2023reu} serve as a general purpose benchmark for low-energy Standard Model processes. These were used to evaluate the effects of the detector acceptance on the final Higgs boson coupling precision, as well as to perform an initial study of the resolution and instrumental effects related to the presence of beam-induced backgrounds.

The search for disappearing tracks~\cite{PhysDect_Capdevilla:2021fmj} is a unique probe to definitively test the WIMP dark matter paradigm at a collider experiment. The signature consists of short tracks formed by a few (3-4) tracking detector hits. As such, it is uniquely sensitive to the detector layout, technology and the levels of beam-induced background.

Furthermore, recent fast simulation studies~\cite{Ruhdorfer:2023uea,PhysDect_Forslund:2023reu} have been crucial in demonstrating the need for not just forward muon tagging, but also the aspirational goal of forward muon measurements and how these measurements affect for example the search for exotic Higgs boson decay modes.

\subsubsection{Planned work}
The effort in view of the evaluation report will be focused on extending the list of preliminary benchmarks mentioned in Section~\ref{subsec:detector_challenges} by assembling a list of representative benchmarks that will allow judging the physics performance more globally. As a part of this effort, a broad survey of measurements and searches will be conducted to identify a list of high-priority benchmarks to be studied in more detail with full simulation. The result of the survey will be a prioritised benchmark list. The primary target of the evaluation will be a collider operating at $\sqrt{s}=10$~TeV. The list should specify for which benchmarks a dedicated assessment for a lower energy stage at $\sqrt{s}=3$~TeV should be performed.

The second key task follows directly from the first and consists in performing the detailed studies for as many of the benchmarks extracted from the prioritised benchmark list as feasible to update to the requirements on the key metrics. As a part of this detailed studies, the CLIC-inspired DELPHES card used for the fast-simulation results will be updated to reflect the expected detector performance in the ``baseline'' and ``aspirational'' scenarios.

The last key task will consist in developing a concept for automating the evaluation of the prioritised benchmarks to efficiently re-evaluate the results for any change in assumed detector design or data-taking conditions.

\subsubsection{Next priority studies and opportunities for additional collaboration}

A precise determination of beam energy is necessary for extracting physics results from the experiments. While it is assumed that techniques foreseen and used at other collider facilities will be compatible with the requirements, additional collaborators interest in working on the design of the instrumentation, the conceptual evaluation of the expected performance and the evaluation of any interplay with the detectors would be welcome.
    
Similarly, only preliminary studies have been performed to study luminosity determination using large-angle Bhabha scattering~\cite{Giraldin:2021gxz}. The feasibility of reaching the required precision needs to be systematically assessed, and possible dedicated detectors need to be designed, if required. New efforts in this direction would make the current assumptions significantly more robust. 

\subsection{Machine-detector interface 
}
\label{sec:Section04_2}

\subsubsection{System overview}
The beam-induced background (BIB) poses a significant challenge for the physics performance of a multi-TeV muon collider. The background is dominated by the decay of stored muons, with additional contributions from incoherent electron-positron pair production and possible beam halo losses on the aperture. These particles interact with surrounding materials and generate a mixed radiation field composed of secondary electrons, positrons, photons, as well as hadrons (photo-nuclear interactions) and muons (Bethe--Heitler pair production). Without dedicated mitigation measures, the beam-induced background would severely impede the reconstruction of collision events in the detector and would lead to significant radiation damage in detector components. 

An optimized design of the machine-detector interface (MDI) is hence critical for minimizing the impact of machine operation on the physics performance reach, and for reducing the ionizing dose and displacement damage in the detector. The MDI design must include an elaborate absorber configuration, consisting of masks inside the final focus region and a conical shielding, which penetrates deeply into the detector region, up to a few centimeters from the interaction point. The optimization of these masks and absorbers must be done jointly with the detector and interaction region (IR) design. A conceptual MDI layout for a 1.5\,TeV muon collider has been devised previously within the MAP collaboration, by means of the MARS Monte Carlo code~\cite{Mokhov2011,Mokhov2012,Alexahin2011}. Based on this configuration, first MDI design studies for higher-energy muon colliders (3~TeV and 10~TeV) were carried out within the IMCC \cite{Calzolari2022,Calzolari2023,Lucchesi2024} using the FLUKA code~\cite{Battistoni2013,Ahdida2022,FLUKAwebsite}.

\subsubsection{Key challenges}

The main challenge for the design of the machine-detector interface is the short lifetime of muons. This intrinsic source of radiation distinguishes a muon collider from other high-energy colliders. One of the key requirements is the design of a sophisticated shielding configuration at the interface between final focus magnets and detector. As shown in previous studies by the MAP collaboration for 1.5~TeV \cite{Mokhov2011,Mokhov2012,Alexahin2011}, the number of secondary particles entering the detector can be suppressed by orders of magnitude by placing massive absorbers in close proximity of the interaction point (IP). The innermost part consists of a nozzle-like shielding, which defines the inner detector envelope and hence the angular acceptance of the detector (10\textdegree \ in MAP). The shape and material budget of the nozzle determines the entry points, directions and energy spectrum of particles reaching the detector. The nozzle must be made of a \mbox{high-Z} material to efficiently shield the electromagnetic showers induced by muon decay products or by beam halo losses on the aperture. In addition, it must embed a layer of borated polyethylene or a similar material to moderate and capture secondary neutrons produced in photo-nuclear interactions. The nozzle must be carefully optimized for different center-of-mass energies. Besides the conceptual design, one also faces engineering challenges including assembly and support of the nozzle, as well as integration of the nozzle inside the detector. In addition, an adequate cooling system must be devised, to evacuate the heat deposited by decay products and beam halo losses.

While the characteristics (e.g.\,spectra) of the decay and halo-induced background are mainly determined by the nozzle, the amount of background particles is also influenced by the interaction region layout, i.e., by the lattice design and the placement of mask-like absorbers inside IR magnets. For example, the previous studies by MAP for a 1.5\,TeV collider suggested that a dipolar component in the final focus configuration can be beneficial for reducing the background flux into the detector \cite{Mokhov2012}. Designing and optimizing the IR layout is one of the key challenges for the collider design, in particular for 10\,TeV  (a 3\,TeV and 6\,TeV IR design was previously devised within MAP \cite{Alexahin2012,Alexahin2016,Alexahin2018}, although no background simulations were published). Furthermore, the flux of secondary particles is also influenced by the presence of a solenoid field in the detector region. This concerns not only the decay-induced background, but also incoherent electron-positron pairs, which are produced in the vicinity of the IP and are not directly intercepted by the nozzle. While the choice of the solenoid field strength is primarily driven by the particle detection efficiency in the detector, it must also take into account the benefits for background suppression.

Even with an optimized MDI and interaction region design, a suitable choice of detector technologies and reconstruction techniques is needed to reduce the effects of the remaining background (see Chapter~\ref{sec:Section05}). The signatures of background particles often exhibit distinct differences with respect to collision products, which can be exploited for background suppression. This concerns, for example, the arrival time of background particles with respect to the bunch crossing, as well as the directions of background particles when they enter the detector.
The shielding requirements and MDI design will hence strongly depend on detector R\&D efforts to suppress the effect of background particles. 


%


\subsubsection{Recent achievements}

A full simulation framework based on the FLUKA Monte Carlo code was developed for computing the beam-induced background entering the detector. First muon decay studies for 1.5\,TeV, replicating the MAP interaction region, exhibited a good agreement with the previous MAP results derived with MARS15 \cite{Collamati2021}. The studies were further extended to 3\,TeV and, in particular, to a higher collision energy of 10\,TeV \cite{Calzolari2022,Calzolari2023,Lucchesi2024}. At the same time various simulation techniques were improved, like the sampling of decays from a fully matched beam phase-space distribution, which provided a more accurate description of the transverse beam tails. While the 1.5\,TeV and 3\,TeV studies were still based on the interaction region lattice from MAP, a new lattice was developed for the 10\,TeV collider, considering a triplet configuration as final focus scheme \cite{Skoufaris2022}. The 10\,TeV lattice allows for $\beta^*$ values of a few millimeters and incorporates an adequate chromatic compensation without sacrificing the physical and dynamic aperture. Using the FLUKA simulation framework, a comparison of decay-induced background for 1.5\,TeV, 3\,TeV and 10\,TeV were performed \cite{Collamati2021,Calzolari2022,Calzolari2023,Lucchesi2024}. As one of the key findings, these studies demonstrated that the secondary spectra and multiplicities of particles entering the detector are not substantially different for a 10\,TeV collider. Figure~\ref{fig:mdibib} illustrates the distribution of arrival times and energy spectra of decay-induced background particles for 3\,TeV and 10\,TeV, respectively \cite{Lucchesi2024}. The results were obtained with the different IR lattices (MAP lattice for 3\,TeV and IMCC lattice for 10\,TeV), as described above. 

\begin{figure}[t]
   \begin{center}
        \includegraphics[width=0.48\textwidth]{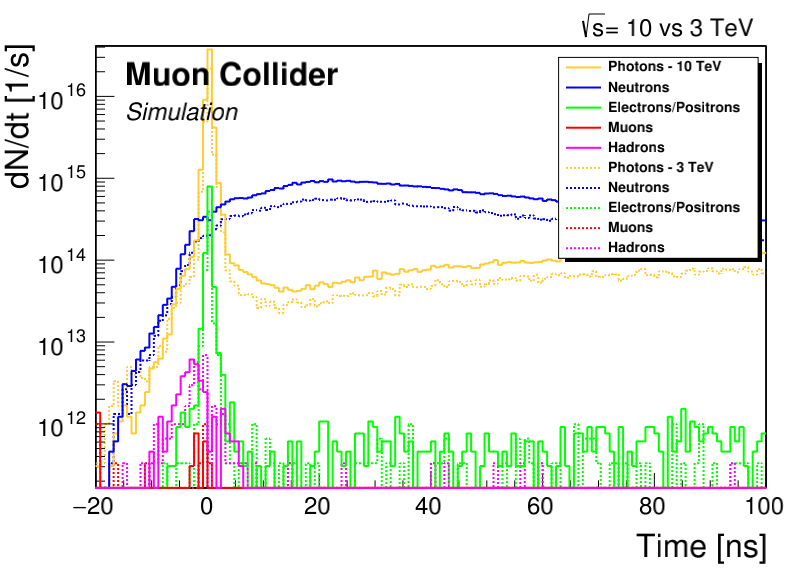}
        \includegraphics[width=0.48\textwidth]{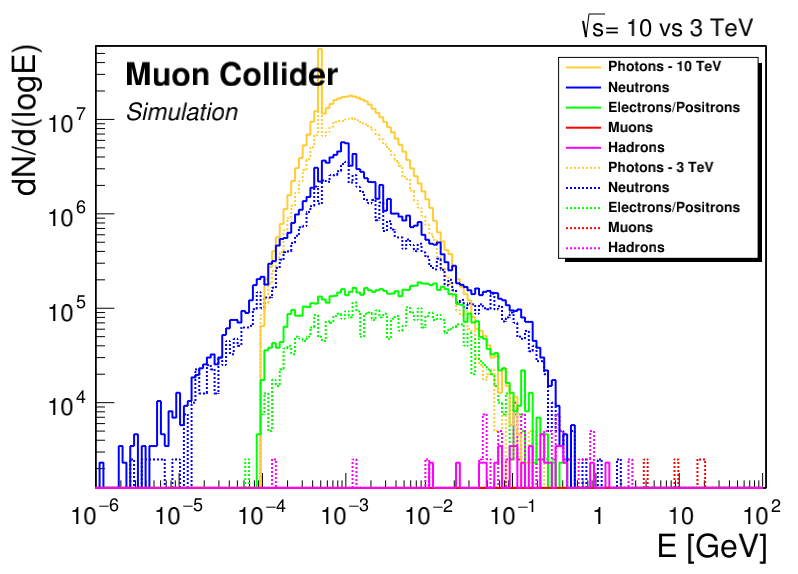}  
    \end{center}
 \caption{Left: Distribution of the arrival time of decay-induced background particles entering the detector. Right: Energy spectra of particles arriving within the time window $[-1:15]$\,ns with respect to the bunch crossing (photons, electrons and positrons below 100\,keV were discarded). The figures were taken from Ref.~\cite{Lucchesi2024}.}
 \label{fig:mdibib}
\end{figure}

In order to explore the impact of lattice design choices on the decay-induced background, a series of studies was performed for 10\,TeV \cite{Calzolari2022}. Three alternative triplet layouts were considered, one based on pure quadrupoles, one based on combined-function dipole-quadrupoles and one using short dipoles placed between triplet magnets. In the two latter cases, the dipolar component strongly alters the azimuthal distribution of e-/e+ impact positions on the vacuum aperture, yet the massive nozzle dilutes any azimuthal dependence of the particle flux into the detector. In general, the studies showed that the dipolar component suppresses only marginally the contribution of decays in the final focus region, but nevertheless is beneficial for reducing the contribution of distant decays. The latter can yield a non-negligible contribution if the final focus is followed by a long drift section. Such a drift might be needed to allow for a a smoother reduction of the beta function at the end of the final focus scheme. The studies showed that in this case, a dipolar component in the triplet can help to reduce the background.

In addition to the above studies, the impact of $L^*$ on the decay-induced background was explored for the 10\,TeV collider \cite{Calzolari2023}. The solenoid in the detector region traps the electrons and positrons, which travel in the beam vacuum until they reach the other side of the interaction region, depositing their energy in the machine components. As a consequence, muon decays between the final focus magnets ($s<L^*$) contribute several orders of magnitudes less to the background than those in final focus quadrupoles. A longer $L^*$ implies larger beta functions in the magnets, therefore increasing the magnet aperture. On the other hand, the quadrupole gradient decreases if the maximum field at the coil aperture is kept the same, thus implying a longer final focus scheme. A comparative study of two different lattices with $L^*=6$ and $10$\,m showed only a moderate background reduction by a few tens of percent, but at the expense of a more complex lattice design. Increasing $L^*$ is hence not considered a viable option for reducing the background. 

All the studies listed above were still based on the MAP nozzle design, which was optimized for 1.5\,TeV. First attempts to adapt the nozzle for 10\,TeV suggest that there is potential to reduce the decay-induced particle flux into the detector \cite{Calzolari2023}. For example, the simulations showed that particle flux into the inner tracker depends on the internal angle of the nozzle tip. The studies also indicated that changing the external shape of the nozzle and adapting the placement of the borated polyethylene layer can reduce the neutron and photon flux. Optimizing the nozzle for 3 and 10\,TeV remains one of the key tasks for future studies.

Preliminary assessments of background sources beyond muon decay were performed for 10\,TeV. In particular, a first comparison of incoherent electron-positron pairs against decay-induced spectra was carried out \cite{Calzolari2023}. Although some electrons and positrons from pair production will be trapped by the solenoid, a non-negligible fraction of particles with energies up to about 1\,GeV is expected to enter the inner tracker. The studies suggest that contribution of incoherent pairs corresponds to a few ten percent of the total number of electrons and positrons entering the detector in the vicinity of the IP (within $\pm$50\,cm). On average, the incoherent pairs have a higher energy than the decay-induced background component since the latter is strongly diluted by the nozzle.

Through the same simulation framework and by using a simplified model of a CLIC-like detector as well as the MAP nozzle (for the detector description, see also Sec.~\ref{sec:Section05}), the cumulative radiation damage in the detector could be calculated for different collider energies (so far for muon decay only). Two quantities were evaluated, the total ionizing dose and the 1~MeV neutron-equivalent fluence in Silicon. The former describes the ionization damage in organic materials and compounds, while the latter is related to the displacement damage. The highest dose values occur around the vertex detector, reaching about $\sim$200\,kGy/year (10\,TeV collider). The 1\,MeV neutron-equivalent fluence is highest in the inner tracker due to increased leakage of neutrons from the nozzle, with a peak value of around $10^{15}$~n/cm$^{2}$/year (10\,TeV collider). Table~\ref{tab:MDI_radiation_detector} provides a summary of the maximum values in different parts of the detector.

\begin{table}[t]
\begin{center}
\caption{Maximum values of the ionizing dose and the 1~MeV neutron-equivalent fluence (Si) in a CLIC-like detector. All values are per year of operation (10\,TeV) and include only the contribution of muon decay.}
\label{tab:MDI_radiation_detector}
\begin{tabular}{lcc}
\hline\hline
& \textbf{Dose} & \textbf{1~MeV neutron-equivalent fluence (Si)} \\
\hline
Vertex detector & 200\,kGy & 3$\times$10$^{14}$ n/cm$^2$ \\
Inner tracker & 10\,kGy & 1$\times$10$^{15}$ n/cm$^2$ \\
ECAL & 2\,kGy & 1$\times$10$^{14}$ n/cm$^2$ \\
\hline\hline
\end{tabular}
\end{center}
\end{table}



\subsubsection{Planned work}

One of the key tasks for the evaluation report is the optimization of the shape and material composition of the nozzle using the FLUKA simulation framework. The studies will be carried out separately for the 3~TeV and 10~TeV machines. The optimization will rely on simple key figures (e.g.\,hit densities in the vertex detector) in order to avoid the need of a full detector simulation for each design iteration. For certain reference configurations, a full evaluation of the background in dedicated detector simulations will be essential, to provide a feedback for the nozzle design. Where applicable, alternative optimization techniques (machine learning) will be explored. The nozzle optimisation will mainly focus on decay-induced beam losses, but the contribution of other background sources (incoherent pair production and beam halo losses) will be assessed through the same simulation framework. This requires input from other codes (GuineaPig) and machine studies (beam dynamics simulations). In this context, it is foreseen to improve the description of incoherent pair production by muons in GuineaPig.

Although a full-scale engineering design of the nozzle is out of the scope of the present studies, it is foreseen to address some of the most important  engineering and integration aspects. This includes in particular the evacuation of the deposited heat by means of a cooling system, the nozzle support structure, the nozzle segmentation and assembly, as well as the integration of the nozzle inside the detector and solenoid. Furthermore, the feasibility of instrumenting the nozzle will be evaluated in close collaboration with detector experts.

Another key task will be the optimization of the interaction region layout, under consideration of the beam-induced background, optics-related constraints, shielding requirement for magnets (heat load to cold mass, radiation damage in the coils), as well as magnet technology limitations (coil aperture versus peak field). The team is prioritizing this optimization for a 10\,TeV machine, but will apply lessons learned to a reoptimization of MAP’s 3 TeV interaction region if more resources become available. 
An important design specification is the minimum acceptable vacuum aperture in the interaction region (in terms of transverse beam sigma). A smaller aperture of the masks embedded in the final focus magnets and the nozzle is beneficial for reducing the decay-induced background, but can lead to enhanced halo losses in the interaction region. It is essential to quantify the permissible level of beam halo losses per bunch crossing, which must remain below the decay-induced background. 
Based on these studies, the interaction region aperture will be revisited and the need of other mitigation measures for halo losses (e.g.\,halo scraping system far from the IP) will be assessed.  

Complementing the background studies, updated estimates for the cumulative radiation damage in the detector will be provided for the optimized interaction region and nozzle layout. These estimates shall include all background contributions, including incoherent pairs and halo losses, which have been neglected so far in the radiation damage studies.


\subsubsection{Important missing effort}

Considering the present resources, the MDI and IR studies described in the previous sub-section will be given priority until the evaluation report planned for 2026.
Although the foreseen design studies will be mainly of a conceptual nature, they will prepare the path towards a fully integrated MDI and IR design. With the present resources, integrated engineering can only be performed for a very limited subset of the systems; however, a detailed design and integration of all technical systems can be elaborated once more resources become available. At that stage, important aspects, like the alignment of the nozzle and correction schemes for possible misalignments, can be addressed in detail. Furthermore, beam instrumentation like beam position or loss monitors, as well as luminosity monitoring, can be studied once resources are increased.

In addition to the MDI and IR design studies, it will be important to design a collimation or scraping system, which reduces the halo-induced background in the detector and also protects the interaction region in case of accidental beam losses. Multi-turn tracking studies will be needed to assess the efficiency of collimation techniques. In addition, it might be necessary to explore novel collimation schemes. The studies described in the previous sub-section will provide the necessary input concerning the acceptable amount of halo losses in the interaction region. It will therefore establish specifications for future design studies of a halo cleaning system, which can be addressed once more resources become available.

\subsection{Neutrino beams}
\label{sec:neutrino_interface}

The collider target parameters 
foresee a single bunch of $N_\pm=1.8\times10^{12}$ muons (and one of anti-muons) injected in the 10~TeV muon collider every $1/f_r=0.2~{\textrm{sec}}$. Assuming $10^7\,{\textrm{sec}}$ of operation per year, this corresponds to a current of $9\times10^{19}$ muons per year. The muons circulate until they decay, and those that decay in a straight section of the collider ring give rise to a collimated beam of neutrinos that could be used for physics measurements in a dedicated detector. The fraction of usable muons is the length $L$ of the straight section divided by the collider circumference, of 10~km, and a region of at least $L=10$~m without fields is required for the installation of the detector close to the Interaction Point (IP). This gives $9\times10^{9}$ neutrinos of each species per second and $9\times10^{16}$ per year at the 10~TeV collider. At the 3~TeV collider, a similar estimate gives $2.4\times10^{10(17)}$ neutrinos per second (year). Notice that our estimate only accounts for the muons that decay in the detector region. The total length of the straight section is about 100~m in the current versions of the collider lattice design, and most of the muons decaying in this longer region should produce usable neutrinos. At the real collider we could thus expect few or 10 times more neutrinos than our estimate.

The neutrinos are very collinear to the decaying muons, with an angle that is typically below $0.1~{\textrm{mrad}}$ already at the 3~TeV MuC, and even smaller at 10~TeV because of the larger muon energy (see e.g.\,Ref.~\cite{PhysDect_King:1997dx}). The muon beam angular divergence is typically much above $0.1~{\textrm{mrad}}$, hence the distribution of the neutrino angle relative to the beam axis is driven by the dynamics of the muon beam and it cannot be estimated without reference to the lattice design, which is still preliminary. Fortunately, the beam angular divergence in the detector region equal to $0.6~{\textrm{mrad}}$, both for the 3 and 10~TeV colliders is a rather robust design parameter because it is directly related to the luminosity at fixed emittance. Our estimate of the neutrino angle (and energy) distribution thus assumes a constant $0.6~{\textrm{mrad}}$ angular spread\footnote{Specifically, we smear $p_{x,y}/p_z$, with $p$ the muon momentum, by independent Gaussians with $\sigma=6\times10^{-4}$.} for the decaying muons, and nominal beam energy of $1.5$ and 5~TeV for the 3 and 10~TeV muon collider, respectively. This simplified simulation of the neutrino angle and energy distribution has been found in qualitative agreement with the complete simulation based on a preliminary design of the lattice.

The relevant parameter for physics studies is the rate of neutrino interactions, which in turn depends on the geometric acceptance and on the mass per unit area of the neutrino target. There are good perspectives to attain order-one geometric acceptance: around 30\% of the neutrinos would intercept a small 10~cm radius cylindrical target placed at a realistically large (200~m) distance from the IP. The target mass is harder to estimate, because it is intertwined with the capabilities of the detector to record the products of the neutrino interactions. Two options are considered in Section~\ref{subse:nu}, and the resulting interaction rates reported in Fig.~\ref{fig:nfl}.
%

Despite the promising potential discussed in Section~\ref{subse:nu}, no resources are currently available within IMCC for the design of dedicated neutrino detectors. We instead plan to refine the estimation of the neutrino flux from updated accelerator lattice designs.

\begin{flushleft}
\interlinepenalty=10000

\end{flushleft}

\clearpage
\section{Detector 
}
\label{sec:Section05}


\subsection{Concepts 
}
\label{sec:Detector_concepts}
\subsubsection{System overview}
    The detector concept for a $\sqrt{s}=3$\,TeV muon collider is based on the detector design proposed by CLIC~\cite{clic}, modified to adapt it to the machine-detector interface. The forward calorimeters (luminometer) have been replaced with the shielding nozzles designed by MAP.
    This structure is constituted by two double-cone made of tungsten and borated polyethylene having an opening angle of $10^{\circ}$, which are necessary to mitigate the impact of the beam-induced background. The nozzles are placed along the beam axis, in the region between 6 and 600\,cm away from the interaction point.
    As a consequence of that, the inner openings in the endcap region of the tracking detector, calorimeter and muon stations have been increased to host the nozzles. In addition to that, the spatial configuration of the vertex detector has been optimised to minimize the occupancy in the region of the tip of the nozzles. Finally, the magnetic field has been set to 3.57 T for consistency with the magnetic field assumed in the MARS15 simulation of the BIB particles.
    The detector is structured as follow.
    The innermost system closest to the beam-pipe is a full-silicon tracking system, that includes a vertex detector made of silicon pixels with double layers, and inner and outer trackers respectively composed of silicon macropixels and microstrips. The tracking system is surrounded by a calorimeter system consisting of an electromagnetic (ECAL) and a hadronic (HCAL) calorimeter, the former composed of alternating layers of tungsten absorber and silicon sensors, and the latter having alternating layers of steel absorber and scintillating pads.
    A solenoid with an inner bore of 3.5\,m radius, provides a magnetic field of 3.57\,T, whose flux is returned by a magnet iron yoke instrumented with muon chambers (resistive plate chambers). A scheme of the full detector is shown in Fig.~\ref{fig:detector}. 
    \begin{figure}[!ht]
        \centering
        \includegraphics[width=0.6\textwidth]{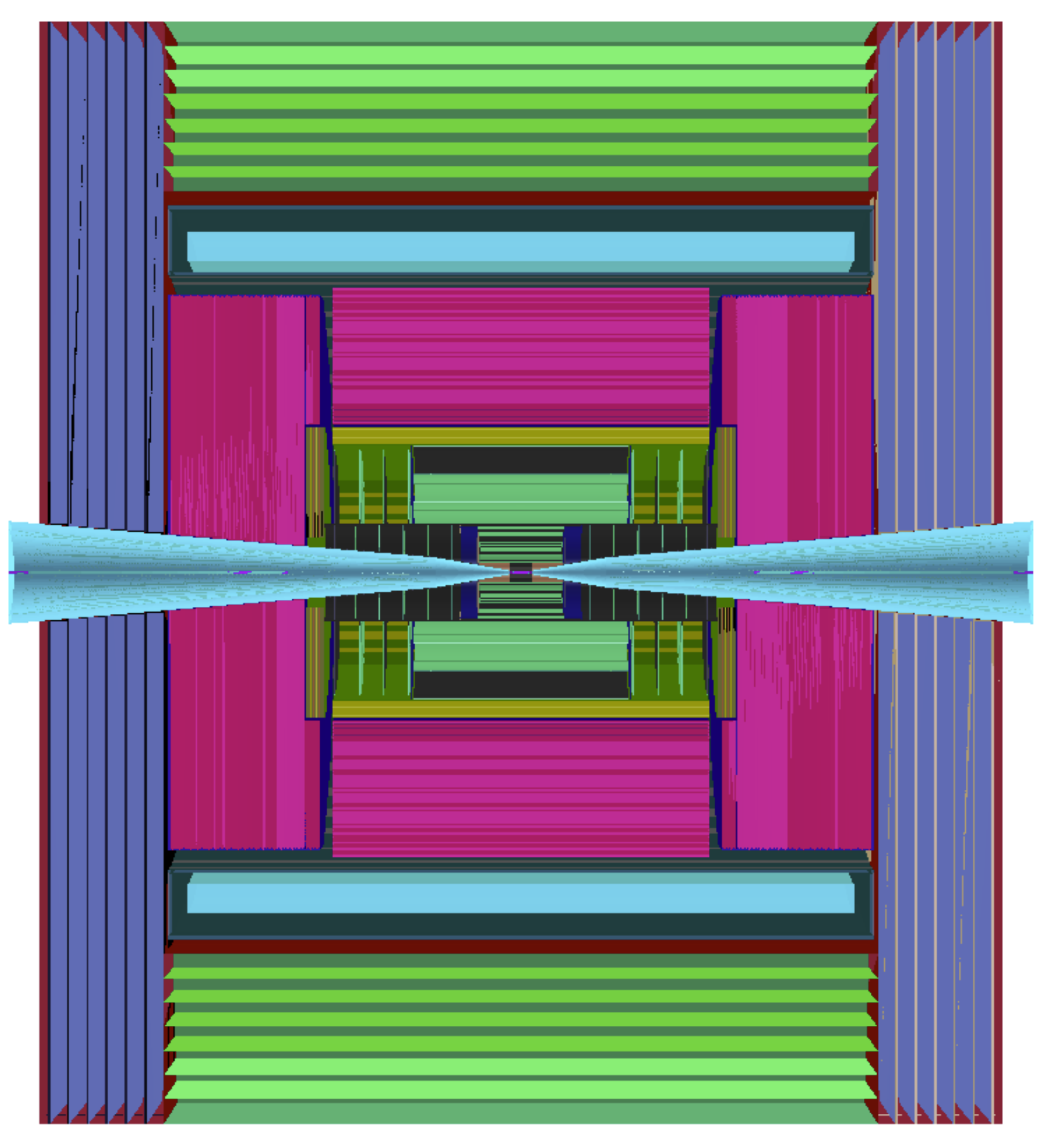}
        \caption{Muon collider detector concept. From the innermost to the outermost regions, it includes the nozzles (cyan), the tracking system (light green), the electromagnetic calorimeter (yellow green), the hadronic calorimeter (red), the superconducting solenoid (light blue), and muon detectors (light green and blue).}
        \label{fig:detector}
    \end{figure}

\subsubsection{Key challenges}
As for the accelerator, the main technological challenges for the detector arise from the short lifetime of muons. In fact the beam-induced background produced by the decay in flight of muons in the beams and subsequent interactions potentially poses a serious limitation to detector operations.
All kinds of particles are produced, like photons, electrons, and neutrons, that can be partially mitigated by a proper design of the machine-detector-interface, as described in Section~\ref{sec:Section04_2}.
Nevertheless, of the order of $10^8$ particles enter the detector at this center of mass (CoM) energy, therefore the detector design, the technology choices, and the reconstruction strategies must primarily take into account the beam-induced-background impact.

It is important to study the features of the beam-induced background by using simulation, to take the proper directions for the detector development. Figure~\ref{fig:bib_detector} shows two key distributions, the arrival time with respect to the bunch crossing time currently set to zero,  and the energy of the beam-induced background particles when they enter the detector. These distributions have been obtained by using MARS15 simulation at $\sqrt{s}=$1.5\,TeV~\cite{mars15}.
It can be seen that a large part of beam-induced background particles are asynchronous with respect to the bunch crossing, and they usually have low energy. Therefore timing measurements and appropriate energy thresholds are important handles to suppress them at the detector level. 

\begin{figure}[!ht]
        \centering
        \includegraphics[width=0.9\textwidth]{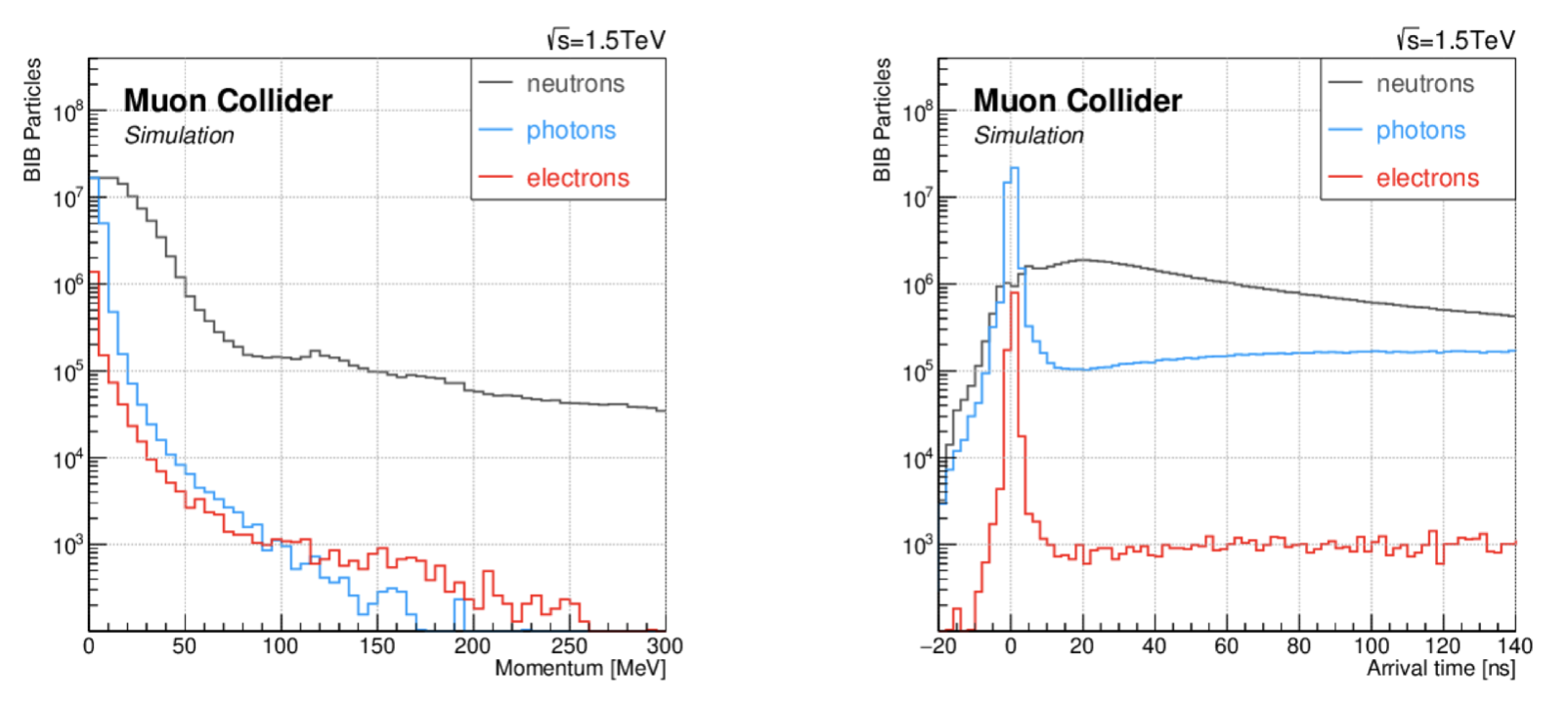}
        \caption{Energy and arrival time, with respect to the bunch crossing time, of the beam-induced background particles when they enter the detector.}
        \label{fig:bib_detector}
    \end{figure}

The following general requirements have been identified for the detector, driven by the necessity of suppressing the beam-induced background and keeping the signal efficiencies as high as possible:
\begin{itemize}
\item time measurements with excellent resolution: at this CoM energy resolutions ranging from 30 to 60\,ps for the tracking system, and around 100\,ps for the calorimeters are evaluated to be appropriate with the detector technologies available at the time of HL-LHC; the time resolution due to the primary vertex uncertainty has not been studied yet, but a resolution below 10\, ps can be easily assumed by considering the beam size in $z$;
\item energy measurements, to apply a proper threshold for rejecting the beam-induced background soft component;
\item high granularity, to reduce the overlaps between beam-induced background and signal hits, and consequently reducing the occupancy;
\item radiation hard devices, since the radiation level is similar to what is expected at the High-Luminosity LHC (HL-LHC)~\cite{muc_forum_report}.
\end{itemize}
The research and development for these technologies will be performed in synergy with the existing programs, from HL-LHC to other future colliders.

The muon collider detector should ensure the physics reaches from low energy (\emph{e.g.}~Higgs and standard model physics) to the highest achievable energy scale for new physics searches (Section~\ref{sec:Section04_1}).
This is particularly challenging in the detector for $\sqrt{s}=10$\,TeV collisions and beyond, where particle momenta from few GeV to several TeV have to be measured. 
The detector layout and design should satisfy physics requirements, and at the same time, it has to accommodate the MDI requirements and geometry. 
In this context, the  magnetic field value plays an important role. High magnetic fields, around  5\,T, could help in improving tracking momentum resolution of high energy particles but it could generate inefficiencies for those with low momentum. In addition, there are significant challenges to achieve the mechanical and electrical stability of the superconductive solenoids needed to reach such values of the field, including the technology transfer from previous HEP projects and the availability of experts.

Another challenge is the simulation of the detector for reconstruction and performance studies. The propagation of millions of beam-induced background particles through the detector volume, done with software based on \textsc{Geant4}~\cite{geant}, is heavily CPU-time consuming. Dedicated software and algorithms that tackle this problem in the proper way have to be developed since the muon collider environment is different from that of other well-known collider machines. Adequate computing resources should also be allocated for the detector studies.
\subsubsection{Work progress since the publication of the European LDG roadmap}
An important progress since the the publication of the European LDG roadmap is the initiation of the studies for the design of the detector at $\sqrt{s}=$10\,TeV.
To properly understand the requirements, three major macro-categories of physics processes for muon collisions at  $\sqrt{s}=$10\,TeV have been identified: 
\begin{itemize}
    \item low energy: typical examples are electroweak (EW) and Higgs boson production, with energy in the range of a few hundreds of GeV; a sample of Higgs bosons decaying to $b \bar b$ jets is used for the studies;
    \item high energy: New physics (NP) processes can create heavy resonances at the order of TeVs; simulated Z$^{\prime}$ boson with a mass of 9.5~TeV is taken as a representative case with physics objects with energies up to  $\approx 5$ TeV;
    \item unconventional signatures such as long-lived particles, disappearing tracks and emerging jets are just a few examples.
\end{itemize}
The transverse momenta, $p_{\mathrm{T}}$, of the decay products of Higgs  and Z$^{\prime}$ bosons are shown in Fig.~\ref{fig:higgszprime10} to illustrate the difference among them.
\begin{figure}[!ht]
    \centering
    \includegraphics[width=0.45\textwidth]{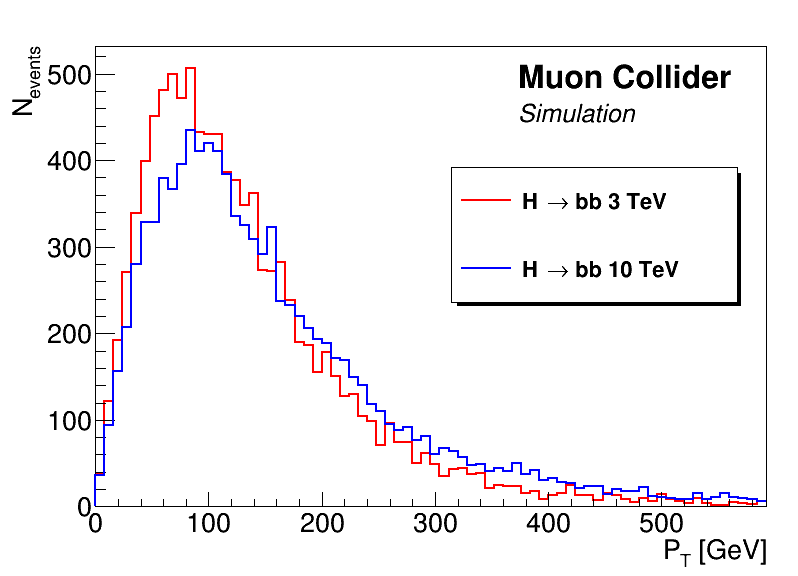}
    \includegraphics[width=0.45\textwidth]{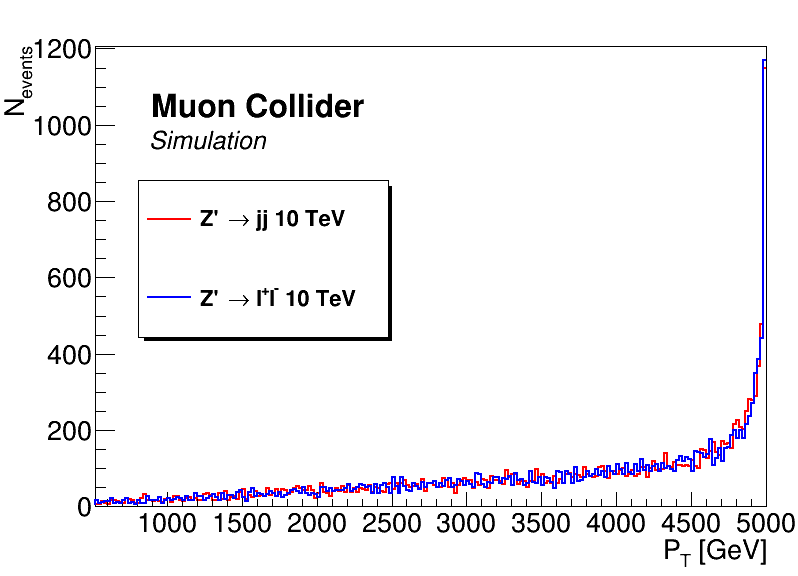}
    \caption{Left: $p_{\mathrm{T}}$ distributions of $b$-quark coming from the $H \rightarrow b \bar{b}$ decays at $\sqrt{s} = 3$ and 10\,TeV, as obtained at generator level. Right: distribution of leptons and jets $p_{\mathrm{T}}$ coming from the decay of a Z$^{\prime}$ boson produced at $\sqrt{s} =10$\,TeV with a mass of 9.5\,TeV, as obtained at generator level. }
    \label{fig:higgszprime10}
\end{figure}
    
The requirements set on the detector by the selected physics benchmarks at $\sqrt{s}=$10\,TeV are determined by optimizing the different sub-detectors.
For the tracking system, a parametric study has been done looking at the relation between the depth of the tracking system and the magnetic field. Figure~\ref{fig:10tev} (left) shows the results of momentum resolution for tracks with a transverse momentum $p_{\mathrm{T}}\sim 5000$\,GeV, as a function of the maximum tracker radius for different magnetic field values where it is evident that a deeper tracker with an high magnetic field value is necessary to keep a good track reconstruction resolution. A preliminary discussion with  experts on detector magnets has been organized with a dedicated workshop~\cite{magnets_workshop}, that set the basis for the studies.
    
Studies with photon/protons/pions particle guns have been performed to determine the calorimeter characteristics.  Deeper calorimeters to contain high energy particles showers are necessary as shown in Fig.~\ref{fig:10tev} (right), where it is visible that a fraction of photon energy (with energies greater than 1\,TeV) leaks into HCAL. For the muon reconstruction, studies on muons with $p_{\mathrm{T}}\sim5000$~GeV have shown that more inclusive algorithms that also use calorimeters information should be used in order to collect the energy released by muons.
    
\begin{figure}[!ht]
    \centering
    \includegraphics[width=0.40\textwidth]{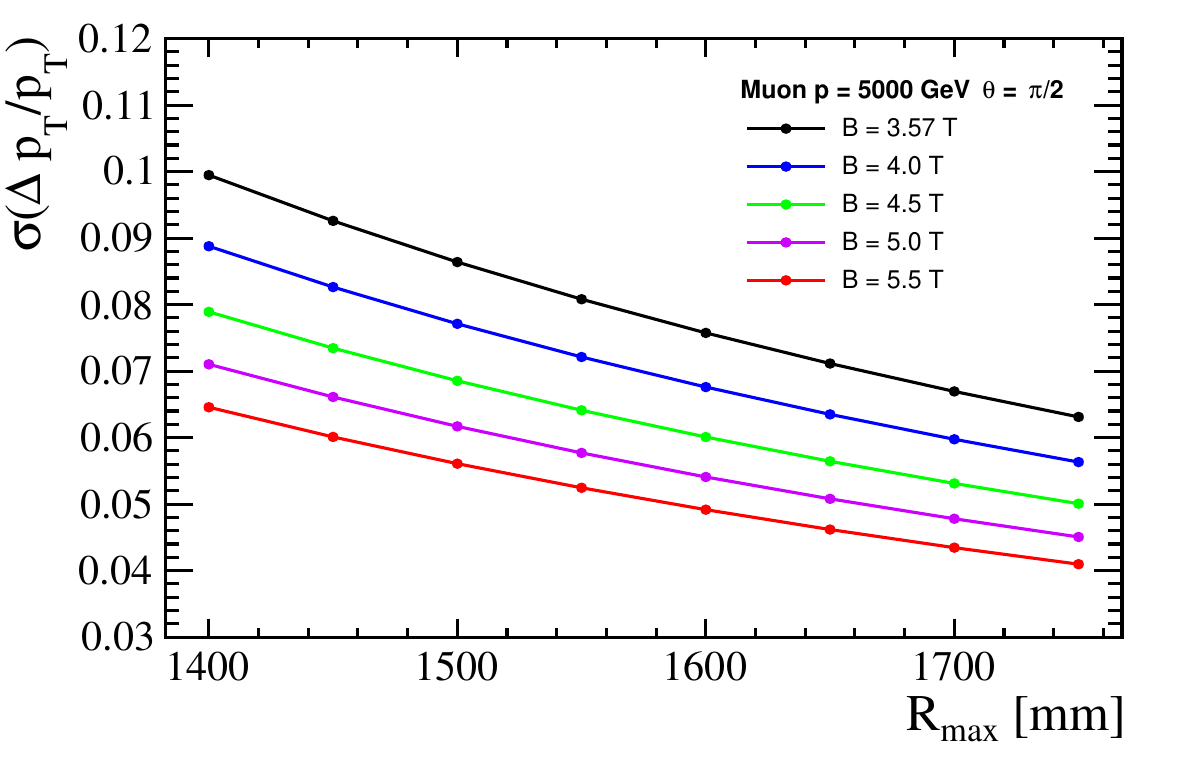}
    \includegraphics[width=0.45\textwidth]{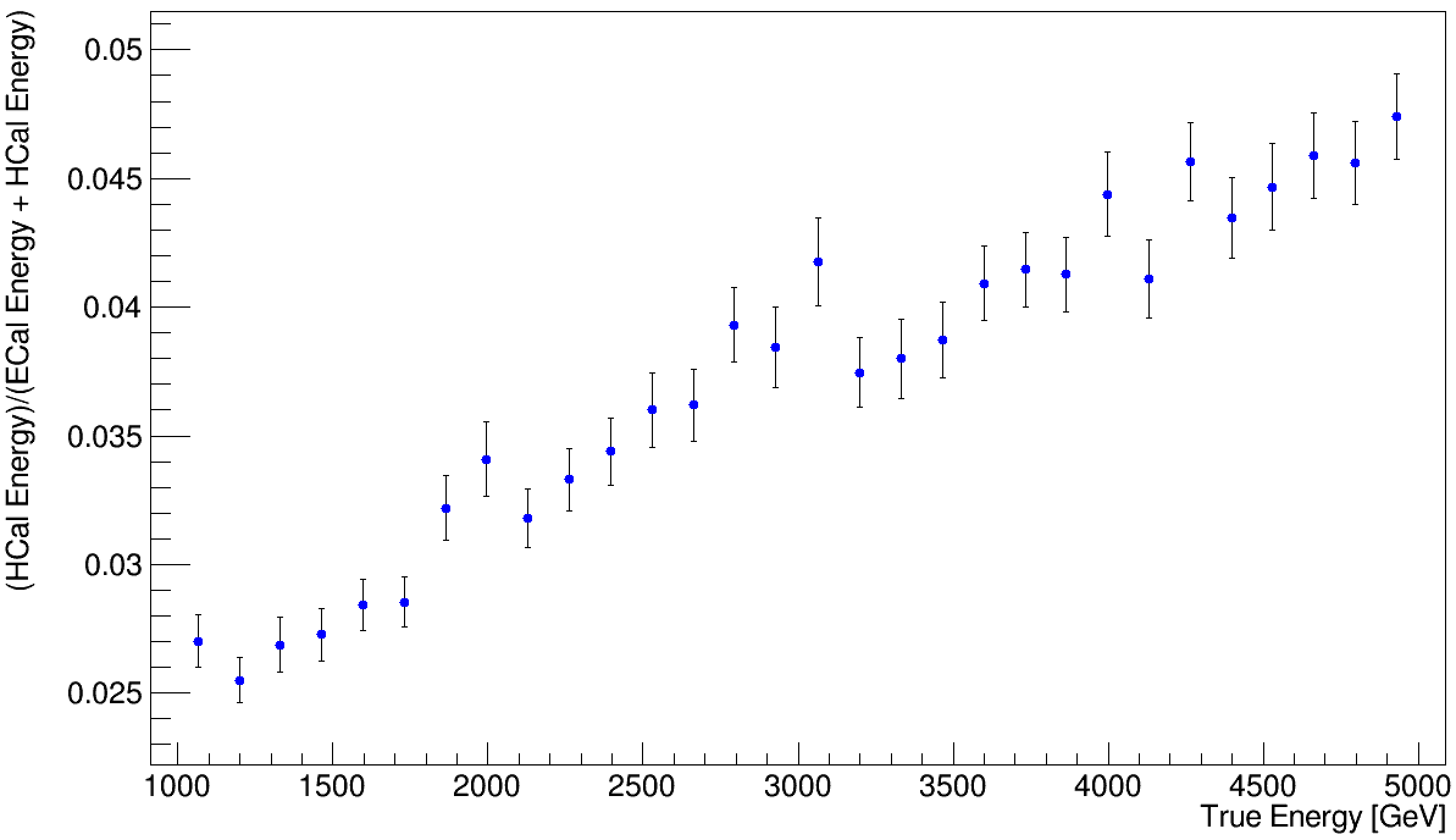}
    \caption{Left: track momentum resolution for muons with $p=5$\,TeV and $\theta = \pi/2$, as a function of the tracker radius ($R_{max}$) and for different values of magnetic field ($B$). Right: fraction of photon energy measured in HCAL with respect to the total energy measured in ECAL+HCAL, as a function of the true photon energy.}
    \label{fig:10tev}
\end{figure}
    
\subsubsection{Work planned for next evaluation report}
The magnet layout and magnetic field intensity will be one of the first items to be tackled for the next activities. This choice will drive the design of the detector sub-systems. The tracking, calorimeter, and muon detector system geometries will be defined, considering the available space (limited by the magnet, MDI, and cavern dimensions). 

The assessment of the beam-induced background effects on the detector at $\sqrt{s}=10$\,TeV is another important activity for the upcoming Evaluation Report.
It will include the study of incoherent electron-positron production, a physics process that could become relevant at high center-of-mass energies, such as 10\,TeV.
Once the simulation samples of the beam-induced background described in Section~\ref{sec:Section04_2} are validated, the studies at the detector level will start.
First, the interactions of the beam-induced background with the detector have to be simulated; a preliminary detector design will be used, considering different configurations and values of the magnetic field.
Several important physics quantities, like detector occupancy and distributions of hit position, time, and energy will be studied, including the determination of the radiation levels.
The detector requirements will be reviewed, to verify if important changes for the layout and the technology choices are necessary, with respect to the $\sqrt{s}=3$\,TeV concept.
The reconstruction performance of the main physics objects will be studied. Initially, using the same algorithms reported in Section~\ref{sec:Detector_performance}, and subsequently, with algorithms tuned specifically for the $\sqrt{s}=10$\,TeV case.
Feedback will be provided to the MDI study group to assess the possibility of improving the detector acceptance (\emph{e.g.}, reducing the nozzle angle) while maintaining excellent reconstruction performance.
The technology choice will be made by considering physics requirements and costs.

A first detector concept description at $\sqrt{s}=10$\,TeV will be delivered and made available in one of the next releases of the muon collider simulation framework and will serve as the starting point for all future studies.
The simulation of the beam-induced background in this detector will be provided, along with a version of the reconstruction algorithms with an initial tune specifically tailored for this concept.

\subsubsection{Next priority activities and collaboration opportunities}

Among the future activities, it will be important to study how to reconstruct and identify very forward muons, with angle below $10^o$ with respect to the beam pipe. These muons are emitted outside the detector acceptance, mainly in the nozzles. Their identification is important in several processes. For example,  muons from $Z^0Z^0$ fusion processes are mainly emitted in these regions, therefore their detection could be employed to tag such events and distinguish them from the $W^+W^-$ fusion.
These muon detectors have to be placed along the beam line, where also the accelerator magnets are placed. The propagation of the very forward muons through the magnetic fields configuration due to the  accelerator components is not trivial, and a dedicated study with FLUKA has to be performed.

\subsection{Performance 
}
\label{sec:Detector_performance}

This section presents an overview of the performance of the 3-TeV detector concept for the reconstruction of the primary physics objects in the presence of beam-induced background.

\subsubsection{Overview of the physics object reconstruction}
\label{sec:objectsReco}
The high levels of beam-induced background in the detector pose unprecedented challenges for the reconstruction
and identification of the particles produced in muon collisions, which are referred to as physics objects.
To fully exploit the physics potential of a muon collider and accomplish its physics program, it
is essential to reconstruct all the physics objects with high efficiency and determine their properties
with the greatest precision, even in the presence of beam-induced background.

A campaign of studies, based on a detailed detector simulation, was carried out to assess the
effects of the background on the detector response and possibly develop the appropriate mitigation measures. The outcome was that the reconstruction algorithms
for all the considered physics objects required revision or fine-tuning.
However, the results indicate that the background effects on the detector response can be minimized
to a degree that does not compromise the detector performance.
This performance was then estimated reconstructing a set of benchmark Higgs boson channels at a 3\,TeV muon collider.


The beam-induced background causes very high hit multiplicities in the detector's tracking system, making the track finding 
process highly challenging, since the number of hit combinations to be considered increases exponentially,
and a uniform diffuse energy deposition in the calorimeters, which makes it difficult to identify and 
accurately measure the energy of particles from collisions.

In general, the physics object reconstruction involves combining information from different detector sub-systems.
Tracks, calorimeter clusters, and muon detector hits are combined to achieve an optimal performance
in terms of identification efficiency, background rejection, and momentum resolution.
For this purpose, a particle flow algorithm is employed, PandoraPFA~\cite{ref:thomson}, which takes
as an input the reconstructed tracks and the hits of the calorimeters and muon detectors.

\begin{figure}[h]
  \centering
  \includegraphics[width=0.45\textwidth]{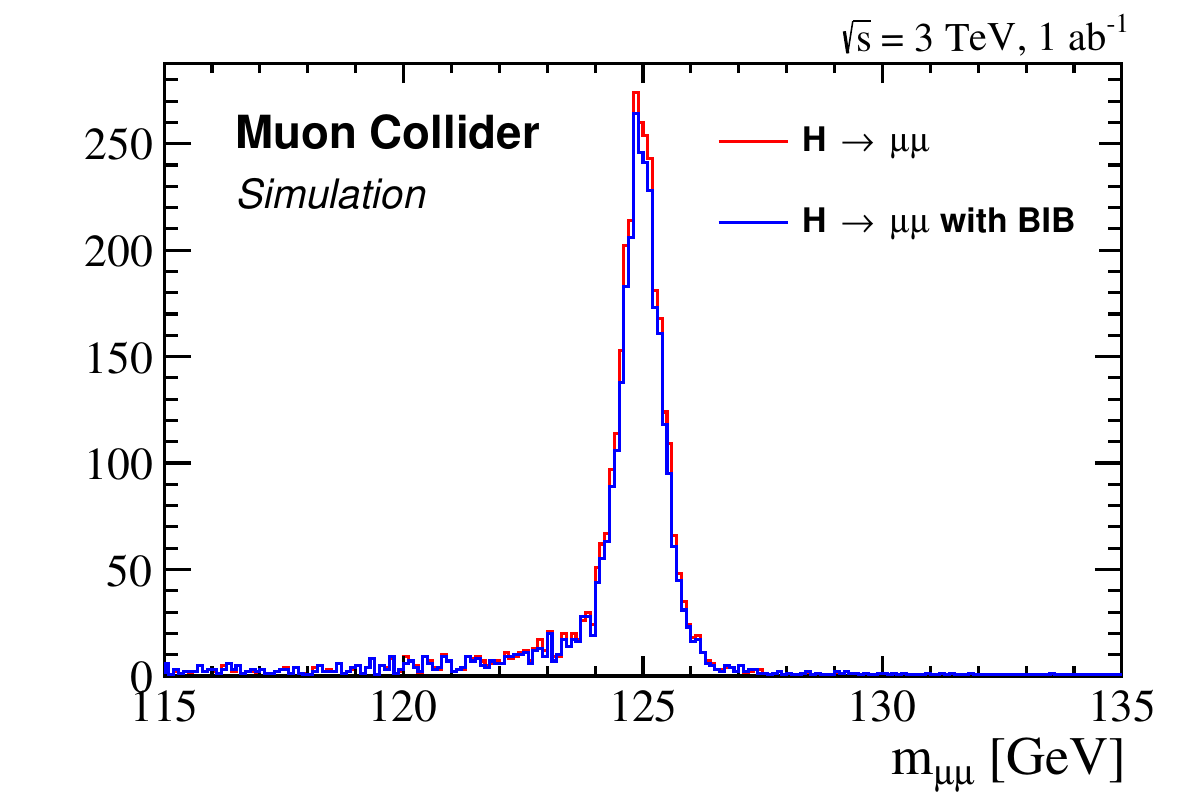}
  \includegraphics[width=0.45\textwidth]{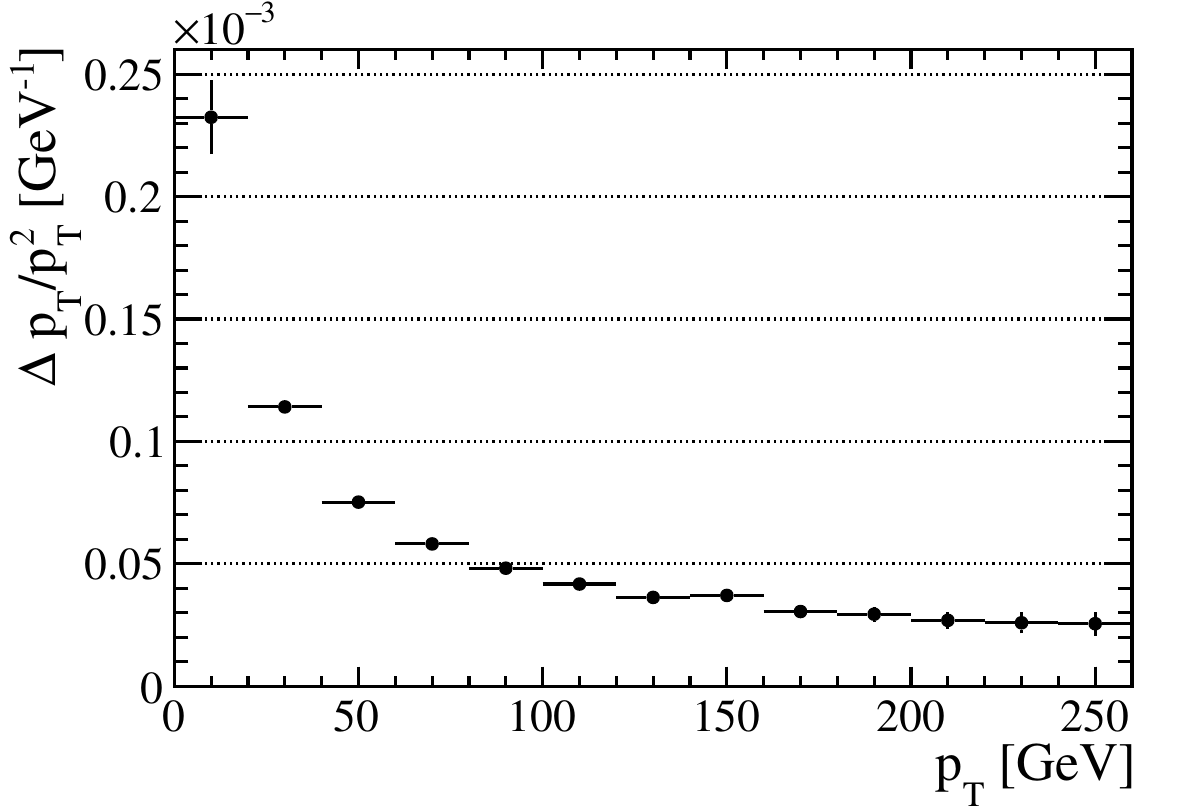}
  \caption{Left: Invariant mass distribution of the muon pairs produced in the
    $\mu^+\mu^-\to H\nu_\mu\bar{\nu}_\mu\to\mu^+\mu^-\nu_\mu\bar{\nu}_\mu$ process without
    the beam-induced background (red line) and with the beam-induced background overlaid to
    the physics events (blue line). Right: Transverse momentum resolution as a function of transverse momentum for the reconstructed muons entering in the plot on the left with the beam-induced background (BIB) overlaid.
  \label{fig:hmumu}}
\end{figure}
The initial focus of the detailed simulation studies was on muons, electrons, photons, and jets.
Muons are identified by matching tracks with hits in the muon detector system. This sub-detector, located far from the beamline, is not significantly affected by the beam-induced background, except for the forward and backward regions with respect to the beam direction.
Electrons and photons are reconstructed by matching tracks with clusters in the electromagnetic calorimeter.
Isolated clusters are classified as photons, while clusters matched with a track are classified as electrons.
The energy of electrons and photons is corrected to take into account inefficiencies and radiation losses.
The reconstruction efficiency of the electromagnetic objects as a function of the energy reaches around 95$\%$ for high energy but drops at low energies due to the beam-induced background.
For the jet reconstruction, tracks and calorimeter hits are filtered to remove the beam-induced background contamination and then are given as input to the PandoraPFA algorithm.
The particles reconstructed by PandoraPFA are used as input to the jet clustering algorithm, which aims to group together particles produced in the fragmentation process of the same quark or gluon, by exploiting their correlations. 
The resulting jet energy is corrected to recover the losses due to particles escaping the detector, detector inefficiencies, and also for beam-induced background contamination. The jet reconstruction algorithm has been tested with simulated samples of different jet flavours: gluons and $u$, $d$, $s$, $c$, and $b$ quarks. Despite the presence of the beam-induced background throughout the detector, the efficiency ranges between 80$\%$ and 95$\%$, with a negligible fake jet probability.
In order to identify the jets originating from heavy quarks, secondary vertices compatible with the decays of $b$- and $c$-hadrons are reconstructed by combining tracks. A jet is identified as a heavy-flavour jet if one of these secondary vertices is reconstructed within the jet cone. The $b$-jet identification probability is approximately 45$\%$ at low $p_{\mathrm{T}}$ (20\,GeV) and increases to 70$\%$ at 120\,GeV, while keeping a misidentification rate of about $20\%$ for $c$-jets and from $1\%$ to $5\%$ for light jets.
This level of performance is similar to what is currently achieved at hadron colliders.
At lepton colliders, a higher efficiency can be expected. The current algorithm is mainly limited by the presence of the beam-induced background. In the future, artificial intelligence-based methods will be implemented to minimize the effects of BIB and further enhance efficiency.

The physics objects described above were used to reconstruct the Higgs boson decay modes into the final states $b\bar{b}$, $\gamma\gamma$, $\mu^+\mu^-$ and derive the reconstruction performance in the momentum range defined by these processes.
The invariant mass depends on the magnitude and the direction of the decay particle momenta.
The peak position at the nominal mass of the Higgs boson indicates a well-calibrated energy of the objects, 
while the peak width reflects the accuracy with which the object properties are measured.
Figure~\ref{fig:hmumu} (left) shows the distribution of Higgs boson reconstructed mass for $\mu^+\mu^-\to H\nu_\mu\bar{\nu}_\mu$ with $H$ decaying into $\mu^+\mu^-$. The reconstruction of high-energy muons produced in this decay, namely above 10 GeV (see Fig.~\ref{fig:hmumu} on the right), is not significantly affected by the beam-induced background, as shown in the figure. The width of the mass peak is 0.4\,GeV, comparable to electron-positron colliders. Figure~\ref{fig:hmumu} (right) demonstrates that relatively high-energy muons are reconstructed with high momentum resolution.
In Fig.~\ref{fig:hgammagamma}, the invariant mass of the $H\to \gamma\gamma$ for the $\mu^+\mu^-\to H\nu_\mu\bar{\nu}_\mu$ process is plotted. As for the muons, high-energy photon reconstruction is not heavily impacted by the beam-induced background. In fact, the di-photon mass distribution has a peak with a width of 3.2\,GeV, with a small tail at higher masses that does not have an impact on the $H \to \gamma \gamma$ measurements. These events are then used to evaluate the energy resolution achieved with the current calorimeter. This range of energy is dominated by the constant term, which is about $3\%$.

\begin{figure}[h]
  \centering
  \includegraphics[width=0.6\textwidth]{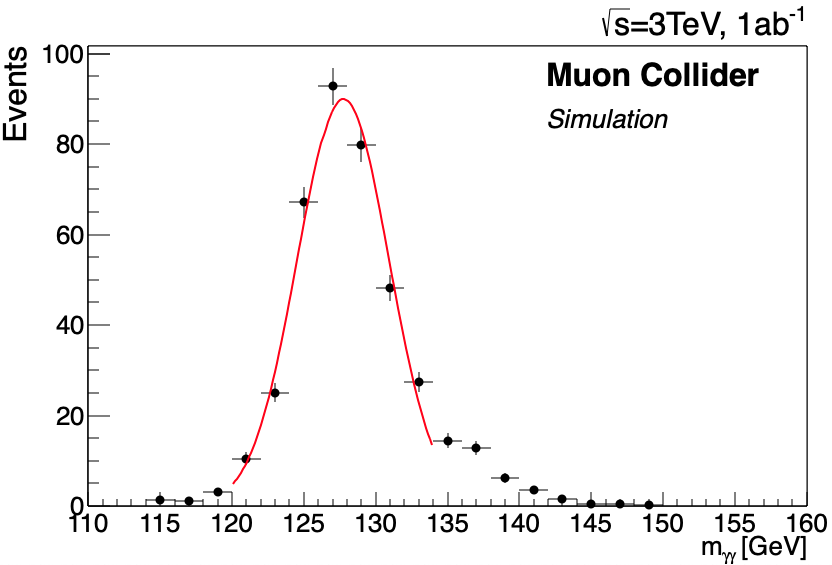}
  \caption{Invariant mass distribution of the photon pairs produced in the
    $\mu^+\mu^-\to H\nu_\mu\bar{\nu}_\mu\to\gamma\gamma\nu_\mu\bar{\nu}_\mu$ process with the beam-induced background overlaid to
    the physics events.
  \label{fig:hgammagamma}}
\end{figure}

The case is different for hadronic jets where an important contribution is given by  low-energy (from 500~MeV to GeV) objects.
The reconstruction performance of low-momentum, around GeV, tracks and low-energy calorimeter clusters is impacted by the beam-induced background.
The di-jet mass resolution is critical, for example, to separate the $H$ and $Z$ peaks as shown in Fig.~\ref{fig:hbb}(left). The current value of the Higgs boson mass resolution is about $18\%$, being dominated by beam-induced background effects, \textit{i.e.} the energy thresholds set on the calorimeter hits. The di-jet mass resolution is sufficient for determining the Higgs and Z bosons yields on a statistical basis, but in the future improving the jet energy resolution will be fundamental for several measurements. This can be achieved by optimizing the filters for removing the beam-induced background contamination, and subsequently by lowering the calorimeter energy thresholds. Fig.~\ref{fig:hbb} (right)~\cite{higgspaper} shows the invariant mass distribution of two jets originating from the leading Higgs boson in a $HH$ samples, compared with that of $q_h \bar{q}_h $ in a $\mu^+\mu^-\to q_h \bar{q}_h q_h \bar{q}_h$  events sample without beam-induced background. It's important to emphasize  that the degraded energy resolution when including beam-induced background is due to the limited performance of the reconstruction algorithm which is currently being improved.

\begin{figure}[h]
  \centering
  \includegraphics[width=0.45\textwidth]{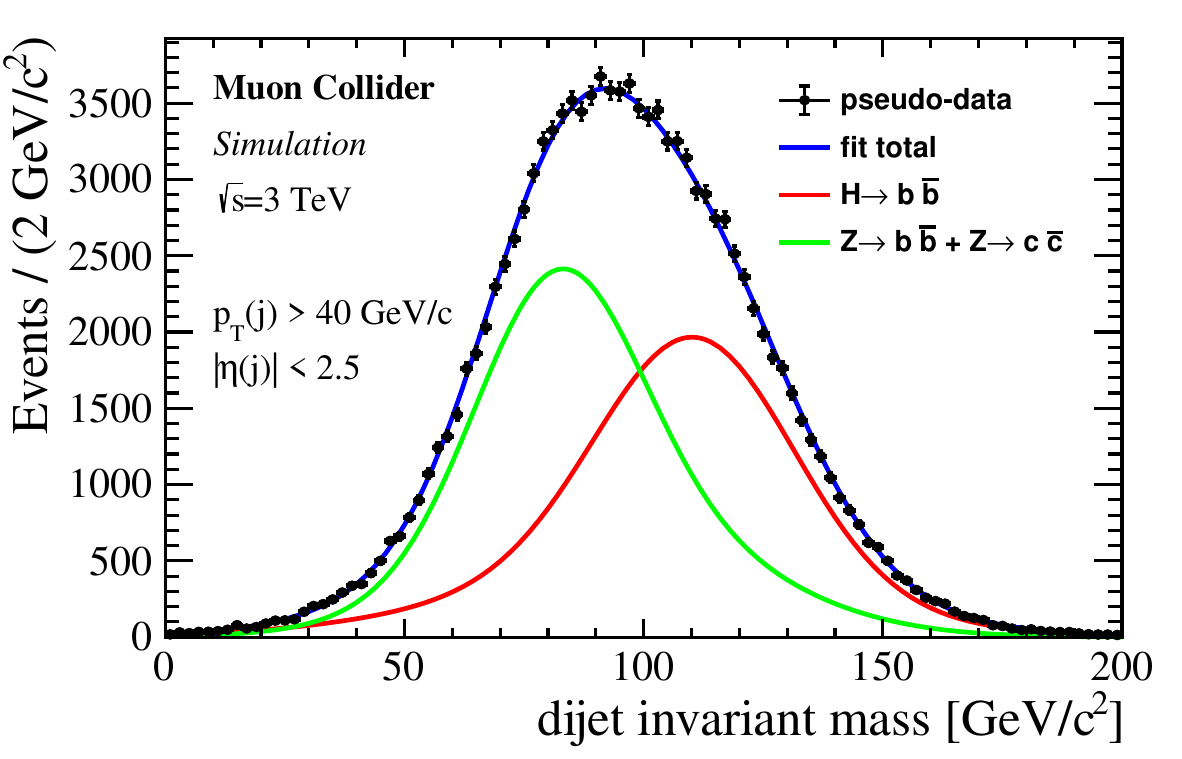}
  \includegraphics[width=0.45\textwidth]{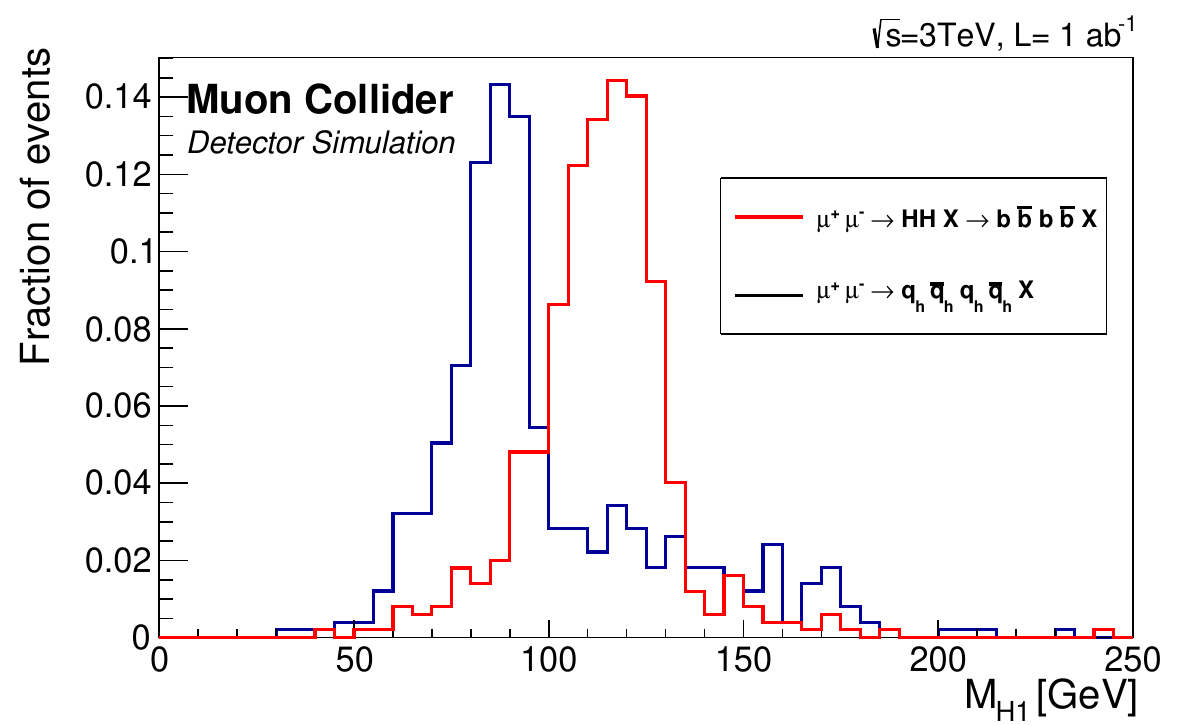}
  \caption{Left: Invariant mass distribution of the jet pairs produced in the
    $\mu^+\mu^-\to H\nu_\mu\bar{\nu}_\mu\to b\bar{b}\nu_\mu\bar{\nu}_\mu$ process
    with the beam-induced background overlaid to the physics events:
    the reconstructed mass peak of the Higgs boson is indicated in red, while the background from $Z\to b\bar{b}$ and $Z\to c\bar{c}$ is in green. Right: Di-jets invariant mass of the leading Higgs candidate in the sample of $HH$ events compared to that one of $q_h \bar{q}_h q_h \bar{q}_h$ events, in samples reconstructed without the BIB.
  \label{fig:hbb}}
\end{figure}

A review of the Higgs precision measurement at a muon collier at $\sqrt{s}=3$ TeV can be found ~\cite{higgspaper}. The analysis of the different decay final states allowed to determine the detector performance in an energy range of few GeV up to hundred of TeV, summarized in Table~\ref{tab:higgsreq}. This performance can be considered the requirements to be competitive on Higgs physics.
\begin{table}[h!]
    \caption{Detector and reconstruction requirements for Higgs physics at the muon collider, determined with the detailed simulation studies. The values have been obtained for objects with $p_T$ of about 100 GeV, that is the average transverse momentum of the particles from the 2-body decay of the Higgs boson with $\sqrt{s}=3$ TeV. The first set of requirements is intended for single Higgs boson measurements, while more stringent requirements are presented for the trilinear coupling determination with the four-$b$-jet final state. Taken from Ref.~\cite{higgspaper}.}
    \centering
    \begin{tabular}{c|c}
    \hline
         \textbf{Object} & Requirements \\
    \hline
    \hline
       muons &  $\frac{\Delta p_T}{p_T} = 0.4 \%$\\
      \hline
       photons & $\frac{\Delta E}{E} = 3\%$ \\
      \hline
       jets & $\frac{\Delta p_T}{p_T} = 15 \%$ \\
      \hline
       $b$-jets & $\frac{\Delta p_T}{p_T} = 15 \%$\\
                        &  $b$ efficiency = 60 \%\\
                        &  $c$ mistag = 20 \%\\
      \hline
      \hline
       $b$-jets & $\frac{\Delta p_T}{p_T} = 10 \%$ \\
       (for $\lambda_3$) &  $b$ efficiency = 76 \%\\
        &  $c$ mistag = 20\%\\
      \hline
      \hline
    \end{tabular}
    \label{tab:higgsreq}
\end{table}

\subsubsection{Key challenges}

The first studies with the detailed detector simulation have shown that the effects of the beam-induced background on high-energy objects in the central region of the detector are negligible.
One of the key challenges in event reconstruction is dealing with physics objects that have low energy and physics objects that are in the forward or backward regions of the detector. 

The beam-induced background mitigation measures must be refined and perfected to improve the reconstruction of the low-energy objects. For instance, the energy thresholds on the calorimeter hits affect the energy resolution of jets and low-energy photons and electrons. 
In the detector forward and backward regions closer to the beamline, the beam-induced background levels are higher.
At multi-TeV collision energies, all the standard model particles are highly boosted in the forward region and their decays may produce particles that are very close to each other in space or even overlapping in the detector.
Dedicated algorithms must be developed for those cases.
%

\subsubsection{Work progress since the publication of the European LDG roadmap}

Since the publication of the European LDG roadmap, an assessment has been performed of the reconstruction performance for tracks, muons, electrons, photons, and jets in the presence of the beam-induced background with a detailed detector
simulation at a 3 TeV collider, as discussed in Section~\ref{sec:objectsReco}.
The reconstruction algorithms of the physics objects were revised and fine-tuned to preserve their reconstruction efficiency and the accuracy in the determination of their properties.

Moreover, such physics objects were used to carry out sensitivity studies on the Higgs boson production cross sections at a 3 TeV collider. 
This paper reports only a small subset of all the results collected in the review by Casarsa \textit{et al.}~\cite{ref:ANR}, where it is demonstrated that the achieved outcomes are competitive compared to the corresponding comprehensive estimates from the CLIC Collaboration at the same collision energy.

\subsubsection{Work planned for next evaluation report}

The work plan for the evaluation report follows two main directions: the completion of the physics objects studies at the 3 TeV center-of-mass energy and a preliminary evaluation of the reconstruction performance for the main physics objects at a $\sqrt{s}=10$ TeV collider.

At 3 TeV center-of-mass energy, a refinement of the electron and photon reconstruction is planned: the energy resolution is currently limited at low energies by the high thresholds set for the electromagnetic calorimeter hits and at high energies by the spillage of the electromagnetic showers into the hadronic calorimeter; electron reconstruction would also benefit from the recovery of the radiated bremsstrahlung photons.
Furthermore, the tau leptons are still missing to complete the set of available physics objects and their reconstruction algorithms must be studied.
In addition, only preliminary studies were conducted on the missing energy determination due to the issues on the calorimeter object reconstruction. The improvements already discussed on this sub-detector will open the possibility also to tackle the missing energy resolution. 

\subsubsection{Next priority activities and collaboration opportunities}

The detector performance has been determined for the most critical and important objects. Additional efforts are needed  to get a complete picture  at 3 TeV and to have a full set of reconstructed objects to be used in physics studies:
\begin{itemize}
\item a study of the reconstruction algorithms for tau leptons with the beam-induced background and an assessment of their performance;
\item an optimization of the jet reconstruction, taking advantage of an improved reconstruction of low-momentum objects and possibly the identification of jet substructure;
\item development of dedicated algorithms to reconstruct and identify boosted physics objects;
\item an improvement of the jet flavour identification, exploiting more sophisticated algorithms based on artificial-intelligence techniques;
\item an assessment of the most suitable particle identification techniques and algorithms in the muon collider environment.
\end{itemize}
%
Neverthless, the next steps will be the study of the detector performance at$\sqrt{s}=10$ TeV following the same priority list used for the $\sqrt{s}=3$ TeV. The algorithms studied for the 3 TeV case will optimized and adapted where needed, to the higher center-of-mass energy. New methods may be developed to reconstruct physics objects that exhibits different properties at high energy.

\subsection{Technologies 
}
\label{sec:Detector_technology}
\subsubsection{System overview}

\noindent
The detector concept described earlier serves as a starting point for evaluating the expected detector performance in different physics scenarios by building an efficient full-simulation workflow to study various beam-induced background mitigation strategies.
At this conceptual stage, only general assumptions about the relevant detector parameters were made, such as dimensions, material composition, granularity, and spatial and time resolution.
This approach simplifies the process of iterative variations towards an optimised detector design for which specific technology details can be implemented at a later stage.
These studies allow the identification of the key features of different sub-detectors at a muon collider experiment, which are summarised in the recent review for Snowmass~2021 on the promising detector technologies and R\&D directions~\cite{promising_rnd}.

From the technology perspective the present $\sqrt{s} = 3$\,TeV detector model can be divided into six sections ordered by increasing distance from the interaction point:
\begin{enumerate}
    \item \textbf{Vertex detector (VXD):} Si sensors with $25\times25$~\textmu m$^2$ pixels and time resolution of $\sigma_\text{t} =30$\,ps arranged in double layers with 2\,mm spacing.
    \item \textbf{Inner (IT) and outer (OT) tracker:} Si sensors with 50\,\textmu m$\,\times~1$~mm (IT) and 50\,\textmu m$\,\times~10$~mm (OT) macro-pixels and time resolution of $\sigma_\text{t} =60$\,ps.
    \item \textbf{Electromagnetic calorimeter (ECAL):} sampling design with \textit{W} absorber and \textit{Si} sensors arranged in $5\times5$~mm$^2$ tiles.
    \item \textbf{Hadronic calorimeter (HCAL):} sampling design with steel absorber and plastic-scintillator tiles arranged in $30\times30$~mm$^2$ cells.
    \item \textbf{Superconducting solenoid:} magnetic field strength of 3.57\,T with a Fe return yoke.
    \item \textbf{Muon detector:} resistive plate chambers (RPC) divided into $30\times30$\,mm$^2$ cells interleaved in the magnet's return yoke.
\end{enumerate}

Description of this geometry is implemented in the DD4hep~\cite{ref:frank} framework, which provides interface to \textsc{Geant4}~\cite{geant} simulation software and to Marlin~\cite{ref:gaede} framework for detector digitization and event reconstruction.
Two generic types of hits are used in the detector simulation:
\begin{itemize}
    \item \textbf{Tracker hit:} corresponds to a single energy deposited by a particle in the sensitive volume, with associated energy, time and position of the deposit within the volume.
    \item \textbf{Calorimeter hit:} corresponds to the integrated energy deposited by one or more particles in the sensitive volume during the fixed integration time, with only the total energy and time assigned to the hit associated with the volume.
\end{itemize}
Tracker-type hits are used for VXD, IT and OT detectors, while calorimeter-type hits are used for the ECAL, HCAL and muon detectors, which have the actual granularity implemented in the geometry.

\subsubsection{Key challenges}

High-intensity beam-induced background that survives after passing through the MDI nozzles poses significant challenges for most of the sub-detectors by inducing high levels of radiation and by creating excessive amount of background hits.
Projected levels of total ionizing dose (TID) and neutron fluence at Muon Collider are comparable to those expected at HL-LHC, while the background hits have signatures unique to Muon Collider and require dedicated treatment in each sub-detector to achieve the necessary physics performance.

In the tracking detectors, the beam-induced background causes very high hit density, up to $5 \times 10^3$~hits/cm$^2$ in the innermost layers of VXD, when integrating over a 15\,ns readout window.
Combinatorial background arising from such a large number of hits makes track reconstruction unfeasible unless the amount of background hits entering the pattern recognition algorithm is significantly reduced.
The primary beam-induced background suppression methods in the tracking detectors are:
\begin{enumerate}
    \item \textbf{Timing:} rejection of hits outside of a very narrow time window of about 100\,ps (exploiting the characteristic time distribution of beam-induced background hits, as shown in Fig.~\ref{fig:trk_timing}.
    \item \textbf{Direction:} rejection of hits inconsistent with pointing at the interaction region by estimating their direction from a stub of two hits in a double-layer.
    \item \textbf{Cluster shape:} rejection of beam-induced background hits represented by wider clusters of pixels due to crossing the sensors at a shallower angle.
    \item \textbf{Goodness of fit:} rejection of fake track candidates due to bad $\chi^{2}$.
\end{enumerate}
Considering these aspects, the great challenge for tracking-detector technology is to provide extreme time resolution and on-detector hit filtering while maintaining low material budget and high radiation tolerance.
A low material budget indirectly requires low power density of the readout electronics such that it can be cooled without introducing additional heat-exchanging materials.
We should also note that these beam-induced background suppression methods could potentially limit the performance of long-lived particles, for which a dedicated strategy should be defined in the future.

\begin{figure}[!ht]
    \centering
    \includegraphics[width=0.6\textwidth]{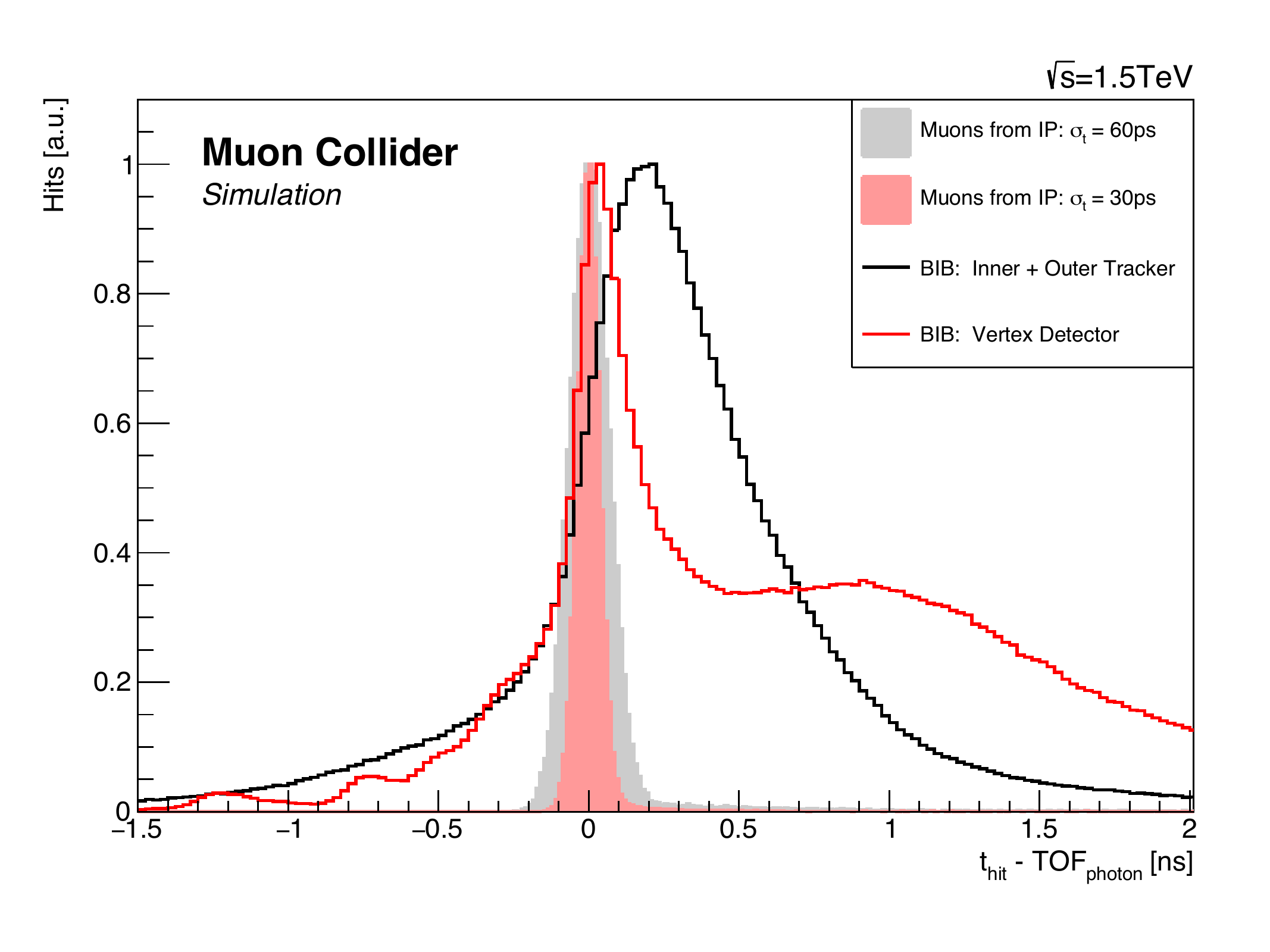}
    \caption{Time distribution of hits in the tracking detector corrected for the expected photon's time of flight from the interaction point. Hits from the beam-induced background (BIB) are shown by solid lines, while hits from single muons smeared by corresponding time resolutions are shown by filled areas.}
    \label{fig:trk_timing}
\end{figure}

In the ECAL and HCAL sub-detectors beam-induced background contribution comes primarily from low-energy photons and neutrons respectively, which have a relatively uniform angular distribution across the detector volume.
Yet, their distribution in time and depth is different from the shower signatures produced by the hard collision, as shown in Fig.~\ref{fig:calo_signatures}.
Given that the beam-induced background contribution partially overlaps with the signal hits, with the current configuration and technology of the calorimeters, it can't be excluded from the readout completely and has to be subtracted instead at a later stage.
To minimise the effect of such beam-induced background subtraction on the resulting energy resolution the corresponding detector should possess high spatial granularity and time resolution, allowing to measure the corresponding time and depth profiles of the shower.
In particular fine depth segmentation implies a large number of sensitive layers in the detector, making the cost of the technology also an important factor.
Thus, the main challenge for the ECAL and HCAL technologies is to provide excellent time resolution in a finely segmented detector at a low cost, while being compatible with the radiation-hardness requirements.

\begin{figure}[!ht]
    \centering
    \includegraphics[width=0.45\textwidth]{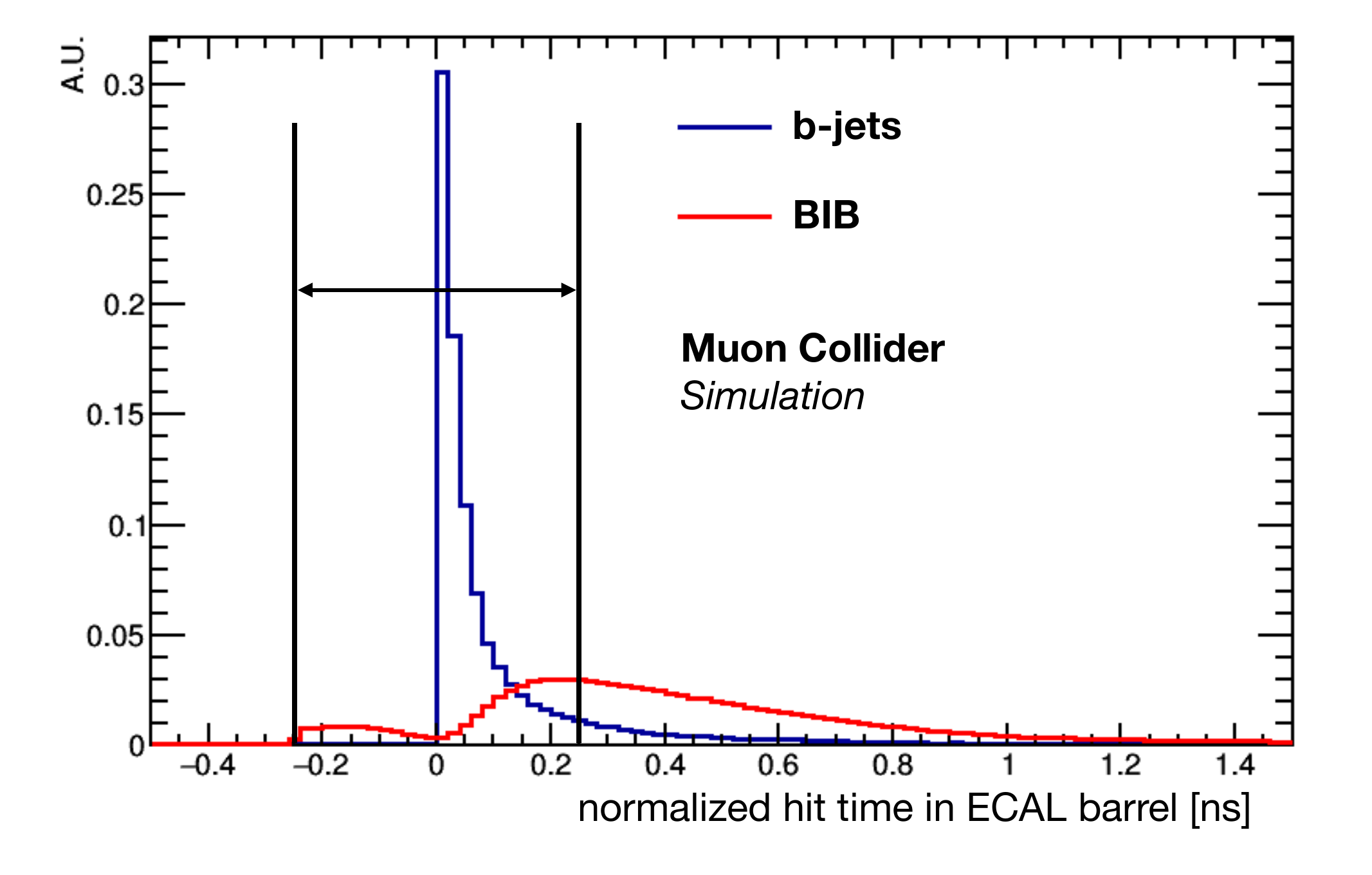}
    \hfill
    \includegraphics[width=0.45\textwidth]{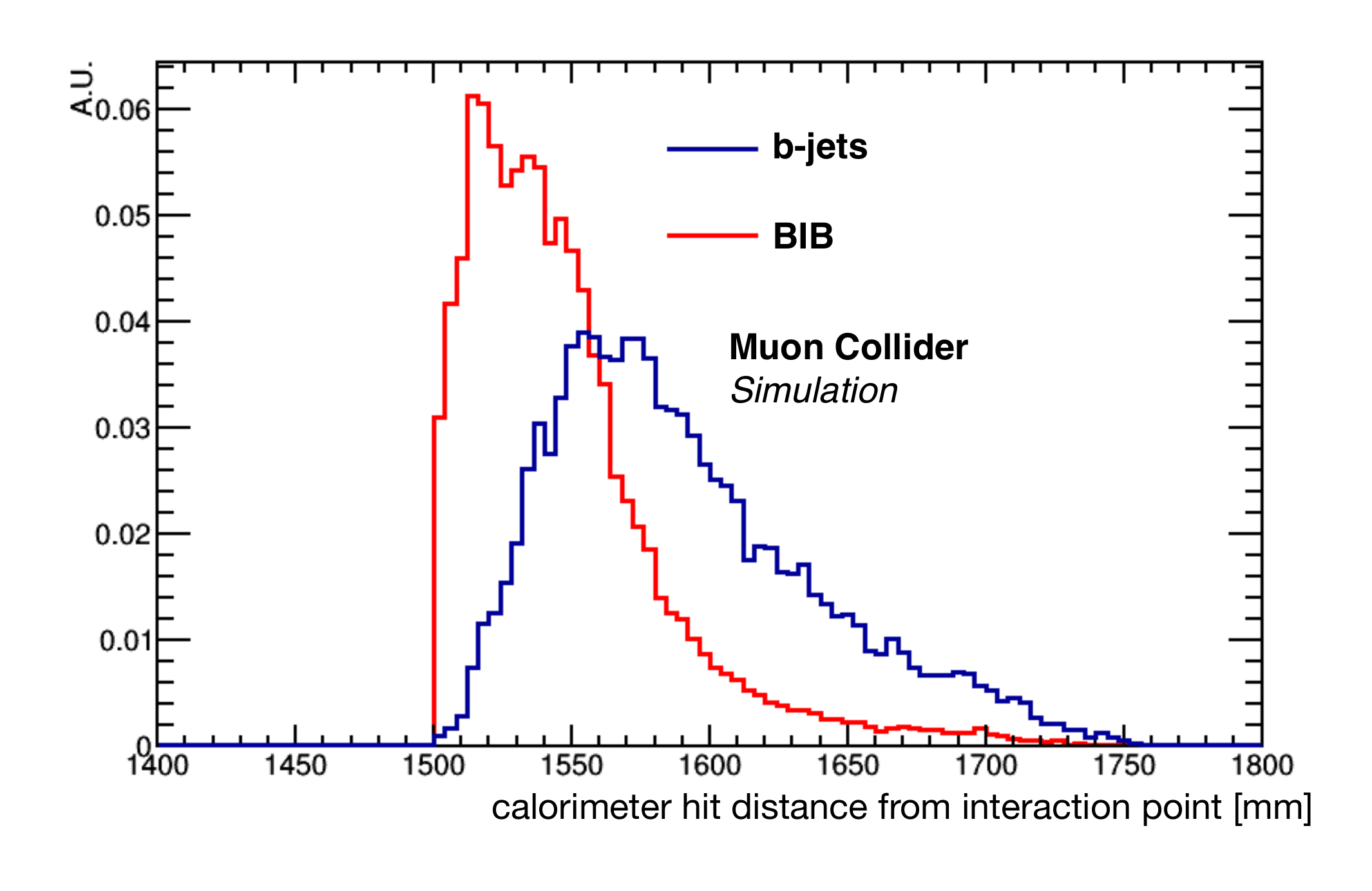}
    \caption{Difference between the time (left) and depth (right) of the hits produced by beam-induced background (BIB) and signal b-jets.}
    \label{fig:calo_signatures}
\end{figure}

The muon system is relatively unaffected by the beam-induced background contribution except for the very forward region, where high-energy neutrons and photons create hits near the MDI nozzles.
While this does not pose any serious challenge for muon detection itself, it is nevertheless important to have high spatial and time resolution to allow matching of standalone muon tracks with hits in the tracking detector.
Such seeding of muon tracks from the muon system would greatly improve muon-reconstruction speed given the high combinatorial background in the tracking detector.

\subsubsection{Work progress since the publication of the European LDG roadmap}

Several developments have been done for different sub-detectors towards a better definition of their technical characteristics.

A realistic digitization algorithm for pixel sensors of the tracking detectors has been implemented and integrated into the Muon Collider simulation software.
With a configurable pixel pitch for the sensor together with other readout parameters like noise and threshold it provides realistic pixel clusters that can be used for beam-induced background suppression.
On the hardware side several technologies have the potential to meet the requirements of the tracking detectors, such as AC- and DC-coupled LGAD, trench-isolated LGAD, as well as MAPS technology thanks to its particularly low material budget.
Estimates of the data rate, mainly due to the beam-induced background hits, underline the importance of fast and radiation-hard data-transfer links, such as LpGBT, and careful design of the data acquisition logic.
Reduction of the amount of data directly in the readout chip is particularly important in this respect.

For ECAL a new conceptual design has been implemented in the simulation based on Crilin technology~\cite{crilin}, which is a semi-homogeneous calorimeter with finely segmented Cherenkov-radiating crystals providing high time resolution at relatively low cost. A semi-homogeneous calorimeter is expected to have a better energy resolution with smaller fluctuations, which can help to minimize the effect of the beam-induced background.
After showing a good performance in Muon Collider simulation studies some physical prototypes have been built and tested to experimentally validate this technology and the expected performance.
The latest beam test \cite{crilin_tipp} of a multi-layer prototype at CERN demonstrated a time resolution of $\mathcal{O}$(20\,ps), as shown in Fig.~\ref{fig:crilin_results}, which gives confidence that reaching $\sim 100$ ps on the final detector would be possible.

\begin{figure}[!ht]
    \centering
    \includegraphics[width=0.4\textwidth]{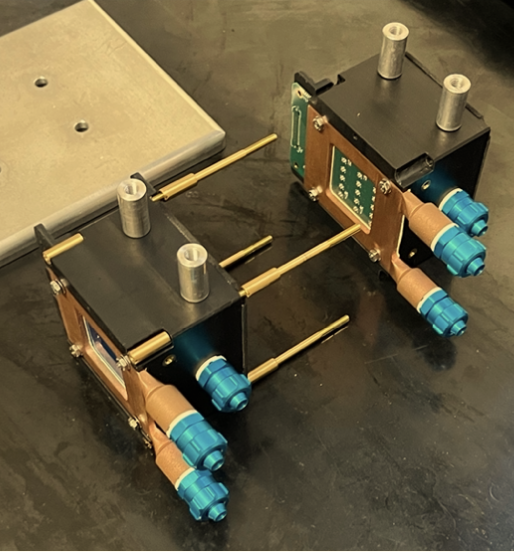}
    \hfill
    \includegraphics[width=0.45\textwidth]{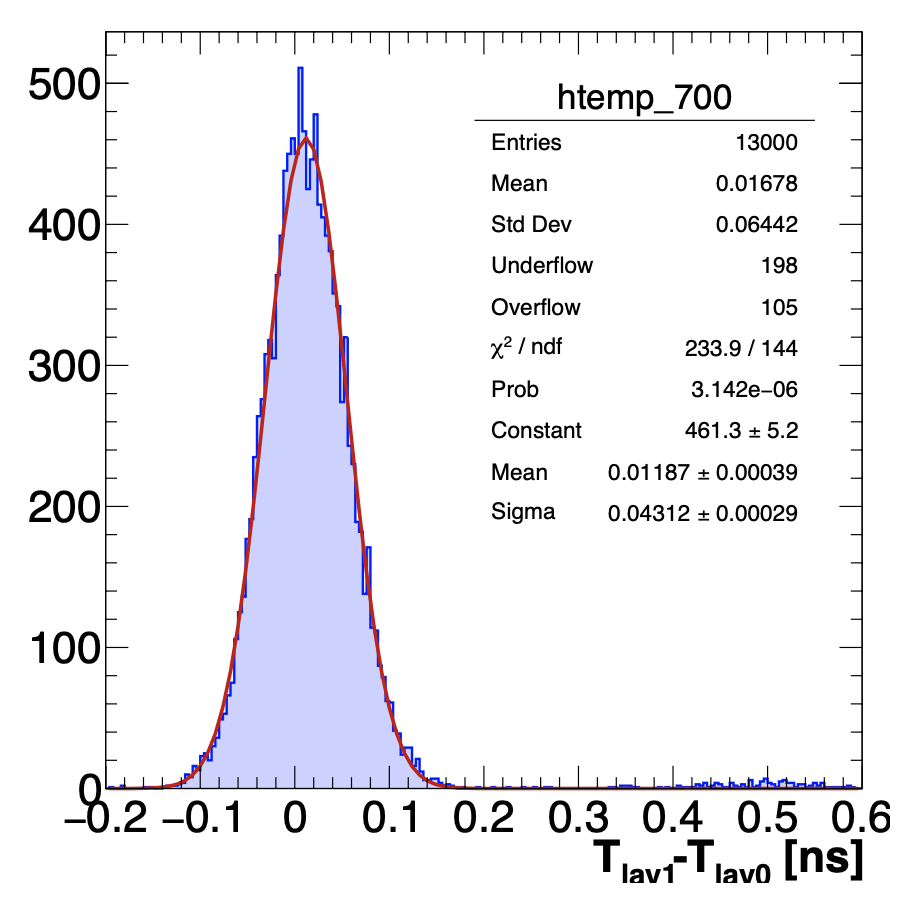}
    \caption{Left: two out of three detection layers of the Crilin prototype proposed for Muon Collider calorimeter  tested at CERN teast beam. Right: Distribution of the time difference between two neighbouring layers of the prototype measured obtained with test beam data.}
    \label{fig:crilin_results}
\end{figure}

For HCAL the use of micro-pattern gas detectors~(MPGD)~\cite{mpgd} as an active layer is under study, which offers good energy and time resolution at high rate and low cost for instrumenting such a large area.
The corresponding design has been implemented in the detector geometry for full-simulation studies, which showed performance comparable to that of the current baseline.
The technologies being considered are \textmu RWell, RPWell and MicroMegas, for which initial digitization logic for digital readout has been implemented in the simulation software.
Physical prototypes for each technology have been also produced and two test-beam campaigns have been carried out at CERN during 2023 (see Fig.~\ref{fig:mpgd_results}), which showed good results in terms of MIP detection efficiency.

\begin{figure}[!ht]
    \centering
    \includegraphics[width=0.55\textwidth]{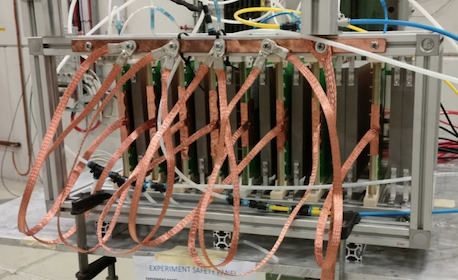}
    \hfill
    \includegraphics[width=0.38\textwidth]{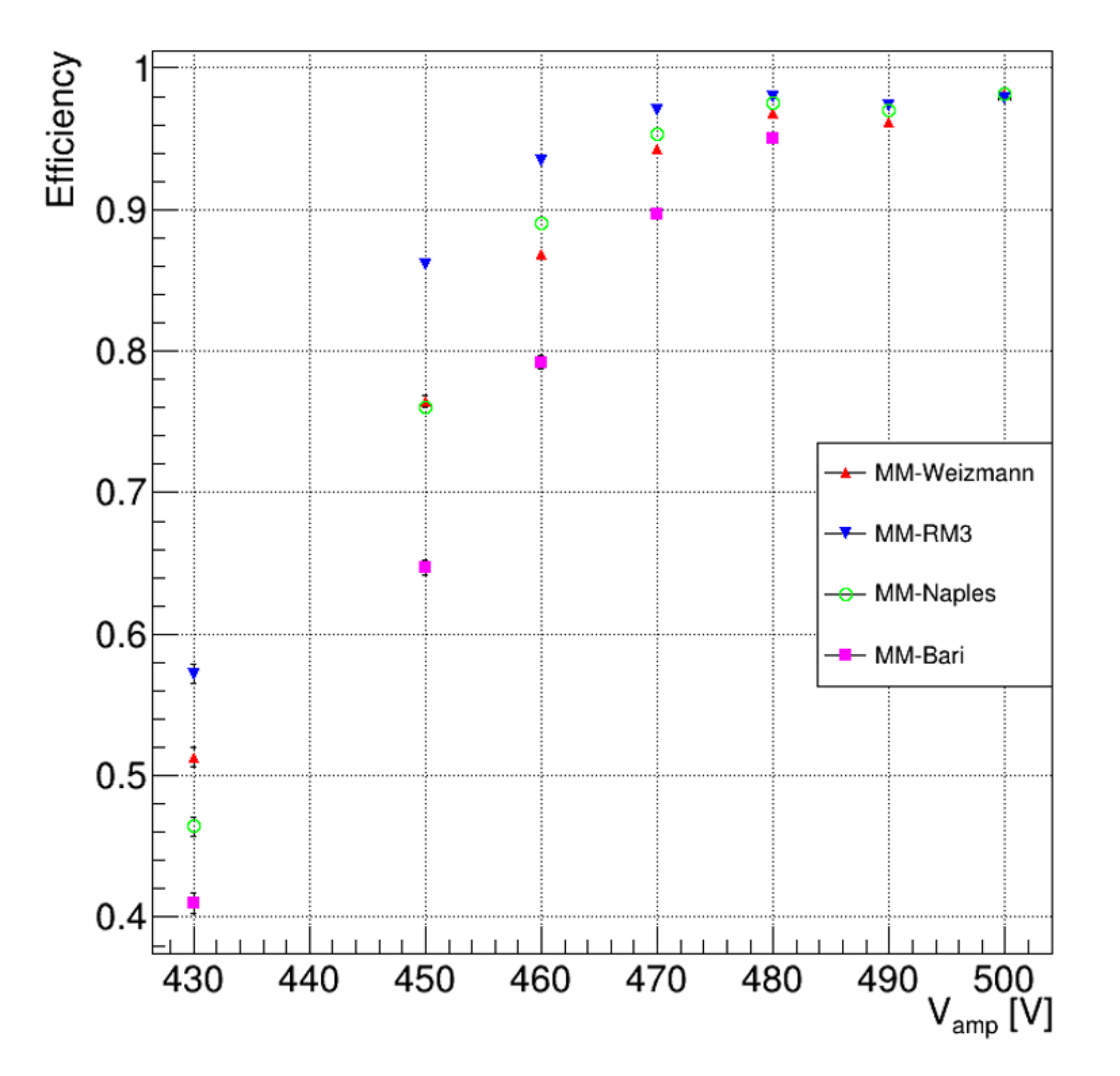}
    \caption{Left: MPGD-based HCAL prototype tested at CERN having multiple layers built with different technologies. Right: measured MIP-detection efficiency of different MicroMegas layers as a function of the applied high voltage.}
    \label{fig:mpgd_results}
\end{figure}

For muon detector, the Picosec technology~\cite{picosec} is being studied as a better-performing alternative to the RPC layers.
It uses a crystal Cherenkov radiator, coupled to a UV-transparent photocathode and a two-stage amplification by Micromegas, providing a time resolution comparable to that of the tracking detector. This would be of great importance for rejecting out-of-time beam-induced background hits, especially in the forward region of the muon system where the rates are high.
Several prototypes have been tested with different photocathode materials and gas mixtures, with very promising preliminary results, as shown in Fig.~\ref{fig:picosec_results}.
Finally, the tested Picosec geometry has been interfaced with the full-simulation framework to study its performance in the forward region in the presence of beam-induced background.

\begin{figure}[!ht]
    \centering
    \includegraphics[width=0.47\textwidth]{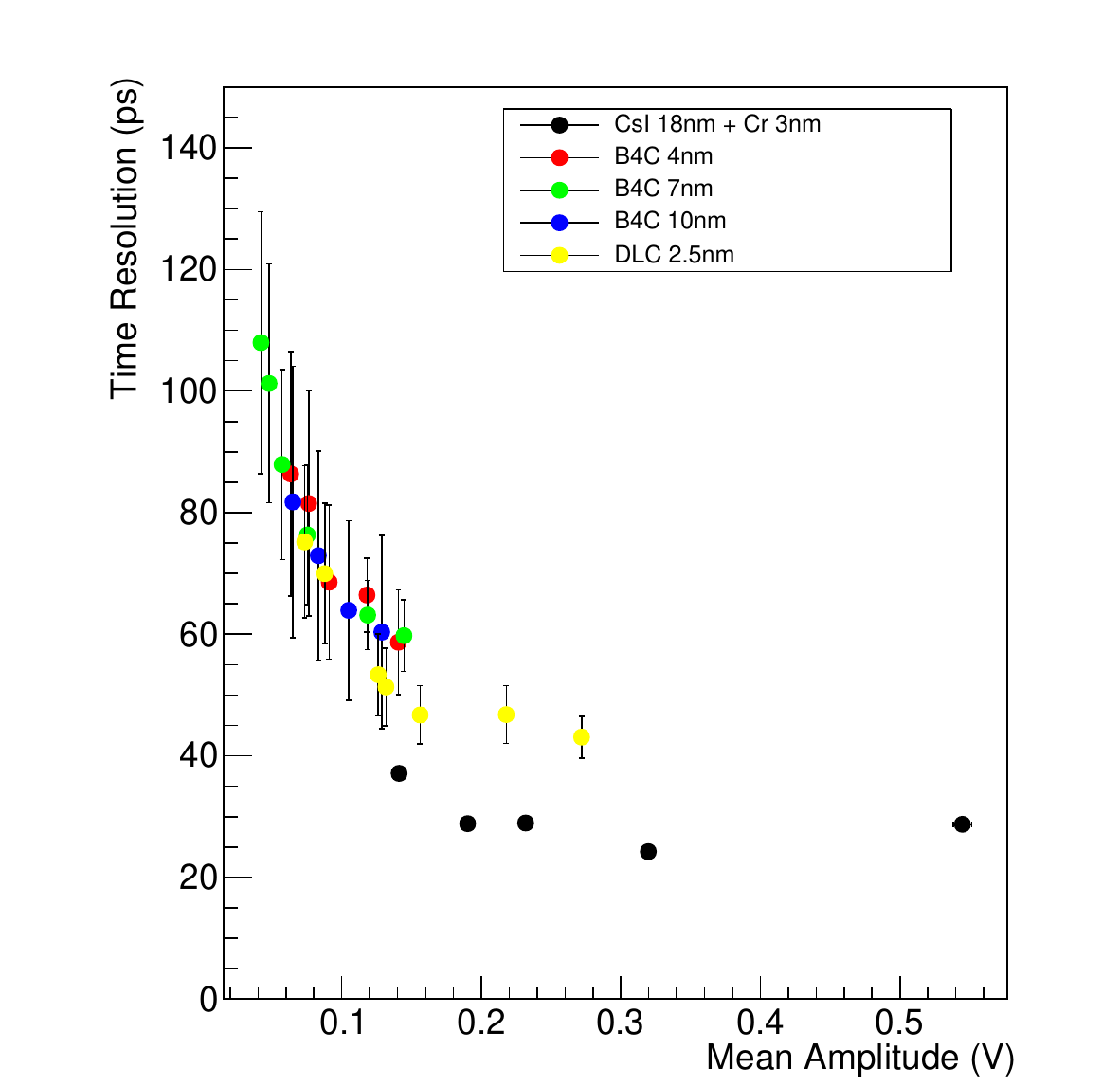}
    \hfill
    \includegraphics[width=0.47\textwidth]{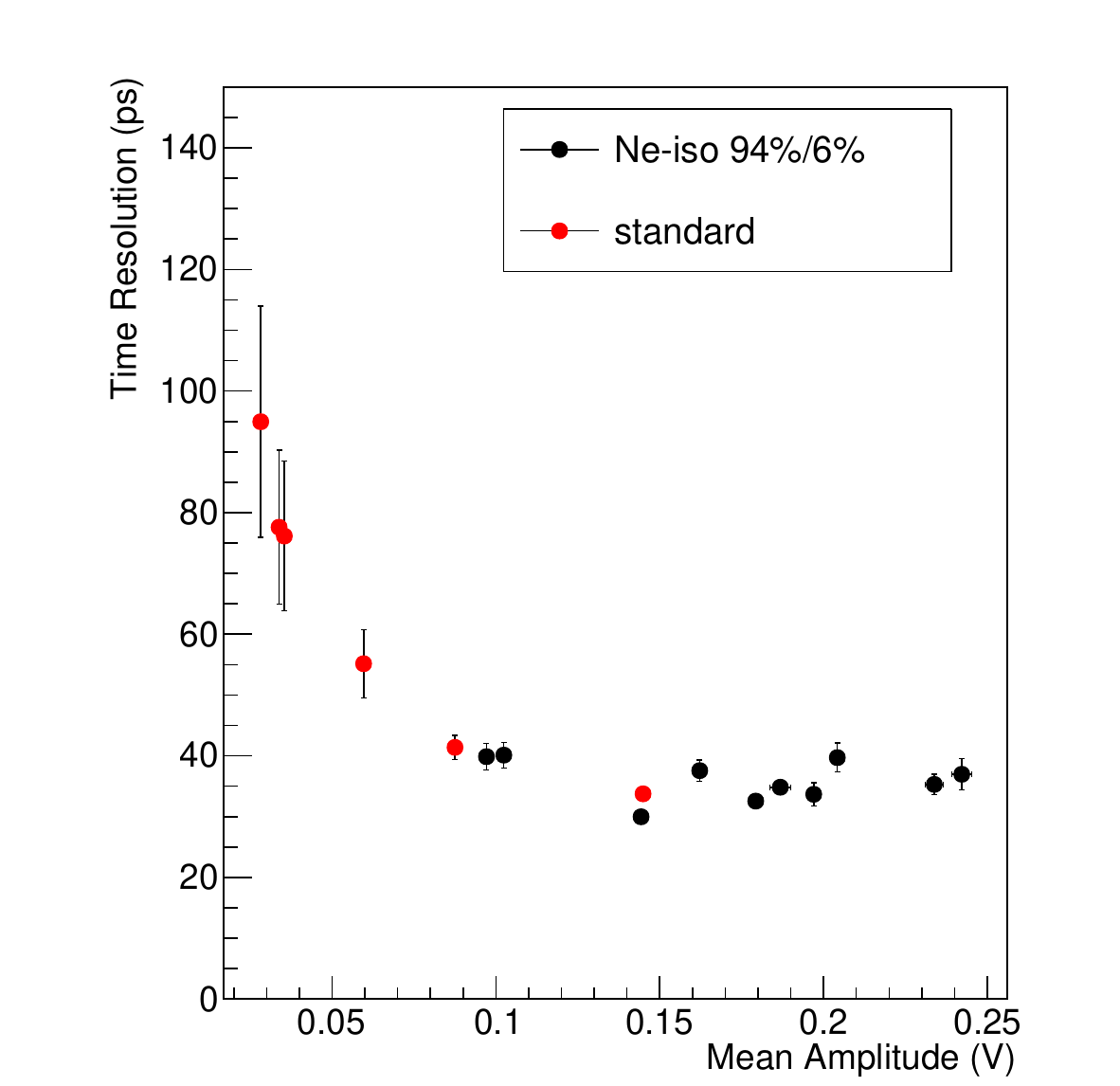}
    \caption{Time resolution obtained by the Picosec prototypes as a function of the mean signal amplitude compared for different photo-cathode materials (left) and gas mixtures (right).}
    \label{fig:picosec_results}
\end{figure}

The community of the IMCC Detector and Physics is deeply involved in the ECFA DRD collaboration. In some of the study, like the calorimeter one, the efforts are lead by the IMCC people. In addition to that, the muon collider community participates to the DRD activities, making sure that all the mentioned technologies relevant for each subdetector are included in the corresponding DRD collaboration roadmaps.
Similar efforts are also taking place within CPAD.
A comprehensive list of all the relevant technologies can be found in Ref.~\cite{promising_rnd}.



\subsubsection{Work planned for next evaluation report}

A number of developments are planned to progress towards a well-defined conceptual detector design.
Important input is expected from experimental tests of the first production batches of DC-RSD sensors~\cite{dc_rsd}, which is a promising technology for the tracking detector.
In particular its ability to achieve low occupancy by controlling the extent of charge sharing will be validated and implemented in the simulation software.
R\&D on the technologies considered for the ECAL, HCAL and muon detector will continue according to their working plans, which are part of the coordinated DRD and CPAD efforts in Europe and U.S. respectively.

\subsubsection{Next priority activities and collaboration opportunities}
Current available technologies allow competitive physics measurement at muon collider if the detector were to be built in the next few years. With the new generation of the detector, the performance can be greatly improved, as the preliminary studies on the calorimeter systems is showing.

The development of an ad hoc tracking detector needs to be better pursued. Here, the technology is at a very early stage and multiple technical aspects need to be addressed.
These include the possibility of having long pixels in the inner and outer trackers, where hit density is lower, which is challenging to implement in DC-RSD technology.
Different possible readout implementations can have great impact on the sensor's efficiency, granularity of the timing information and the resulting power consumption.
An integrated approach to the tracker design that also includes the mechanics and cooling will be very important in the future.

The integration of all sub-detectors, currently missing, will be studied once the technology for each individual sub-detector is defined.
\subsection{Software and computing: Concepts}
\label{sec:Detector_software}

\subsubsection{System overview}

The muon collider software framework~\cite{ref:bartosik} is based on CLIC's iLCSoft. It includes the DD4hep toolkit~\cite{ref:frank} to model the detector geometry, the \textsc{Geant4}~\cite{geant} program to simulate the detector response, and the Marlin package~\cite{ref:gaede} for the event digitization and reconstruction.
Stable particles are generated by external packages and given as input to the aforementioned software. Signal and physics backgrounds events are produced by users with common packages like \textsc{Whizard}~\cite{whizard} and \textsc{MadGraph}~\cite{madgraph}. The beam-induced background bunch-crossings, instead, are produced by using the FLUKA~\cite{fluka} and FlukaLineBuilder~\cite{flukaLB}, as described in Section~\ref{sec:Section04_2}.
The programs workflow proceeds as:
\begin{itemize}
    \item generation of $\mu^+\mu^-$ interactions and beam-induced background bunch crossings;
    \item simulation of the interaction of the generated particles with the passive and active elements of the detector; the results are stored as {\textsc SimHits};
    \item overlay of beam-induced background {\textsc SimHits} to the $\mu^+\mu^-$ interaction events;
    \item reconstruction of the energy deposits in each sub-detectors, i.e.~tracker, calorimeters, muon system producing  {\textsc RecHits};
    \item reconstruction of high-level physics objects e.g.~charged particle tracks, jets, secondary vertices, muons, electrons, etc.
\end{itemize}
The physics objects are then used in the analysis measurements, as presented in Section~\ref{sec:objectsReco}. 
The software infrastructure is currently supported and maintained by two INFN experts.

The computing infrastructure is formed by: 
\begin{itemize}
    \item CPU: accessed via cloud services provided by INFN and CERN;
    \item storage: disk space of about 600 TB at INFN sites and about 100 TB at CERN.
\end{itemize}

\subsubsection{Key challenges}
The major challenges identified so far are the following:
\begin{itemize}
    \item production of the beam-induced background bunch-crossing events, see Section~\ref{sec:Section04_2} for a description;
    \item simulation of the beam-induced background particles in the detector; this task requires a significant amount of CPU time, in Ref.~\cite{ref:bartosik} such a time is quantified depending on the configuration used to track particles throughout the material;
    \item the overlay of the beam-induced background hits with the $\mu^+\mu^-$ interactions particle hits; this process demands a significant amount of computing memory (RAM) per event, for a detailed discussion see Ref.~\cite{ref:bartosik};
    \item tracks and calorimeter object reconstruction from {\textsc RecHits} including the beam-induced background; it can be performed only under particular conditions, for example by filtering tracker hits depending on their arrival time on the sensor and subtracting a tuned amount of energy in the calorimeter cells.
\end{itemize}

\subsubsection{Work progress since the publication of the European LDG roadmap}
The software has been improved significantly since the publication of the roadmap to reduce the amount of computing resources and the time necessary to simulate and reconstruct a complete bunch-crossing event. A complete description of the improvements can be found in Ref.~\cite{ref:bartosik}. 
The notable optimisations include the early exclusion of irrelevant beam-induced background particles from \textsc{Geant4} simulation by filtering tracker-hit collections based on the arrival time on the sensor, as well as the integration of ACTS (A Common Tracking Software) track-reconstruction framework~\cite{acts_paper}.

Another important step forward is the transition to the common Key4hep software stack that is almost completed, but not yet in production.

From the point of view of the code distribution to the users, two major software releases have been done since the publication of the roadmap, in June 2022 and April 2023. The software is available with its distribution via docker images and CERN CVMFS.

\subsubsection{Work planned for the evaluation report}
The major activities currently in progress and likely to be ready at the time of the evaluation report are:
\begin{itemize}
    \item complete transition to Key4hep and distribution of a production release, depending on the person-power available;
    \item definition of the release validation workflow;
    \item definition of a user support, with voluntary personnel, to help newcomers to use the software structure;
    \item expansion of computing infrastructure to include additional European and US sites.
\end{itemize}

\subsubsection{Next priority activities and collaboration opportunities}
The computing infrastructure is distributed in different sites, a federation is needed with an authentication and an authorization method common to the IMCC. A similar problem is present on data management, where different architectures are used by different sites and within IMCC there is no data management system in place. Both these activities are not pursued due to the lack of person-power.

The usage of advanced artificial intelligence methods for physics object reconstruction like tracks and jets and their identification could significantly improve the performance by helping to minimize the impact of the beam-induced background. These activities require a significant increase of the person-power.

\begin{flushleft}
\interlinepenalty=10000

\end{flushleft}

\clearpage
\section{Accelerator design 
}
\label{sec:Section06}



This chapter describes the whole accelerator chain of the proposed muon collider complex. The challenge of the proton complex is to provide the few MW proton with very short bunch lengths of a few ns and low repetition rate of 5\,Hz. The muon production cooling part consists of a target inside an as high as possible magnetic field and standing the drive beam power, a decay channel with decreasing magnetic fields and comprising a chicane to remove unwanted particles, a section for longitudinal capture of the muons and, finally, several ionization cooling channels interleaved with a section merging several muon bunches into one bunch. The acceleration to high energy has to be fast requiring large average RF gradients at all stages; the baseline solution consists of a sequence of recirculating linacs followed be very rapid cycling synchrotrons accelerating the beam within a few tens of turns. Finally, the muons are injected into the muon collider ring, which has to provide sufficient luminosity leading to a challenging design with $\beta^*$ and rms bunch length of a few mm and large rms momentum spread of around $1 \cdot 10^{-3}$.

\subsection{Proton complex 
}
\label{sec:Section6_2}

\subsubsection{System overview}

The proton complex is the first piece in the muon collider complex. It comprises of a high power acceleration section; an accumulator and compressor; and a target delivery section. The high power acceleration section can take many shapes as shown in Fig.~\ref{fig:proton_complex_layout}. In our baseline design we will study a final energy linac. This section is responsible for delivering high power, high intensity bunches to the accumulator and compressor section. At the end of the high power acceleration section the pulse total charge is distributed among a high number of bunches with a small energy spread. Such a long train of bunches do not reach the desired instantaneous power for intense pion production on the target.

\begin{figure}[h]
\centering
\includegraphics[width=0.8\textwidth]{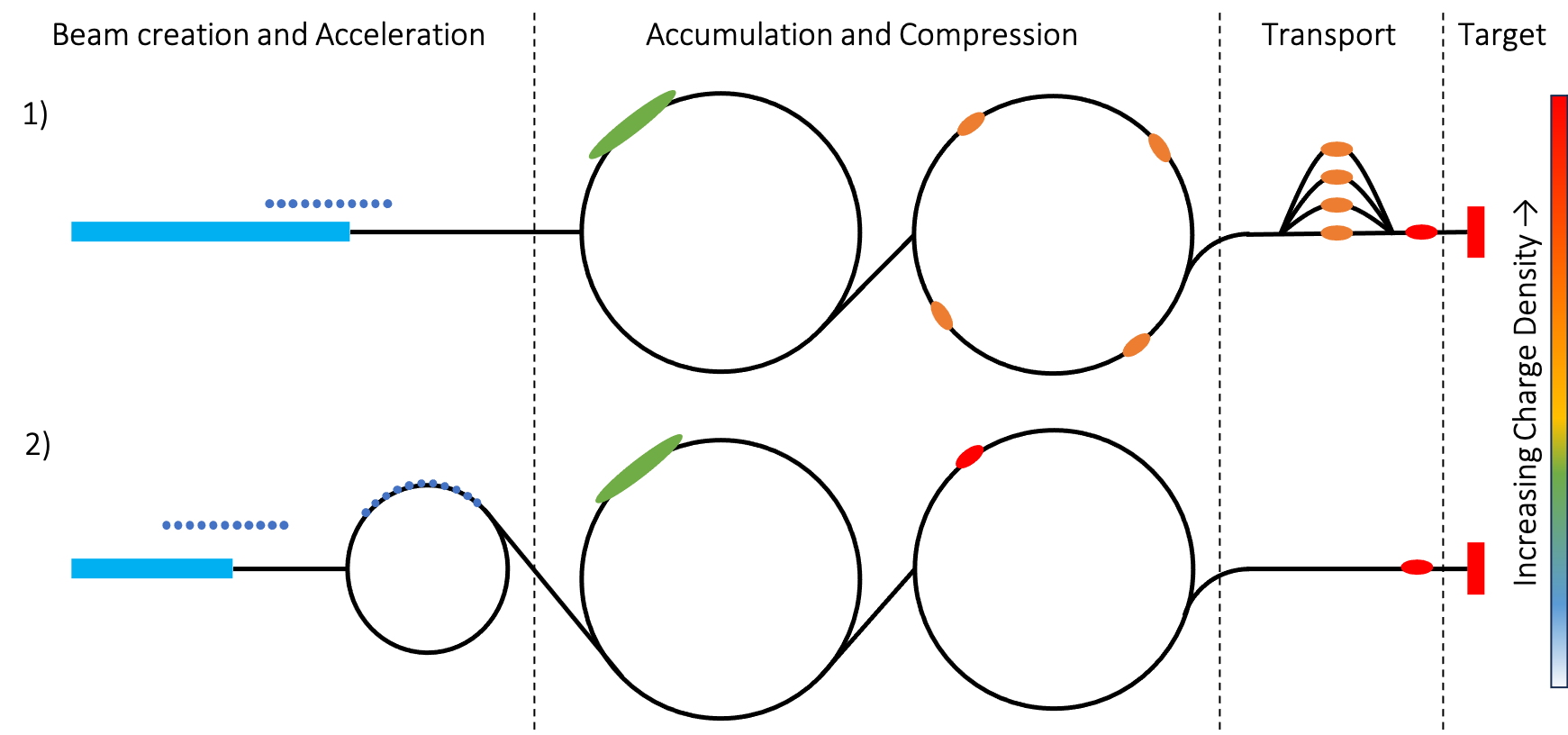}
\caption{Schematic layouts for the proton complex section. Our baseline design is the schematic 1 on the figure. The bunch density as the proton travels thought the complex is also depicted. The closer to the target the highest the bunch charge density and at the point where is collides with the target. To reach high densities is one the the main challenges in the design of the part of the muon collider. \label{fig:proton_complex_layout}}
\end{figure}

In order to achieve the high proton intensities in a short pulse, an Accumulator and Compressor ring will be used to combine and compress the protons into very short bunch, with a length of the order of $\sim 2$\,ns rms.  The first storage ring, the Accumulator, accumulates the protons via charge stripping of the incoming H$^−$ beam from the linac section, preserving the energy spread. The incoming beam is chopped to allow a clean injection into the ring circumference. The second storage ring, the Compressor, accepts between 1 and 6 intense bunches  from the Accumulator and performs a 90\textdegree-bunch rotation in longitudinal phase space, shortening the bunches to the limit of the space-charge tune shift just before extraction. The Compressor ring must have a large momentum acceptance to allow the storage of the beam with a momentum spread of a few percent, which arises as a consequence of the bunch rotation. 

The target delivery section transports the final intense short bunches from the compressor to the target, ensuring that neither the transverse nor longitudinal properties of the beam are compromised in the process. To achieve optimal luminosity, the final collider will work in single bunch colliding mode. Therefore, in the multi-bunch solution, short bunches extracted from the Compressor must be recombined and hit the pion production target simultaneously.

The primary requirement of the proton complex is to enable the production of a high number of useful muons at the end of the decay channel after the target. The production rate, to good approximation, is proportional to the primary proton beam power, and (within the 5–15\,GeV range) only weakly dependent on the proton beam energy. Considering a conversion efficiency of about 0.013 muons per proton-GeV, a proton beam in the 1–4\,MW power range at an energy between 5\,GeV and 10\,GeV will provide the sufficient number of muons required.

\subsubsection{Key challenges}

The critical challenges for the proton complex include the determination of proton final energy on the target with significant implications for particle creation, the layout of the high power acceleration section, and the choice of target material and size. Two final proton beam energies, 5 GeV and 10 GeV, will be investigated during the length of the grant.

A comprehensive study is required to identify key parameters for the Proton Complex, including a compression scheme and the final number of bunches, thus influencing the layout of compressor ring lattice and the required number of turns for compression. The minimum number of bunches needed will affect the layout of the {\em ring-to-target} delivery design.
The beam parameters on the target surface will impact the design of the expense beam dump, which can be challenging due to target geometries and constraints.

Determining the linac chopping scheme is another key parameter, which will directly influence the linac pulse length and current, and subsequently, the accumulator and further accelerator design.

Other important challenges include the design of the accumulator lattice for both energies to be studied, investigation of the instability thresholds in the accumulator and compressor rings, error analysis across the entire complex with a focus on bunch recombination at the target delivery system, and continuous development of the H$^{-}$ source to meet the needed pulse current and length for the muon collider's requirements, including potential upgrades.

\subsubsection{Recent achievements}

A baseline lattice for the accumulator and compressor for the 5\,GeV case, based on the studies for the neutrino factory at CERN, was revised and ported to XSuite and PyOrbit. A campaign to cross benchmark of both codes, specially for high space charge conditions, is ongoing.  Basic beam physics parameters are being collected and compiled and a parametric study is underway. A first look at the transport between ring and target of a 2\,ns high current bunch has started. A compilation of the literature on past projects, with special attention to US Muon Acceleration Program (MAP) \cite{Berg,MAP2} results, was done.

Since both the accumulator and compressor rings are strongly space charge limited machines, an evaluation of available codes for space charge dominated beam simulation is under way. In order to complement this evaluation a collection of data from CERN PS bunch rotation and Booster bunch recombination, both with high density bunches, was carried out. Data analyses and further plans are underway.

\subsubsection{Planned work}

The majority of the planned work for the following year is concentrated on the compressor lattice design, refinement, and target delivery parameters. We plan to explore possibilities at a beam energy of 5\,GeV, including the definition of minimum number of bunches possible and a rotation scheme. Also for the 5\,GeV case, we plan to narrow down the beam parameters that are possible to be delivered to the target. A second step in the process will be the design of a 10\,GeV compressor ring and the re-design of the target delivery section, in case the 5\,GeV solution is unable to achieve the required beam quality.

Other planned work regarding the proton complex includes:
\begin{itemize}
    \item Compilation of linac technologies and main parameters
    \item Study on H$^{-}$ source capabilities (current and pulse length)
    \item Study on H$^{-}$ stripping for injection into the Accumulator
    \item Definition of a possible chopping scheme for the linac and final accumulator setup
    \item Definition of RF requirements for bunch rotation
    \item Initial study of main instabilities in the accumulator and compressor rings
\end{itemize}

\subsubsection{Next priority studies and opportunities for additional collaboration}

Several studies with lower priority are at present not planned and could represent great opportunities for future collaborative work. Notably, the study proposal does not encompass work on the accumulator or further acceleration (beyond the linac final energy of 5\,GeV), alternative rotation schemes and in-depth error and instability studies in various sections of the complex. All these topics would be of interest to further optimise the proton complex scheme and could be envisaged with additional resources and collaborators. On a more technical side, integrated engineering activities, as well as the detailed RF design for the rotation scheme, will be lacking. In depth work on the linac lattice design and many R\&D topics regarding H$^{-}$ source development, H$^{-}$ stripping, high efficient acceleration and design of  high-gradient super conducting (SC) cavities will be either partially covered or not at all developed at this stage, and we will use already proven schemes and setups for the SC linac in the baseline design. Other area that represents further opportunities for knowledge exchange and collaboration are tests in a established machine that can achieve space charge levels similar the the final bunches in the proton complex, such measurements can further our knowledge of the extreme space charge effects and help benchmarks the codes used for design and simulations.

\subsection{Muon production and cooling 
}
\label{sec:Section6_3}

\subsubsection{System overview}

\begin{figure}
\includegraphics[width=\textwidth, trim={0 5cm 0 4cm}, clip]{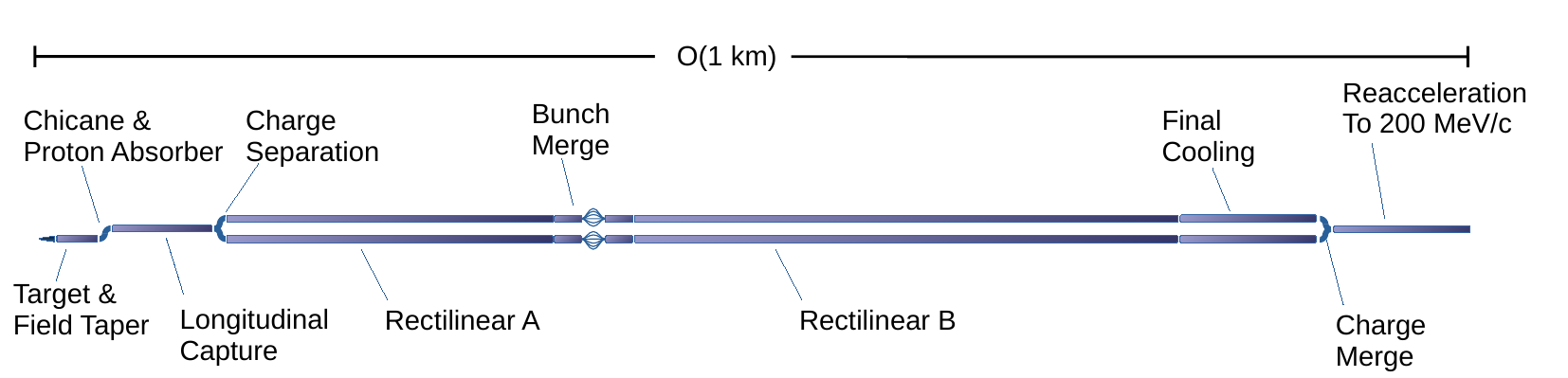}
\caption{Schematic of the muon production system. Protons are incident onto the target on the left of the figure, while the cooled muon beam leaves on the right of the figure, entering into the low energy acceleration system \label{fig:production_schematic}}
\end{figure}

The muon production and cooling system comprises several subsystems, as shown in Fig.~\ref{fig:production_schematic}. The protons provided by the proton complex intersect a target to produce pions. The target is immersed in a 20\,T solenoid field, yielding a high pion flux. The field is rapidly tapered to 1.5\,T. Pions and muons traverse a solenoid-focused chicane where high momentum beam impurities are removed followed by a Beryllium absorber which ranges out remnant low energy protons. The remaining beam passes through a longitudinal drift where any remaining pions decay. The beam is then captured longitudinally. A multi-frequency RF system that captures the beam into 21 bunches is required owing to the initially large longitudinal beam emittance. Another solenoid chicane system splits the beam by charge species, in preparation for the ionisation cooling system.

The ionisation cooling system comprises a series of solenoids, which focus the beam onto energy absorbers where the beam momentum is removed. The momentum is restored longitudinally using RF cavities resulting in a reduction in the beam's emittance (volume in position-momentum space). An initial rectilinear cooling system, Rectilinear A, reduces the beam emittance sufficiently that the many initial bunches can be merged into one single bunch. The beam is then cooled further to the final emittance, using a continuation of the rectilinear cooling system, Rectilinear B, followed by a sequence of high field solenoids operated ultimately with a low momentum (non-relativistic) beam. The beam is finally reaccelerated to semi-relativistic speeds so that it can be delivered to the acceleration system.

\subsubsection{Key challenges}
Critical challenges have been identified in the muon production and cooling system.
\begin{itemize}
\item An outline cooling system was prepared by the US Muon Accelerator Program (MAP) \cite{MAP2} that yielded an emittance that did not meet our requirement by a factor of two. The MAP system used conservative equipment parameters, but more demanding parameters may yield improved performance while remaining practical. Significant risk was identified associated with heating in the final cooling absorbers. The MAP system must be assessed and optimised to get a better understanding of the potential system performance including thorough assessment of engineering challenges.
\item A detailed design for the pion production target, solenoid and surrounding infrastructure must be performed. This target is as demanding as state-of-the-art horn-based pion production targets. Solenoid focusing at high beam powers in this environment has not yet been done in practice. Assessment of the associated issues must be performed, building on the work done by MAP, including detailed assessment of solenoid feasibility accounting for radiation and thermal loads in liaison with the magnet team. Handling of the spent proton beam must be studied in detail.
\end{itemize}
Other important challenges have been identified.
\begin{itemize}
\item A system to separate the charge species has not been designed. In the baseline design the system must accommodate very large emittance which is not readily manageable using a conventional system, while maintaining the bunch structure.
\item A design for the longitudinal capture system exists. In order to enable pion yield optimisation studies to continue, the sensitivity to incident beam parameters must be understood.
\item A design for the bunch merge system was done but the lattice is no longer available. A new lattice must be generated and sensitivity to parameters must be performed.
\item It is very likely that the overall system performance may be greatly increased by looking at alternative designs to the baseline described above. This could dramatically improve the facility performance.
\end{itemize}

\subsubsection{Recent achievements}
\begin{figure}
\includegraphics[width=\textwidth]{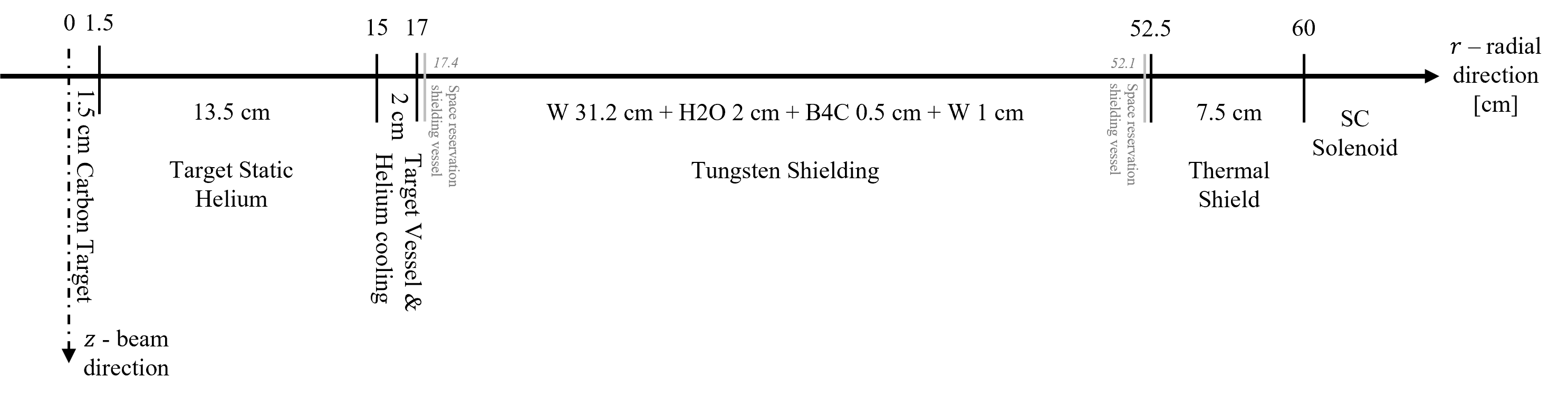}
\caption{Target assembly radial build showing different azimuthal regions from the target the axis up to the SC solenoid coils. \label{fig:pion_production_target}}
\end{figure}
The target team have designed a baseline radial build for the pion production target, taking into account radiation flux into the surrounding solenoid, heat load on the graphite target itself and appropriate cooling systems for the target and shielding (detailed in Sections~\ref{sec:Section7_4} and \ref{sec:Section7_5}). The radial build is shown in Fig.~\ref{fig:pion_production_target}. A 3D FLUKA model for the target area has been constructed and the pion and muon yield through the target system has been assessed. The magnet team have identified HTS technology as having greater tolerance to radiation issues. The target team have updated the design with reduced shielding. Preliminary concepts for the spent proton beam handling have been studied.

\begin{figure}
\includegraphics[width=\textwidth]{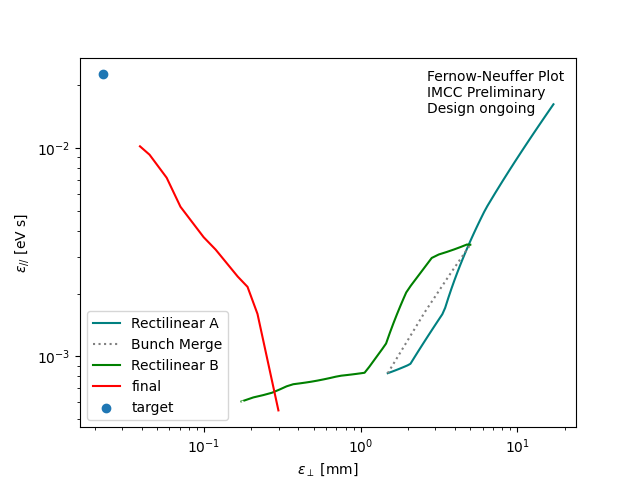}
\caption{System performance of the muon cooling system, including an updated rectilinear and final cooling system. Optimisation continues; in particular the rectilinear cooling system final emittance has not been integrated with the final cooling system. \label{fig:fernow_neuffer_plot}}
\end{figure}

Beam physics designs have been made for the ionisation cooling systems. Two cooling systems have been studied, the rectilinear cooling system, used at high and intermediate emittances, and the final cooling system, used for low emittances. The optical parameters of the solenoid system have been assessed and parameter dependencies established, including non-linear optics dependencies on momentum and transverse emittance. Possible lattices using different RF technologies have been assessed. A preliminary system optimisation for both lattices has been performed yielding potentially improved performance. The current performance estimate for the ionisation cooling system is shown in Fig.~\ref{fig:fernow_neuffer_plot}. The performance of both the rectilinear and final cooling systems have been improved yielding significantly smaller transverse and longitudinal emittance compared to previous baselines. Optimisation continues and it is expected that the simulated performance will improve further.

\subsubsection{Planned work}
The target design will continue with increasing detail. Systems for cooling of the shielding, moderation and absorption of neutrons will be refined. Improved models for the incident proton beam will be considered and the design for the upstream beam window will be updated. Preliminary studies for the handling of the spent proton beam have been envisaged but a detailed design does not exist. The existing beam physics design for the chicane and proton absorber will be refined and an engineering design for handling of the significant spent proton beam in this region will be developed. The 1.5\,T solenoid field in this region will likely require superconducting technology. An appropriate collimation and shielding system will be designed to protect the solenoids in this region.

The intermediate emittance rectilinear cooling system design will be further optimised. The optimisation will be extended to include the high emittance rectilinear cooling system. The optimisation will be refined to include a more detailed assessment of the limitations of solenoid and RF fields, taking into account the required field profiles and highly compact nature of the lattice. An integrated design will be completed for a small subset of the cooling lattices that are proposed.

The final cooling system design will be further optimised. The lattice design will be refined in collaboration with magnet and RF experts. Absorber heating increases rapidly at lower beam emittance and lower momentum characteristic of the downstream end of this system. Schemes to mitigate the impact on the liquid hydrogen will be assessed. Impact of misalignment and field errors will be assessed. Other collective effects will be assessed. The final cooling system will be integrated with the final emittance of the rectilinear cooling system. A system to reaccelerate the beam to relativistic speeds will be studied.

Limitations arising from space charge and beam loading will be assessed to estimate a maximum beam current that can traverse the cooling system.  A preliminary estimate of tolerances to alignment will be performed. A preliminary design for the absorber will be made including a preliminary heat load assessment. A preliminary assessment of radiation load from muon decays will be performed and collimation systems will be considered. Beam instrumentation requirements will be considered. Simulation in alternate codes will be performed to validate the performance estimates.

For all systems, a future R\&D path will be developed including an assessment of necessary equipment tests. Detailed studies of a rectilinear cooling channel cell for a cooling Demonstrator are described below.

\subsubsection{Next priority studies and opportunities for additional collaboration}
Several studies with lower priority will not be carried out, assuming current resource levels continue. In the target region, effects of misalignments and field errors will not be studied. Mitigation schemes, to handle the case that the baseline design can not accommodate the required beam power or the requirements change, will not be studied in detail. Equipment tests will not be possible but a plan will be developed.

Integrated engineering will be performed for only a representative subset of the rectilinear cooling cells. Beam instrumentation will not be studied in detail, but is expected to be feasible based on existing facilities. Correction schemes for misalignments will not be studied. Low frequency acceleration systems such as induction linacs, that may be advantageous for the final cooling system, will not be studied.

Integration systems, key for a start-to-end simulation, will not be developed. Consequently the muon production system will carry significant performance, risk, cost and power uncertainty. These integration systems are unconventional and technically challenging. In particular, the charge separation system, which has no current design, will not be developed. This system must separate muon charge species before cooling, where the beam that has transverse and longitudinal emittances that are expected to be too large for a conventional dipole-based scheme to manage without significant beam degradation.

The longitudinal capture and bunch merge system, which have beam physics designs, will not be developed further. Both of these systems have complex arrangements of RF cavities, operating at several frequencies, for simultaneous manipulation of several bunches. The schemes would benefit from analysis of the challenges in the RF systems. The bunch merge system additionally incorporates a challenging transverse funnel.

The baseline subsystems were chosen as they have the most mature design. Alternative schemes are likely to improve performance significantly, but have lower Technology Readiness Level. For example an emittance exchange scheme, employing a wedge absorber and dipole, could yield better performance compared to the final cooling scheme with simpler technology. A combined `HFoFo' channel would be capable of cooling two charge species simultaneously, potentially yielding a more cost- and power-efficient cooling system. Frictional cooling, ring coolers, Parametric Ionisation Cooling channels and helical cooling channels may all yield improved cost, power or luminosity performance. None of these alternatives will be studied.

\subsection{Acceleration 
}
\label{sec:Section6_4}
\subsubsection{Low-energy acceleration 
}
\paragraph*{System overview} 
The  low energy section  involves three superconducting linacs operating at 352\,MHz and 1056\,MHz: a single pass linear pre-accelerator (PA) followed by a pair of multi-pass ‘Dogbone’ recirculating linacs (RLA).  In the presented scenario, acceleration starts after final cooling at 255\,MeV/c and proceeds to 63\,GeV, where the beam is going to be injected into a first rapid cycling synchrotron. A schematic of the low energy section is shown in Fig.~\ref{fig:lowEnSchematic}. The 352 MHz linac, which has a large aperture, is sufficient to transmit a beam that has received relatively little cooling. The 1056 MHz cavities linearise the RF waveform to minimise the growth of uncorrelated energy spread in the beam.


\begin{figure}[htb]
    \centering
    \includegraphics[width=\textwidth]{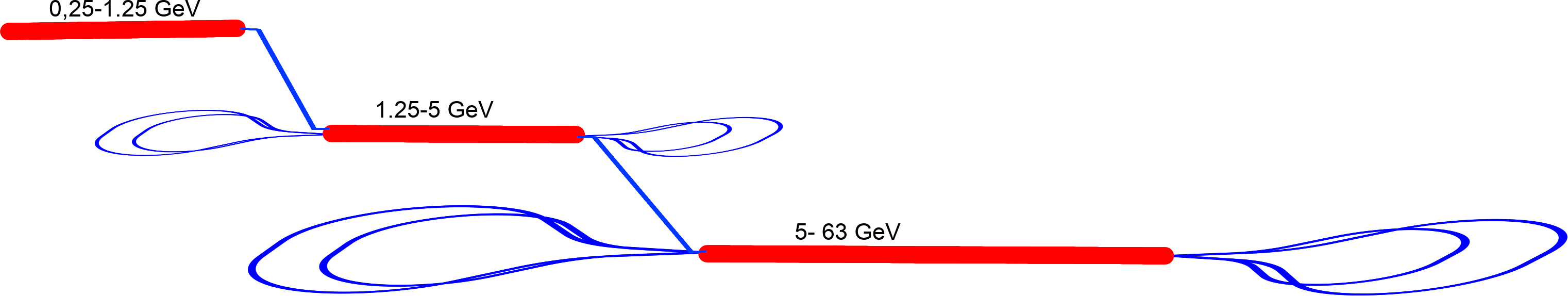}
    \caption{ Layout of a two-step-Dogbone RLA complex. Pre-accelerator, Dogbone I and Dogbone II are stacked up vertically; $\mu^{\pm}$ beam can be transferred between the accelerator sections  by the vertical dogleg.}
    \label{fig:lowEnSchematic}
\end{figure}

\paragraph*{Key challenges}
To ensure that the survival rates of  muons are sufficient, the acceleration must be done at high average gradient. Since muons are generated as a secondary beam they occupy large phase-space volume. In addition to providing high average gradient, the accelerator must have very large transverse and longitudinal accelerator acceptances. 

For the given longitudinal emittance, in order to accelerate of the muon beam within the given transverse and longitudinal emittance tolerances, the beamline must be designed to minimize transverse chromatic effects, thus tight focusing in bending plane with weak quadrupoles. In addition to preservation of the longitudinal emittance, the bunch length needs to be precisely controlled  in the arcs. 

The weak FODO channel chosen to minimize chromatic effects for first passages (where the beam energy is low) will lead to large betatron amplitudes at high energies, making the linac vulnerable to transverse wakefield effects. Considering the high bunch charge to be accelerated, the initial offset of the beam needs to be controlled precisely. 

Another key challenge is the beam loading effect due to the high bunch charge. The following bunch will always gain less energy due to the beam loading effect of the the preceding bunch. Since both bunches will be recirculating in the same arc, the energy difference will lead to a transverse orbit deviation from the reference orbit in the arc, which will in turn lead to a transverse beam offset in the following linac pass. This offset will create a wakefield in the linac that will amplify the beam offset and potentially lead to beam loss or degradation. Therefore the beam loading effect needs to be controlled either by exchanging bunch positions or some other methods. 

\begin{figure}[h!]
    \centering
    \includegraphics[width=\textwidth]{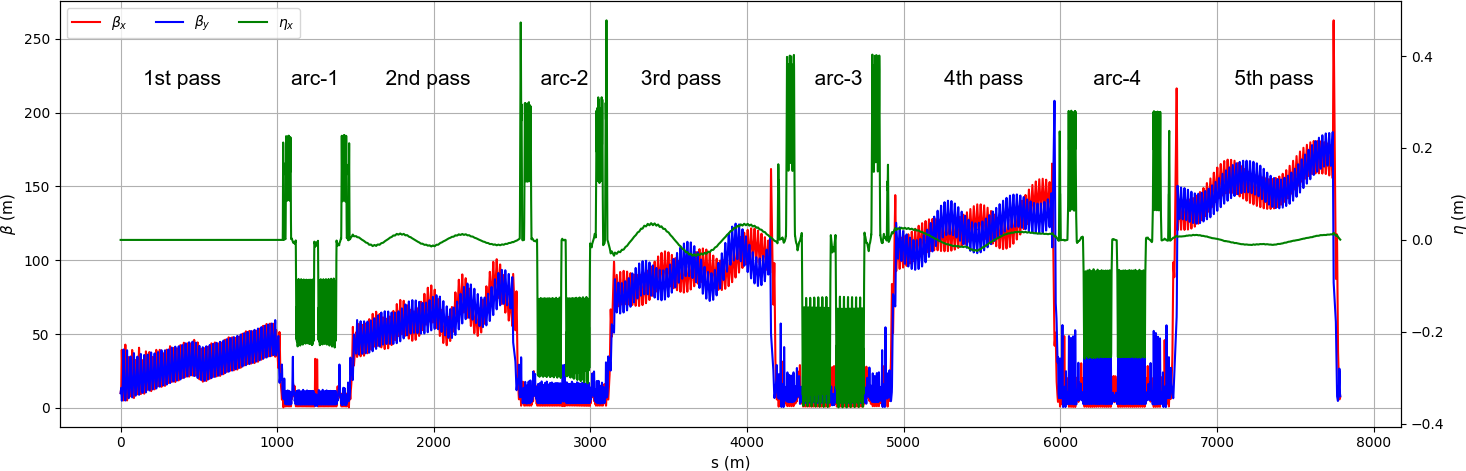}
    \caption{ Start-to-end Twiss functions along RLA2}
    \label{fig:s2e_twissRLA2}
\end{figure}

\paragraph*{Work progress since the publication of the European LDG roadmap}
We created a preliminary lattice design for RLA2. We used a FODO focusing structure in the linac to reduce chromatic effects at low energy.  Several lattice configurations were considered for the recirculating arcs including FODO, Triplet, Theoretical Minimum Emittance (TME) and Double Bend Achromat (DBA). The DBA showed better performance. An initial bunch with 100\,mm.rad longitudinal and 40\,mm.mrad transverse emittance could be accelerated from 5\,GeV to 63\,GeV while preserving longitudinal emittance, a 25\% increase in transverse plane, and a better than $90$\% muon survival rate. Figure~\ref{fig:s2e_twissRLA2} shows the linear lattice  functions along the RLA2. Because the quadrupole gradients in the linac are the same for each linac pass, the increasing beam energy in later passes leads to large beta functions in later passes. Table~\ref{tab:RLA} contains the main parameters for the simulated RLA2 and the predicted RLA1.
 \begin{table}[h]
     \centering
          \caption{Preliminary parameters for low energy acceleration.}
     \begin{tabular}{ccccccc}
     \hline\hline
     \textbf{Parameter} & \textbf{Unit} & \textbf{PA} & \multicolumn{2}{c}{\textbf{RLA1}}  &  \multicolumn{2}{c}{\textbf{RLA2}}\\
     \hline
         Injection energy & [GeV] & 0.255 & \multicolumn{2}{c} {1.25} & \multicolumn{2}{c} {5} \\
         Ejection energy & [GeV] & 1.25 & \multicolumn{2}{c} {5} & \multicolumn{2}{c} {63} \\
         RF Frequency & [MHz] & 352 & 352 & 1056 & 352 & 1056\\
         Number of cavities & \# & 30 & 25 & 2 & 320 &56 \\
         RF gradient & [MV/m] & 25 &  25 & 30 & 25 &30 \\
         Energy gain / pass & [GeV] & 0.955 &  \multicolumn{2}{c}{0.8} & \multicolumn{2}{c}{11.6} \\
         Passes & \# & 1&  \multicolumn{2}{c} {5} & \multicolumn{2}{c} {5} \\
     \hline\hline
     \end{tabular}
     \label{tab:RLA}
 \end{table}

\paragraph*{Work planned for the evaluation report}
The following works are planned as a next step as voluntary effort and provided resources can be made available.
\begin{itemize}
    \item No work has been performed for PA and RLA1. Our next step will be to create designs for these lower energy parts.  The large acceptance of especially in PA, requires large apertures and tight focusing. Combined with moderate beam energies, this favors solenoid rather than quadrupole focusing for the entire PA linac. 
    \item  We have made significant progress with the lattice design in RLA2. However, the design still does not meet our performance requirements (see above) even without machine imperfections. We will therefore continue to improve this design until it meets our requirements.
    \item The gradients estimated in simulations are somewhat optimistic. All simulations will be repeated for lower acceleration gradients: ~16\,MV/m for 352\,MHz, 30\,MV/m for 1056\,MHz. 
    \item The impact of machine imperfections and operation strategies  balancing contradicting requirements will be implemented in simulations
\end{itemize}

\paragraph*{Next priority studies and opportunities for additional collaboration}
The expected gradient and balancing beam loading effect for the accelerating gradient plays a main role on the muon survival rate in the low energy section. 

An initial 6D phase space distribution is essential for performing more realistic simulations in the low energy acceleration chain of muon collider. Our simulations have thus far assumed a Gaussian distribution, but the upstream systems could deliver a beam very different from that. Simulations that provide that distribution will not be available during this study, if for no other reason than the systems upstream of the PA will not be fully designed.

\subsubsection{High-energy acceleration 
}
\paragraph*{System overview 
}
For the high-energy acceleration from \SI{63}{\GeV} to the TeV range, a chain of rapid cycling synchrotrons~(RCS) is foreseen to accelerate the two counter-rotating bunches with a repetition rate of \SI{5}{\Hz}. 
One option, inspired by the US Muon Acceleration Program (MAP) \cite{Berg,MAP2}, includes intermediate stages of \SI{0.30}{\TeV}, \SI{0.75}{\TeV} and \SI{1.5}{\TeV} to finally reach \SI{5}{\TeV} before injecting the muons into the separate collider ring. This scenario is illustrated in Fig.~\ref{fig:RCSs}. 
\begin{figure}[ht!]
\centering
{\includegraphics[width=0.85\columnwidth]{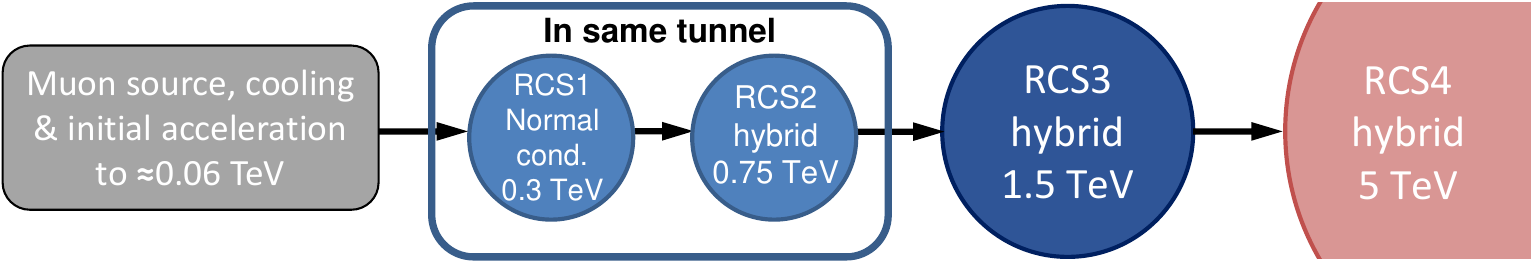}}
\caption{Sketch of the chain of rapid cycling-synchrotrons for the high-energy acceleration complex. From Ref.~\cite{ipac2023Batsch}.}
\label{fig:RCSs}
\vspace*{-1\baselineskip}
\end{figure} 
The first two RCS share the same tunnel, meaning that they have the same circumference and layout~\cite{Berg2}. The bending in the first RCS is provided by normal conducting magnets. The RCS2 to RCS4 are planned as hybrid RCSs where strong fixed-field, superconducting magnets are interleaved with normal conducting magnets cycling from $-B_{\mathrm{nc}}$ to $+B_{\mathrm{nc}}$. 
This combination allows for a large energy swing combined with a high average bending field to minimize the radius and thus the travel time and decay of the muons. 

\paragraph*{Key challenges}
To keep decays acceptably low, the RCSs must have high average gradients $G_\text{avg}$, as high as  \SI{2.4}{\mega\volt/\metre}. The average gradient is the energy gain in one turn divided by the circumference; since most of the ring does not contain RF, the gradient in the RF sections must be significantly higher than that average gradient. Assuming a survival rate of 90\% per RCS, ultra-fast acceleration with tens of GeV energy gain per turn is required. 
The RF voltage has therefore to be provided by hundreds of superconducting cavities. To both provide large gradients
, ILC-like \SI{1.3}{\GHz} cavity structures, also known as TESLA cavities~\cite{TESLA}, have been assumed for the simulations.
As a result of the tens of GV of RF voltage, the number of synchrotron oscillations per turn is much larger than the conventional stability limit for stable synchrotron oscillations and phase focusing of $1/\pi$~\cite{Pi} in a synchrotron with one or few localized RF sections. As a consequence, the RF system must be distributed over the entire RCSs in order to reduce the synchrotron tune between two consecutive RF stations to a value much smaller than that instability limit.

As a direct result of the decay and ultra-fast acceleration, the acceleration times are in the millisecond range and the number of turns in each RCS is between 17 and 66. In the RCS chain, the ramp rate of the normal conducting magnets is on the so far unprecedented order of kT/s, a particular challenge. Further, due to the hybrid structure of the three downstream RCSs, beam orbits and radii are not constant anymore, but change locally, as depicted in Fig.~\ref{fig:RCS_orbit_RCS2}.
\begin{figure}[htb]
  \begin{center}
    \includegraphics[width=0.5\textwidth]{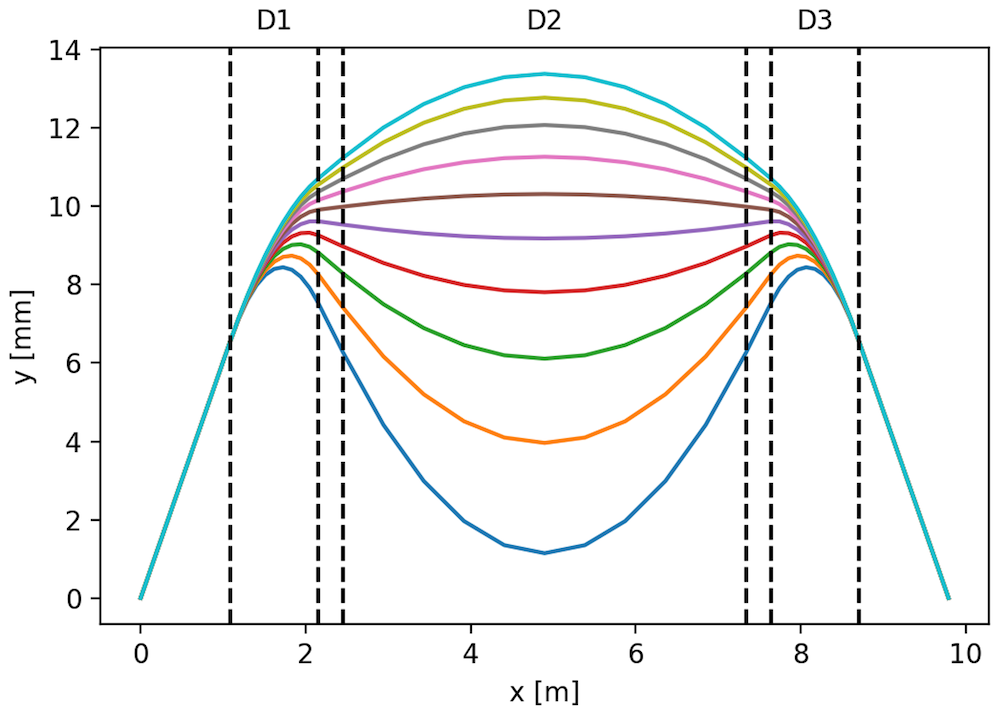}
  \end{center}
\caption{Trajectories in a half-cell of RCS2. The injection/extraction orbits
are respectively in the inner/outer side (in blue/cyan). From Ref.~\cite{ipacAntoine}.}
\label{fig:RCS_orbit_RCS2}
\end{figure}
Special attention is therefore being paid to the lattice design in order to minimize the orbit length change during acceleration. Fast cavity detuning on the kHz scale to match the RF frequency with the changing revolution frequency have to be developed.

With the special beam structure of counter-rotating, intense single \textmu$^+$ and \textmu$^-$ bunches, large transient beam loading for both fundamental and higher-order modes must be expected.

\paragraph*{Recent achievements}
To simulate longitudinal beam dynamics, we use the longitudinal macroparticle tracking code \mbox{BLonD~\cite{Blond,Blond2}}. The code was successfully extended to model multi-turn wakefields in the multiple RF stations per ring. A plot of the longitudinal phase space at injection for RCS1 
is displayed in Fig.~\ref{fig:RCS_emittance_growth}(a).

Tracking simulations on the influence of the number of RF stations on the longitudinal emittance and beam stability have been performed for all RCSs. As shown in Fig.~\ref{fig:RCS_emittance_growth}, their minimum number is around 32~RF~stations for RCS1 and RCS4, and 24~stations for RCS2 and RCS3~\cite{ipac2023Batsch}. 
\begin{figure}[htbp]
  \begin{center}
  \includegraphics[width=0.45\textwidth]{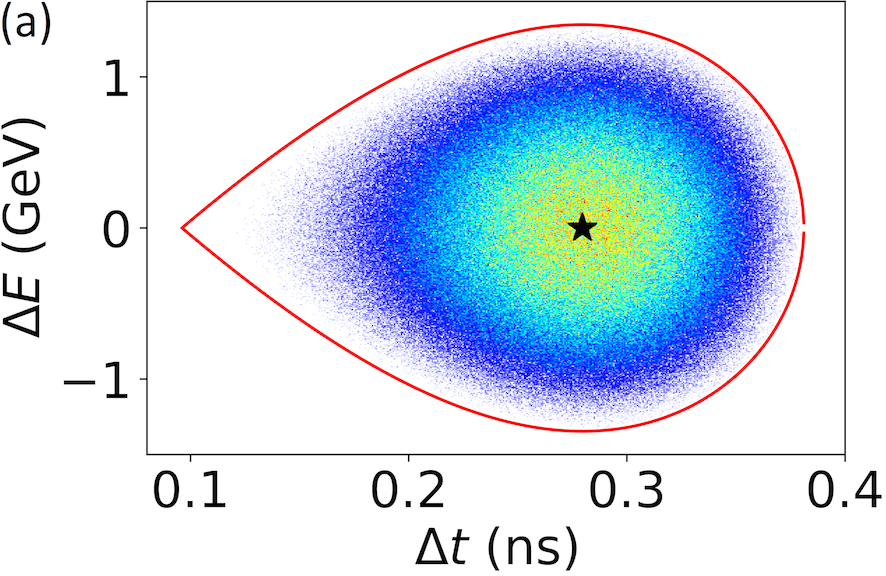}
  \includegraphics[width=0.45\textwidth]{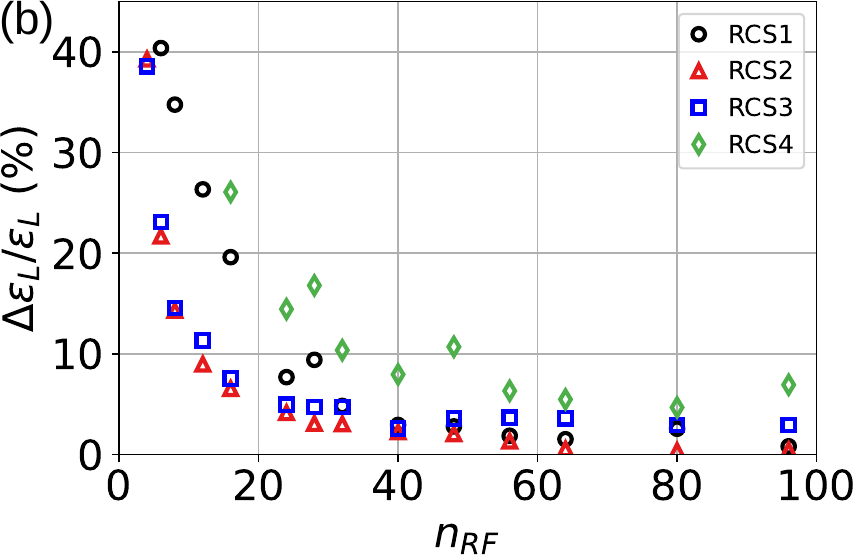}
  \end{center}
\caption{(a) Longitudinal phase space for a $\mu^+$ bunch in RCS1 for $n_{\mathrm{RF}}=32$ at injection. The red line indicates the separatrix (with continuous energy approximation) for $f_{\mathrm{RF}}=\SI{1.3}{\GHz}$. \\
(b) Relative emittance growth at the end of the cycle of each RCS with respect to the emittance at injection versus $n_{\mathrm{RF}}$ for RCS1 to RCS4 from simulations without intensity effects.}
\label{fig:RCS_emittance_growth}
\end{figure}
This allows to keep the longitudinal emittance growth close to the required 5\% level (not taking into account intensity effects nor phase and energy errors at injection). 

A first lattice of the RCS2 has been developed with Xsuite~\cite{bib:xsuite}. In the current version (see Fig.~\ref{fig:RCS_lattice_RCS2}, the RCS2 consists of 26 arcs. 24 out of the 26 insertions between the arcs host some cavities. The two remaining insertions will be dedicated to injection and extraction. The arc half-cells house three dipoles: two superconducting dipoles (\SI{10}{T}) on the outside, and a pulsed normal conductor with a peak field of \SI{1.8}{T} in the middle.

\begin{figure}[htb]
  \begin{center}
    \includegraphics[width=0.8\textwidth]{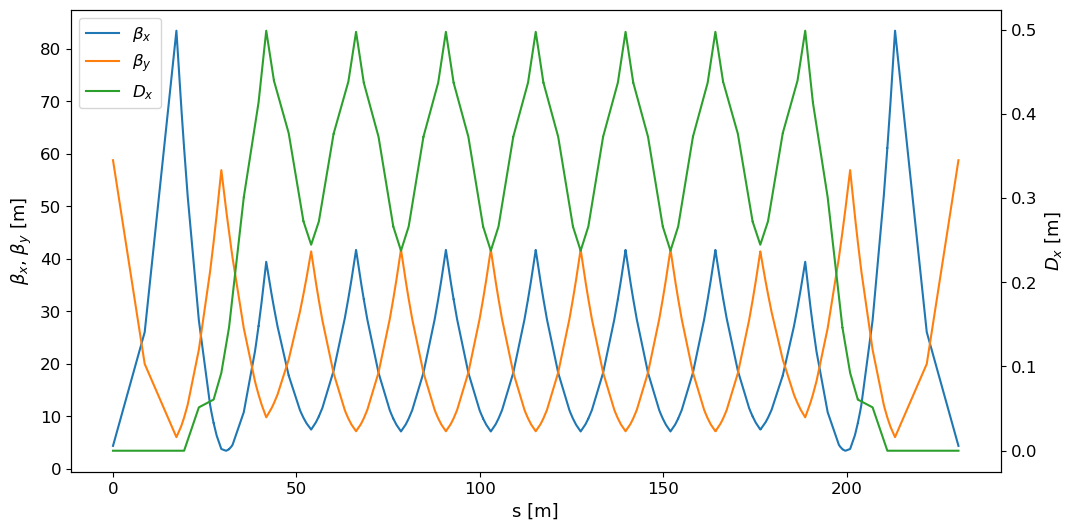}
  \end{center}
\caption{Twiss functions in one out of 26 arcs of RCS2.}
\label{fig:RCS_lattice_RCS2}
\end{figure}

Simulations that include the intensity effects of the single but intense muon bunch show the effect of short-range and long-range induced voltages. The short range wakefields, calculated once following the theory of K.~Bane~\cite{Bane} and once using a resonator model for all fundamental and higher-order modes of an ILC-like cavity, show induced voltages on the order of \SI{1}{\mega\volt} per meter (and per cavity), i.e., stay below 10\% of the cavity voltage without beam. The effect of multi-turn wakefields due to HOMs (higher-order modes) on the longitudinal stability was investigated and the power which is induced into these HOMs was studied as a function of their quality factors~\cite{HB23Batsch}. 

The orbit length variation $\Delta \mathcal{C}$ has been minimized by optimizing the layout of the FODO cells~\cite{ipacAntoine} so as to reduce the range of tuning required for the RF cavities and their power sources. With the current design, values reach $\Delta \mathcal{C}/\mathcal{C}\cdot f_\text{RF}\approx\SI{2}{\kHz}$ in RCS2, resulting in a tuning requirement of $\Delta \mathcal{C}/\mathcal{C}\approx\num{1.5e-6}$, equivalent to a peak tuning speed of approximately \SI{10}{\MHz/\s}. 

At this state, giving some tolerances on the magnets and thus $\Delta \mathcal{C}$ becomes very challenging. Due to the fast acceleration, it becomes very difficult to correct any dynamic errors like a jitter on the powering of the magnets. The lattice should enable some passive mitigation of this jitter. The task is even more challenging because it should work also for a counter-propagating beam. A main concern is the eddy currents generated in the pulsed magnets like the normal conducting dipoles. 

\paragraph*{Planned work}

The longitudinal tracking and RF studies are well advanced for a single beam, but the transient beam loading in the RF cavities due to the counter-rotating beams remains to be investigated. The impact of fundamental and higher-order mode voltages induced in the RF cavities on the longitudinal beam dynamics will be studied in detail. We will not place RF cavities at points where the two opposite sign muon beams cross to prevent the short range wakefield induced in the cavities by one beam from affecting the other one.

The present parameter set for the RCS chain (Table~\ref{tab:ch07:RCS_RFpars}) has been derived assuming the same muon survival rate for each of the stages. This results in very different requirements for the total RF voltages. While the first RCS must be equipped with almost \SI{21}{\giga\volt}, the RF voltage is smaller for the following two stages, despite the fact that RCS3 is significantly larger than the first two accelerators. Parameter optimization of the sequence of RCSs, in particular with respect to survival rate and the energy levels at which beam is transferred, will be performed to reduce the overall RF voltage installed in the entire chain and to maintain the same average accelerating gradient through the entire RCS chain.

The required RF voltage for a given transmission is smallest for a perfectly linear energy ramp. This would require that the current driving the pulsed magnets in the RCS varies linearly with time. While a close approximation to this is possible, that would require a very costly power source for the magnet. It would be less costly to drive the magnets with a current pulse that is closer to a portion of a sine wave. However, a nonlinear rise of the field must be compensated by additional RF voltage to reach the same muon survival. First discussions on cost models for the RF system and the bending magnets with their power converter have started, but there are not yet at the level of detail and consistency to allow cost optimization. This effort will continue in 2024, adding also software tools to extract cost estimates.

The work on RCS lattice, started only recently, will be continued and extended to cover the whole RCS chain and possible limitations. Special emphasis will be to ensure a sufficient energy acceptance and a careful evaluation of possible synchro-betatron resonances
The effect of deviations in the magnetic field strength of fast ramping magnets on the beam and possible mitigation method have to be studied. Code development will be pushed to enable start-to-end simulations and integrate the magnetic ramp. Such simulations will give a first estimate of the emittance growth during the acceleration.


\paragraph*{Next priority studies and opportunities for additional collaboration}
While the RCS injection chain accelerates the bunch to the energy required for the collider ring, its current design does not deliver the longitudinal bunch parameters required. The short bunches required pose tight constraints on the longitudinal emittance growth in the entire RCS chain. 
Scenarios where the longitudinal bunch parameters for the collider are not achieved during the normal course of acceleration in the final RCS should be designed and simulated (for example, a longitudinal phase space rotation could be performed on the bunch in less than one turn).

Furthermore, the present RF frequency for the muon acceleration of \SI{1.3}{\GHz} has been taken as an assumption, as detailed cavity designs are available for the International Linear Collider (ILC). Although the frequencies around \SI{1.3}{\GHz} are well suited in terms of achievable gradient, further studies considering bucket area, beam loading and wakefields could force to move to a lower RF frequency. Once the suitable frequency range has been fixed, the exact RF frequency remains to be determined based on arguments like the matching in frequency with the muon pre-acceleration and cooling stages.

While the focus has been on the optimization of separate beam dynamics simulations for the transverse and the longitudinal planes, full simulations of the beam dynamics in six dimensions (6D) have yet to be performed and it remains to be clarified which tracking tool is suitable or could be extended to the parameters of the muon RCS chain.

The strong transient beam loading with the high-intensity single bunches also poses challenges to the low-level RF feedback systems. A gated 1-turn delay feedback is being considered, but the conceptual design must be made to estimate its benefit and RF power requirement.

The extremely fast acceleration of muons in RCSs extends the synchrotron concept in uncharted territory in terms of parameters, which are beyond those of present accelerators. Already the conventional design tools had to be carefully scrutinized and extended to make sure that they remain applicable for the muon RCSs. No existing RCS has a comparable circumference, accelerates similarly fast and with such a large synchrotron tune. In addition the concept of interleaving super- and normal-conducting magnets is novel.

We could use existing accelerators to test operation with parameters at least beginning to approach the parameters required for the muon RCSs. For example, synchrotron tunes as large as $0.1$ could be tested with ions at injection energy in the CERN SPS, to benchmark the consequences of discretised synchrotron motion. Further beam tests may be possible in the existing RCS at JPARC or the \SI{8}{\GeV} Booster synchrotron at Fermilab, both accelerating proton beams within few $10^4$~turns.

\subsection{Collider ring 
}
\label{sec:Section6_5}

Present work on the muon collider ring design focuses on a 10\,TeV center of mass machine.

\subsubsection{Collider design overview}
The optics design of the collider design comprises:
\begin{itemize}
  \item The interaction regions (IR) with a straight section housing the detector and an inner focusing triplet. This is followed by a longer (almost) straight section possibly with weak bendings and/or quadrupoles. In the present scenario, injection, extraction for beam dumping, RF and other systems will use the straights available in this region.
  \item A chromatic compensation section (CCS) correcting chromatic aberrations generated due to the small $\beta^*$ required to achieve sufficient luminosity. 
  \item A matching section (MS) connecting the CCS to the arc cells. Present efforts aim at designing an MS with sufficient tuning margin for the betatron phase advances to allow adjustments of the working point.
  \item The arc will be made of negative momentum compaction cells required to tune the momentum slip factor of the whole collider ring to very small values required to keep the bunches short ($\approx 1.5$\,mm) over about 1000 turns contributing to luminosity.
\end{itemize}

The following effects related to muon decays have to be mitigated:
\begin{itemize}
  \item Electrons and positrons lead to showers either by synchrotron radiation generated by them towards the outside of the ring or where they impact on the inside of the ring. Tungsten absorbers installed inside the magnet aperture together with residual heat load and radiation to the magnet are described in Section~\ref{sec:Section7_5}. The cryogenic system is dimensioned to cope with the residual heat load from muon decays and is described in Section~\ref{sec:Section7_7}. 
  \item Neutrinos leave the ring almost tangential to the design trajectory and reach Earth's surface close to the plane defined by the collider ring with hot spots in the direction of straight sections. Dose levels generated by these muons reaching Earth's surface or rather showers created by them are described in Section~\ref{sec:Section7_10}. Measures to ensure that dose levels remain at negligible levels are described and comprise to limit the length of straight sections outside the IR and mechanical system described in Section~\ref{sec:Section7_12} to periodically deform the machine in vertical direction. 
\end{itemize}

\begin{figure}[htbp]
  \begin{center}
  \includegraphics[width=12cm]
  {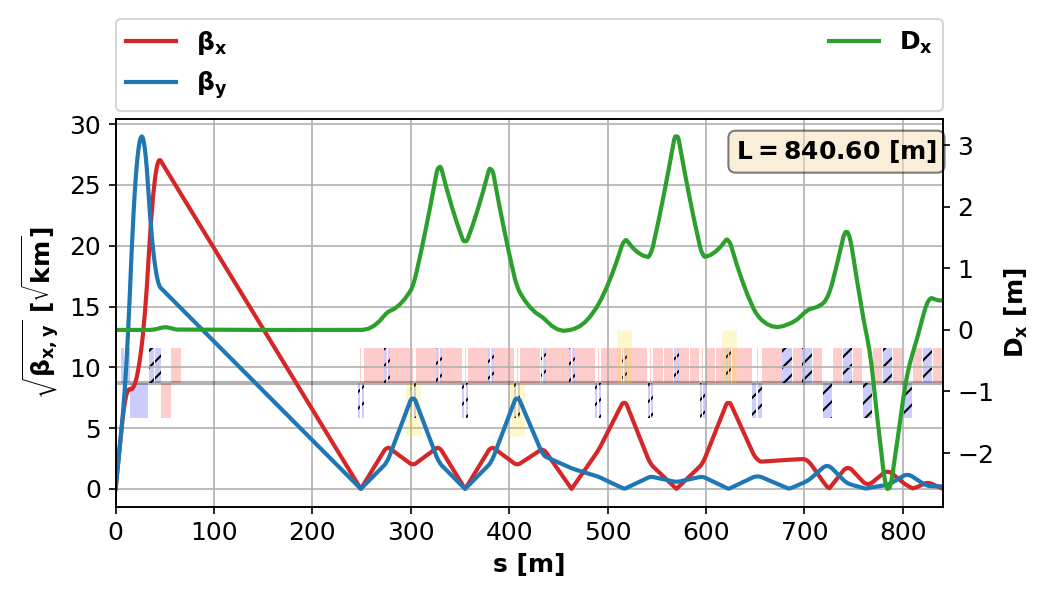}
  \end{center}
\caption{Twiss functions in the collider ring from the IP through the final focusing, the CC section up the beginning of the matching section.}
\label{fig:ColliderIPCC}
\end{figure}

\subsubsection{Key challenges}
A major challenge of the optics design are the chromatic aberrations caused by the small Twiss betatron function at the interaction point $\beta^* = 1.5 \mbox{ mm}$ and the large rms momentum spread of $\sigma_\delta = 10^{-3}$ together with the high beam rigidity and maximum quadrupole gradients. Twiss parameters in the most challenging part of the muon collider ring are shown in Fig.~\ref{fig:ColliderIPCC}. The design of the the local CC section to correct these aberrations and, in particular, the control of higher order effects is underway. The small $\beta^*$ and the large beam rigidity unavoidably leads to large Twiss betatron functions at location of strong magnets around the final focusing structure and, to a lesser extent, in the CC section. A consequence is stringent requirements on unwanted multipolar components to be evaluated once a working lattice for a perfect machine is available. The rms bunch length has to be kept short (rms $\approx 1.5 \mbox{mm}$) for about 1000 revolutions contributing to luminosity production. This leads to stringent requirements for the momentum compaction factor including higher order contributions and the need to control the (average) dependence of the path length from the betatron oscillation amplitudes using sextupoles. It has not yet been decided whether a high gradient high frequency RF system will be used to slightly mitigate these constraints. 

Pushing the luminosity requires to push the  fields of the magnets (see Section~\ref{sec:Section7_1}) to the maximum possible with existing technologies or, if feasible, to develop new concepts. This is necessary to reduce the machine circumference allowing to increase the number of revolutions contributing to luminosity production and for a compact design of the final focusing structure.

The dose levels, where neutrinos reach Earth's surface, will be kept at negligible  levels with the help of mitigation measures as installation of the collider ring deep underground, avoidance of straight sections in regions outside the straights housing the experiments by adding dipolar components to quadrupoles and higher order multipoles (becoming combined function magnets) and wobbling, i.e.~periodic deformation of the ring in the vertical direction as described in Section~\ref{sec:Section7_12}, of the machine. Challenges are the required high precision movement system and to ensure that adverse effects on the machine optics and performance can be controlled.

Challenges generated by muon decay products (electrons and positrons and synchrotron radiation generated by them) and showers generated by them and their interception by tungsten absorbers is described in Section~\ref{sec:Section7_5}. Challenges due to the heat load from residual shower particles not intercepted by the absorber are described in Section~\ref{sec:Section7_7} on the cryogenic system.

The machine-detector-interface (MDI) with tungsten absorbers and nozzles to minimize beam induced background seen by the detector is described in Section~\ref{sec:Section04_2}.

\subsubsection{Work since the publication of the European LDG roadmap}

Substantial progress has been made on the design of a lattice for a 10\,TeV center of mass collider \cite{EPS2023}. Despite significant improvements of the transverse acceptance for off-momentum particles, a lattice working for nominal beam parameters is not yet available even for a perfect machine. Nevertheless, relaxing slightly requirements (increasing $\beta^*$ and the bunch length and decreasing the momentum spread by the same factor) would allow to find such a lattice.

Progress of Beam Induced Backgrounds (BIB) studies and Machine Detector Interface (MDI) design optimizations are described in Section~\ref{sec:Section04_2}. BIB mitigation strategies elaborated with the collider lattice design are described there.

Progress on studies of radiation at Earth's surface caused by neutrinos generated by muon decays comprise FLUKA studies, an improved "source term" describing radiation from a decaying muon beam without divergence and a procedure to take the properties of the lattice into account. A detailed description is given in Section~\ref{sec:Section7_10}.

\subsubsection{Work planned for the evaluation report}

Optimization aiming at designing a lattice with sufficient transverse acceptance over the whole relevant momentum with the nominal $\beta^*$ will be continued. If no such lattice is identified, parameters may have to be slightly relaxed (increase of $\beta^*$ and bunch length leading to smaller momentum spread).

Starting with a working lattice for a perfect machine without machine imperfections, realistic scenarios will be investigated. Imperfections to be added comprise component misalignments, magnet strength errors (possibly caused by limited precision of power converters) and unwanted multipolar components of magnets. Initially, a lattice without $\beta^*$ squeeze will be assumed. Afterwards, correction of imperfections based on observations with the beam will be simulated. As a result of these simulations, specifications for component alignment tolerances, maximum magnet strength uncertainties and maximum unwanted multipolar components will be obtained. In case some of these conditions, e.g., maximum unwanted higher magnetic field multipolar components cannot be met in practice, an optimized lattice design may allow to improve and parameters may have to be relaxed.

The impact of periodic machine deformations to mitigate radiation from neutrinos reaching Earth's surface described in Sections~\ref{sec:Section7_10} and \ref{sec:Section7_12} on beam dynamics will be studied. Machine deformations introduce by design vertical dispersion inside the challenging CC section with large betatron functions and strong sextupoles possibly leading to a substantial impact on the dynamic apertures and machine performance.

\subsubsection{Next priority studies and opportunities for additional collaboration}

Present studies concentrate on a \SI{10}{\tera\eV} center of mass collider and no further investigations on a lower energy machine are underway. In case a staged implementation of muon collider facility with a lower energy collider as first stage is envisaged, thorough studies in addition to the existing \SI{3}{\tera\eV} center of mass muon collider design from MAP\cite{MAP2} are necessary.

Interplay of machine nonlinearities and beam-beam effect have not yet been looked into in detail but drive the design and alignment tolerances to ensure that the expected performance is reached. Further studies on beam-beam effect have to be clarified.

\subsection{Collective effects 
}
\label{sec:Section6_6}

\subsubsection{System overview}

Collective effects are a general concern over the whole muon collider, from the proton driver to the collider ring.
Different mechanisms will dominate the system, depending on the particle's type, their energy, the bunch emittance and intensity.

In the proton system, the high-intensity and medium energy of the proton bunches will make them prone to collective instabilities from space-charge effects and interplay with impedance, in particular in the accumulator and compressor rings.
In the muon cooling system, collective effects created by beam-matter interaction in the ionisation cooling system could make the muon bunches unstable or lead to an increase of the emittance.
These effects will be investigated as part of muon production and cooling described in Section~\ref{sec:Section6_3}.
In the muon acceleration system, the two high-intensity, counter-rotating muon and anti-muon bunches will undergo fast acceleration to meet the survival rate goal.
The numerous RF cavities required for acceleration will generate strong wakefields that could degrade the bunch emittance and lead to intensity losses.
Beam-beam effects will also be present in the droplet arcs of the recirculating linacs, at two crossing points in the rapid cycling synchrotrons (RCS) and at the two interaction points in the collider ring.
In the latter, the beam chamber will generate resistive-wall wakefields whose magnitude will be determined by the chamber radius and materials used.

\subsubsection{Key challenges}

In the first RCS, an energy gain per turn of \SI{14.7}{\GeV} is required to reach the \SI{90}{\percent} survival rate of the muon beams.
To obtain such large accelerating gradient, $O(700)$ TESLA superconducting RF cavities operating at \SI{1.3}{\GHz}~\cite{bib:TESLA_cavity} and with an individual accelerating gradient of \SI{30}{\mega\volt\per\m} will be needed, as described in Section~\ref{sec:Section6_4}.
The cavities high-order modes (HOMs) will generate short and long-range wakefields.
Combined with the high bunch intensities of \num{2.7e12}, \num{2.4e12}, \num{2.2e12} and \num{2.0e12} muons per bunch at injection of RCS1, RCS2, RCS3 and RCS4 respectively, the strong wakefields could disrupt the beam motion and lead to emittance blow-up and particle losses.
Moreover the fast acceleration the two bunches will undergo in the RCS (only 17 turns for RCS1) would require the transverse damper to have a fast response to efficiently mitigate the coherent beam instabilities.
Sextupoles magnets could be introduced to correct and control the RCS lattice chromaticity and mitigate potential head-tail instabilities, but this would however require further space in the RCS lattice.

In the \SI{3}{\tera\eV} and \SI{10}{\tera\eV} collider rings the beam chamber will generate short-range resistive-wall wakefields.
To reach the highest possible magnetic field in the dipole, the magnet aperture should be as small as possible and must host a \SI{4}{\centi\metre} thick tungsten shielding to intercept muon decay products, as described in Sections~\ref{sec:Section7_1} and~\ref{sec:Section7_5} for the \SI{10}{\tera\eV} collider.
Because of the absence of synchrotron radiation damping mechanism, the transverse normalised beam emittance of \SI{25}{\micro\metre\radian} must be preserved over the storage time to reach the target luminosity.
The inner radius of the tungsten shield and its material properties, such as the operating temperature or the use of a low electrical resistivity coating, will influence the resistive-wall wakefield and determine the coherent stability threshold.
At the interaction points, strong beam-beam effects will arise from the high bunch intensity of \num{1.8e12} muons per bunch at injection and the small $\beta^*$ target of \SI{1.5}{\milli\metre}.
The short bunch length of $\sigma_z = \SI{1.5}{\milli\metre}$ could also lead to beam induced heating in some devices.
However the strength of collective effects will reduce over the beam storage time in the collider thanks to the muon decay.
With a \SI{5}{\tera\eV} beam, the Lorentz factor $\gamma=47323$ and the muon lifetime in the laboratory frame $\tau$ is $\tau = \gamma \tau_0 = 47323 \cdot \SI{2.2}{\micro\second}=\SI{104}{\milli\second}$, equivalent to $3121$ turns in the \SI{10}{\kilo\metre} long ring.

\subsubsection{Recent achievements}
\label{subsec:Section6_5_3}
A transverse impedance model for the RCS1 was created using the PyWIT framework~\cite{bib:pywit}, assuming the accelerator comprises 670 superconducting TESLA type RF cavities.
The HOMs used are those of the low loss type TESLA cavities described in Ref.~\cite{bib:LL_cavities}.
The three modes with the largest shunt impedance $R_s$ are reported in Table~\ref{tab:rcs_impedance_cavity_HOMs}.
Figure~\ref{fig:rcs1_impedance_model_LL_TESLA_HOMs} shows the real part of the transverse horizontal dipolar beam coupling impedance obtained with this model.
The vertical impedance is assumed to be identical to the horizontal one.

\begin{table}[htb]
\begin{center}
\caption{List of transverse HOMs from the Low Loss TESLA cavity, as described in Ref.~\cite{bib:LL_cavities}.}
\label{tab:rcs_impedance_cavity_HOMs}
\begin{tabular}{ccc}
\hline\hline
  {\textbf{Frequency $f_{\mathrm{res}}$ [MHz]}} & {\textbf{Shunt Impedance $R_s$ [M$\Omega$/m]}} & {\textbf{Q factor [$10^4$]}} \\
  1927.1    & 3865.11    &  1.5\\
  2451.07   & 6155.63    &  10.0\\
  2457.04   & 4314.02    &  5.0\\
\hline\hline
\end{tabular}
\end{center}
\end{table}

\begin{figure}[thb]
  \begin{center}
    \includegraphics[width=0.5\textwidth]{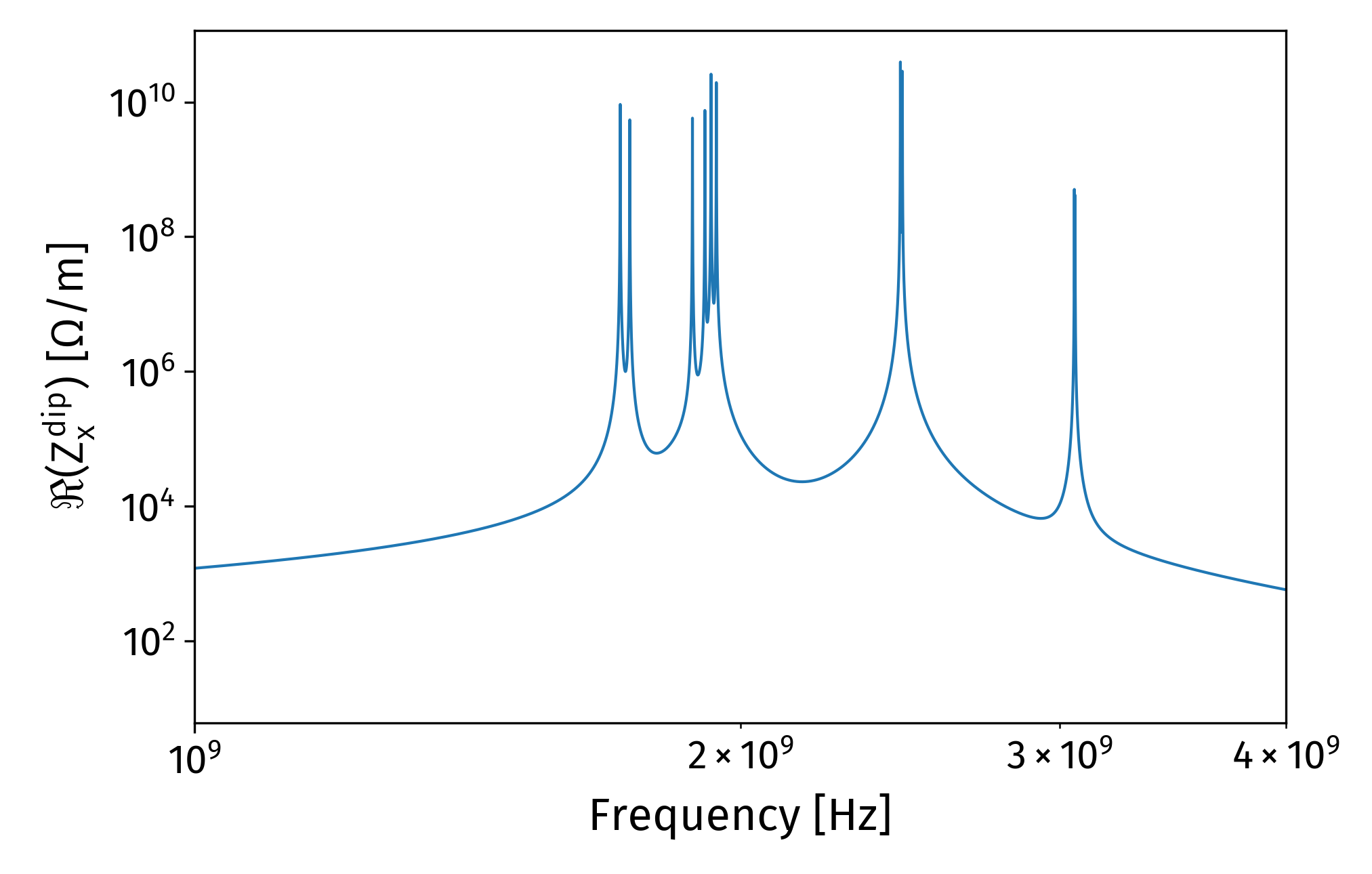}
  \end{center}
\caption{Real part of the horizontal dipolar impedance for the RCS1, including the HOMs created by 670 TESLA type RF cavities.
The plot is zoomed on the \SIrange{1}{4}{\giga\hertz} range to highlight the HOMs.
}
\label{fig:rcs1_impedance_model_LL_TESLA_HOMs}
\end{figure}

Transverse beam dynamics simulations were performed with the PyHEADTAIL~\cite{bib:pyheadtail} macroparticle tracking code.
To handle the large acceleration gradient present in the RCS, the RF cavities must be distributed over multiple RF stations along the ring as detailed in Section~\ref{sec:Section6_4}.
Specific modules were developed to handle in PyHEADTAIL simulations of this particular configuration of the RF cavities~\cite{bib:amorim_rcs_imcc_annual_meeting_2022, bib:Amorim:ipac2023-wepl186, bib:Amorim:hb2023-thbp17}.

A first stability criterion was establish through tracking simulations to find the maximum single high-order mode shunt impedance $R_s$, resonance frequency $f_{\mathrm{res}}$ and quality factor $Q$ admissible in a cavity.
To guarantee single bunch stability with a single HOM, in single-turn wakefield regime $R_s < \SI{100}{\mega\ohm\per\metre} Q / f_{\mathrm{res}}^2$ where $f_{\mathrm{res}}$ is expressed in \si{\GHz} whereas in multi-turn regime $R_s < \SI{10}{\tera\ohm\per\metre}$.
When all the cavity HOMs are included in the impedance model, tracking simulations show that there is a factor of two margin on the total transverse shunt impedance $R_{s, \mathrm{tot}}$ of the modes, assuming that all modes shunt impedance can be summed as we remain in the single-turn wakefield regime.

These simulations were performed assuming that the chromaticity $Q^{\prime}$ is corrected to zero (i.e.~using sextupoles).
Assuming the RCS lattice will be a FODO type and an average Twiss beta function of $\overline{\beta}_{x, y} = \SI{50}{\metre}$, the transverse tunes can be approximated to  $Q_{x, y} \approx C_0 / (2\pi\overline{\beta}_{x, y}) = 19$ with $C_0 = \SI{5990}{\metre}$ the machine circumference.
The chromaticity $Q_{x, y}^{\prime}$ was scanned from \numrange[retain-explicit-plus]{-19}{+19}.
With the impedance model pictured in Fig.~\ref{fig:rcs1_impedance_model_LL_TESLA_HOMs}, no transverse emittance blow-up was found after 17 turns of acceleration, both with $Q^{\prime} = -19$ and $Q^{\prime}=0$.
Applying a factor two on the impedance, a large emittance blow-up appears with $Q^{\prime} = -19$ and $Q^{\prime} = +19$.
With smaller chromaticity values such as $Q^{\prime} = -4$ or $Q^{\prime} = +4$, the transverse blow-up is still present but limited to a few percent.
Simulations including a small initial transverse offset of the beam were also started in order to assess the impact of injection errors on transverse coherent stability, scanning the initial horizontal beam offset in the \SIrange{0.1}{1000}{\micro\metre} range.
A transverse damper located in the second RF station after the injection point would be required to damp the bunch centroid motion.
For larger offset such as \SI{1}{\mm}, either the damper strength should be quite strong (in the 4-turn range) or multiple damper units should be used.

In the \SI{10}{\tera\eV} collider ring, the resistive-wall impedance from the dipole magnets vacuum chambers would be the main source of wakefields.
Impedance and beam stability simulation results performed for the \SI{10}{\tera\eV} collider are described afterwards.
Similar studies were also done for the \SI{3}{\tera\eV} collider and can be found in Ref.~\cite{bib:amorim_collider_imcc_annual_meeting_2022}.
Assuming that the innermost surface seen by the beam is the tungsten shield, impedance models were created for a range of radii between \SI{10}{\milli\metre} and \SI{40}{\milli\metre}.
To reduce the chamber impedance, the inner surface of the tungsten shield would be copper-coated.
The material temperatures are assumed to be \SI{300}{\kelvin}, as described in Section~\ref{sec:Section7_7}.
A scan on the copper coating thickness was also performed, from \SI{0.1}{\micro\metre} to \SI{100}{\micro\metre} to find the minimum thickness required to shield the tungsten material.
The impedance models were generated using the ImpedanceWake2D code~\cite{bib:iw2d_website}.
The chamber and material parameters used in simulations are reported in Table~\ref{tab:coll10tev_impedance_parameters}.
Figure~\ref{fig:coll10tev_impedance_model_20mm_chamber} shows the transverse dipolar impedance for a \SI{23}{\milli\metre} radius chamber, and for different copper coating thicknesses.

\begin{table}[!hbt]
   \centering
   \caption{Chamber parameters used in simulations for the \SI{10}{\tera\eV} collider.}
   \begin{tabular}{lccc}
\hline\hline
           Chamber length and geometry                    &                  &                         &    \SI{10}{\kilo\metre}, circular\\
           Copper coating thickness            &             &   \si{\micro\metre}                 &   0.1, 1, 10, 100 $+$ no coating $+$ infinite\\
           Copper resistivity at \SI{300}{K}   & $\rho_{\mathrm{Cu}, 300\,\mathrm{K}}$&  \si{\nano\ohm\metre}   &    \num{17.9}      \\
           Tungsten resistivity at \SI{300}{K}  & $\rho_{\mathrm{W}, 300\,\mathrm{K}}$&  \si{\nano\ohm\metre}   &  \num{54.4}   \\
\hline\hline
   \end{tabular}
   \label{tab:coll10tev_impedance_parameters}
\end{table}

\begin{figure}[htb]
  \begin{center}
    \includegraphics[width=0.8\textwidth]{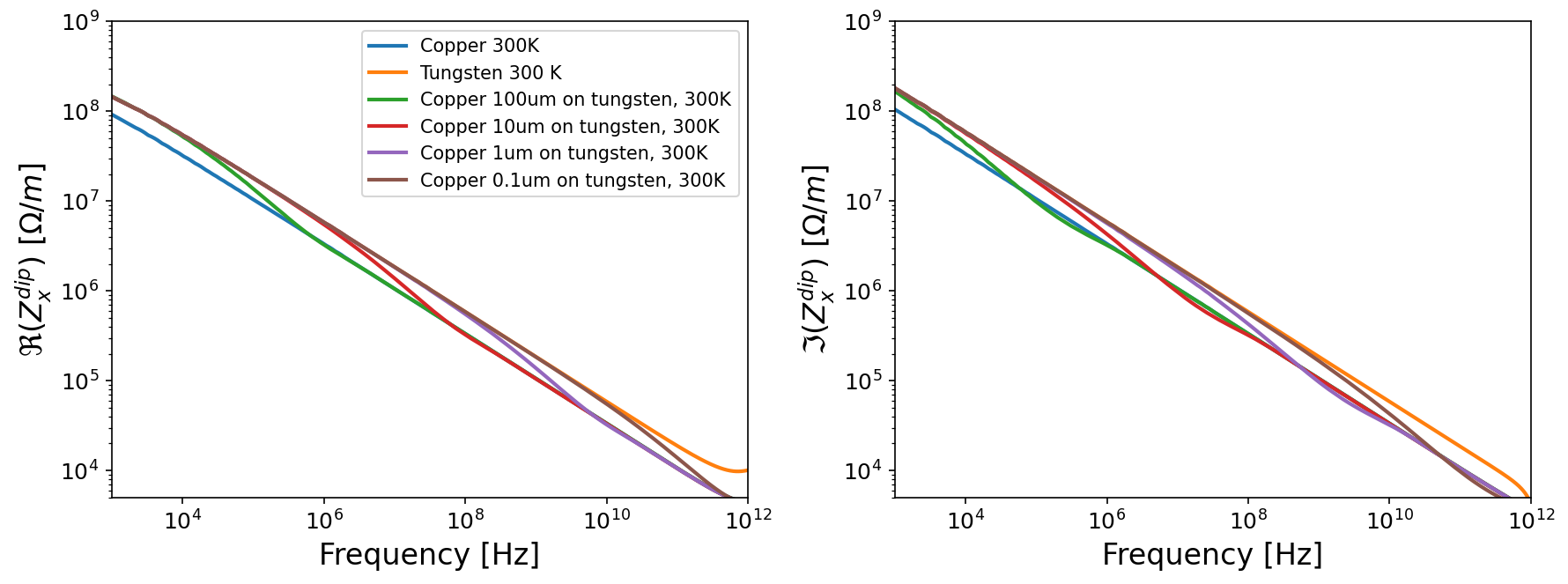}
  \end{center}
\caption{Transverse dipolar impedance for a \SI{23}{\milli\metre} radius tungsten shield in the \SI{10}{\TeV} collider.
Left plot shows the real part, right plot the imaginary part.
Different copper coating thickness are assumed, and are represented alongside two models assuming an infinite copper thickness (orange curve) or tungsten thickness (blue curve).
}
\label{fig:coll10tev_impedance_model_20mm_chamber}
\end{figure}

Tracking simulations using PyHEADTAIL were performed with these impedance models~\cite{bib:Amorim:ipac2023-wepl185, bib:Amorim:hb2023-thbp17}.
They allowed to find the minimum chamber radius needed to ensure transverse coherent beam stability, as well as the copper coating thickness that should be used to reduce impedance effects.
The muon decay effect has to be included in simulations since the bunch intensity reduction over time will reduce the coherent effects strength.
A PyHEADTAIL module was therefore implemented to reproduce this effect in simulations.

Figure~\ref{fig:coll10tev_minimum_chamber_radius} summarises the results of the tracking simulations.
The minimum chamber radius required to keep the beam stable is plotted as a function of the transverse damper strength.
Each curve corresponds to a different copper coating thickness.
For coating thickness larger than \SI{1}{\micro\metre}, the curves are superimposed on the pure copper case.
It shows that a copper coating of at least \SI{1}{\micro\metre} is required to shield the tungsten.
Results of simulations showed that if a tungsten chamber with a \SI{10}{\micro\metre} copper coating is used, with a 50-turn transverse damper gain, the minimum chamber radius required is \SI{13}{\milli\metre}, smaller than the \SI{23.5}{\milli\metre} currently assumed for radiation shielding studies.

\begin{figure}[htb]
  \begin{center}
    \includegraphics[width=0.4\textwidth]{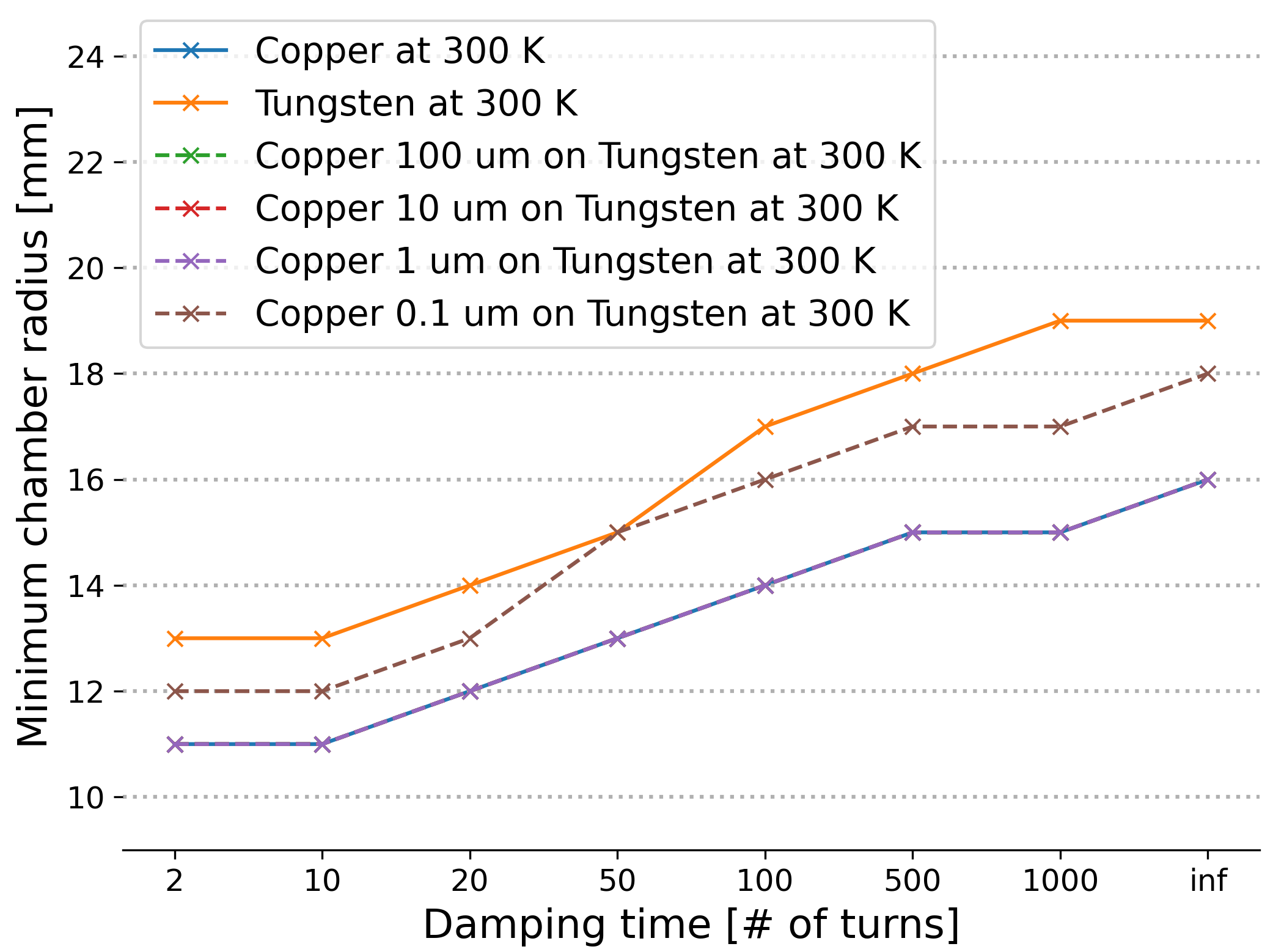}
  \end{center}
\caption{Minimum chamber radius required to keep the emittance growth below \SI{10}{\percent} after 3000 turns, for the different copper coating thickness considered and pure copper or pure tungsten (no copper coating).
}
\label{fig:coll10tev_minimum_chamber_radius}
\end{figure}

For both RCS and collider rings the impedance models and beam dynamics codes are made available on a dedicated Git repositories and can be found at \url{https://gitlab.cern.ch/muc-impedance/muc-impedance}.
A web page was also developed to visualise the developed impedance models and is available at \url{https://muc-impedance.docs.cern.ch/}.

\subsubsection{Planned work}

Impedance models for the RCS2, RCS3 and RCS4 will be created, following the machine parameters reported in Section~\ref{sec:Section7_3}.
Simple models for the resistive-wall impedance of the RCS are also currently being investigated, in particular for the pulsed magnet chambers.
Because of the fast magnet ramping, in the order of $\dot{B} = \SI{4200}{\tesla\per\second}$ for the RCS1 for example, ceramic chambers with outer RF shielding stripes such as those used in the J-PARC RCS~\cite{bib:jparc_chamber} must be considered.
The tracking simulations will be ported to the XSuite~\cite{bib:xsuite} framework.
This will allow to perform start-to-end tracking simulation in the different RCS as well as in the collider ring, without and with collective effects.
The study of the effect of an initial transverse offset of the beam will be refined to better assess the required strength and location of the transverse damper.
Beam-beam effects will also be included in simulations performed within this framework.

The tracking simulation framework for the collider ring will also be migrated to XSuite.
This will allow to investigate the coherent beam-beam effects in \SI{10}{\tera\eV} collider, and their possible interplay with impedance effects.
Similarly to the RCS studies, the effect of an initial bunch offset at injection will be studied through simulations.
The impact on coherent stability of the quasi-isochronicity of the ring caused by the very small momentum compaction factor $\alpha_p$ will also be investigated, as well as the effect of the higher-order momentum compaction factors.

Collective effects studies for the proton driver part will also be started, in association with the Linac, Accumulator and Compressor ring developments exposed in Section~\ref{sec:Section6_2}, as well as the collective effects during the ionisation cooling developed in Section~\ref{sec:Section6_3}.
In particular the effect of space-charge, and its interplay with impedance, on the coherent beam instability thresholds should be investigated.

\subsubsection{Next priority studies and opportunities for additional collaboration}

The transverse tracking simulations for the RCS are for now limited to a simple linear optics model.
The development of a detailed lattice model would allow to refine the coherent beam stability simulations in the RCS and ensure through start-to-end simulations the choice of beam parameters for the RCS chain.
In particular collective effects in the hybrid synchrotrons, which will see large horizontal beam excursions in certain sections of the ring, could be investigated.

In collaboration with the vacuum and the magnet design teams, the technology options for the RCS normal and super-conducting dipole vacuum chamber should be investigated to have a better estimate of the resistive-wall impedance in the RCS.

In the collider ring, an RF system design should be investigated in collaboration with the lattice design and RF design teams in order to evaluate its impact on the coherent beam stability.

Experimental study of space-charge could be performed in machine development sessions in the CERN PS to investigate beam dynamics effects that could arise in the proton driver Accumulator and Compressor rings.
Moreover these sessions could also be used to study coherent beam dynamic effects close to transition for isochronous rings.

\begin{flushleft}
\interlinepenalty=10000

\end{flushleft}

\clearpage
\section{Accelerator technologies
}
\label{sec:Section07}


\subsection{Magnets
}
\label{sec:Section7_1}
Prior to the IMCC, the US Muon Accelerator Program (MAP) study \cite{palmer2014, palmer2015} provided the most complete concept of a MuC, including an overview of the magnet requirements. The IMCC has hence taken MAP as a starting point, identified the key magnet challenges and technology options, and evolved the magnet configurations to provide a current set of main magnet performance parameters. These parameters extend the feasible performance space by considering recent advances in superconductor and magnet technology. 

Work carried out so far to evolve the MAP configurations has a strong focus on HTS. A driving reason for this are the higher field reaches possible with HTS. But, equally important are considerations of efficient cryogenic operation and helium inventory. Operation in helium gas at temperatures in the range of 10\,K to 20\,K can provide a coefficient of performance up to a factor four higher than a cryoplant producing liquid helium at 4.2\,K. This improved energy efficiency will reduce the impact of energy depositions from muon decay and triggered cascades, which applies throughout the MuC complex, and vital to the sustainability of the accelerator complex. 

The sections below describe the magnet systems for the main parts of the MuC  complex, namely the muon beam production (target, decay and capture channel), muon cooling, acceleration and collision. We will expand on the concept and magnet engineering, making reference to specific technologies such as Low Temperature Superconductors (LTS, Nb-Ti or Nb$_3$Sn), High Temperature Superconductors (HTS, mainly considering REBCO) as well as normal conducting (NC) iron dominated magnets. More extensive descriptions can be found in the publications cited. We also bring the attention to the internal review of magnet technology options that was organized during the 2023 IMCC Annual Meeting (\url{https://indico.cern.ch/event/1250075}), with a final report presented at the Accelerator Design Meeting of 30 October 2023 (\url{https://indico.cern.ch/event/1337221/}).

\subsubsection{Magnet system for muon beam production}
\label{sec:Section7_1_1}

\paragraph*{System overview}
Muons are produced from the decay of pions generated from the collision of a short, high intensity proton pulse with a target. The target, varying in outer diameter dimensions from 150\,mm to 250\,mm depending on technology, is placed within a steady-state, high field “target solenoid”, whose purpose is to capture and guide pions into a “decay and capture channel”, also embedded in solenoid magnets. An approximately 18\,m long magnetic field profile from  the target to the end of the capture channel enables the capture of pions and has a characteristic shape with a peak field of 20\,T on the target and an adiabatic decay to roughly 2\,T at the exit of the channel \cite{HKSayed}. 

To prevent heating and radiation damage from the interaction of the multi-MW proton beam with the target, a radiation shield is necessary for the target, decay, and capture solenoids. This shield comprises a combination of a dense metal such as tungsten (to intercept highly energetic photons generated by the decay electrons) and a moderator like water (to reduce neutron fluence). The shield thickness dictates the magnet bore and is chosen to reduce to acceptable levels the radiation damage and heat load on the solenoid coils and surrounding infrastructure. Calculations show that with a shield approximately 500\,mm thick, radiation induced heating in the coils would fall within the range of 5\,kW, local radiation dose in the range of 80\,MGy, and a yearly peak DPA in the range of $10^{-3}$. These values, and especially the radiation damage, are at the limit of present magnet technology, thus setting a minimum magnet bore dimension of 1200\,mm used for the design of the target, decay and capture solenoids. A larger bore dimension implies higher stored magnetic energy, influencing electromagnetic forces, magnet protection, and cost.

To achieve the peak 20\,T field profile at the target, MAP considered a hybrid normal conducting (NC) and superconducting (SC) target solenoid configuration, consisting of a rad-hard resistive NC insert (approximately 5\, T with a 150\,mm bore), within a large bore Low Temperature Superconducting (LTS) magnet (approximately 15\,T with a 2400\,mm bore) \cite{Weggel2001, Weggel2011, Weggel2014}. While this concept remains a possibility and can be built based on the extrapolation of known technology, the system's stored energy (approximately 3\,GJ), coil mass (about 200\,tons), and wall-plug power consumption (around 12\,MW, dominated by the resistive insert) are substantial.

\paragraph*{Key challenges}
The main challenge in our proposed design (see below) is radiation damage of the superconductor and insulation, both requiring further analysis and measurements, and especially advances in understanding the damage mechanism in HTS. Other challenges include magnet engineering (field performance, mechanics, stored energy and protection) and infrastructure and operating cost (power and cooling), including integration/maintenance in a high-radiation environment. These challenges appear manageable, and overlap broadly with those of magnets for high magnetic field science (e.g.~user facilities based on hybrid SC and NC solenoids and all-SC solenoids), as well as magnets for fusion devices (e.g.~the central solenoid magnets for a Tokamak) \cite{Tsuji2001}.

\paragraph*{Recent achievements}
Following recent advancements in HTS magnets for fusion \cite{MITpress}, we have proposed an alternative design involving an HTS cable operating at 20 K. The cable considered is shown in Fig.~\ref{fig:MITcable}, an internally cooled conductor inspired by the VIPER developed at MIT \cite{MITcable}. The analysis performed thus far indicates the potential to eliminate the resistive insert, thus reducing the magnet bore to the minimum of 1200\,mm, roughly half the size of the US-MAP LTS coil, while still maintaining the desired field profile for efficient muon capture. If compared to LTS, operating at a temperature higher than liquid helium diminishes the need for radiation heat shielding and maintains good overall energy efficiency.
The proposed system is shown schematically in Fig.~\ref{fig:TargetSolenoid}. It has a stored energy of $\sim$1\,GJ, a coil mass of around 100 tons, and a wall-plug power consumption of approximately 1\,MW. This represents a significant reduction compared to the previously suggested hybrid solution from the MAP study. Our future focus will be on identifying specific HTS conductor and winding technologies suitable for magnets of this class. The details of the electro-magnetic and mechanical design can be found in Ref.~\cite{portoneMT28}, while the cooling and quench analysis can be found in Ref.~\cite{botturaAPS2023}.

\begin{figure}
\centering
\includegraphics[width=0.7\textwidth]{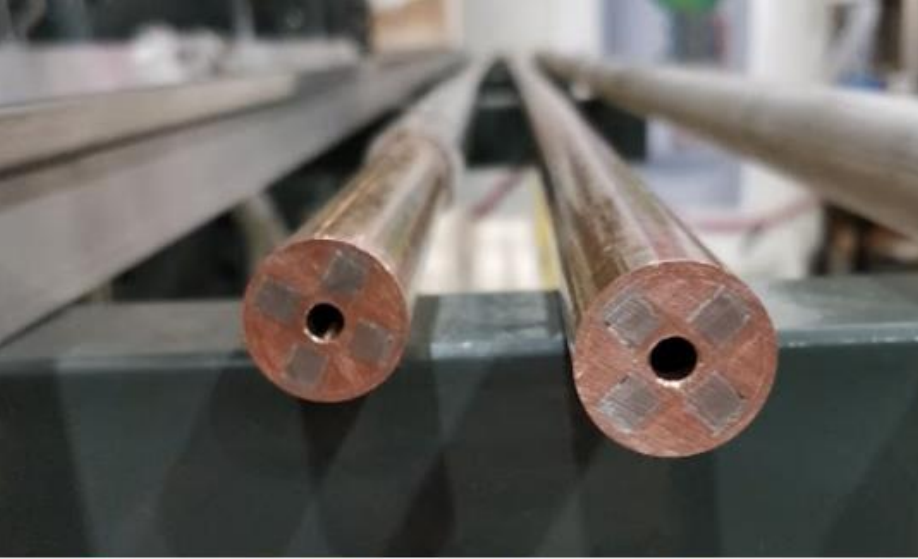}
\caption{MIT VIPER cable, image taken from Ref.~\cite{MITcable}.} \label{fig:MITcable}
\end{figure}

\begin{figure}
\centering
\includegraphics[width=0.7\textwidth]{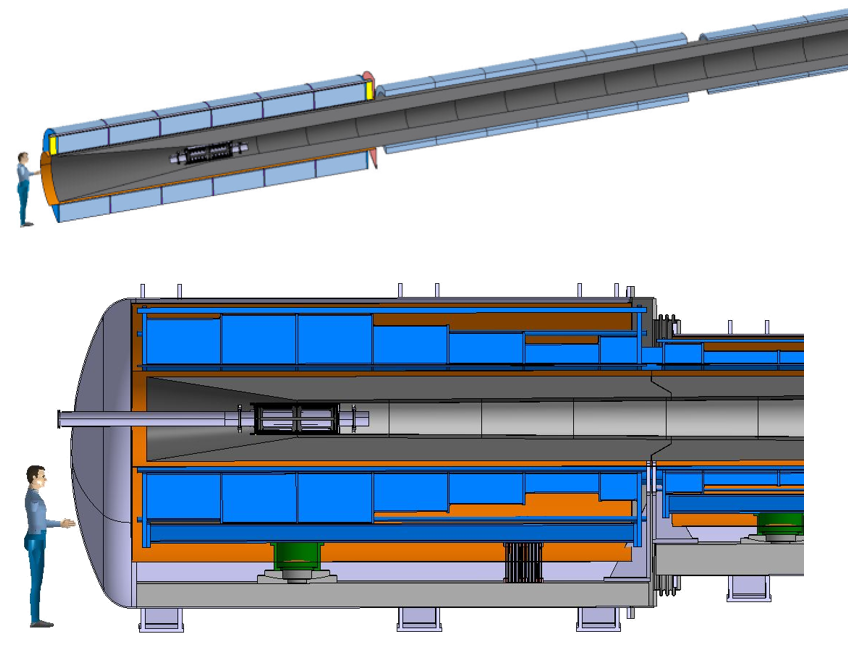}
\caption{Schematic view of the HTS target, decay and capture channel proposed (top), with a detail of the high field solenoids around the proton target area (bottom).} \label{fig:TargetSolenoid}
\end{figure}

\paragraph*{Planned work}

After completing a first iteration on the electro-magnetic, mechanic, cooling and quench protection designs, we plan now to iterate on the free bore to resolve known mechanical interference with target and shield, and mitigate radiation heat load and damage. Alternative cable designs and coil geometries are being considered, to see whether a better cost optimum can be found. We will then integrate the design in an engineering concept, including considerations of manufacturing, installation and maintenance. This effort can be completed within the planned timeline. It will then be important to plan for technology demonstrators, from cable to coil, including testing efforts. Given the size of this magnet, this work will require considerable material and personnel investment.

\paragraph*{Next priority studies and opportunities for additional collaboration}

Besides the technology demonstrators already quoted earlier, an important study not planned with available resources is the characterization and understanding of radiation damage in HTS. This is a recent field of research, data is scarce and we still lack a sound understanding of the mechanism by which particle radiation induces degradation of the critical properties of the superconductor. Such understanding is mandatory to extrapolate measurements to the conditions experienced by the magnets in a MuC, and thus reliably predict the lifetime.

\subsubsection{Magnet system for muon cooling }
\label{sec:Section7_1_2}

\paragraph*{System overview}
Muons exiting the target, decay and capture channel have a large spatial and energy spread, and thus must be cooled in both dimensions to reach values suitable for a collider. This 6D “beam cooling” process occurs over a long (almost 1\,km) sequence of tightly integrated absorbers (consisting of light elements such as hydrogen), solenoids, and RF cavities. The solenoids create a broadly increasing oscillating magnetic field in the beam direction, that when combined with absorbers and re-acceleration (RF cavities), cools the beam in the six dimensions of the beam phase space (position and momentum). The final emittance of the muon beam is inversely proportional to the strength of the final cooling solenoids. To achieve maximum cooling efficiency and minimize the emittance there is hence a clear interest in very high field final cooling solenoids.

The MAP design study provided a baseline of magnets (geometry and number) over the entire cooling chain (up to the final cooling solenoids) which could produce the desired on-axis field profile \cite{Stratakis2015}. However, it is important to note that these solenoids were created purely to best fit a desired field profile, rather than from a magnet engineering point of view, and therefore present un-optimized parameters (stresses, forces, volume of conductor) which pose significant challenges. In the MAP version, in total there are approximately 2400 solenoids in cooling channels of approximate 1\,km length (excluding the final cooling), which are divided into a total of 826 cooling cells. There are 12 types of cooling cells (labeled A1 to A4, B1 to B8) composed of an absorber, one or two RF modules, and two, four or six solenoids in split pairs (depending on cell type). 

For the final cooling solenoids, the MAP scheme consisted of 17 solenoids with a bore field up to 30\,T. With this field, an emittance of about 50\,\textmu m, roughly a factor of two greater than the transverse emittance goal (25\,\textmu m), can be achieved \cite{Stratakis2015, Sayed2015}. However, other analyses \cite{Palmer2011} show that higher fields (in the range of 50\,T) offer further gains in the final emittance and beam brightness. 

\paragraph*{Key challenges}
As detailed in the next section, individual solenoids and cells based on the MAP configuration pose challenges in terms of peak fields, high forces and stresses, and quench management. The range of specifications is however within what can be manufactured already with present technology. Indeed, it is rather the integration of solenoids within a cell containing RF cavities and absorbers, as well as in a series (where the field of one cell affects neighboring ones), which poses the primary difficulty in this aspect of the overall MuC. Furthermore, the large number of solenoids in the 6D cooling section will require standardization and effective industrialization.

For the final cooling, in the scope of improving upon the design by the US-MAP study, the fundamental challenge will be developing ultra high-field solenoids (as explained in the next section) with magnetic fields reaching beyond the current state of the art, more than 40\,T in a 50\,mm bore. The number of these solenoids, which are nearly free standing, is limited. We hence expect that in this case it will be the peak field that will pose the main challenge.

\paragraph*{Recent achievements}
Starting from the MAP baseline, the first step has been to characterize the parameters of all of the solenoids during operation, as shown in Table~\ref{tab:coolingcells}. 
\begin{table}[ht]
\begin{center}
\caption{Summary of main cooling solenoid characteristics for the 6D cooling cells of a muon collider, based on the US-MAP configuration. For a given cell type, $B_{\mathrm{peak}}$ is the peak on axis field, $\sigma_{\mathrm{Hoop}}$ is the maximum hoop stress seen by a solenoid, and $\sigma_{\mathrm{Radial}}$ shows the minimum/maximum radial hoop stress seen by a solenoid.}
\begin{tabular}{lcccccc}
\hline\hline
\textbf{Cell} & $J_E$ (A/mm$^2$) & B$_{\mathrm{peak}}$ (T) & E$_{\mathrm{Mag}}$ (MJ) & e$_{\mathrm{Mag}}$ (MJ/m$^3$) & $\sigma_{\mathrm{Hoop}}$ (MPa) & $\sigma_{\mathrm{Radial}}$ (MPa) \\
\midrule
\textbf{A1} & 63.25 & 4.1 & 5.4 & 20.5 & 34 & $-$ 4.6/0.0 \\
\textbf{A2} & 126.6 & 9.5 & 15.4 & 76.3 & 137 & $-$ 28.3/0.0 \\
\textbf{A3} & 165 & 9.4 & 7.2 & 72.8 & 138 & $-$ 28.5/0.0 \\
\textbf{A4} & 195 & 11.6 & 8.4 & 91.5 & 196 & $-$ 49.4/0.0 \\
\textbf{B1} & 69.8 & 6.9 & 44.5 & 55.9 & 95 & $-$ 13.5/0.0 \\
\textbf{B2} & 90 & 8.4 & 24.1 & 61.8 & 114 & $-$ 20.1/0.0 \\
\textbf{B3} & 123 & 11.2 & 29.8 & 88.1 & 174 & $-$ 36.6/0.0 \\
\textbf{B4} & 94 & 9.2 & 24.4 & 42.4 & 231 & $-$ 23.5/19.7 \\
\textbf{B5} & 168 & 13.9 & 12 & 86.3 & 336 & $-$ 55.7/21.1 \\
\textbf{B6} & 185 & 14.2 & 8.2 & 68.3 & 314 & $-$ 43.1/22.3 \\
\textbf{B7} & 198 & 14.3 & 5.7 & 59.6 & 244 & $-$ 37.4/20.7 \\
\textbf{B8} & 220 & 15.1 & 1.4 & 20.3 & 119 & $-$ 22.9/22.1 \\
\hline\hline
\end{tabular}
\label{tab:coolingcells}
\end{center}
\end{table}

The solenoids from the MAP optics exhibit a diverse range of parameters, from large-bore variants with a diameter exceeding 1.5\,m and a modest on-axis field of 2.6\,T, to smaller-bore with a 90\,mm diameter and a high on-axis field of 13.6\,T. This diversity features substantial stored magnetic energy (up to 44\,MJ in the single cell B1), notable associated stresses (hoop stress up to 300\,MPa and radial tensile stress reaching 20\,MPa in cells B5 and B6), and quench management challenges (energy density approaching 100\,MJ/m$^3$ of a coil in cell A4). 

While the MAP study provides a baseline, the magnets were created from a beam physics point of view (find solenoid geometries which best fit a desired on-axis field) rather than from a magnet engineering point of view. This is evident in Table \ref{tab:coolingcells}, i.e.~large volumes of conductor (and therefore larger cost) are paired with lower current density, when larger current densities/lower volumes can be considered. Therefore, significant work has gone into creating an optimization program to revise the coil configurations to address identified challenges while fitting the desired on-axis field to within some error percentage. This is currently a work in progress, and will incorporate advancements from beam optics studies. 

For the final cooling solenoids, to improve upon the results obtained by the US-MAP study, we are considering a solenoid design with the potential to reach and exceed 40\,T, a clear bore of 50\,mm, a magnet length of 500\,mm, and compact enough in size to reduce overall footprint, mass, and cost. See Fig.~\ref{fig:FinalCooling}. We are studying in particular a non-insulated HTS winding solution, where the cooling solenoid is built as a stack of soft-soldered pancakes. To reduce the coil size, forces, and stored energy, and meet the field target, the operating current density target is high, 650\,A/mm$^2$.  The coil size is exceptionally small, with a 180\,mm outer diameter. 

\begin{figure}
\centering
\includegraphics[width=0.34\textwidth]{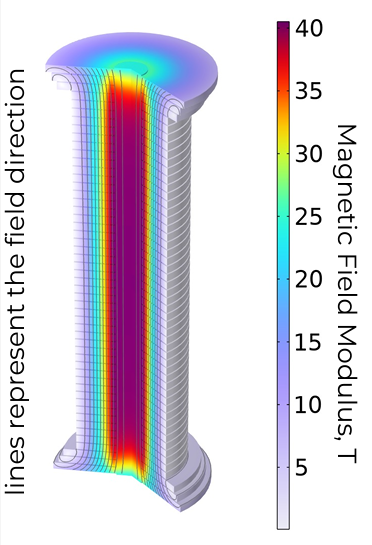}
\includegraphics[width=0.47\textwidth]{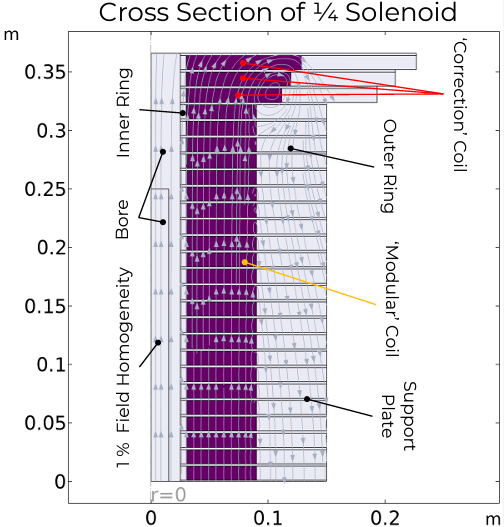}
\caption{The current 40\,T and beyond conceptual design for the final cooling solenoids. It consists of 46 identical 'modular' and six 'correction' pancakes, with a bore aperture of 50\,mm. Figures courtesy of B. Bordini.} \label{fig:FinalCooling}
\end{figure}

Two critical parameters were identified in the conceptual design phase of a final cooling solenoid, the stress state and the selection and control of the transverse resistance which must be a good balance between the required ramp-rate and quench management. The mechanics design features a maximum hoop stress of 650\,MPa, ensuring no tensile stress under any operating condition. To achieve this, the wound and soldered pancakes undergo radial loading from a rigid external ring, providing a radial pre-compression of 200\,MPa at room temperature. This radial pre-compression is carefully chosen to nearly offset the outward electromagnetic stress at 40\,T, a field range identified as the stress limit for a single pancake. Beyond this point, higher fields would necessitate segmenting the coil into concentric layers with independent support, introducing added complexity and size.

Efforts are underway to address the transverse resistance, with ongoing work focused on reducing values typically associated with soldering tapes. Given that most current transfer in REBCO tapes occurs through the copper coating on the tape sides, it has been observed that minimizing or eliminating this copper component significantly enhances transverse resistance \cite{Bordini2024}. Our objective is to achieve protection through a low transverse resistance (potentially with mechanisms to actively trigger a quench), while still enabling a full ramp in less than six hours and maintaining field stability at a flat-top better than 10\,pm/s. Further details on the concept, design studies, and recent advancements are available in Ref.~\cite{Bordini2024}. 

\paragraph*{Planned work}
For the evaluation of the MAP baseline 6D cooling solenoids, one important aspect is ongoing---the quench protection analysis. In parallel, optimization and revision of all magnets from an engineering perspective (stress, cost, protection, etc.) as mentioned above is underway. Tight collaboration with beam physicists is necessary to incorporate new magnet configurations to fit evolving on-axis field profiles. Specific manufacturing options will be explored, and the possibility of standardization will also be considered to manage the substantial number of solenoids slated for construction.

The final cooling solenoid is a case by itself. Besides the design and analysis work, which will include multi-physics evaluation of electrical, magnetic (screening currents), mechanical and thermal (quench) performance, we have initiated supporting experimental activity. Tests are planned on HTS tapes and pancakes to provide an experimental basis for the electro-mechanical performance limits (stress and strain, as well as in-field measurement of delamination) and prove the feasibility of concepts. Stacks and windings are being manufactured to establish mechanical properties, and eventually perform tests in field to probe performance in the range of 20\,T to 40\,T.

\paragraph*{Next priority studies and opportunities for additional collaboration}
The cryogenic cooling system needs to be defined both in concept, as well as in practical engineering solutions. For various reasons, such as paramagnetic effects on helium, cooling by a bath of helium is not practical. Direct and indirect cooling need to be considered to find the optimal baseline. A second matter that will require attention is to explore manufacturing routes for the large number of solenoids needed by the cooling channel. Finally, a full-size demonstration of a 6D cooling cell integration, as well as the final cooling solenoid, will be required to validate the designs.

\subsubsection{Magnet system for muon acceleration to TeV energies}
\label{sec:Section7_1_3}

\paragraph*{System overview}
To extend their lifetime in the laboratory reference frame, the muons must be accelerated rapidly to relativistic momentum. The acceleration begins with an initial linear accelerator and sequence of re-circulating linear accelerators followed by a sequence of rapid-cycling synchrotron (RCS) and hybrid cycling synchrotrons (HCS)~\cite{Batsch2023, Chance2023}. The RCS employs fast-ramped normal conducting (NC) magnets, swinging from an injection to an extraction field level. In a HCS, static superconducting (SC) magnets establish a field offset, while NC fast-ramping magnets are powered from peak negative to peak positive field, enabling a full field swing. This design results in a more compact accelerator compared to an equivalent RCS. 

Design concepts of NC fast-ramped magnets with a peak operating field of 1.5\,T were developed by US-MAP \cite{Berg2016}. SC dipoles for HCS were not studied in depth, besides setting target value for bore field and magnet length.

In the current IMCC baseline configuration, the NC dipoles in the initial RCS need to sweep from 0.36\,T to 1.8\,T within 0.35\,ms (equating to a rate of 4\,kT/s). In the concluding HCS, the dipoles sweep from $-$ 1.8\,T to 1.8\,T in 6.37\,ms (equating to a rate of approximately 560\,T/s). Studies are ongoing to evaluate the possibility of realizing the final slower HCS pulsating magnets with HTS REBCO based tapes and ferric core. The SC dipoles in the first two HCS have a nominal field of 10\,T, while in the final HCS the assumption is of 16\,T field. All dipole magnets, NC and SC, have a rectangular nominal aperture of 30\,mm (vertical) $\times$ 100\,mm (horizontal).

\paragraph*{Key challenges}
Beyond magnet engineering, a fundamental challenge in constructing an accelerator ring of the required dimensions is the substantial stored energy, reaching several tens of MJ. Efficiently powering at a high-pulse rate with effective energy recovery between pulses necessitates expertise in managing peak power in the range of tens of GW. Resonant circuits combined with energy storage systems appear to be the only viable solution. This is described in detail in the next section, devoted to power converters.

A second challenge will be mitigating the losses originating from the iron hysteresis, eddy currents in the lamination and coils, as well as any other metallic component exposed to field changes. For example, the vacuum chamber is inserted into a gap where the $\frac{dB}{dt}$ is of the order of few thousands of Tesla per second. Standard metallic vacuum chambers cannot be used, as the induced losses would be not sustainable leading to severe overheating or, in the worst case, melting of the chamber. Alternative concepts are needed to reduce the impact of these losses.

\paragraph*{Recent achievements}
As noted in the previous section, the primary challenge of the RCS and HCS magnets lies in effectively managing the multi-GW power necessary to pulse them while maintaining precise control of the field ramp shape, homogeneity, and maximizing energy efficiency. The power required for pulsing, given a specified ramp time and shape dictated by beam design and RF limitations, is directly proportional to the magnetic energy stored in the ramped magnets. A lower bound for stored energy is the magnetic energy within the beam aperture, nominally defined as 30\,mm (gap) $\times$ 100\,mm (width). To limit saturation, which can impact losses and field quality, an upper design field limit of 1.8\,T has been set for the resistive magnets. This corresponds to a magnetic energy of 3.9\,kJ/m in the beam aperture, with the energy stored in the magnet being higher. Analysis of various resistive magnet configurations, considering different iron cross-sections, materials, coil designs, and current densities, indicates that the lowest magnet stored energy is in the range of 5.4\,kJ/m, representing a 1.4-fold increase compared to the magnetic energy in the beam aperture \cite{Breschi2024}.

The present strategy is to realize the dipoles as single turn magnets without return connections and let the power converters connect the ends and starts for the coils from subsequent magnets (see section on power converters). Two type of conductors are analysed, either flat bars cooled at the ends or hollow direct cooled conductors. 

A secondary concern involves the magnitude of resistive, eddy current, and hysteresis losses — power drawn from the grid and dissipated. Although a definitive target has not been established, a suitable range is considered to be 200 to 500\,J/m per pulse. After analyzing multiple configurations, it has been determined that the best compromise between stored energy, losses, and field quality is achieved with "H" and "Hourglass" shaped iron cores \cite{Berg2016}. These configurations will be used in further magnetic analysis, including 3D and end effects. It is important to note that the design of the power converter is intricately linked with magnet design and analysis, aiming for an optimal cost solution (CAPEX+OPEX), while considering beam dynamics and RF. The exploration of various powering configurations has given way to identifying two main cost drivers: capacitor-based energy storage and the active power converters necessary for controlling ramp linearity and reproducibility.

Finally, the steady state SC dipole magnets of the HCS’s are only in the early stages of development. While specification of these magnets is still an active discussion, it is accepted that the aperture will be similar to that of the pulsed resistive magnets, i.e.~30\,mm $\times$ 100\,mm rectangular, necessitating non-conventional windings. Rejecting the cos-theta coil geometry due to its inefficiency for a rectangular aperture, we have shifted our conceptual focus toward flat racetrack coils. These coils are simpler and more suitable for winding HTS tapes. It appears feasible to achieve a target magnetic field of around 10\,T while operating at temperatures significantly above liquid helium (10\,to 20\,K) in case HTS is used. This approach would be highly advantageous to help with operating margin and efficiency requirements in HCS’s, which tend to experience higher beam losses and operational challenges.

\paragraph*{Planned work}
Further detailed magnetic analysis, including 3D and end effects, of the possible NC magnet configurations (“H” and “Hourglass” shaped iron cores) is ongoing. This will lead to a final optimized design. Tests are planned on soft magnetic steel in a range of frequency centered around 1\,kHz, and field up to saturation, where the experimental database is sparse. These tests will provide input for the NC magnet design.

An initial design study of magnet configurations and parameters of the SC dipole magnets is underway, with 2D and 3D analysis. This study combined with consideration of more complex geometries, quench protection analysis, and engineering (mechanics, cryogenics, etc.) will provide the basis for selecting the most suitable configuration.

\paragraph*{Next priority studies and opportunities for additional collaboration}

Besides the design and analysis work, and the limited experimental activity, also in the case of the fast pulsed accelerators the program would need a validation of the integrated system made of magnet and power converter. Given the modular nature of the design, one such test could be performed on a single powering cell. The configuration for this test is not yet defined, but as a minimum it should consist of NC dipoles and power converters, including the energy storage units. Such a set-up would allow measurement with accurate field tracking, including field quality, losses, as well as overall efficiency of powering and energy storage.

\subsubsection{Magnet system for muon collision}
\label{sec:Section7_1_4}

\paragraph*{System overview}
The final stage of the muon accelerator complex is the collider ring, where two counter-rotating bunched beams of positive and negative muons collide. A critical design parameter for the collider ring is to maximize collisions of stored muon beams in their limited lifetime. This is achieved by minimizing the collider ring circumference \cite{AccTech_Skoufaris2022}, translating into a requirement of the highest possible field in the arc dipoles. At the same time, luminosity depends critically on the strength and aperture of the quadrupoles in the Interaction Region (IR). In fact, not only high gradients and aperture are important for luminosity, but also a short IR length reduces the cumulative neutrino flux generated by muon decays in the straight sections around the experiments and reduces chromatic aberrations. 
Effective radiation shielding is essential to protect against substantial radiation and heat loads generated by muon decay and collisions. To give representative figures, the heat load originating from muon decay (electrons) and synchrotron radiation in the arc, initially at 500\,W/m at the level of the beam chamber, is reduced to below 5\,W/m at the level of the coil, with a radiation dose below 40\,MGy, by inserting a 40\,mm thick tungsten shield. The need of thick shielding results in exceptional demands on magnet aperture \cite{Alexahin2022, Kashikhin}. 

Assumptions in the present study of the 10 TeV collider optics include the generation of a steady-state magnetic field up to 16\,T within a 160\,mm aperture by the main arc magnets. To minimize straight sections and address effects from the high neutrino flux, these arc magnets are assumed to serve combined functions (e.g., dipole/quadrupole and dipole/sextupole) \cite{AccTech_Skoufaris2022}. The latest optics necessitate dipole fields in the range of 10\,T and gradients of the order of 300\,T/m. While these field requirements combined with aperture constraints are part of an initial assessment, they currently exceed practical limits and will necessitate iteration. As for the IR quadrupole magnets, the assumption from optics studies is a peak field of 20\,T, also associated with large apertures, up to 300\,mm.

It should be noted that the US-MAP study investigated a reduced collider energy of 3\,TeV, which could stand as a fast-track option leveraging Nb$_3$Sn technology nearing demonstration.

\paragraph*{Key challenges}
As detailed in the next section, the present optics demands high field and large bore combined function arc and IR magnets. While this study is evolving and practical limits being developed for these magnets, they must be designed to withstand large stresses (mechanics), with large stored energy (quench), and deal with significant energy deposition and radiation dose from muon beam decay and interaction debris.

\paragraph*{Recent achievements}
To advance the study, specifying the performance limits of accelerator dipole and quadrupole magnets is crucial. We have initiated this effort through analytical evaluations of operating margin, peak stress, hot-spot temperature, and magnet cost, assuming a sector coil geometry \cite{Novelli2023}. While the results are specific to this coil, extending them to other coil geometries is possible and would not significantly alter the main outcomes.

The analytical evaluation was utilized to create design charts with maximum magnet aperture (A) versus bore field (B) (see Fig.~\ref{fig:ABchart_dipole}), a format convenient for iteration with the beam optics. These A-B charts were generated for various superconductors and operating points, including Nb-Ti at 1.9\,K, Nb$_3$Sn at 4.5\,K, and REBCO at either 4.5\,K or 20\,K. For a 10\,TeV collider, Nb-Ti at 1.9\,K appears sub-optimal due to a low operating margin (considering the substantial energy deposition), cryoplant efficiency, and energy consumption concerns. Similarly, Nb$_3$Sn at 4.5\,K falls short of the required field performance for arc magnets, providing feasible solutions only up to 14\,T, being limited by mechanics when considering the necessary 160\,mm aperture. Our initial evaluation of REBCO indicates that also in this case the available design space does not meet the required performance. However, for REBCO, operating margin is not a concern, and operation in the range of 10\,K to 20\,K could be considered. The main limitations for REBCO arise from the cost of the superconductor and quench protection. Cost considerations drive the current density in an all-HTS coil toward high values where standard detect-and-dump protection strategies are not sufficiently fast. Therefore, alternative protection schemes need to be devised to leverage the large current carrying capacity and margin of present REBCO conductors. Assuming a reduction in the cost per meter of REBCO tape by a factor of three to four, and relaxing the need for very high values of current density, a suitable design range for the arc magnets can be defined. This range spans from a nominal aperture of 160\,mm at a reduced bore field of 12\,T to a nominal bore field of 16\,T but with a reduced aperture of 100\,mm. The entire range can be achieved with REBCO at 4.5\,K and 20\,K, while the low field range can also be reached with Nb$_3$Sn at 4.5\,K, providing at least two technology options.

\begin{figure}
\centering
\includegraphics[width=0.47\textwidth]{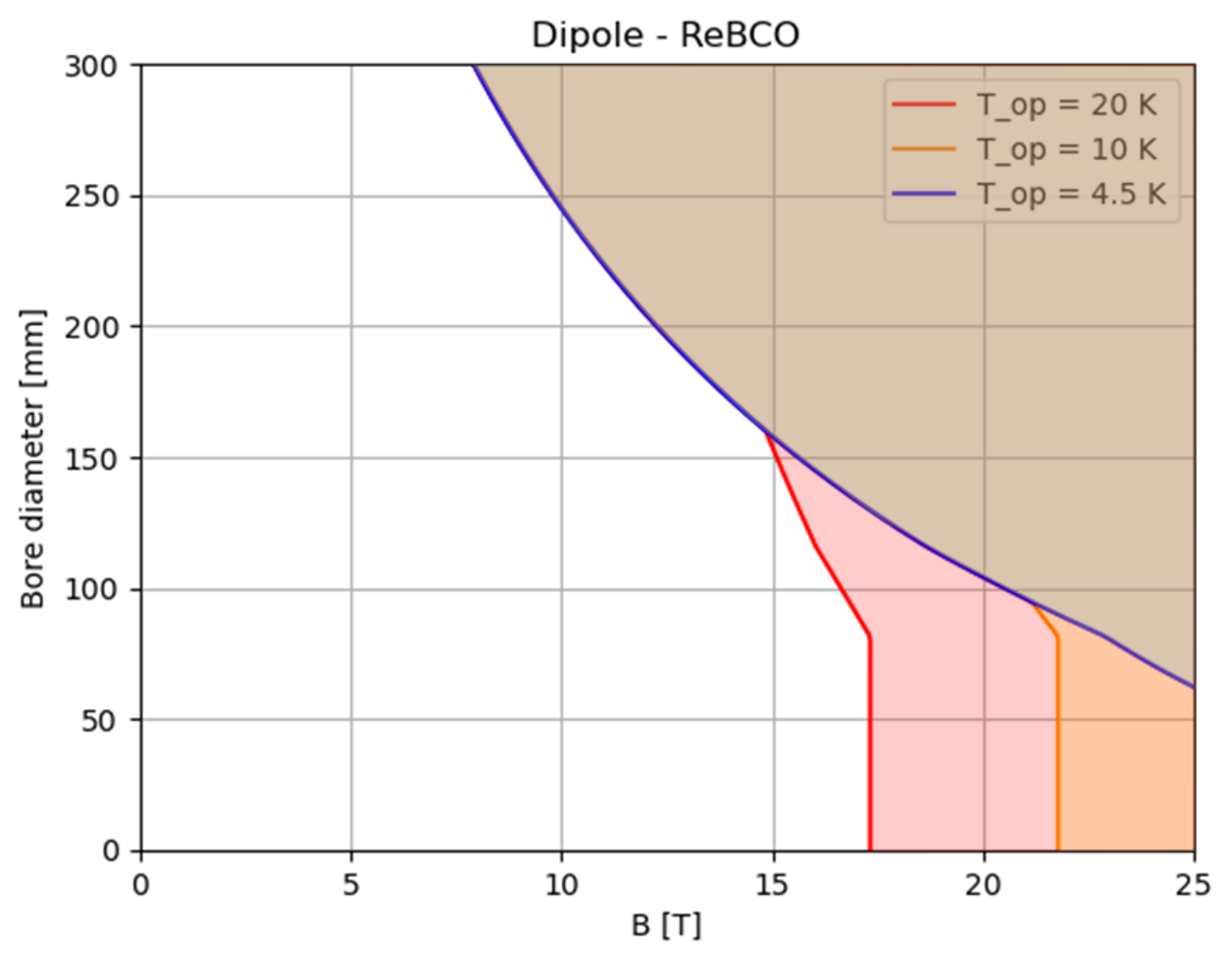}
\includegraphics[width=0.47\textwidth]{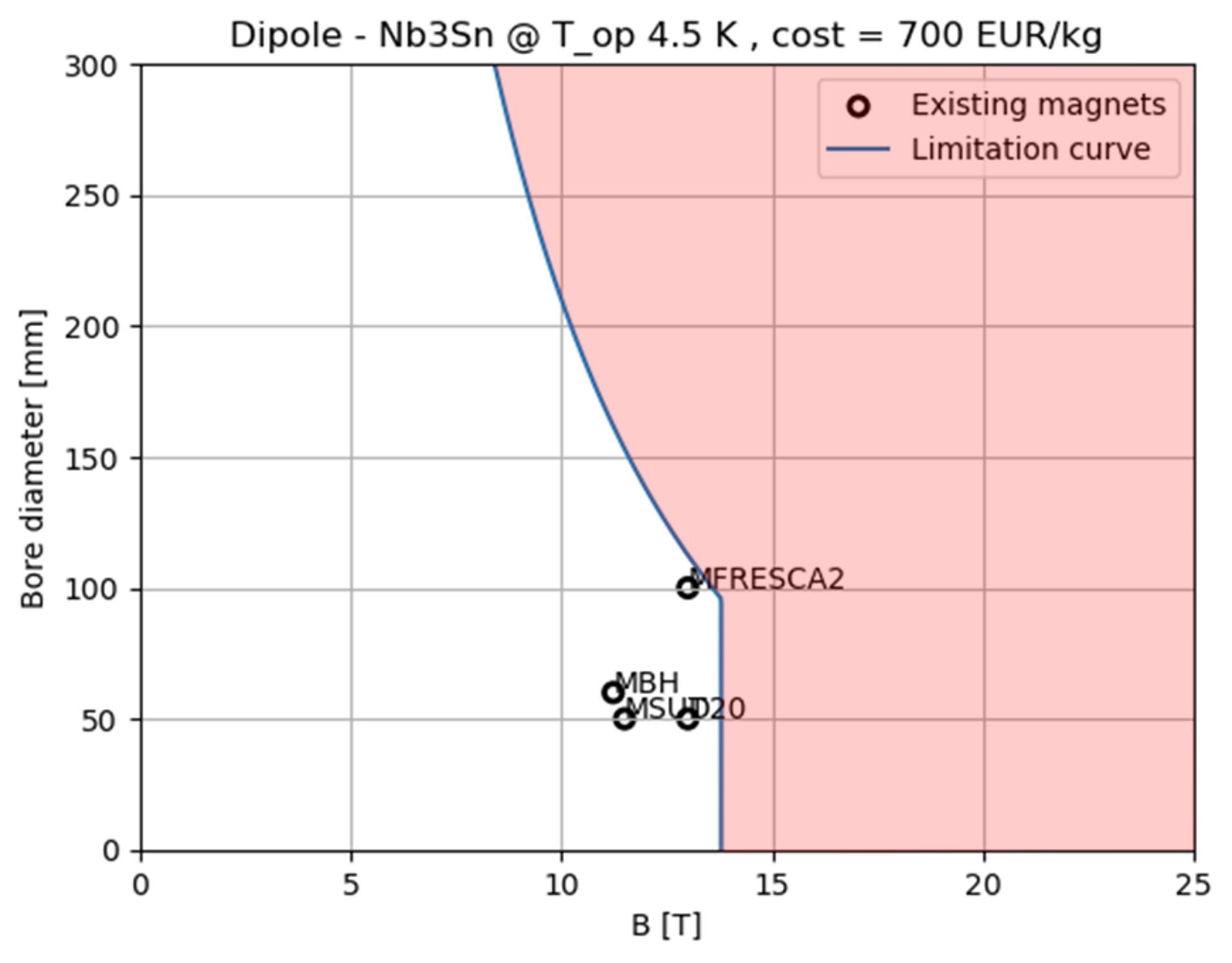}
\caption{Dipole limitation curves as a function of stress, protection, and margin shown against bore diameter vs. magnetic field for different conductor material and temperature. Left: ReBCO at different operating temperatures (Top), assuming an aspirational cost of 2500\,EUR/Kg, a peak stress of 300\,MPa, and a hot spot temperature of 200\,K. For example, the red line represents the limit assuming an operating temperature of 20\,K plus 2.5\,K margin. Right: Nb3Sn at 4.5\,K, assuming an aspirational cost of 700\,EUR/Kg, a peak stress of 150\,MPa, and a hot spot temperature of 350\,K. Plots are advanced versions from ref. \cite{Novelli2023}.} \label{fig:ABchart_dipole}
\end{figure}

The analysis for quadrupoles, using the same method outlined above and in \cite{Novelli2023}, is presently in progress (see Fig.~\ref{fig:ABchart_quad}). The range of feasible designs in this case covers gradients of 300\,T/m, with an aperture of 100\,mm, down to gradients of 100\,T/m, and aperture of 250\,mm. This is only marginally short of the present requirements for the IR optics, which indicates that the arc dipoles are actually the most demanding magnets.

\begin{figure}
\centering
\includegraphics[width=0.47\textwidth]{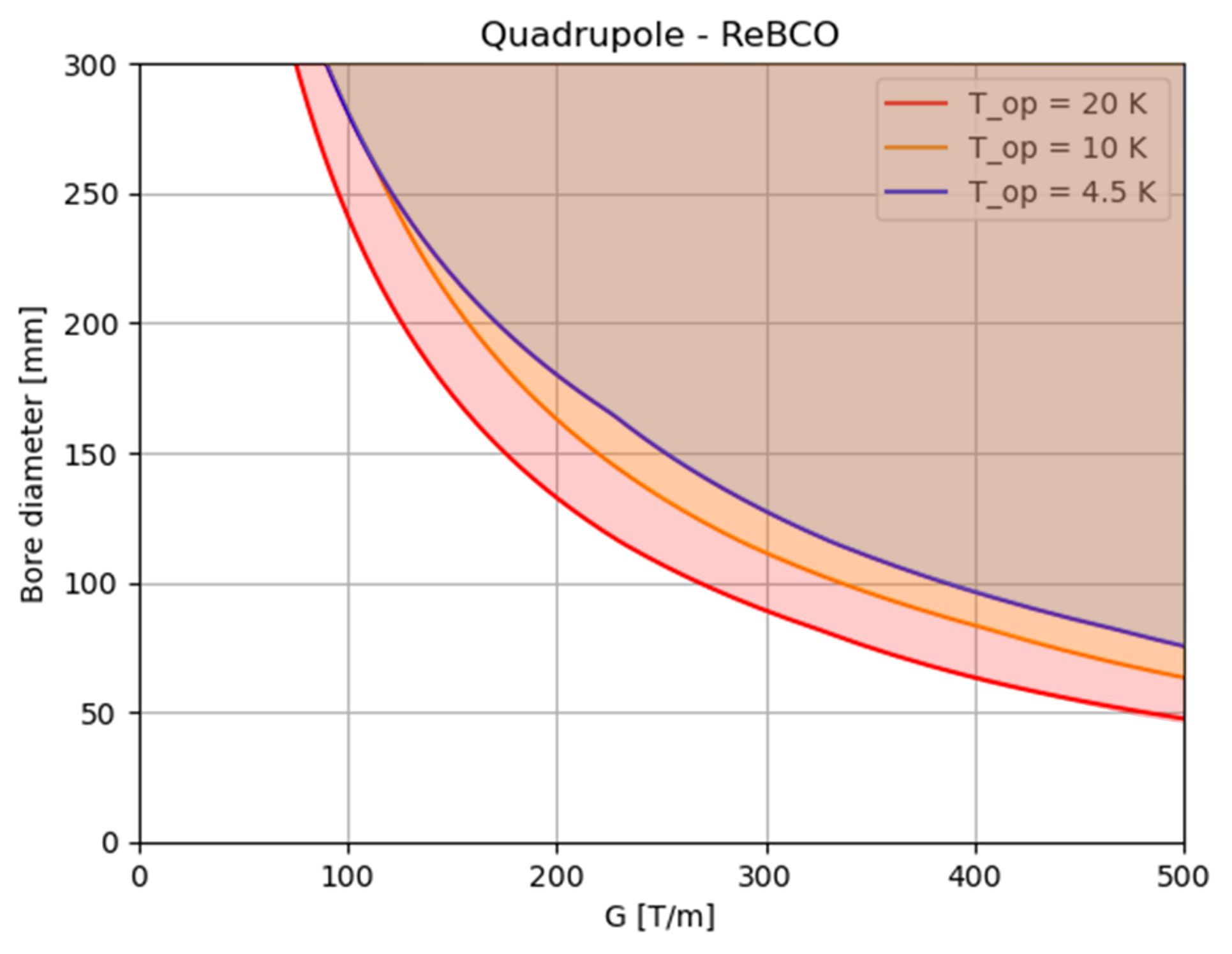}
\includegraphics[width=0.47\textwidth]{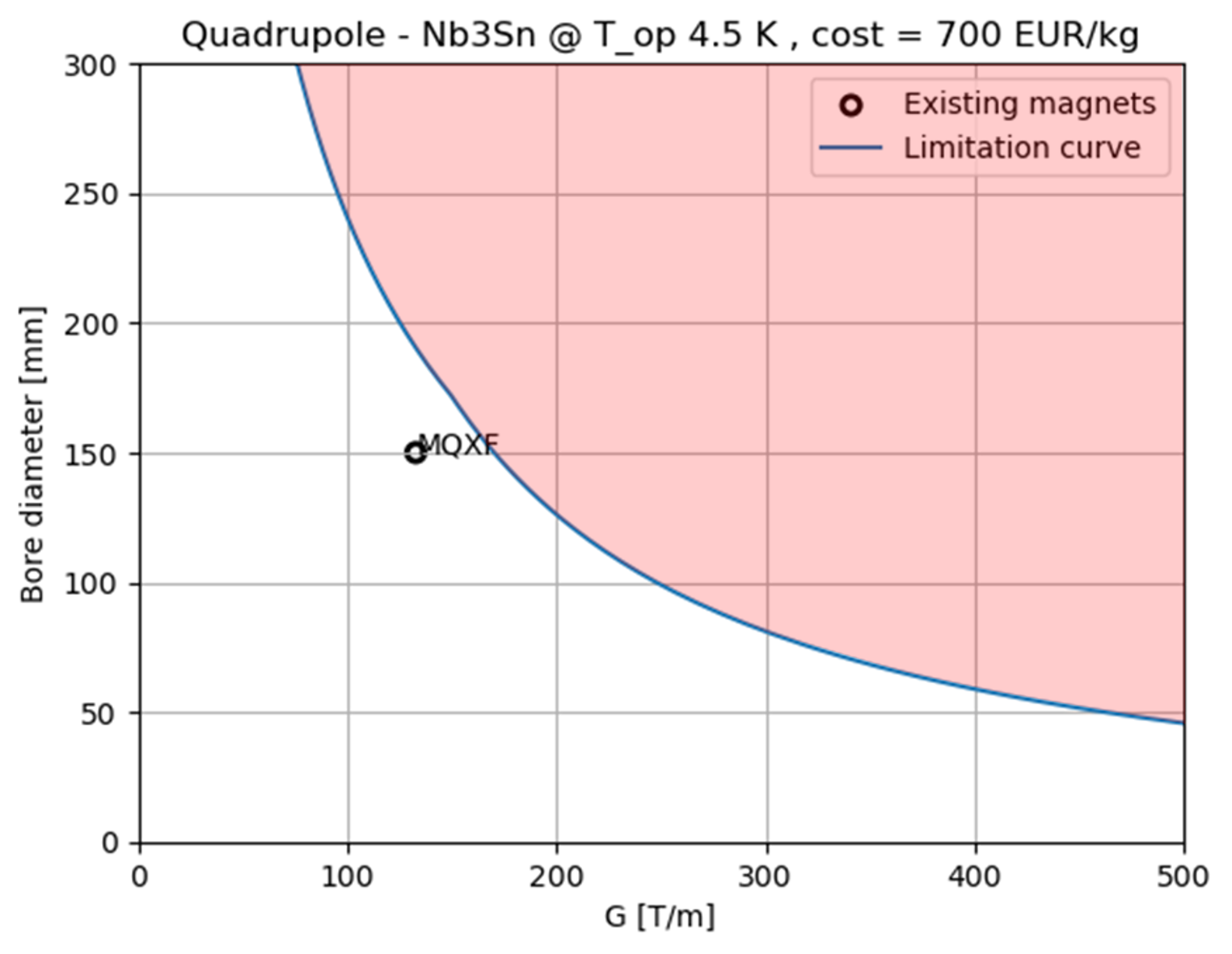}
\caption{Limitation curves as a function of stress and margin shown against bore diameter vs. magnetic field gradient of collider quadrupoles consisting of ReBCO at different operating temperatures (left) or Nb$_3$Sn at 4.5\,K (right). This is a current work in progress, with final design targets still being decided and protection limits created. Plots courtesy of D.\,Novelli.} \label{fig:ABchart_quad}
\end{figure}

\paragraph*{Planned work}
We will pursue the analysis of quadrupole performance limits (for the design of the interaction regions) and combined function magnets, necessitated by neutrino flux mitigation. We plan to finalize the "A-B" charts for the Superconducting (SC) dipoles and quadrupoles in the collider, incorporating an assessment of the feasible range of combined function gradient versus dipole strength. The main difficulty will be to extend the analysis to combined function magnets. We will most likely need to resort on parametric Finite Element Analysis (FEA) for this. 

Following an iteration on the beam optics, we will then advance to the examination of engineering designs for the collider arc dipole and combined function magnets. This study will encompass stress management strategies aimed at enhancing robustness and compensating for inevitable stress peaking factors. Additionally, it will be of interest to explore a fast-track line for a reduced collider energy, specifically in the range of 3\,TeV, as investigated by US-MAP, leveraging the Nb$_3$Sn technology nearing demonstration.

\paragraph*{Next priority studies and opportunities for additional collaboration}

The engineering design of the arc and IR magnets would require a considerable amount of personnel resources, well beyond what is presently allocated in the study. Although we will focus on the most challenging elements (arc dipoles and IR quadrupoles), other magnets may need due attention. One striking example is the Nb$_3$Sn option that could be used to fast-track the construction of a first collider at 3\,TeV. This design is presently beyond the foreseen scope.

As noted in Section~\ref{sec:Section7_1_1}, the effect of radiation on HTS can be an issue. The radiation effects of the harsh and unique environment of a Muon Collider cannot be reproduced and tested in an existing installation, thus an evaluation of material response will need an understanding and scaling of existing and future data. For this specific issue we plan to rely on advances in material characterization for magnetically confined fusion.

\subsection{Power converters for the muon acceleration to TeV energies
}
\paragraph*{System overview}
The quick acceleration times can only be achieved with considerable electric power provided by the power converter. The fast $\frac{dB}{dt}$ also carries huge voltages with it. In order to deal with both huge power and voltages (Table~\ref{tab: PowerTab1}), the power converters of the RCS are based on resonant circuits and must be divided into several series/parallel connected power cells. A power cell is the basic power converter block. Many such power cells are connected to power one sector of the accelerator. One sector is formed by series of connected magnets, i.e.~with the same current flowing trough. The RCS is divided into a given number of sectors, each supplied by an independent power converter.
Inside one Sector, the power cells can be connected in Cell-Load cascade or Unified Cell to load connection mode, as illustrated in Fig.~\ref{fig:PowerFig1}. The magnets and power converters will need to be designed such that they possess the same number of independent circuits and series connected power cells and magnets. Figure~\ref{fig:PowerFig1} shows the case of one sector with four separated magnets each possessing four independent circuits and four series connected power cells per circuit.

\begin{figure}[H]
\centering
\includegraphics[width=\textwidth]{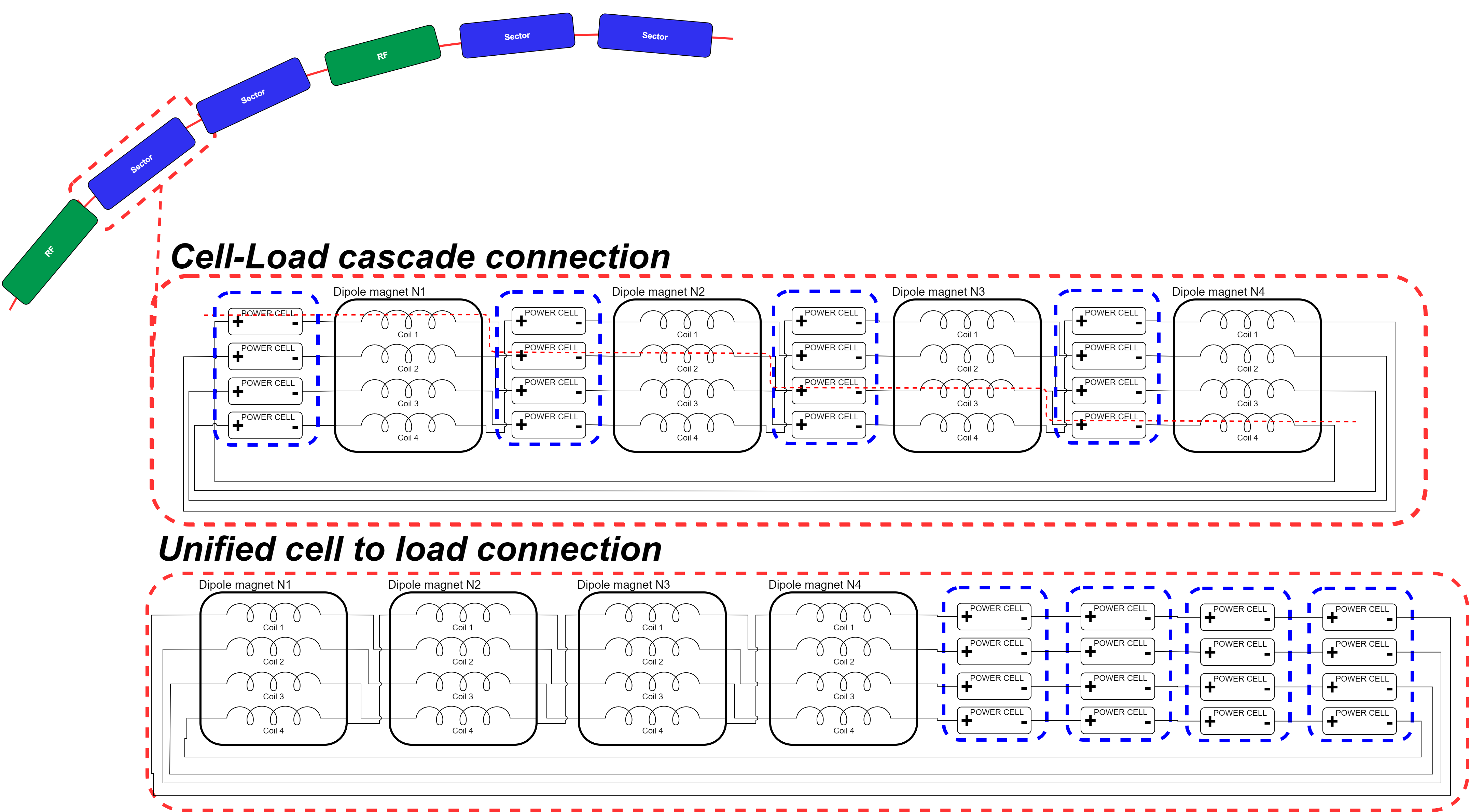}
\caption{Connection of the power cells with the magnets.}
\label{fig:PowerFig1}
\end{figure}

The choice of the connection depends upon the voltage insulation level of the magnets and the location of the power cells. The Cell-Load cascade connection allows a much smaller voltage insulation to ground but only makes sense if the power cells are located close to the magnets i.e.~in the tunnel. The Unified cell to load alternative, has higher insulation requirements but power cells can be installed on the surface without having too much cable routing.

The power cell design is based on a discontinued resonance approach where the energy is stored in one or more capacitor banks and then discharged on the magnets with a combination of several semiconductor switches. Two different topologies have been identified and are shown in Fig.~\ref{fig:PowerFig2}. In both cases, the LC resonance starts by bringing the current (Field) on the negative polarity at the injection and continues the oscillation towards positive values during the real particle acceleration phase. It then returns back to zero where the resonance is actively stopped by the semiconductor switch.
Capacitors are then recharged to the initial value and the power converter waits to be restarted for the next pulse.The concept is illustrated in Fig.~\ref{fig:PowerFig3}.

\begin{figure}[ht]
  \begin{minipage}{0.5\textwidth}
    \centering
    \includegraphics[width=\linewidth]{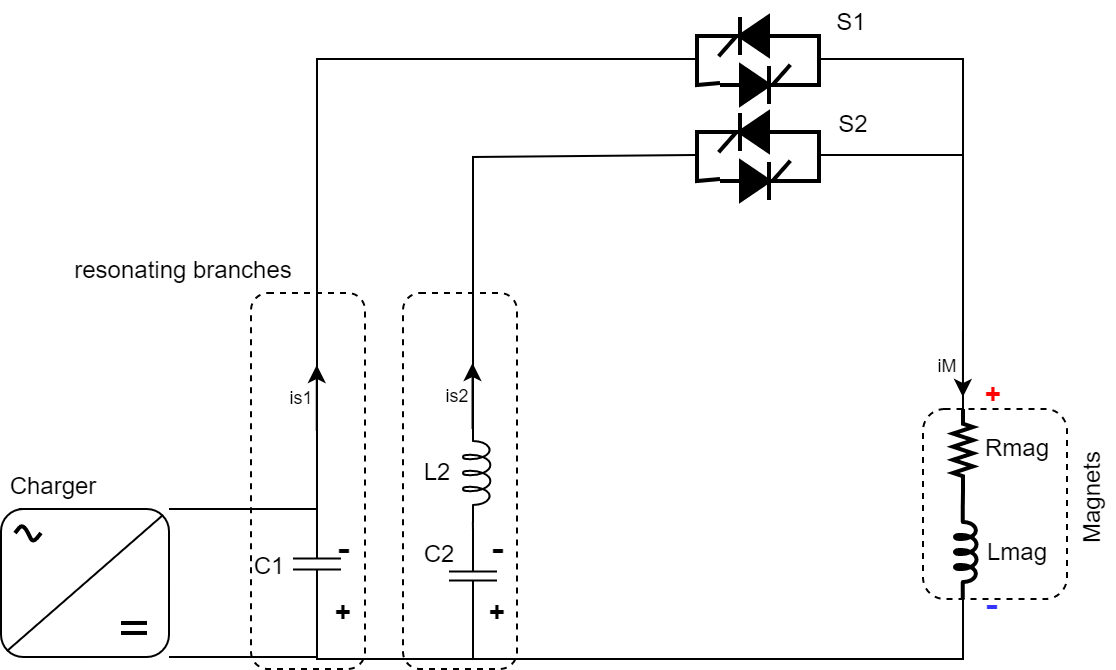}
  \end{minipage}
   \begin{minipage}{0.5\textwidth}
    \centering
    \includegraphics[width=\linewidth]{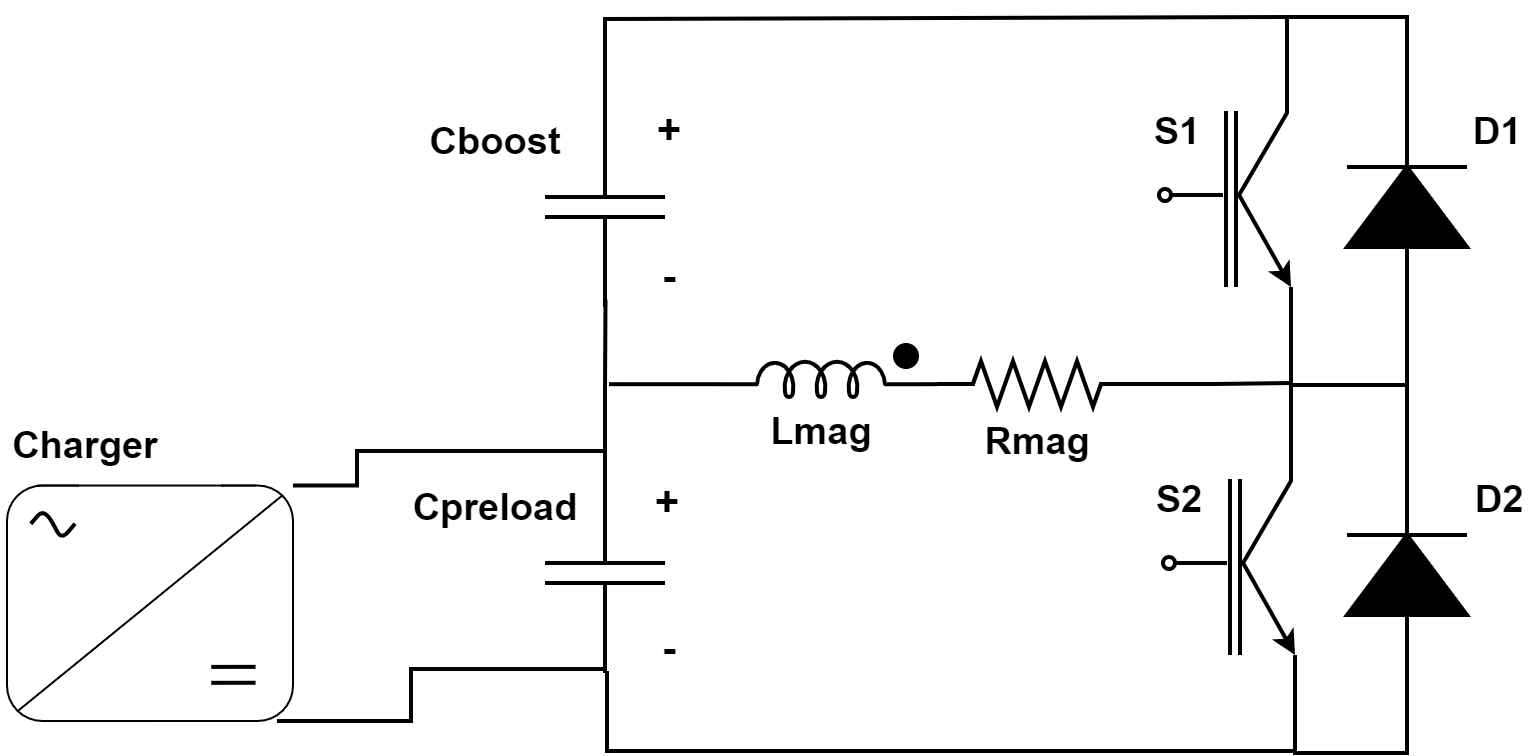}
  \end{minipage}
  \caption{Full wave (left) and switched (right) discontinued resonance principles}
  \label{fig:PowerFig2}
\end{figure}

\begin{figure}
\includegraphics[width=\textwidth]{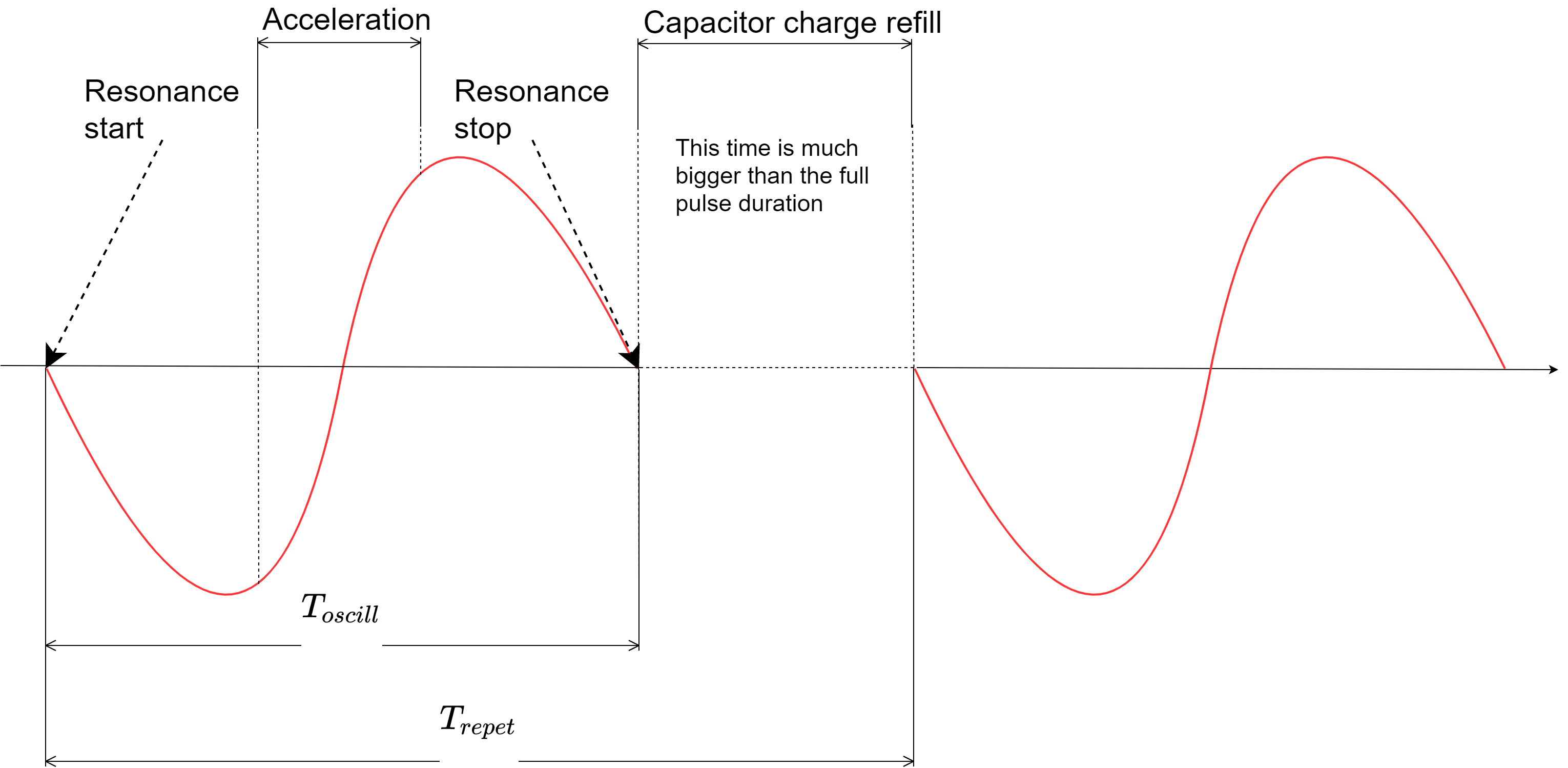}
\caption{Discontinued resonance approach \label{fig:PowerFig3}}
\end{figure}

\paragraph{Key challenges}
The quick acceleration requires extremely high peak power sources that can only be economically guaranteed by making large use of resonance circuits such as the one proposed in the previous paragraph.
On the other side, resonant circuits are not very easily controllable and it is difficult to assure high precision among the different sectors(see Fig.~\ref{fig:PowerFig1}) and high repeatability from pulse to pulse.
In addition, LC resonance doesn't match with linear acceleration profiles as required to minimize the RF requested power.
The above parameters will have to be carefully analysed to avoid overrating of the power converters.
\begin{table}[!hbt]
   \centering
   \small
   \caption{RCS peak power and voltages for a linear Bref ramp.} 
   \begin{tabular}{lccccc}
       \toprule
       \textbf{} &\textbf{RCS1} & \textbf{RCS2} & \textbf{RCS3} & \textbf{RCS4}\\
       \midrule
       Inj Energy {[GeV]} & \SI{63}{} & \SI{314}{} & \SI{750}{} & \SI{1500}{} \\
       Acc. length {[km]} & \SI{5.99}{} & \SI{5.99}{} & \SI{10.7}{} & \SI{35.0}{} \\
       Res. mags length {[km]} & \SI{3.65}{} & \SI{2.54}{} & \SI{4.37}{} & \SI{20.38}{} \\
       Binj in gap {[T]} & \SI{0.36}{} & \SI{-1.8}{} & \SI{-1.8}{} & \SI{-1.8}{} \\
       Bextr in gap {[T]} & \SI{1.8}{} & \SI{1.8}{} & \SI{1.8}{} & \SI{1.8}{} \\
       B ramp time Tramp {[ms]} & \SI{0.35}{} & \SI{1.10}{} & \SI{2.37}{} & \SI{6.37}{} \\
       Dipoles Pmax {[GW]} & \SI{111}{} & \SI{54}{} & \SI{43}{} & \SI{74}{} \\
       Dipoles Vmax {[MV]} & \SI{2.4}{} & \SI{1.1}{} & \SI{0.9}{} & \SI{1.5}{} \\
       
       \bottomrule
   \end{tabular}
   \label{tab: PowerTab1}
\end{table}
\paragraph{Recent achievements}
The two resonating circuits reported in Fig.~\ref{fig:PowerFig2} are a concept generalization of two possible approaches for the power converters. 
The full wave resonator is the approach by which it is possible to get close to the desired linear acceleration profile, by adding two or more resonant circuit. The most efficient approach is to use the first and second harmonics (Fig.~\ref{fig:PowerFig4}), but other approaches are possible. The circuit is very simple and, in theory, the cost of the power electronic switch, (two thyristors) is limited.

\begin{figure}[ht]
  \begin{minipage}{0.5\textwidth}
    \centering
    \includegraphics[width=\linewidth]{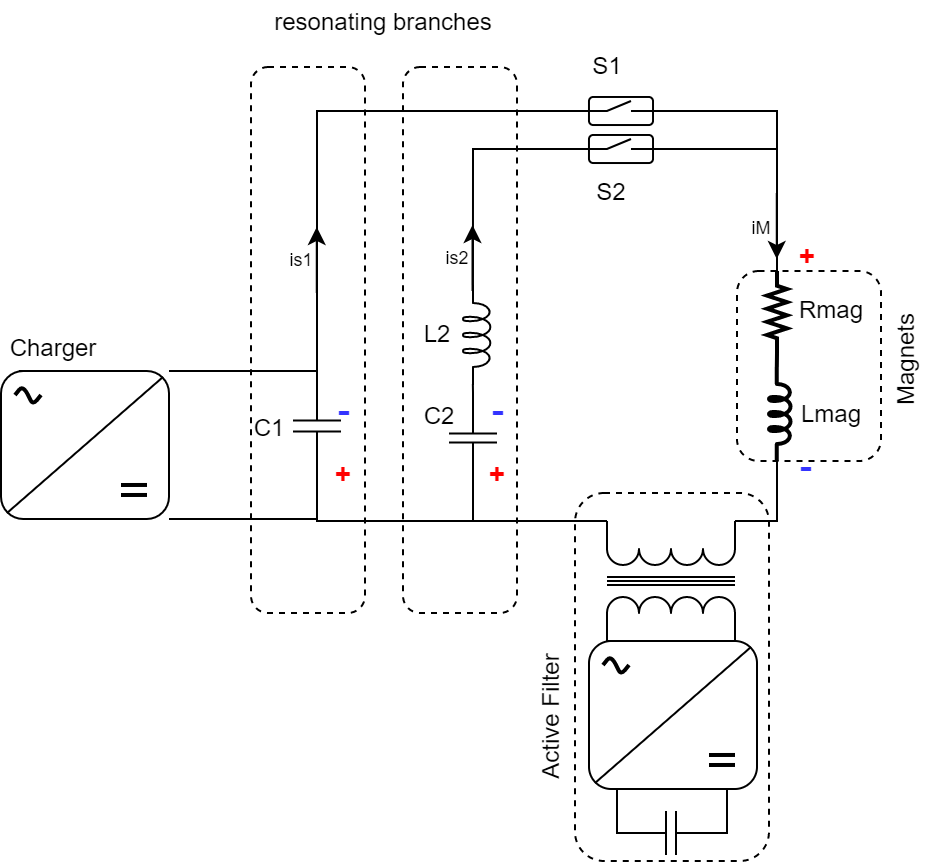}
  \end{minipage}
   \begin{minipage}{0.5\textwidth}
    \centering
    \includegraphics[width=\linewidth]{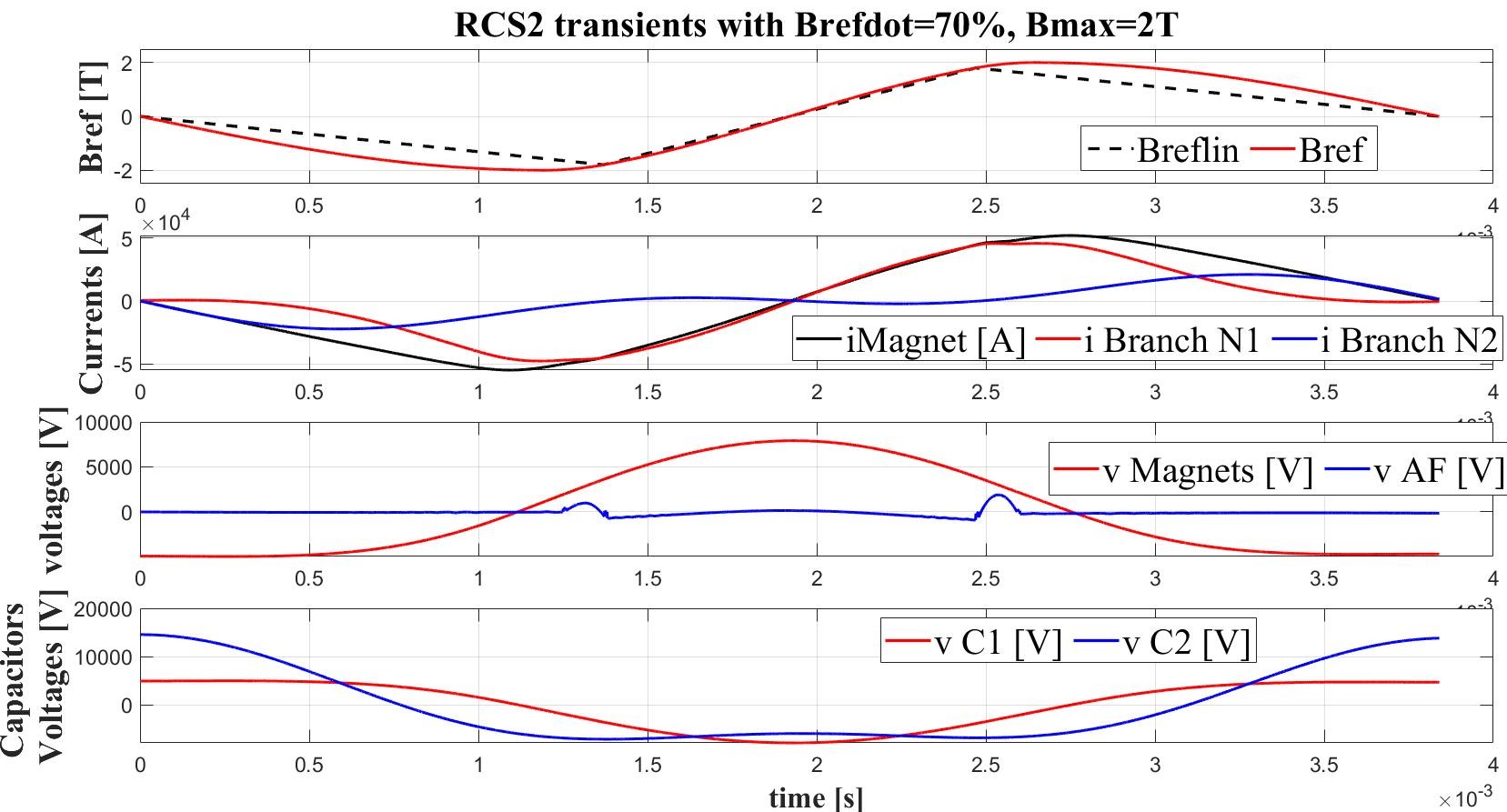}
  \end{minipage}
  \caption{Full wave resonance with active filter contribution}
  \label{fig:PowerFig4}
\end{figure}

However, the circuit doesn't offer any clear possibility to make a regulation and therefore to control the current among the different sectors of the accelerator. To overcome this limitation, additional hardware would have to be installed (Active filter) whose cost can be high.
In addition, energy storage capacitors oscillate with bipolar voltage thus leading to a considerable less energy dense design (and therefore higher cost).
The introduction of the switched resonance circuit makes the circuit more complicated and costly from the power electronics point of view, but no additional magnetic components (L2 in Fig.~\ref{fig:PowerFig2}) are required and energy storage capacitors only oscillate on positive voltage so that a more favorable design is possible for these (Fig.~\ref{fig:PowerFig5}).

\begin{figure}
\includegraphics[width=\textwidth]{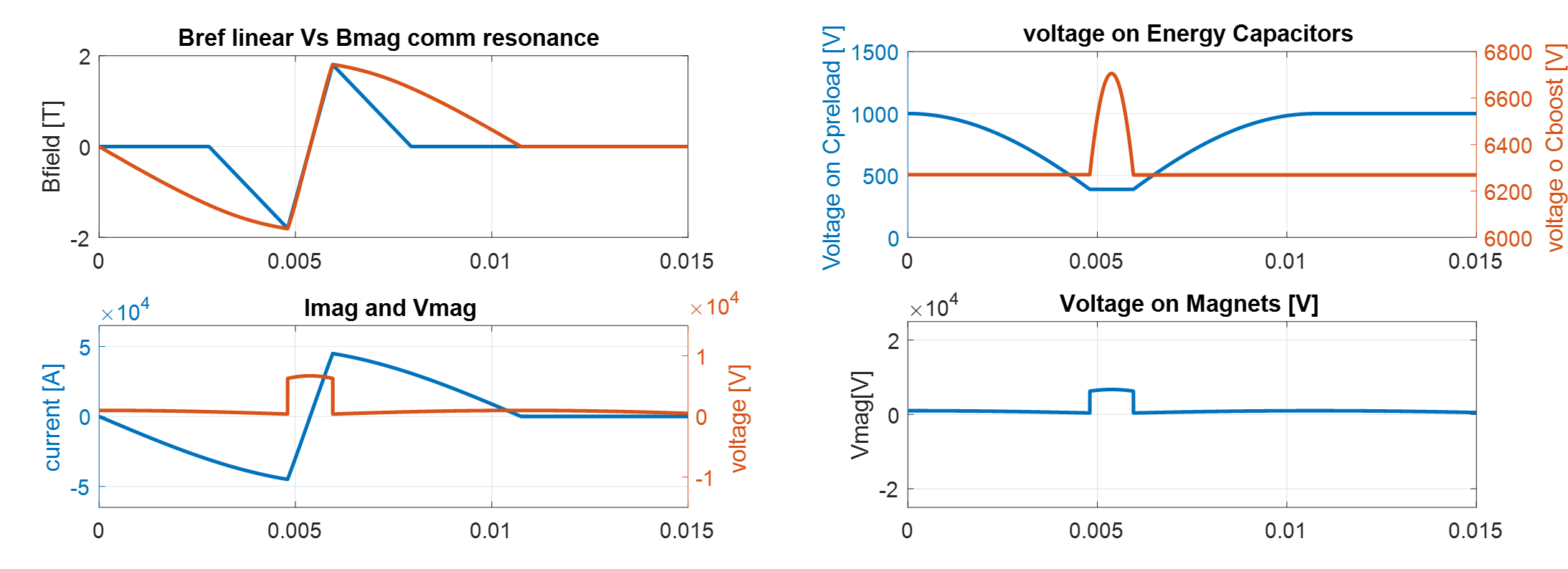}
\caption{Simulation of a switched resonance case \label{fig:PowerFig5}}
\end{figure}
Even more important, it is possible to act on the switching on-off time instants to play with the equivalent output voltage. When this is done using several tens of power cells (as in the configuration shown in Fig.~\ref{fig:PowerFig1}) one realizes that it is possible to have basic control of the currents in each sector and possibly avoid using an active filter. The control concept would be based on an "Iterative Learning Controller" approach, which acts on the reference voltage of each power cell to achieve matching of the current profile across the different sectors. The study of the control algorithm feasibility is part of the important missing efforts listed in the next paragraphs.

\paragraph*{Planned work}
Because of the strong links with the design of the resistive magnets and with other technological areas of the accelerator, it is difficult to determine the design of the power converters alone. Instead a combined python based design suite is being developed. A preliminary sketch is reported in Fig.~\ref{fig:PowerFig6}. 
\begin{figure}
\includegraphics[width=\textwidth]{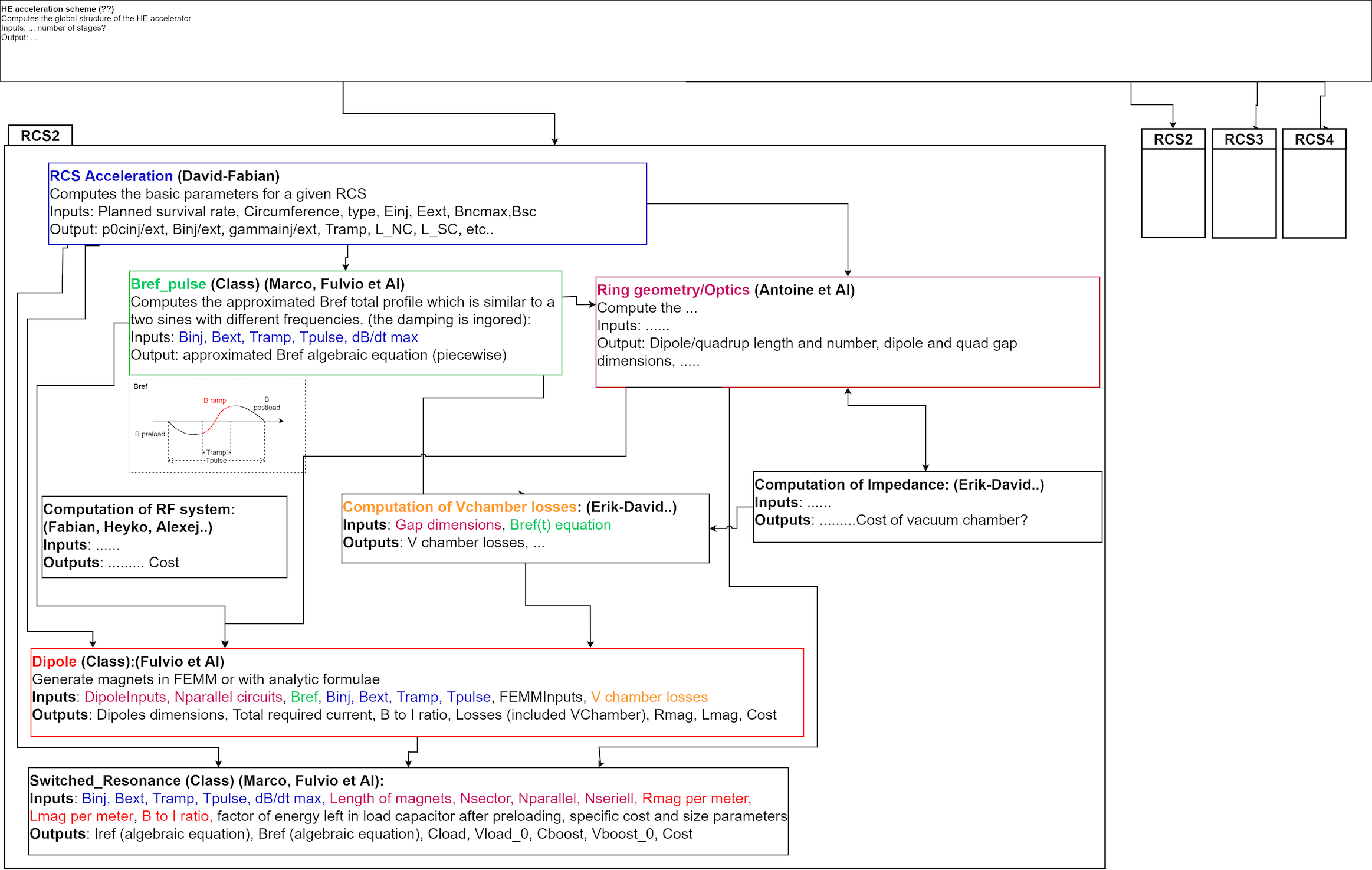}
\caption{Python Accelerator simulation suite schematic \label{fig:PowerFig6}}
\end{figure}

Different Python classes shall perform the necessary computations to design the relative parts and compute the cost of it. An integrated optimization can then lead to a cost effective design of the accelerator as a whole.

Another very important part in the design of the power converters is with the study of the ILC controller to verify if such an approach would allow controlling the current among sectors with the required accuracy. 

\paragraph{Next priority studies and opportunities for additional collaboration}
Throughout our analysis we have assumed many design values that have a remarkable impact on the cost and carbon footprint of the final design. As the application is very peculiar, our assumption cannot be justified by existing similar applications.
In particular the limits of the IGBTs could be different from that assumed in the design. As the systems only works for few ms every 200, the DC maximum voltage and peak switched current values may be higher than normally assumed with "standard" power converters, thus reducing the equivalent IGBT number.
Also the energy density design value of the capacitors may be higher (or lower) than assumed.
Thus, the optimisation of the system will require to build (after the design optimization effort) a full scale power cell and study the limits of the different components.
In addition, the system behaviour and control of several connected power cells has to be demonstrated with a string test using a limited number of full scale power cells.

\subsection{RF
}

\label{sec:Section7_3}
\subsubsection{RF system for muon cooling
}
\paragraph*{System overview}
The RF system for the muon cooling complex consists of approximately 8000 RF cavities operating in the frequency range starting from a few hundred of MHz. The majority of the cavities operate at 352 and 704\,MHz in the initial and 6D-cooling channels. The main peculiarity of the RF system is that the RF cavities are interleaved with strong solenoidal magnetic fields, which enhance the cavity breakdown rate. It has been proven to affect the high gradient operation of the RF cavities necessary to re-accelerate muons as fast as possible due to their short life time. This also limits the choice of the RF cavities to normal conducting ones since SRF cavities cannot operate in magnetic field. In order to reach high gradient on the order of a few tens of MV/m in normal conducting cavities, high peak RF power on the order of few MWs per one single cell cavity is needed. This results in very high peak power requirements of a few tens of GW. On the other hand, the single (or few) bunch operation mode results in relatively short pulse lengths of $\approx 10$ \textmu s and relatively low duty factor of $\approx10^{-4}$.
\paragraph*{Key challenges}
Several experiments on breakdown rates in cavities under high magnetic field have been carried out in the frame of dedicated programs (International Muon Ionization Cooling Experiment (MICE) \cite{mice2020}, Muon Accelerator Program at Fermilab (MAP) \cite{palmer2015}). Higher breakdown rates leading to lower achievable accelerating gradients were observed in copper cavities operating at 805\,MHz. A model based on the "beamlet" principle obtained good correlation with experiments \cite{bowring2020}. That model  predicts the appearance of breakdown  when the local temperature rise exceeds a threshold beyond, which plastic deformation and surface damage may affect cavity behavior. Additionally, it also relates the local temperature rise to the magnetic field. The same study predicted an improvement for cavities made of beryllium, which is corroborated by measurements: a maximal gradient of nearly 50\,MV/m is kept in presence of 3\,T, while for copper the gradient drops from 24\,MV/m at 0\,T to the 13\,MV/m at 3\,T \cite{bowring2020}.

Integration of the RF cavities together with high-field superconducting solenoids and absorbers in the common vacuum vessel is another engineering challenge. The superconducting solenoids operate at cryogenic temperature whereas RF cavities and the associated RF network operate at room temperature. Combining all the subsystem in a compact way is a key challenge in the engineering design of the muon cooling channel.

\paragraph*{Recent achievements}
A simplification of the "beamlet" model for short pulses (< 10\,\textmu s) has been developed, where the heat diffusion in the cavity walls can be neglected. It showed that reducing the pulse duration should strongly increase the maximal acceptable B field, for the same accelerating field. Moreover, materials other than copper (beryllium, aluminum, copper alloys such as CuAg) with different mechanical properties could give other solutions to mitigate this effect. Lower temperatures could also mitigate this effect, as it affects the field emission of the cavity, as well as increasing the surface hardness and the thermal conductivity . Further, the external magnetic field may affect elliptical cavities less than pillbox cavities.

\paragraph*{Planned work}
Further simulations involving thermal studies must be performed in order to estimate the local thermal diffusion and its potential effect on the breakdown limit in different configurations. These simulations must include multiple physics, including RF, particle tracking, and thermodynamics. A dedicated test stand is needed for studying these possibilities, which must include a high-power (some MW) RF source, one or several cavities and a strong magnet (some Tesla). See Section~\ref{sec:Section09_2} for proposed test stands.

On the design side, based on the input from the muon cooling beam dynamics design, a consistent set of parameters for all of the RF cavities and their associated RF systems will be elaborated creating the backbone of the conceptual design of the RF system for the muon cooling complex. For certain specific cases, starting with the cavity which will be used for the demonstrator, more advanced engineering design will be performed including 3D RF design and thermo-mechanical simulations, as well as integration into the cooling cells for operation at very high gradients in strong solenoidal magnetic fields. The impact of the beam loading on the muon energy spread will have to be investigated and appropriate mitigation measures to be proposed.

\paragraph*{Next priority studies and opportunities for additional collaboration}
The conceptual design of all of the cavities for the whole moun cooling complex including front end, rectilinear initial and 6D-cooling, bunch merge and final cooling will not be covered due to the large number of different cavities. Design and construction of prototype cavities at nominal RF frequency as well as testing at high gradients in strong magnetic fields is required as next step.

\subsubsection{RF system for acceleration
} 
\paragraph*{System overview}
After the cooling channel, a linear accelerator and two Recirculating Linear Accelerators (RLAs) provide an initial acceleration up to 63\,GeV. 
The linacs operate at the RF frequency of 352 MHz and accelerating gradient of 25 MV/m.
Downstream from the RLAs, four Rapid Cycling Synchrotrons (RCSs) will gradually increase the energy of the two muon bunches within a few tens of turns each up to the collision energy of 5\,TeV. 
During the transition from the second RLA to the first RCS, the bunches are split and continue as counter-rotating bunches. 
The design choices for the RF system will be guided by the requirements resulting from the beam dynamics simulations and the short muon lifetime. 
As a result, a high RF voltage per turn are required, supplied by hundreds of superconducting 1.3 GHz cavities operating at accelerating gradient of 30 MV/m per RCS. 
A more detailed description of the acceleration chain can be found in Section~\ref{sec:Section6_4}. 
\paragraph*{Key challenges}
All accelerators following the first RCS will be implemented as hybrid RCSs with both normal- and superconducting magnets.  
Due to the nature of this hybrid magnet design, the orbit length and, thus, the revolution period changes during the acceleration, leading to the necessity of fast frequency tuning capabilities for the cavities. \\
The large bunch charge of up to $2.7\times10^{12}$ muons per bunch in the RCS chain will lead to significant transient beam loading and HOM-induced power within the cavities. 
While the instantaneous requirements for cavity powering and HOM power extraction are high, the machine's duty cycle is low. 
The cavities will, therefore, be operated in a pulsed mode. 
The requirements for powering and HOM extraction further increase in the RLAs. 
The number of passes is  $\approx$ 5, while the beam current is higher than in the first RCS due to the higher bunch charge and lower travel time between cavity passages. 
\\
Additionally, the impact of the beam loading will not be consistent in all cavities in an RCS due to the differing time between the passages of the counter-rotating $\mu^+$ and $\mu^-$ bunches.
The same challenge applies to the extraction of the HOM power and cavity powering. 
In the RLAs, the HOM power and cavity powering need to be investigated in detail, as the bunches are planned to pass in buckets directly after each other, which might significantly impact HOM power extraction requirements. In the cavity shape design, the cavity HOMs performance has to be carefully balanced against the fundamental mode performance, which might lead to different designs for each ring due to the different beam currents. 
In the cavity design for the low-energy acceleration, one also has to take the particle speed into account, as some of the accelerators operate in an energy regime where the particles are not ultra-relativistic.

\paragraph*{Recent achievements}
A first approximation of the power requirements for the RCS chain has been performed using the ILC cavities, cryomodules, and powering infrastructures \cite{ch12:ILC-tdr} as a baseline. The results of which can be found in Table~\ref{tab:ch07:RCS_RFpars}. The 9-cell ILC cavity operates at a fundamental frequency of 1.3\,GHz with an accelerating gradient of 30\,MV/m. 
While the ILC features the same repetition rate (5\,Hz) as the muon collider, the beam current and bunch structure differ significantly.
The requirements do not consider transient beam loading, HOM power contributions, cryogenic loss, impact of orbit change detuning, and counter-rotating beams.
\begin{table}[ht]
\begin{center}
\caption{RF parameters for the RCS chain. The average and peak RF power includes losses from the cavity to the klystron, while the wall plug power also includes the klystron efficiency.}
\label{tab:ch07:RCS_RFpars}
\begin{tabular}{lcccccc}
\hline\hline
& Unit & {\textbf{RCS1}} & {\textbf{RCS2}} & {\textbf{RCS3}} & {\textbf{RCS4}} & {\textbf{All}}\\
\midrule
Synchronous phase & [\textdegree] & 135.0 & 135.0 & 135.0 & 135.0 & - \\
Number of bunches/species & - & 1 & 1 & 1 & 1 & - \\
Combined beam current ($\mu^+$ and $\mu^-$) & [mA] & 43.3 & 39.0 & 19.8 & 5.49 & - \\
Total RF voltage & [GV] & 20.9 & 11.2 & 16.1 & 90.0 & 138.2 \\
Total number of cavities & - & 700 & 380 & 540 & 3000 & 4620 \\
Total number of cryomodules & -  & 78 & 42 & 60 & 334 & 514 \\
Total RF section length & [m] & 990 & 530 & 760 & 4230 & 6510 \\
\midrule
Combined peak beam power ($\mu^+$ and $\mu^-$) & [MW] & 640 & 310 & 225 & 350 & - \\
External Q-factor & [$1\times 10^{6}$]& 0.94 & 1.04 & 2.05 & 7.42 & - \\
Cavity detuning for beam loading comp. & [kHz]& $-$0.69 & $-$0.62 & $-$0.32 & $-$0.09 & - \\
Max. detuning due to orbit length change & [kHz]& 0 & 2.0 & 0.33 & 0.20 & - \\
Total number of klystrons & - & 88 & 42 & 30 & 47 & 207 \\
Cavities per klystron & - & 8 & 9 & 18 & 64 & - \\
\midrule
Beam acceleration time & [ms] & 0.34 & 1.1 & 2.37 & 6.37 & - \\
Cavity filling time & [ms] & 0.23 & 0.25 & 0.50 & 1.81 & - \\
RF pulse length & [ms] & 0.57 & 1.35 & 2.87 & 8.18 & - \\
RF duty factor & [\%]& 0.29 & 0.68 & 1.44 & 4.09 & - \\
Peak cavity power & [kW] & 910 & 820 & 420 & 120 & - \\
Total peak RF power & [MW] & 850 & 410 & 300 & 460 & - \\ 
Average RF power & [MW] & 2.42 & 2.76 & 4.27 & 19.0 & 28.4 \\
Average wall plug power for RF System & [MW] & 3.72 & 4.25 & 6.56 & 29.2 & 43.7 \\
HOM power losses per cavity & [kW] & 9.75 & 12.14 &  6.74 & 2.77 & - \\
Total HOM power losses  & [MW] &  6.83 & 4.61 & 3.64 & 8.3 & 23.38\\
\midrule\midrule 
\end{tabular}
\end{center}
\end{table}


\paragraph*{Planned work}
Within the work package, optimised cavity geometries for the specific cases of the muon collider RCSs and the low-energy acceleration complex are planned to be designed. 
Due to the different beam- and machine parameters in the low- and high-energy acceleration, the impact of this optimisation will be significant. 
The target parameters, such as the fundamental mode frequency or HOM stability requirements, will be discussed with the other working groups involved. 
For the final choice of the frequency, the power requirements of the acceleration and cryomodules might play a significant role. \\
During the acceleration, muons will constantly decay, resulting in many seed particles for multipacting, which might lead to breakdowns. 
The magnitude of this effect needs to be studied in detail and be included in the shape optimisation. \\
For the chosen cavity geometry, the development of an HOM damping scheme as well as HOM coupler and FPCs (Fundamental Power Couplers) design will be conducted according to the requirements of the collective effects and beam dynamic studies. 
Depending on the powering and beam dynamics requirements, it might be necessary to adjust the number of cells per cavity to stay within the power limits of the couplers.
In addition to the design of the cavities, an uncertainty quantification is planned to be conducted in order to investigate the impact of fabrication imperfections on the performance of the cavity. 
The results of this analysis will serve as a requirement for the fabrication precision in the cavities. \\
The power consumption of the system will be determined for the different frequencies under consideration. 
To achieve a good compromise between power consumption and voltage stability under the different conditions in each accelerator, simulations will be used to determine an optimal value for the loaded quality factor $Q_L$ and the cavity detuning $\Delta\omega$. 

\paragraph*{Next priority studies and opportunities for additional collaboration}
The work package will focus on the design of the cavities and couplers for both the high- and low-energy acceleration. The integration of the cavities into cryomodules is not part of the work package. 
Possible improvements to the fabrication procedure could enhance the accelerating gradient for the accelerating cavities and should be examined, as these could significantly reduce the number of required cavities.

\subsubsection{RF power sources
}
\paragraph*{System overview}
The muon cooling system will require a significant amount of RF power to supply around 8000 cavities each with around 3\,MW, summing up to around 10\,GW of peak power. The frequencies required are likely to be around 352\,MHz and 704\,MHz. This will be driven by klystrons at this power and frequency range.

\paragraph*{Key challenges}
Klystrons at 352 or 704\,MHz are rather large devices which are typically located in a surface building. Using a standard commercially available 3\,MW klystron for each cavity will require a surface building on the order of 180\,m by 30\,m, as well as requiring shafts for large waveguides. A key requirement is to reduce the size of the physical footprint. This can be achieved either by using fewer higher power klystrons to feed multiple cavities or by making the klystrons more compact using improved technologies. For example, the E37503 Multibeam Klystron is a 1\,GHz, 20\,MW klystron has an RF circuit length of 1.5\,m, Including the collector and gun, it would have a total length in excess of 2\,m. A 352\,MHz klystron would be significantly longer than this. 
Another issue is the wall-plug electrical power required to drive the klystrons. Klystrons are typically around 60\% efficient at their saturated output power, however, they are typically operated at lower power to provide a low-level RF overhead to cope with RF transients, which further reduces the efficiency. The duty cycle of a muon collider is likely to be around 0.015\%, and including HVDC system and modulator efficiency and RF losses the total power is close to 2\,MW of average power. This could be halved to under 1\,MW if the operating efficiency could be increased to closer to 80\%. An advantage of having a higher efficiency is also a reduction in the power handling of the collector further reducing space and cooling requirements.

\paragraph*{Recent achievements}
There has been significant developments in high efficiency, compact klystrons for LHC, FCC \& CLIC at CERN and Lancaster University. The use of two stage klystrons where the gun is operated at a lower voltage and beam power making the bunching process occur in a much shorter distance. This in turn allows better collection of particles into a well focused bunch improving the efficiency, while at the same time reducing the length of the klystron. A two stage (TS) multi-beam klystron has been designed for the CLIC drive beam that operates at 1\,GHz and an output power of 24\,MW, which has an efficiency of 82\% and an RF circuit length of only 0.9\,m.

\paragraph*{Planned work}
Specific design work on 352\,MHz and 704\,MHz klystrons for MuCol will start in 2024 at Lancaster University. This will most likely be a two-stage klystron to minimise the size. The number of cavities fed by each klystron will be optimised to minimise the size of the surface building while also increasing the efficiency.

\paragraph*{Next priority studies and opportunities for additional collaboration}
A prototype two-stage klystron should be constructed and demonstrated, however this could be done for LHC, FCC or CLIC. 

\subsection{Target
} 
\label{sec:Section7_4}

\paragraph*{System overview} 
The production of muons in the muon collider front-end involves the collision of a proton beam with a target material. This interaction triggers deep inelastic reactions that generate kaons and pions, subsequently decaying into muons. To effectively capture these particles and maintain control over their emittance, a robust solenoidal magnetic field is crucial. This magnetic field confines the charged particles along helical trajectories within the production target and subsequent beamline. To sustain the megawatt-class production target, active cooling is essential, requiring separation from the primary vacuum through beam windows. Moving downstream through the tapering sector before reaching the muon-cooling section, provision for an extraction channel for the unspent proton beam is required to accommodate a high-power dump absorber.

Initially, a graphite target is being considered as the primary option due to its suitability. Graphite allows operation at high temperatures and exhibits remarkable resistance to thermal shock. Its extensive usage in various facilities attests to its reliability for \SI{2}{MW} operation. However, two other distinct technological alternatives are currently under investigation. The first alternative involves a heavy-liquid metal (HLM) target made of pure lead. This option addresses concerns regarding radiation damage and could potentially eliminate the necessity for integrated active target cooling within the cryostat of the superconducting (SC) solenoid. The second alternative explores a fluidized tungsten target that circulates micro-scale spheres in a closed loop. Notable features include high thermal-shock resistance and reduced susceptibility to cavitation, corrosion, and radiation damage. Both options present a possible pathway for even higher beam power delivery.

\paragraph*{Key challenges} 
The MAP study has laid the groundwork for a muon collider but has left certain crucial aspects unexplored. One significant gap is the lack of an integrated system design and optimization for the target systems for production of a high-charge, high-quality muon beam essential for achieving the desired luminosity. Similarly, detailed specifications for other front-end beam-absorbers, such as beam windows and a proton dump, were not addressed until now within the IMCC. 

The integration of a MW-class production target, radiation shielding, and their respective cooling systems within the cryostat of the SC Solenoid poses significant challenges. For instance, in a \SI{2}{MW} facility utilizing a graphite muon production target, around \SI{110}{kW} of thermal power are deposited on the target. Additionally, up to 34\% of the stored beam energy is deposited onto tungsten shielding near the target, necessitating efficient heat extraction and isolation from the surrounding solenoid. Addressing these issues requires dedicated vessels and the routing of multiple services. The ability to align and perform target exchanges independently, without removing the surrounding shielding of over 20 tonnes, needs consideration. Furthermore, residual radiation doses on the target systems necessitate designs compatible with remote handling.

Whether employing a carbon, HLM, or fluidized tungsten target, the high-intensity \SI{2}{ns} pulse on target results in a very high adiabatic temperature rise of the bulk material. This temperature rise can lead to dynamic stress-strain responses in solid targets, potentially surpassing material limits. HLM targets, though free of structural concerns, may induce extreme loads on the containment vessel due to pressure waves in the liquid, necessitating in-depth design and modelling of this complex multiphase problem. Understanding erosion management and powder handling for a fluidized tungsten target is imperative, particularly due to limited operational experience in present facilities.

The substantial levels of radiation exposure, encompassing hadronic and electromagnetic showers, as outlined in Section~\ref{sec:Section7_5}, pose a significant challenge for the target system and its adjacent components. A comprehensive study of target and beam window materials against radiation damage is required. High integrated fluence of protons can induce atomic changes in the crystalline structure of the target materials, resulting in conductivity loss and increased brittleness, potentially reducing the lifespan of the device. For graphite, radiation annealing via operation of the target at high-temperature, coupled with larger beam sizes on target, is considered a mitigation strategy. Proton beam windows, separating the primary vacuum from the target confinement atmosphere, benefit from low Z materials such as beryllium, in order to reduce their integrated dpa (displacement per atom).

Effective cooling is essential to maintain reasonable temperature conditions for the entire system and minimize heat dissipation towards the superconducting magnet. Simultaneously, designing a sustainable facility requires optimal sizing of auxiliary services. 

Lastly, a major challenge involves designing a proton absorber capable of fully absorbing the entire beam if necessary at a regular dumping power, alongside the intricate design of a beam extraction channel.

\paragraph*{Recent achievements} 
The culmination of current studies has resulted in a baseline target design featuring an \SI{80}{cm}-long isostatic graphite rod of \SI{30}{mm} in diameter, enclosed within a titanium vessel for hermetic confinement. A carbon composite frame holds the rod in the central position of the vessel. Internally, static helium gas at \SI{1}{bar} prevents sublimation of the graphite during expected high-temperature operation (around 2000$^\circ$C). Additionally, it facilitates heat dissipation through natural convection, as opposed to a radiation-only cooled target. While direct cooling (in contact with the rod) could lower the graphite temperature, the baseline solution was preferred due to its capacity for enhancing radiation defect annealing and reducing erosion concerns.

Active cooling of the inner vessel utilizes a helicoidal helium cooling circuit operating at room temperature and \SI{20}{bar}. Although this setup offers a lower heat transfer coefficient compared to a liquid coolant, it mitigates the formation of pressure waves due to energy deposition on the coolant and concerns regarding water radiolysis. Positioned within the bore of the shielding assembly, the target vessel is planned to be supported on the cryostat's upstream side, allowing easy replacement of the target without interfering with the shielding assembly.

The assembly's protection from radiation has been developed in detail (see also Section~\ref{sec:Section7_5}). A tungsten shielding surrounding the pion capture solenoid is required. To address the dissipation of heat (\SI{680}{kW} for a \SI{2}{MW} facility), the approach shifted from contemplating water cooling, which posed complexities and handling limitations, towards utilizing helium gas cooling. This transition aims to resolve technological challenges associated with corrosion, erosion, and hydrogen embrittlement on the shielding material. The tungsten shielding design has evolved to adopt pie-shaped segments, perforated for improved cooling efficiency, together with the engineering design of the shielding vessel. 

The IMCC studies have been focused on refining the target design and optimizing the pion/muon yield towards the cooling section. After careful consideration, a beam size of \si{5}{mm} ($1\sigma$) was determined as the optimal compromise to ensure that the graphite's dynamic response remains within acceptable limits. This solution aligns with the requirements of a \SI{2}{MW} production target facility and is compatible with a proton injection of \SI{5}{GeV} and \SI{2}{ns} ($1\sigma$) pulses at a frequency of \SI{5}{Hz}.

The proton beam window, currently envisioned to be made of beryllium due to its low density and reduced interaction with the primary beam, will be located outside the cryostat. This component will undergo extensive radiation damage and necessitates dedicated cooling. Initially, materials such as titanium were considered for windows but were dismissed due to an expected accumulation of 10s of displacement per atom (DPA) per year. After several iterations, the decision was made to position it outside the solenoid assembly, ensuring improved accessibility for inspections, potential replacements, and increased availability of service space.

Focused research on an HLM lead target emphasized defining a high-level concept while addressing critical obstacles. An initial concept involving flow in a pipe was discarded due to high magnetohydrodynamic (MHD) losses and concerns regarding shock waves on the retaining target pipe. The current model under consideration involves a free-flow of lead in the vertical direction (similar to a curtain), aiming to eliminate the concerns posed by the initial concept. Ongoing work involves further modelling and development of this concept.

\paragraph*{Planned work} 
The plan includes defining a layout for the proton dump extraction along with its conceptual design. Additionally, there will be a comprehensive focus on fully integrating the target assembly system inside the cryostat, encompassing detailed aspects of its supporting structure. Concurrently, the design of the proton beam window, especially its cooling system, will be developed in conjunction with a high-level integration design study. Careful considerations will be made regarding the installation, dismantling, and maintenance of the frontend target systems equipment. Lastly, comprehensive physics simulations, spanning from muon production to the cooling section, are scheduled to be conducted.

\paragraph*{Next priority studies and opportunities for additional collaboration} 
The ongoing conceptual studies lack vital engineering data essential for the expected operational conditions in a muon collider. There is a need for extensive material characterization and beam tests to complement these studies adequately. For instance, prototyping key components and carrying out beam tests in facilities such as HiRadMat at CERN would provide valuable feedback. Additionally, it is imperative to include the dimensioning of auxiliary services along with respective high-level costing.

\subsection{Radiation shielding
}
\label{sec:Section7_5}

\paragraph*{System overview}
Radiation to equipment poses a significant challenge for all stages of the muon collider complex, from the front-end up to the collider ring. Dedicated shielding configurations must be designed for the different machines in order to dissipate the radiation-induced heat and protect equipment against long-term radiation damage. This concerns in particular the protection of superconducting magnets in different parts of the complex, like the target and capture solenoids of the muon source as well as the superconducting magnets in the RCS and collider rings. The radiation shielding must prevent beam-induced magnet quenches, reduce the thermal load to the cryogenic system, and prevent magnet failures due to the ionizing dose in organic materials (e.g.~insulation, spacers) and due to atomic displacements in the superconductors. In most cases, dedicated absorbers need to be embedded inside the magnet aperture in order to attenuate secondary radiation showers before they can reach the coils. 

The proton-driven MW-class muon source is one of the areas in the muon complex where massive shielding elements are needed for equipment protection. The solenoids near the production target are exposed to the secondary radiation showers generated by inelastic collision products in the target. The front-end design elaborated within MAP considered a combination of resistive copper magnet inserts and larger-aperture solenoids made of low-temperature superconductors. The latter are more sensitive to radiation than resistive magnets and were foreseen to be shielded by a thick layer of helium gas-cooled tungsten beads. A new solenoid configuration, based entirely on high-temperature superconductors (HTS), has been developed within the IMCC, combined with a new arrangement of tungsten shielding inserts for absorbing electromagnetic and hadronic showers. 

\begin{table}[t]
\begin{center}
\caption{Muon decays in the collider ring (one bunch per beam), assuming an injection frequency of 5\,Hz and 1.2$\times$10$^{7}$\,s (=139 days) of operation per year. }
\label{tab:parameters_radcollider}
\begin{tabular}{lcc}
\hline\hline
& \textbf{3~TeV} & \textbf{10~TeV} \\
\hline
Muons per bunch & 2.2$\times$10$^{12}$ & 1.8$\times$10$^{12}$ \\
Circumference & 4.5\,km & 10\,km\\
Muon decay rate per unit length & 4.9$\times$10$^{9}$~m$^{-1}$s$^{-1}$ &  1.8$\times$10$^{9}$~m$^{-1}$s$^{-1}$\\
Power per unit length carried by decay $e^\pm$ & 0.411\,kW/m & 0.505\,kW/m\\
Total decays per unit length and per year & 5.87$\times$10$^{16}$~m$^{-1}$ & 2.16$\times$10$^{16}$~m$^{-1}$\\
\hline\hline
\end{tabular}
\end{center}
\end{table}

Dedicated shielding configurations are also needed in the accelerators and the collider, protecting equipment against the radiation load due to muon decay. The power carried by decay electrons and positrons is on average about 35\% of the energy of decaying muons (the rest is carried away by neutrinos, which are irrelevant for the radiation load to the machine). With the presently assumed beam parameters, this amounts to about 500\,W/m dissipated in the collider ring; the collider parameters are chosen such that the power load is about the same at 3\,TeV and 10\,TeV (see Table~\ref{tab:parameters_radcollider}).
The decay electrons and positrons can have TeV energies and emit synchrotron radiation while travelling inside the magnets. The energy is then dissipated through electromagnetic showers in surrounding materials. In addition, secondary hadrons can be produced in photo-nuclear interactions, in particular neutrons, which dominate the displacement damage in magnet coils. Shielding studies for muon colliders have been previously carried out within MAP, indicating that a continuous liner (few centimeters of tungsten) is needed inside magnets and magnet interconnects. The shielding requirements for collider energies up to 10\,TeV have been studied more recently within the IMCC.

Absorbers and shielding configurations might also be needed for the protection of other accelerator systems like RF cavities. In addition, massive absorbers are also required for the Machine-Detector Interface (MDI), in order to reduce the beam-induced background and radiation damage in the detector. The MDI shielding is discussed in more detail in Section~\ref{sec:Section04_2} and will not be covered here. 

\paragraph*{Key challenges}
Designing shielding configurations for a harsh radiation environment like in the muon collider target complex requires a good understanding of radiation damage limits for equipment components like superconductors, insulation materials, etc. The shielding requirements will strongly depend on R\&D efforts towards more radiation-hard solenoids and accelerator magnets. Radiation damage effects in superconductors, in particular in HTS, are also of critical interest for other applications like fusion reactors. The irradiation tests and theoretical studies carried out by the fusion community are highly beneficial for the muon collider design study. One of the key challenges is to derive a relation between radiation damage quantities, e.g., displacement per atom (DPA), and relevant macroscopic material properties like the critical temperature. This is crucial for establishing a relationship between irradiation experiments and radiation environments in the muon collider complex. Despite the still existing uncertainties, it is expected that the understanding of radiation effects in HTS magnets, including the possible benefits of annealing, will significantly improve in the coming decade. 

The shielding blocks inside the superconducting solenoids around the pion/muon production target need to be tens of centimeters thick (radially), in order to sufficiently reduce the heat load and radiation damage in the magnets. The shielding has to be efficiently cooled since it has to dissipate a significant fraction (about 35\%) of proton drive-beam power (2\,MW). Other important aspects are the machining, handling, assembly and support of the shielding blocks. Optimizing the shielding configuration around the target and in the downstream capture section is crucial, since the shielding will determine the final aperture requirements for the solenoids. Another challenge is the overall integration of the heavy shielding inside the cryostats. Furthermore, the shielding needs to be thermally insulated from the cryostats, considering the significant heat dissipation in the shielding.  

One of the key challenges for the collider shielding design is the overall optimization of geometrical aspects (beam aperture, shielding thickness, coil aperture) and thermal aspects (shielding and coil temperature, thermal insulation, cooling scheme). This optimization critically depends on technology choices, e.g., low-temperature versus high-temperature superconductors, and requires a multi-disciplinary design approach including radiation and beam physics, magnet engineering, cryogenics and vacuum.
A careful optimization of the absorber thickness is important since the absorbers significantly impact the aperture requirements for magnets.
An essential design parameter for the shielding is the maximum allowed heat deposition in the cold mass of collider magnets and the resulting cooling requirements, which will be an important cost factor for facility operation due to the associated power consumption (see Section~\ref{sec:Section7_7} for details). Besides the conceptual shielding design, one also faces different engineering challenges. The shielding needs to be equipped with a cooling circuit in order to extract the heat deposited by the particle showers. In addition, the vacuum chamber needs to support the weight of the shielding absorbers ($>$ 100\,kg/m). Another important aspect is the raw material and manufacturing cost of the shielding. Considering the length of the collider (10\,km for the 10\,TeV machine), the shielding cross section hence needs to be carefully optimized.  

\paragraph*{Recent achievements}

A conceptual radiation shielding design was conceived for the target area and the downstream capture section and chicane. Dedicated shielding studies were performed with the FLUKA Monte Carlo code, in order to quantify the heat load and radiation damage in the solenoids. The shielding was assumed to be composed of helium gas-cooled tungsten segments. In order to thermalize and capture neutrons, the shielding around the target was assumed to embed a layer of water and boron carbide. This can reduce the cumulative displacement damage in the coils by about a factor of two. Table~\ref{tab:dpadosetargetsolenoid} shows the annual DPA and dose in the target solenoid for different tungsten shielding thicknesses. The studies indicate that a thickness of about 40\,cm is needed for reducing the DPA in the coils to 10$^{-3}$\,DPA/year and the dose to about 5\,MGy/year. Considering the beam aperture requirements, the target vessel size and the space required for supports and thermal insulation, the minimum coil aperture (radius) for the target solenoid is hence expected to be about 60--70\,cm. The maximum allowed DPA in the coils per year of operation needs to be further assessed, and will depend on the expected effect of annealing cycles. A dose of 5\,MGy/year could possibly be acceptable for insulation materials.

\begin{table}[t!]
\caption{Radiation load on the target solenoid in terms of the maximum displacement per atom (DPA) and the maximum absorbed dose per year of operation for various shielding configurations. The studies considered a graphite target. The inner aperture of the shielding around the target vessel was 17.8\,cm. A gap of 7.5\,cm was assumed between the shielding and the coils (supports, thermal insulation).}
\label{tab:dpadosetargetsolenoid}
\centering\resizebox{1.0\textwidth}{!}{%
\begin{tabular}{cccc}
\hline\hline
Inner radius of the magnet coils & Shielding thickness around the target & DPA/year [$10^{-3}$] & Dose [MGy/year] \\
\hline
\SI{60}{cm} & {W \SI{31.2}{cm} + H\textsubscript{2}O  \SI{2}{cm} + B\textsubscript{4}C \SI{0.5}{cm} + W \SI{1}{cm}} & $1.70  \pm 0.02$ &  $10.0 \pm 0.3$\\
\SI{65}{cm} & W \SI{36.2}{cm} + H\textsubscript{2}O  \SI{2}{cm} + B\textsubscript{4}C \SI{0.5}{cm} + W \SI{1}{cm} &  $0.90 \pm 0.02$ &  $5.6 \pm 0.2$\\
\SI{70}{cm} & W \SI{41.2}{cm} + H\textsubscript{2}O  \SI{2}{cm} + B\textsubscript{4}C \SI{0.5}{cm} + W \SI{1}{cm} & $0.49 \pm 0.01$ &  $3.1\pm 0.1$\\
\SI{75}{cm} & W \SI{46.2}{cm} + H\textsubscript{2}O  \SI{2}{cm} + B\textsubscript{4}C \SI{0.5}{cm} + W \SI{1}{cm} & $0.29 \pm 0.01$ & $1.9 \pm 0.1$ \\
\SI{80}{cm} & W \SI{51.2}{cm} + H\textsubscript{2}O  \SI{2}{cm} + B\textsubscript{4}C \SI{0.5}{cm} + W \SI{1}{cm} &  $0.16 \pm 0.01$ &  $1.0 \pm 0.1$ \\
\SI{85}{cm} & W \SI{56.2}{cm} + H\textsubscript{2}O  \SI{2}{cm} + B\textsubscript{4}C \SI{0.5}{cm} + W \SI{1}{cm} & $0.09 \pm 0.01$ &  $0.6 \pm  0.1$ \\
\hline\hline
\end{tabular}
}
\end{table}

In order to estimate the general shielding requirements for superconducting arc magnets in the 3\,TeV and 10\,TeV colliders, generic shielding studies were performed with FLUKA. The main focus of these studies was on muon decay, whereas other source terms like beam halo losses still have to be addressed in the future. As in the MAP studies, tungsten was assumed as absorber material due to its high atomic number and density. For engineering reasons, pure tungsten may be substituted by tungsten-based alloys without significantly affecting the shielding efficiency if the alloy has a similar material density. The simulations indicated that the total power penetrating through the shielding is similar for all collider energies, despite the harder decay spectrum at 10\,TeV. On the other hand, the maximum power density and dose in the coils was found to be a factor of 1.5--2 higher at 10\,TeV than at 3\,TeV. 

Table~\ref{tab:parameters_radloadcollider} summarizes the calculated power load and radiation damage in collider arc magnets as a function of the radial tungsten absorber thickness (for the 10\,TeV collider option). The power penetrating the absorber, mostly in the form of electromagnetic showers, amounts to almost 20\,W/m in the case of a 2\,cm shielding, and decreases to 4\,W/m in the case of a 4\,cm shielding. Most of this power is deposited in the cold bore and cold mass of the superconducting magnets. 
As a design criterion, the beam-induced heat load due to muon decay shall not exceed 5\,W/m, considering the additional static heat load in the magnet cold mass and the resulting cooling requirements and power consumption (see Section~\ref{sec:Section7_7}). This suggests that the tungsten shielding thickness needs to be at least 4\,cm. With such a thickness, the cumulative radiation damage in magnets (dose and DPA) is expected to be acceptable, even for 10~years of operation. This shows that the heat load in the cold mass remains the driving factor for the shielding design. Together with the required beam aperture of 2.35\,cm (see Section~\ref{sec:Section6_5}) and other space requirements (supports etc.), the performed shielding studies indicate that the inner coil aperture (radius) of the collider arc magnets needs to be about 8\,cm, which is a key figure for the magnet design.  

\begin{table}[t]
\begin{center}
\caption{Power load and radiation damage in collider ring arc magnets (10\,TeV) as a function of the radial tungsten absorber thickness. The power penetrating the shielding does not include neutrinos, since they are not relevant for the radiation load to the machine; the percentage values are given with respect to the power carried by decay electrons and positrons. The results include the contribution of both counter-rotating beams.}
\label{tab:parameters_radloadcollider}
\begin{tabular}{lcc c}
\hline\hline
        &  \textbf{2~cm} & \textbf{3~cm} & \textbf{4~cm} \\
\hline
Beam aperture (radius) & 23.5~mm & 23.5~mm & 23.5~mm\\
Outer shielding radius & 43.5~mm & 53.5~mm & 63.5~mm\\
Inner coil aperture (radius) & 59~mm & 69~mm & 79~mm \\
\hline
Power penetrating tungsten absorber & 19.1~W/m (3.8\%) & 8.2~W/m (1.6\%) & 4.1~W/m (0.8\%)\\
Peak power density in coils & 6.5~mW/cm$^3$ & 2.1~mW/cm$^3$ & 0.7~mW/cm$^3$ \\
Peak dose in Kapton (5/10 years) & 56/112~MGy & 18/36~MGy & 7/14~MGy\\
Peak dose in coils (5/10 years) & 45/90~MGy & 15/30~MGy & 5/10~MGy \\
Peak DPA in coils (5/10 years) & 8/16$\times$10$^{-5}$ DPA& 6/12$\times$10$^{-5}$ DPA& 5/10$\times$10$^{-5}$ DPA\\
\hline\hline
\end{tabular}
\end{center}
\end{table}

\paragraph*{Planned work}

The shielding studies for collider magnets performed so far were mostly of conceptual nature, with the goal to establish key parameters like the required shielding thickness. It is foreseen to progress towards a more realistic absorber design, by optimizing the transverse absorber cross section for dipoles and quadrupoles. Besides the contribution of muon decay, also beam halo losses have to be studied. In addition, it is planned to repeat the shielding studies for a wobbled machine configuration, which is essential for mitigating the environmental impact of neutrinos. Furthermore, material and engineering aspects have to be studied, including the choice of absorber material (pure tungsten versus other heavy tungsten-alloys), machining and handling, as well as the design of supports. Another important aspect is the definition of the operating temperature and the design of the cooling circuit, which requires a close collaboration between various experts (cryogenics,  vacuum).  So far, the studies focused mainly on the collider arcs, but they have to be extended to the experimental insertions, in particular the final focus regions. Besides the studies for the collider, shielding studies are also foreseen for the accelerators.

\paragraph*{Next priority studies and opportunities for additional collaboration}

In several cases, a detailed engineering design of the shielding configurations will not be possible. In addition, it is out of the scope of the present activity to proceed with the prototyping of radiation absorbers or mock-ups of entire shielding configurations. Furthermore, with the present resources it is not feasible to perform dedicated irradiation tests for probing the radiation effects in superconductors like HTS. However, possible synergies with other activities, for example within the fusion community, can be exploited. 

\subsection{Muon cooling cell
}
\label{sec:Section7_6}
The Cooling cell is the building block of a Muon Cooling section as described in Section~\ref{sec:Section6_3} and is considered one of the highest priority for prototyping efforts towards the demonstration of the feasibility of a Muon Collider. The cooling of muons requires in fact a very compact assembly comprising room temperature RF cavities, superconducting solenoids and either Liquid Hydrogen or Lithium Hydride absorbers (more details in Section~\ref{sec:Section7_8}). All those components will require space for services, power feeds and other ancillaries which will make the mechanical integration extremely complex. Within the European Project MuCol it has therefore been decided to have a dedicated work package (WP8) to produce a complete 3D model of a representative cell among the several types that will be necessary for the cooling sections. In this chapter we describe the strategy on how we intend to proceed in phases towards the full integration of a cell, and on the rationale behind the choice of the cooling cell type that will be implemented. We expect to validate the choice at the annual meeting of the IMCC collaboration that will be held in March 2024. 

\paragraph*{System overview}
A cooling cell, in its simplest version, is defined as the assembly of an RF structure made of several cells, surrounded by two or more superconducting solenoids, and one absorber at each end of the cavity (see Fig.~\ref{fig:Cooling_Cell_Terminology} for a conceptual drawing).

\begin{figure}[!ht]
    \centering
    \includegraphics[width=1\textwidth]{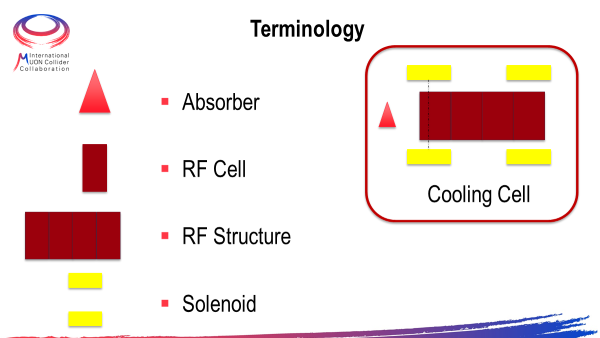}
    \hfill
    \caption{Terminology adopted for the different components of the Cooling Cell}
    \label{fig:Cooling_Cell_Terminology}
\end{figure}

Different types of cooling cells will be necessary to cope with the muon beam characteristics along the cooling section of the complex. A description of the layout of a possible implementation of the 6D cooling section can be found for instance in the cooling section and in papers from the MAP programme   \cite{Stratakis2015, Sayed2015}. Such layout will be further optimised as discussed in Section 6.3. We will choose the cell that can yield

\begin{itemize}
  \item an excellent cooling performance; and
  \item an appropriate level of technical challenge.
\end{itemize}

In order to acquire the necessary experience to properly design a cooling cell, it was agreed that the WP8 design team would be in charge as well of the design of the "Radio Frequency cavity in a Magnetic Field Test Facility" (RFMFTF) which encompasses a pillbox type cavity inside a superconducting solenoid. This is considered a somewhat simpler integration exercise yet including most of the difficulties presented by the cooling cells.

\paragraph*{Key challenges}
The main challenges for the integration of a cooling cell are related to the limited space available in order to fulfil the requirements. The design will have to cope with the following constraints:

\begin{itemize}
  \item have a large magnetic field on axis (order of a few Tesla),
  \item have a large electric field (25\,MV/m) in the RF cavity,
  \item decide the technology to be used to limit the breakdown rate in the RF cavity under high magnetic field,
   \item decide whether all solenoids are in a single cryostat or in separate cryostats,
  \item decide whether services and ancillaries  for the RF cavity (cooling, RF power couplers,  HOM damping couplers will come radially or longitudinally ,
  \item in case the absorber, or the mitigation strategy for high breakdown rate in the RF cavity requires the use of Liquid or Gaseous Hydrogen, understand the safety issues and study a proper distribution network around the cell,
  \item provide the shortest cell to cell coupling space.
\end{itemize}

\paragraph*{Recent achievements}

A preliminary conceptual design of the RFMFTF has been presented and discussed, and will be the object of further studies. The parameters of such test facility are listed in Table~\ref{tab:ch07:RFMFTF_parameters}: 

\begin{table}[ht]
\caption{Parameters for the RFMFTF }
\label{tab:ch07:RFMFTF_parameters}
\centering
\begin{tabular}{p{4.5cm}ccp{6cm}}
\hline\hline
& Unit & {\textbf{Parameter}} & {\textbf{Comment}} \\
\midrule
Magnetic Field on Axis & [T] & 7 & -  \\
RF Frequency & [MHz] & 3000  & - \\ 
Number of solenoids & - & 2 & In order to allow working in phase or antiphase \\
Number of cryostats & - & 1 & Both solenoids will be in the same cryostat \\
Position of RF power coupler & - & Longitudinal & RF Power Coupler will come from the front as all the services \\

\midrule\midrule 
\end{tabular}
\end{table}

\paragraph*{Planned work}
In the next few months we are going to complete the design of the RFMFTF that is planned to be mostly built by INFN and placed at INFN - LASA in Milan. An initial funding has been granted by INFN, although a full estimate of the cost has to be established taking into account that some parts might come from  in kind contributions (e.g.~cryogenics, power converters and RF power source) of some of the labs participating to the collaboration. On January 18 and 19, 2024, a workshop on the cooling cell technology has been organised at CERN. We presented to the community and some experts in the field the choices taken by the team and  collected feedback, in order to prepare a proposal for the Governing board of MuCol about the type of cell for which the detailed design will be performed. After the endorsement of the GB, the design work will start at full speed. 

\paragraph*{Next priority studies and opportunities for additional collaboration}

The next, not yet planned step, is the actual construction of a cooling cell. The cost is estimated to be between 5 and 10 MCHF for the first prototype.

\subsection{Cryogenics
}
\label{sec:Section7_7}

\paragraph*{Overview}
A cryogenic infrastructure that enables the operation of superconducting magnets and radio-frequency (RF) cavities is an essential part of the Muon Collider complex. Systems requiring cryogenics will be present at most stages of the complex, including but not limited to:
\begin{itemize}
    \item Proton Driver: cooling of high-Q RF cavities;
    \item Front End: cooling of the target solenoid;
    \item Cooling Channel: cooling of the magnet array, operation of the H$_2$ absorbers;
    \item Accelerator and Collider rings: cooling of superconducting magnets.
\end{itemize}

With sustainability in mind, one of the main drivers for the design of the cryogenic system will be the overall electrical power consumption. The various challenges of each part of the complex shall be tackled while keeping the highest possible thermodynamic efficiency (\textit{i.e.} temperature levels as high as feasible) and the complexity to a minimum.

\paragraph*{Key challenges}
Several key challenges have been identified at the level of the Cooling Channel, and the Accelerator and Collider rings. For the moment, the Proton Driver and the Front End systems have not been assessed from a Cryogenics standpoint.

\par
\textbf{Cooling channel:} the aim is to economise the overall cryogenic system investment and operation by harmonizing the demands of the various magnets that make up the channel. The most demanding magnets of the chain are the final cooling solenoids, where a substantial heat load needs to be extracted from each structure. Here, the challenge is rather on designing the local heat extraction rather than coping with the overall generated heat load. Solutions are strongly dependent on chosen time constant and ramp scheme for the magnets. The heat load profile and peak heat load generated during ramping varies significantly with these two parameters; a lower time constant reduces the heat load during ramp, but also reduces thermal and electrical stability during operation.

\par
\textbf{Accelerator ring:} the focus for the accelerator ring is to investigate sustainable cryogenic system possibilities in a heterogeneous magnet environment such as the accelerator ring, which will have interleaved normal conducting (warm) and superconducting (cold) magnets, and to design a cooling layout to cope with such arrangement. The fast ramping magnets will most likely generate significant ramping losses, and the sheer number of cold-to-warm transitions calls for careful design to keep heat in-leaks manageable. Another challenge is that of magnet cooling over long distances: with the aim of minimizing cost and complexity, one should strive to maximize the distance required between the accelerator's continuous-cryostat's cryogenic-service feed points and re-cooling stations. Considering the rather short field-free regions between magnets (\textit{i.e.} interconnects), the internal routing of cryogens in the cryostat while satisfying minimum magnet interconnect distances needs careful consideration.

\par
\textbf{Collider ring:} Convergence on a radial build that can be achieved from both the magnet point of view (aperture \textit{vs.} field) and cryogenic/thermal design is crucial. It has to support appropriate mitigation strategies for the heat load deposited on the coil, such as a heat intercept between the coil and absorber and the shielding thickness of the absorber so that the total heat load stays below 10\,W/m. As with the accelerator ring, magnet cooling over long distances needs to be optimised: with the aim of minimizing cost and complexity one should strive to maximize the distance required between the collider's continuous-cryostat's cryogenic-service feed points and re-cooling stations. Optimization of very short magnet interconnects compliant with magnet movement for neutrino flux mitigation is necessary: the internal routing of cryogens in the continuous collider cryostat and the integration of cryogenic-service feed-points has to be solved while satisfying minimum magnet interconnect distances and allowing for magnet movement.

\paragraph*{Recent achievements}
Since the definition of the road map, the principal achievement has been the identification of the main drivers for the cryogenic design and operation of the Muon Collider complex. The collider ring has been the subject of a more focused study, due to the stringent conditions regarding power deposition on the superconducting magnets. An estimation of the heat loads has been carried out for collider ring magnets as function of operating temperature and absorber thickness, including static heat loads from radiation and conduction via mechanical supports. This assessment will be completed with the presently missing contribution from resistive splices, current leads and ramping losses. The heat load estimation evidenced the need for a heat intercept (thermal shield) in the radial build between the absorber and the coils. This intercept is a necessary component to reduce the heat in-leak from thermal radiation to manageable levels. Also from the heat load estimation, the electrical power-consumption was estimated for the collider ring at the refrigerator interface as a function of the temperature level of the magnets and absorber circuits. Boundaries were defined for operating temperature, cryo-related power consumption and heat load to coil for collider-type magnets. These cryogenic, sustainability-driven, operational-landscape mappings contribute to facilitating the magnet/coil-cable design options.

\paragraph*{Planned work}
Most of the planned work will focus on the accelerator ring, namely to carry out an exercise similar to what has been done for the collider ring: heat load estimation and corresponding proposals of cryogenic options for the accelerator ring, considering the overall cryogenic electrical consumption.
Concerning the Proton Driver, Front End and Cooling Channel, work will focus on heat load estimation and identification of potential showstoppers/challenging magnets.
A common thread while working on the several systems that make up the complex will be the cryogenic distribution and the exergetic analysis of the accelerator and collider rings, with the overall sustainability of the cryogenic system in mind.

\paragraph*{Next priority studies and opportunities for additional collaboration}
At the moment, resources are insufficient to carry out an in-depth assessment of the cooling solution for the radiation absorbers in the collider ring, namely to investigate the trade-off between water and CO$_2$ as coolant. 
It should be discussed if hydrogen as a coolant for the collider ring's superconducting magnets is an option worth investigating, and if so a preliminary evaluation into safety considerations for hydrogen use should be started.
In view of the considerable heat load deposited onto the collider magnets via the absorber and cold mass supports, a thermal design optimisation of the supports along with appropriate integration of heat intercepts (thermal shield) will reduce  the incoming heat loads to the temperature level of the magnets, which can have a significant impact on the overall power consumption of the 10\,km collider ring.
At the moment resources are also insufficient to carry out an individual thermal design optimisation of the extensive Muon Collider magnet family.

\subsection{Vacuum system}
\label{sec:Section7_8}

\subsubsection{Overview}
\label{sec:Section7_8_1}
The vacuum technology of a muon collider will encompasses the vacuum system of the different elements of the accelerator complex: proton driver, cooling cells, beam acceleration and collider. Each element of the accelerator complex will have different vacuum requirements and specific constrains. Due to the short storage lifetime of the muons the pressure requirement of a muon collider will be more relaxed than in existing electron and proton storage rings \cite{v_gallardo}.\\
The accelerator complex will require a vacuum system for beam transmission, including superconducting cavities with specific dust-free requirements, thermal insulation for the superconducting magnets, and several barriers between vacuum and absorbers in the cooling cells.\\
Starting at the proton driver, the pressure requirement would be $<10^{-9}$\,mbar, assuming a similar concept as the SPL machine \cite{SPL}. To achieve this pressure level, only conventional vacuum equipment is required and no specific challenges are expected in this area. The presence of superconducting cavities would require dust management during the assembly and clean pumpdown and venting procedures.\\
After the muon production at the target, the cooling section will be dominated by the presence of drift pipes, absorbers and RF cavities inside strong solenoidal fields. Despite the pressure requirement may not be challenging, the configuration and integration of the cooling cells will be very demanding. The presence of strong magnetic fields will limit the use of mechanical pumps and difficult the integration of ion pumps and vacuum instrumentation. This area can be pumped using chemical pumping, i.e., with NEG coatings or cartridges, cryopumping (taking advantage of the availability of cryogenics for the superconducting magnet cooling), or ion pumps in special configurations like shown in \cite{knaster}, taking advantage of the magnetic fields available in the area.\\
The cooling section will need to integrate either lithium hydride or high density hydrogen (liquid or gas) as absorber to reach the target emittance (see Section~\ref{sec:Section7_6}). Multiple windows will be required to separate the absorbers from the vacuum of the beam line along the cooling channel.\\
Finally the beam acceleration and the collider will also integrate conventional vacuum equipment to reach a modest vacuum, as showed in Ref.~\cite{v_gallardo}. The vacuum chamber of the collider would be integrated with a thick tungsten shield (see Section~\ref{sec:Section7_5}) that have to dissipate a significant amount of the muon decay power. Assuming an aperture of 47\,mm, as shown in Table~\ref{tab:parameters_radloadcollider}, the optimal pump distribution requires 10\,l/s pumps spaced 90\,m to reach an average pressure  <10$^{-6}$\,mbar, not considering beam induced desorption.
\subsubsection{Key challenges}
\label{sec:Section7_8_2}
The main challenges that have been identified are the production of very thin windows for the final stages of the cooling, the integration of liquid hydrogen as absorber in contact with thin windows, and the integration of the collider beam screen to intercept the power produced by the muon decay, having low resistive wall impedance (see Section~\ref{subsec:Section6_5_3}).\\
The kinetic energy of the muon beam at the final cooling will be as low as 20\,MeV at the entrance of some stages \cite{kamal}, reaching a few MeV at the exit. At this energy, the typical thickness used for accelerator windows (tenths of mm) will significantly perturb the beam. 1.3\,mm of beryllium are enough to stop the muon beam completely at 5\,MeV. The development of thin windows ideally below 10\,\textmu m and able to survive high intensity beams is required.\\
The presence of liquid hydrogen in those last stages poses another challenge. A cooling cell where the beam kinetic energy inside the absorber is reduced from 20 to 5\,MeV involves the deposition of 9.6\,J. For a small beam of $\sigma_\textrm{RMS}$ equal to 0.6\,mm, it will produce a pressure excursion inside the hydrogen of several hundred bars. Under these conditions the integrity of a thin window is questionable. If the muon beam has a repetition rate of 5\,Hz, the power deposited in the liquid hydrogen is approximately 50\,W. The evacuation of this power will be also challenging, considering the very limited bore radius of the solenoid (see Section~\ref{sec:Section7_1_2}).\\
The superconducting dipole magnets of the muon collider will require a thick tungsten shielding to protect the coils from the muon decay products. The vacuum chamber should integrate this shielding, it should be able to evacuate a significant heat load from the muon decay (in the order of 500\,W/m), and have an acceptable resistive-wall impedance. The high heat load excludes running this beam screen at cryogenic temperature. On the other hand, the modest vacuum required is achievable with distributed pumping with lumped pumps separated by tenths of meters.
\subsubsection{Recent achievements}
\label{sec:Section7_8_3}
Different materials used for the production of thin windows for x-ray transmission, like Be, C, and Si$_3$N$_4$ have been evaluated. Among them, silicon nitride is easily available with several suppliers. A setup for mechanical characterization of silicon nitride windows at different temperatures, from 77\,K to 500\,K, by bulge testing has been build and commissioned. The measurement campaign has started. Several windows of 1\,\textmu m thick Si$_3$N$_4$ with a 6 by 6\,mm aperture on a silicon frame of 10 by 10\,mm were tested at room temperature and at liquid nitrogen temperature. All the membranes were leak tight and survived a differential pressure of more than 5\,bar at room temperature and more than 2\,bar at cryogenic temperature.\\
One silicon nitride window has been irradiated with HiRadMat beams at SPS. The objective of the experiment was to prove the window could survive to proton intensities equivalent to those of the final cooling with a mechanical load of 1\,bar of differential pressure. Assuming all the power remains in the window (using Bethe-Bloch equation) $4\times10^{12}$ muons at 5\,MeV and $\sigma_{\mathrm{RMS}}$ 0.6\,mm is equivalent to 440\,GeV/c protons with 0.25\,mm beam size and $3\times10^{12}$ protons. The membrane survived intensities 20~times higher than this threshold with a significant deformation, but still leak tight (see Fig.~\ref{fig:Waves}).\\
\begin{figure}[!ht]
    \centering
    \includegraphics[width=1\textwidth]{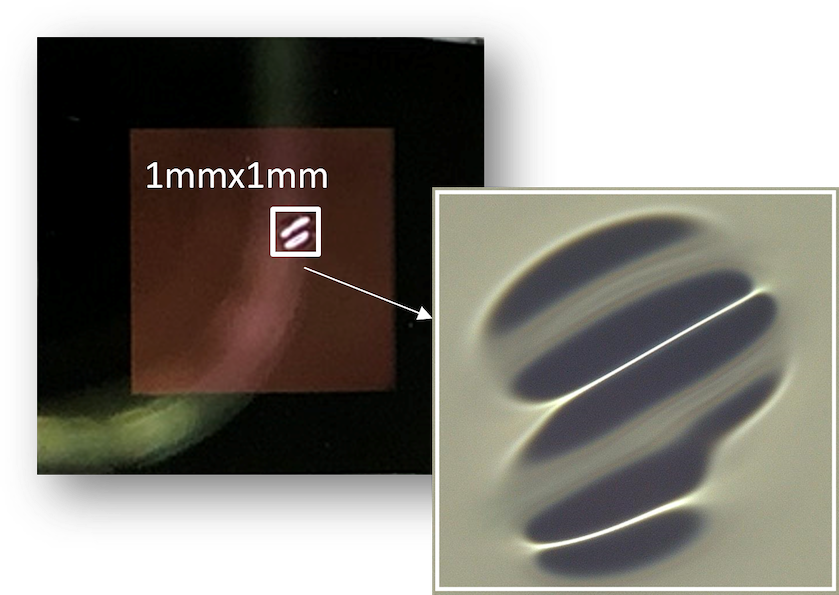}
    \hfill
    \caption{Deformation of Si$_3$N$_4$ membrane after exposure to several pulses of a proton beam at 440\,GeV/c and $>5\times10^{14}$ p$^+$/mm$^2$.}
    \label{fig:Waves}
\end{figure}
\subsubsection{Planned work}
\label{sec:Section7_8_4}
The main efforts are being focused on the development of very thin windows for the muon cooling. The mechanical characterization of the Si$_3$N$_4$ membranes will continue. The setup is shown in Fig.~\ref{fig:experimental_setup}. The setup consist on a window holder made of copper that can be heated or immerse in liquid nitrogen, a cover flange that allows to create a vacuum for thermal insulation and an confocal optical sensor to measure the deflection of the membrane when pressurised. With this setup will be possible to measure the mechanical performance of the membrane at different temperatures.\\
A HiRadMat experiment called SMAUG2 has been approved for 2024, and will allow to irradiate six membranes with different proton beams. The aim of the experiment is to gain a better understanding of the membrane behaviour during irradiation and the mechanism that generate the strong deformation shown in Fig.~\ref{fig:Waves} and test the performance at maximum beam brightness.\\
On top of the window characterization studies, the work will include the proposal of new concepts that could allow the integration of hydrogen as absorber for low emittance and low energy beams. Another topic that will progress in the coming months and years is the definition of the beam screen of the collider including the cooling scheme and possible coatings for reduction of the resistive wall impedance.\\
\begin{figure}[!ht]
    \centering
    \includegraphics[width=1\textwidth]{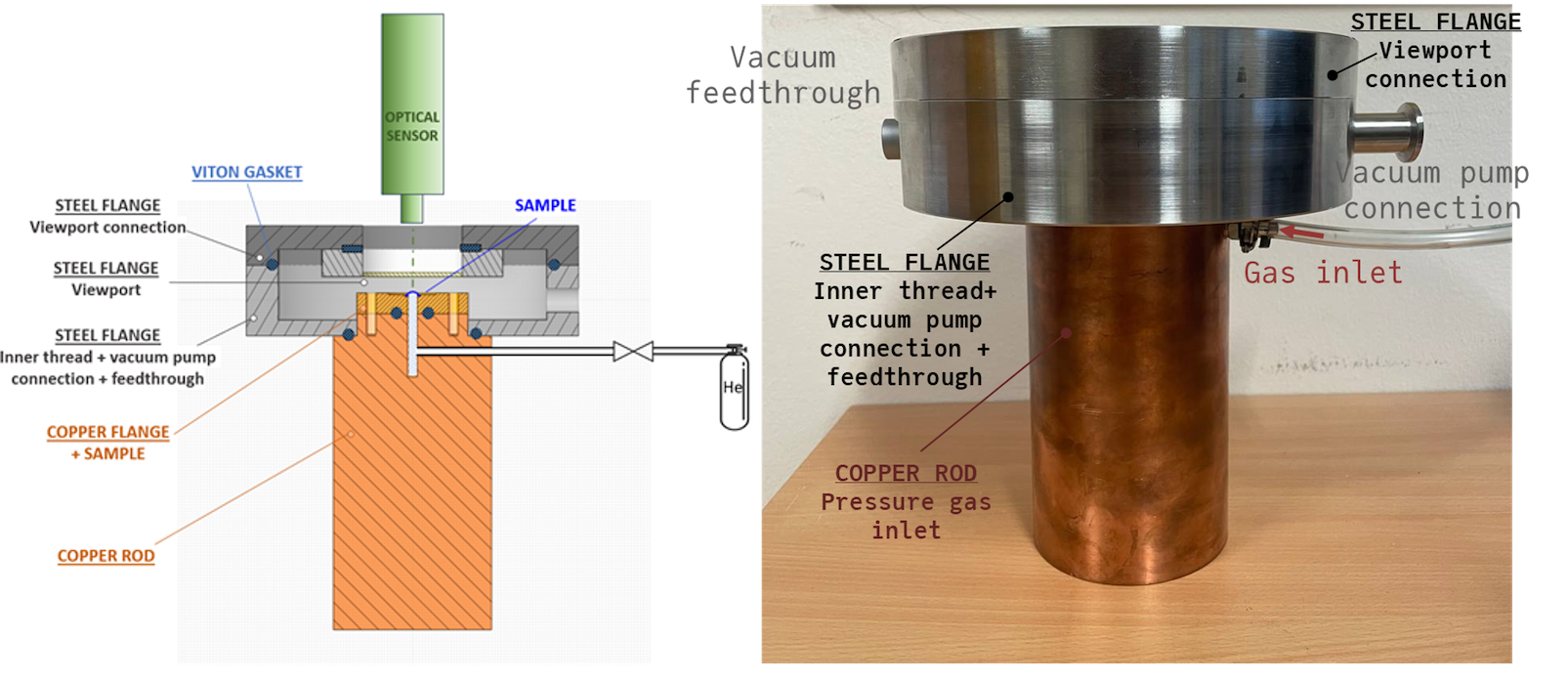}
    \hfill
    \caption{Experimental setup for mechanical characterization of Si$_3$N$_4$ by bulge testing}
    \label{fig:experimental_setup}
\end{figure}
\subsubsection{Next priority studies and opportunities for additional collaboration}
\label{sec:Section7_8_5}
The present amount of resources limits the studies of thin windows to commercial available Si$_3$N$_4$ membranes. No development or optimization of the membrane properties will be carried out. Other candidate materials like silicon carbide, carbon, etc.~are not included in the study either. No investigation about the mechanical integration of the window is being carried out. All the tests are carried out gluing the silicon frame to the support. A future application would require to develop different radiation hard joining techniques and/or  a way to remove the membrane from the silicon frame\\
One of the driver specifications of the final cooling window is the absorber material and pressure excursions after each beam passage. The integration of the absorber in the final cooling will not be pursued more than at conceptual level.

\subsection{Instrumentation
}
\label{sec:Section7_9}

No specific studies have been launched to address the design  of beam instrumentation for the Muon Collider since the expected intensities do not seem unfeasible with the presently available technologies. In particular instrumentation already used in existing muon beam facilities as well as other type of test beams are considered to be adaptable to the requirements as they are presently understood. Once the layout will be frozen, studies will be performed to confirm such hypothesis. The main difficulties will be related to the space available (very limited in the Ionisation Cooling channels), and to the radiation levels and background due to muons decay. The IMCC will strive to provide at least a roadmap for the most challenging instruments, with detailed specifications and corresponding proposed devices

\subsection{Radiation protection
}
\label{sec:Section7_10}

\paragraph*{Overview}
The Radiation Protection (RP) aspects of the Muon Collider can be separated into two main, intrinsically different topics. One is related to the conventional RP aspects connected to prompt and residual radiation, air, water and soil activation, as well as radioactive waste produced in the whole facility starting from the proton driver up to the collider ring. These aspects are principally well understood and can be mitigated to acceptable levels as long as they are addressed in the design phase. The second topic is related to the 
flux of high energy neutrinos emitted from the Muon Collider that have a 
very small probability to interact far away in material near the Earth’s surface producing secondary particle showers. This aspect is 
most relevant for the collider ring.

\paragraph*{Key challenge}
    
Muons circulating in the muon collider decay and generate neutrinos within a small solid angle that may lead to a large local flux of neutrinos in the plane of the collider ring. Due to their particularly low interaction cross section, the attenuation of TeV-scale neutrinos traversing the Earth’s matter is very low (e.g.~0.01\% for \SI{100}{km}). A large flux of neutrinos may therefore reach the Earth’s surface even far away from the collider, where rare single events of these high-energetic neutrinos create secondary particle showers. 
One of the challenges of a high energy muon collider is therefore to reduce the small amount of radiation created by these showers so that it is negligible.

A refined neutrino dose model has been developed. 
It comprises FLUKA Monte Carlo simulations that allow study of the expected neutrino and secondary-particle fluxes and evaluation of the main parameters for the effective dose predictions. The results are then folded with a realistic neutrino source term taking into account the given real collider lattice. The findings are further used to develop a map in which the dose is projected on the Earth’s surface subsequently used for the final dose assessment. Next to the dose assessment also the uncertainties are to be studied and methods allowing to demonstrate compliance. This is an iterative process where optimization is addressed at the various working steps.


The goal is to ensure that the neutrino-induced effect does not entail any noticeable addition to natural radioactivity and that the environmental impact of the muon collider is negligible, i.e.~an effective dose of the order of 10\,\textmu Sv/year, similar, for instance, to the impact from the LHC. Below this constraint, the optimisation requirement is considered as fulfilled and public acceptance can be expected.

\paragraph*{Recent achievements: neutrino dose model}

Exhaustive FLUKA simulations were performed to evaluate the effective dose in soil, resulting from the interaction of the neutrinos emerging from the decay of 1.5\,TeV and 5\,TeV mono-directional muons. The angular and energy distributions of the neutrinos were sampled according to the respective distributions expected from the muon decays. 
It was found that after only a few meters of path in soil the sampled neutrino-induced showers reach a plateau condition, where the effective dose saturates and remains constant farther ahead. This condition is considered the conservative worst-case scenario since the neutrino-induced showers rapidly decrease in the transition between soil and air, as further demonstrated with additional simulations. 
The effective dose kernel has been evaluated on the basis of a Gaussian fit to the radial projection of the saturated effective dose. 
These kernels, characterised by a peak dose and a lateral width, were evaluated for a wide range of neutrino interaction distances from the decay point ($5, 10, 15, 20, 40, 60, 80, 100~\mathrm{km}$) and further interpolated to obtain a kernel shape for any distance between $5$ and \SI{100}{km}. It was found that the widening of the neutrino radiation cone due to the lateral extension of the shower is very small compared to the effect of the neutrino flux angular divergence.  The results additionally allowed for a general verification of simplified analytical expressions showing good agreement.


In order to evaluate the effective dose due to neutrinos 
reaching the Earth's surface, the ”source” term given by the effective dose kernel needs to be further combined with the properties of the accelerator lattice taking the divergence of the muon beam into account. 
This has been performed based on numerical evaluations for half an arc cell of the latest version of the 10\,TeV center-of-mass energy collider. The results show that sharp dose peaks are expected to be caused by short straight sections of 0.3\,m length between magnets. Focusing and chromaticity correction based on combined function magnets helps avoiding even higher peaks. Even though the length of all straight sections is identical, the height of the peaks are lower at the beginning of the cell than towards the end of the cell. The dipolar magnetic field of combined function magnets is lower than the one of pure dipoles leading to slightly increased radiation levels. The variations of dose from different positions along the quadrupoles is caused by variations of the beam divergence.

Obvious mitigation measures are to minimize the length of straight sections in regions outside the long straight section housing the experiments and to install the device deep underground leading to large neutrino distance from the Earth's surface so that the divergence of the neutrino flux leads to dilution of the resultant shower. Careful lattice design avoiding straight sections without horizontal beam divergence due to dispersion derivative may allow the radiation dose peaks from the arcs to be mitigated. The present plan for the 10\,TeV collider is to implement an additional mitigation method, which implies that the beamline components in the arcs are placed on movers to deform the ring periodically in small steps such that the muon beam direction would change over time (for details see Section~\ref{sec:Section7_12}). It has been found that such vertical deformation of the beam within $\pm$1\,mrad by the movers would result in a reduction of the dose kernels by a factor 80 to 90 for distances between 5-100\,km.

Higher neutrino fluxes arising from the long straight sections are unavoidable. A tool to determine the location on the surface corresponding to the neutrino flux position and to easily adjust the positioning of the machine is being developed. This tool can be used to optimize the position of the collider ring such that higher radiation levels are localized in regions such as rock face, which may be owned by the laboratory and fenced, or at the sea.  
A first Proof of Concept (POC) to identify the intersection points with the Earth's surface with such a tool has been achieved by simulating straight line exit points in the case SPS \& LHC. 
It was a static simulation to validate the methodology and the potential results, but without allowing dynamic change of the position of the accelerator. Following this POC, it was decided to add this functionality to ’Geoprofiler’, a decision-support web application for future accelerator localisation study. This functionality allows users to move dynamically the position of the collider via 2D placement with specification of the altitude as well as the ability to "tilt" the collider around two axes, 
and to then perform an analysis of the main radiation lines of the collider for this position. The objective was to show the intersection between these lines and the surface of the Earth (Digital Elevation Model). Due to the large range of action of the radiation lines, we chose to use European Space Agency Copernicus DEM, which covers the whole European territory. At this point, because of performance issues, it was decided as a first step, to restrict the number of radiation lines to the two main ones. An analysis of the positioning with the tool allowed us to identify at least one potential option in the local CERN area that would direct the neutrino flux of the IPs to a limited non-built-up area in the mountains on one side and a far away area in the Mediterranean Sea on the other side as shown in Fig.~\ref{fig:Geoprofiler_Muon_Overview_RadLines}. In addition, this configuration has the advantage of having a single exit point in the mountains from each of the two IP's located with a steep exit angle thanks to the collider inclination and ground profile.

    
    \begin{figure}
        \centering
        \includegraphics[width=0.75\linewidth]{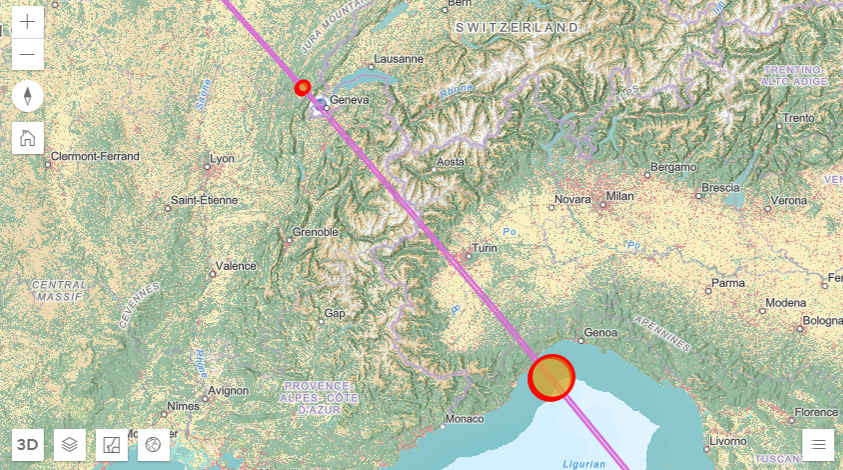}
        \caption{Overview of the exit points of the potential collider placement option in the local CERN area. The radii of the exit points were arbitrarily chosen for illustration purposes.}
        \label{fig:Geoprofiler_Muon_Overview_RadLines}
    \end{figure}


Generally, for the dose assessment all possible exposure pathways have to be taken into account. In the case of the neutrino induced radiation far away from the collider, it has been found that the only pathway to consider is the external exposure directly from the secondary particle showers induced by the neutrinos. It has been demonstrated with FLUKA simulations that the exposure due to activated or contaminated soil and air can be considered negligible.  
FLUKA simulations were furthermore used to investigate the composition of the secondary particle showers created by the neutrino interactions, which is relevant for identifying methods to demonstrate compliance. It was found that the effective dose stemming from the neutrino interactions is dominated by neutrons and the electromagnetic component. Moreover, various dosimetric quantities were compared to each other allowing to evaluate the field-specific conversions factors.
Finally, it was concluded that the exposures resulting to biota, i.e.~animals and plants, are expected to be orders of magnitude below the respective reference levels.

\paragraph*{Recent achievements: conventional RP aspects}

A first investigation of the conventional RP aspects of the collider ring arcs was performed based on FLUKA simulations. 
The studies comprised the evaluation of the residual ambient dose equivalent levels in the arc dipoles region for different cool-down times. The radiation levels after 5~years of beam operation were found to be better than in the inner triplet region of the HL-LHC and are thus considered acceptable assuming optimization of interventions. 
The residual dose rates after the commissioning phase, assumed to be of three months with 20\% beam intensity, were found to be more than one order of magnitude lower and complying with a Supervised Radiation Area ($< 15$\,\textmu Sv/h) after one week of cool-down, thus facilitating for interventions. Furthermore, the radioactive soil thickness around the tunnel was investigated and found to be similar to that of the LHC. With suitable collider placement deep underground the level of soil activation is expected to be acceptable.

\paragraph*{Planned work: neutrino dose model}
For the dose kernel evaluation the plan is to investigate more realistic neutrino interaction scenarios, taking into account different exit angles of the neutrino flux at the Earth's surface, interactions in various materials and more realistic geometries. Additionally, various sensitivity studies are foreseen to be conducted in a systematic manner. 
For the folding with the collider lattice, further studies will be carried out to investigate the feasibility of the lattice, understand the impact of the machine ”wobbling” on beam dynamics and to refine the model around transitions between magnets and straight sections.
The Geoprofiler tool is planned to be further improved allowing a more in-depth analysis of the areas impacted by the neutrinos, including a height/depth clearance taking into account the respective neutrino flux width. 
Next to that, the feasibility of the potential option in the local CERN area will be investigated more closely. 
For the dose assessment the evaluation of more realistic exposure scenarios and potential dose modification factors that are location dependant will be further investigated.


\paragraph*{Planned work---conventional RP aspects}

Depending on the advancement of the target area and the test facility designs, an RP assessment may be conducted in the future to evaluate shielding requirements and provide input relevant for the optimization of the conventional RP aspects.
    

\paragraph*{Next priority studies and opportunities for additional collaboration}

 The full evaluation of the feasibility of a collider option will not be possible in view of its dependence on the approval by the authorities of the impacted States. Furthermore, a study of high interest in case additional resources become available the evaluation of the potential to reuse the LHC.  

\subsection{Civil engineering
}
\label{sec:Section7_11}

\subsubsection{Overview}

Civil Engineering (CE)  represents a significant proportion of the implementation budget for tunneling projects such as the Muon Collider at CERN. As a result, CE studies are of critical importance to ensure a viable and cost-efficient conceptual design from the beginning.

\subsubsection{Layout and placement study}
The civil engineering design and planning is a key factor in establishing the feasibility of the Muon Collider.  
Experience from previous construction projects at CERN has shown that the existing sedimentary rock in the Geneva Basin, known as molasse, provides favourable tunneling conditions.  Regions of limestone geology are also present, and cause more challenging tunneling conditions. During the excavation of the LEP tunnel, water ingress from the limestone formations caused significant problems. For this reason, the underground structures should be located as much as possible in the molasse and avoid the limestone.
Another priority is to orientate the tunnel to minimise the depth of the shafts and reduce the overburden pressure on the underground structures.

\paragraph*{Recent achievements}

To plan the tunnel alignment and analyse the geological constraints, an application called ‘Geoprofiler’ has been developed as seen in Fig.~\ref{fig:Geoprofiler}. This application provides geographical visualisation of the accelerator’s footprint alongside detailed geological profiles. Within the application, user can modify the location of the tunnel alignment as well as its depth and tilt. To ensure a productive user experience, the resulting geology of the accelerator placement is be displayed in real-time.

\paragraph*{Key challenges}

A key requirement of the application is to be able to freely modify the footprint of the accelerator and determine the most favourable position in terms of risk and cost. Additionally, radiation analysis requires high precision to position the accelerator and was added to the tool as well.
Many key advancements were made during the development of the Geoprofiler tool:

\begin{itemize}
\item Integration of the alignment data. Although currently there is no official alignment, it is likely that a MAD-X file format will be provided. Therefore, it has been made feasible to upload a MAD-X file to ‘Geoprofiler’.

\item Several geodetic challenges were encountered. The available data in the MAD-X file first needs to be converted from a Cartesian coordinate system to a geographic one. Once the accelerator is in a geographic system, the coordinates must remain accurate even when the alignment is moved or rotated.

\item The geological data used for the tool was collected from UNIGE (University of Geneva). There are two main components in the application to assess the geology:
\begin{itemize}
    \item Firstly, the profile of the alignment, which displays the geological data. This includes the depth of each geological layer, the depth of the accelerator and the surface height.
    \item Secondly, the geology surrounding the shafts is displayed, from which the depths of each geological layer can be determined. 
\end{itemize}
\begin{figure}
    \centering
    \includegraphics[width=1\linewidth]{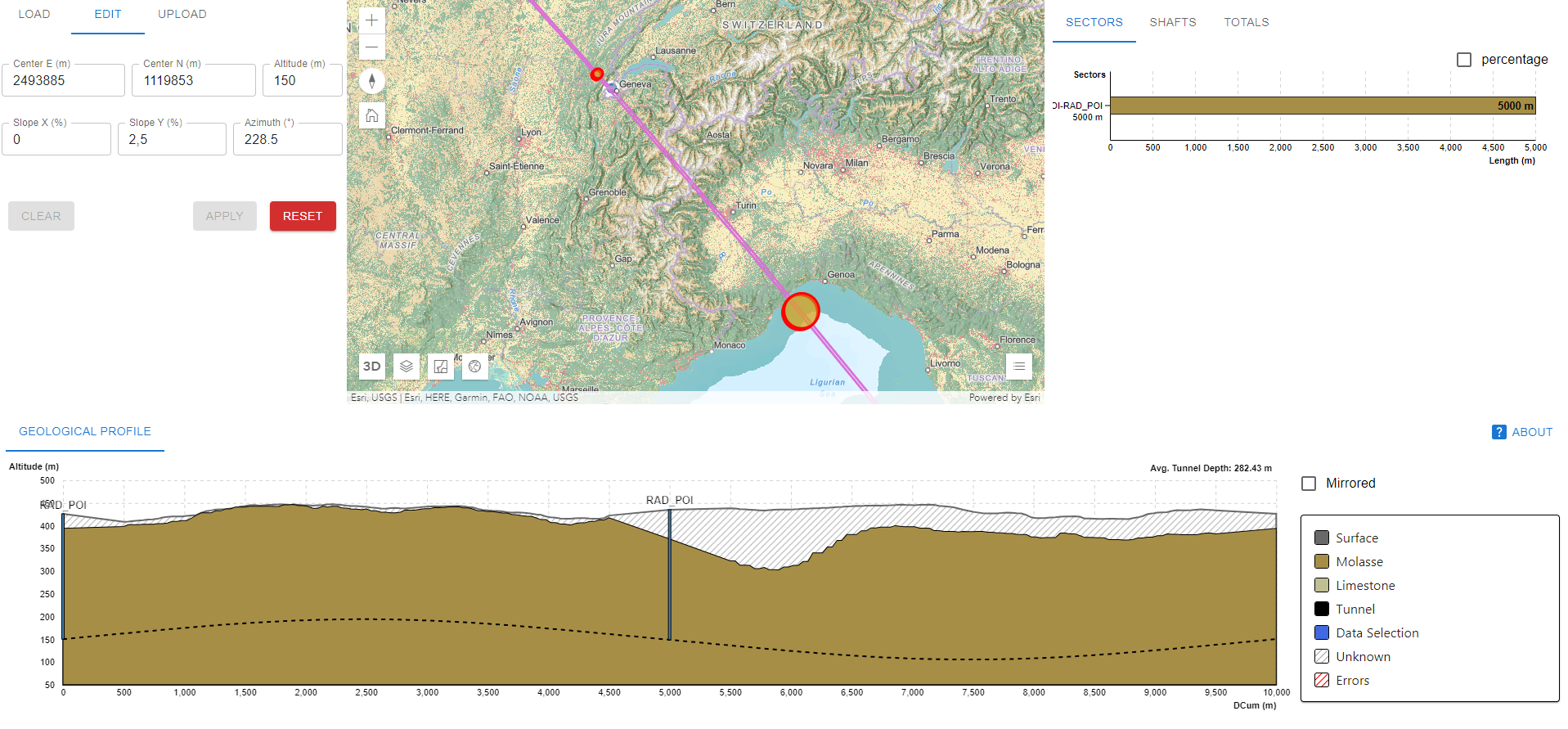}
    \caption{Geology analysis results from Geoprofiler for a given position.}
    \label{fig:Geoprofiler}
\end{figure}

\item For radiological analysis, a feature to compute the intersection of radiation lines with earth surface was added.
\end{itemize}

\subsubsection{Muon Test Facility}
A beam test facility was proposed to demonstrate the feasibility of muon ionisation cooling in order to prove that the required luminosity could be achieved in the Muon Collider. 
A preliminary civil engineering study has been carried out, investigating the possibility of constructing the Test Facility at the two different locations.

\paragraph*{TT10 option}
The first option is to use the existing TT10 tunnel, an existing transfer line to SPS. From TT10 a new beamline would be extracted via a  tunnel to the proposed Muon Collider Demonstrator Facility. 
A  conceptual layout of the facility includes a target hall, cooling tunnel, service gallery and access shafts, as shown in Fig.~\ref{fig:CE_1}. 

\begin{figure}[htb]
    \centering
    \includegraphics[width=0.9\textwidth]{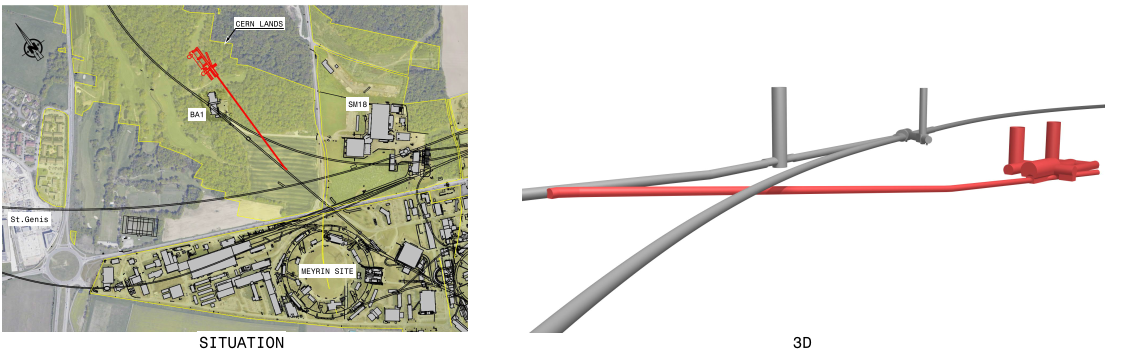}
    \caption{General layout of the proposed test facility at TT10 area.}
    \label{fig:CE_1}
\end{figure}

Due to the complexity and excessive cost of the required works a second option was proposed and studied implementing the proposed facility in an already existing facility at CERN.  

\paragraph*{TT7 option}
A second option proposes the extension of the existing TT7 transfer tunnel up to building 181. The new surface test facility building would  be connected to the underground tunnel through waveguides located every 10\,m, as shown on the Fig.~\ref{fig:CE_2}.

\begin{figure}[htb]
    \centering
    \includegraphics[width=0.9\textwidth]{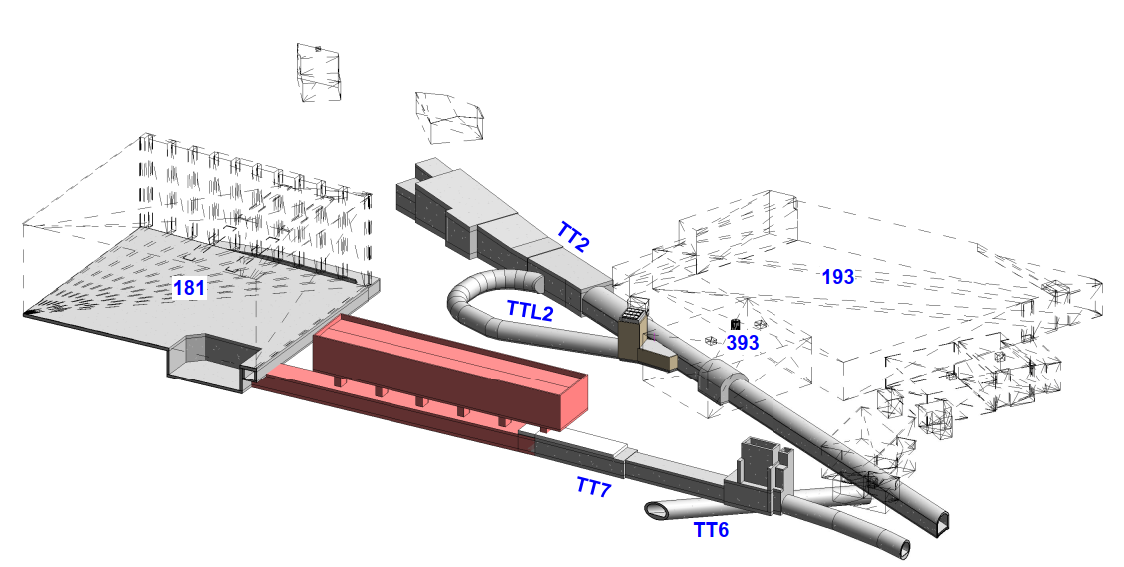}
    \caption{General layout of the proposed test facility at TT7 area.}
    \label{fig:CE_2}
\end{figure}

Currently, existing steel shielding blocks can be found at the end of TT7 tunnel. The blocks are proposed to be  either fully or partially removed during the construction of the facility, and if possible, reused at CERN. 
Preliminary civil engineering cost estimates have been prepared for the two scenarios. 

\subsubsection{Planned work}

The main goal for the following year is to choose a suitable position for the machine, that satisfies multiple constraints, such as the geological situation, territorial, and environmental aspects.

A more detailed understanding of the experimental requirements and integration studies is required, to feed into the civil engineering design and define tunnel/cavern dimensions. The aim is to have a typical cross section for the main tunnels and access shafts, to allow the feasibility studies to advance. 

\subsection{Movers
}
\label{sec:Section7_12}

The neutrino flux created by the decaying muon beam in the collider is to be distributed on the surface by deforming the collider in vertical direction periodically, say every 12\,hours. The collider deformation patterns are a combination of pieces of parabola with opposite curvature depicted in Fig.~\ref{fig:Mover-Patt}. The wavelength and the period of the deformation pattern are adjusted such that the slope of the tangent varies by $\pm 1$\,mrad allowing to reduce the peak doses due to neutrinos reaching earth's surface by almost two orders of magnitude.

\begin{figure}[htb]
    \centering
    \includegraphics[width=10cm]{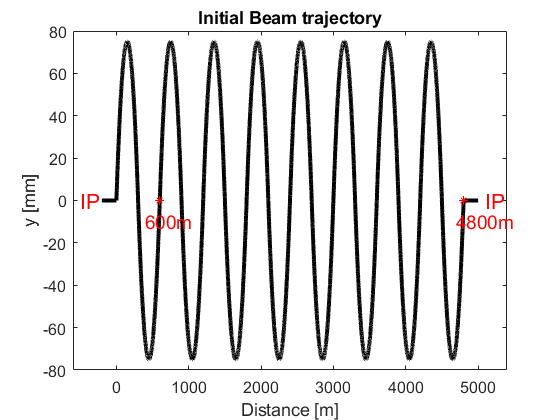}
    \caption{Typical muon collider deformation pattern to mitigate radiation from neutrinos reaching Earth's surface.}
    \label{fig:Mover-Patt}
\end{figure}

During operation, the vertical position and the pitch of all collider magnets (pure bendings and combined function magnets) must be changed to vary the orientation of the tangent to the beam trajectory, that represent the cone of radiation induced by the neutrino flux. Several collider ring vertical deformation patterns, each to be used during typically 12\,hours, are depicted in Fig.~\ref{fig:Mover-Combs}.

\begin{figure}[htb]
    \centering
    \includegraphics[width=10cm]{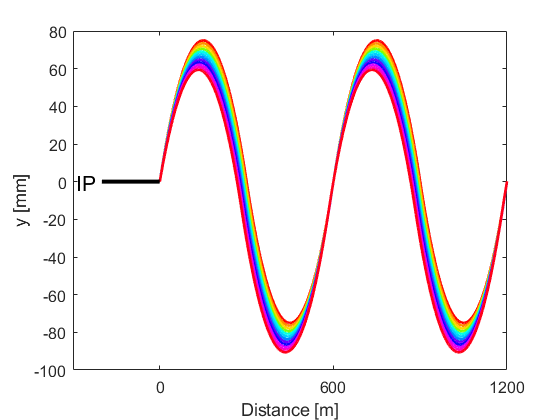}
    \caption{Several collider ring vertical deformation patterns.}
    \label{fig:Mover-Combs}
\end{figure}

Several machine deformation "wobbling" patterns resulting in $\pm 1$\,mrad variations of the slope of the tangent are plotted in Fig.~\ref{fig:DeformationsVars}.
To keep the radiation level at an acceptable level, the tangent to the beam trajectory should change between -1\,mrad and +1\,mrad. 
If we assume a period of 100\,m, thus correspond to vertical movement required of the magnets is $\pm25$\,mm with respect to the horizontal beam trajectory. Two adjacent magnets cannot be allowed to have opposite maximum deviation in order to protect the interconnection bellows. It should also be noted that increase of the required movement of the magnets will lead into a larger tunnel diameter.

\begin{figure}[htb]
    \centering
    \includegraphics[width=10cm]{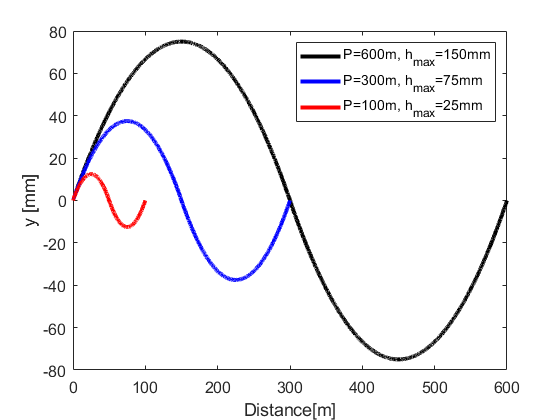}
    \caption{Proposed deformation patterns with different amplitudes and periods.}
    \label{fig:DeformationsVars}
\end{figure}

\subsubsection{Key challenges}
Specifications of positioning accuracy will become available only after thorough beam dynamics studies based on a working based on the collider lattice. The required positioning accuracies are expected to be stringent, in particular, for the chromatic correction and matching sections. For the present studies, positioning tolerances of $\pm0.1$\,mm with respect to the theoretical position are assumed.

The control system of the system may need both sensors on the magnet moving components (i.e.~jacks, gearboxes) and on the magnets themselves to confirm the position of the magnets after each movement. This will lead to controlling of several hundreds of electrical motors in closed loop feedback from up to couple of thousands of sensors.

The time required for the change from one magnet lay-out to another is dependent on the maximum number of motors that can be driven in parallel. The parallel movement of all magnets would require even hundreds kW of power – its realistic scenario for the movement operation, and thus time impact, remains to be studied in detail.

The collider magnets will be served by a cryogenic supply system. We have used the existing scheme of the LHC machine as reference. This involves a connection between the static cryogenic and magnet assemblies at approximately every 100\,m. The beam trajectory of the muon collider does not allow a rigid connection between the cryogenic supply and the magnets at the same interval; thus, development is required.

\begin{figure}[htb]
    \centering
    \includegraphics[width=10cm]{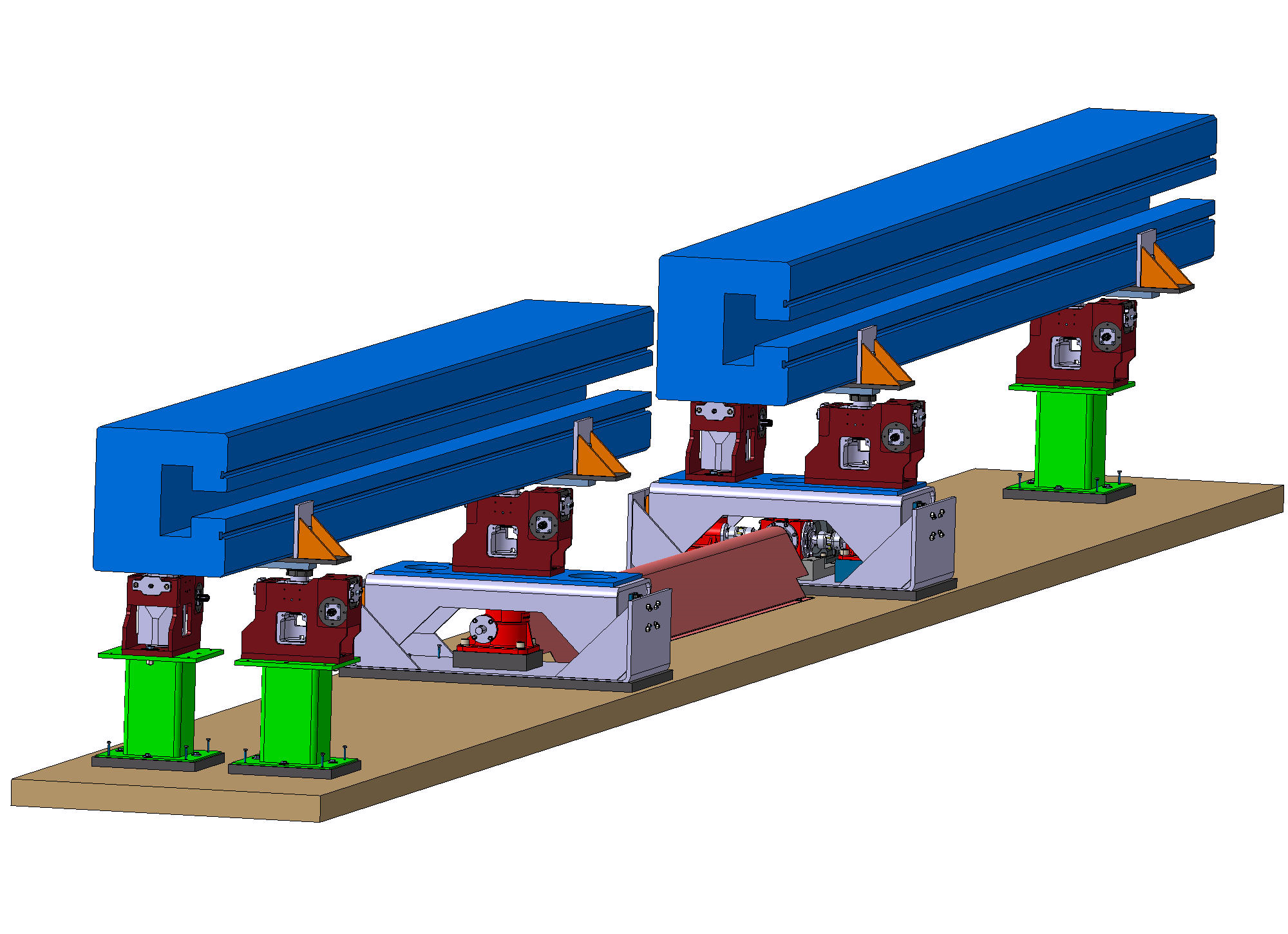}
    \caption{Mock-up of the proposed magnet movement system.}
    \label{fig:MockUp}
\end{figure}

\subsubsection{Planned work}
A mock-up of the first concept is planned to study the movement system, Fig.~\ref{fig:MockUp} shows the mock-up 3D model. The plan is to test it over one interconnection. This involves two dummy magnets, the LHC type jacks for their alignment and jacking system for the re-positioning.

\subsection{Infrastructure
}
\label{sec:Section7_13}
The study of the technical Infrastructures of the Muon Collider complex has not started yet as the configuration of the machines is still in evolution. A survey of expected energy consumption will be performed during 2024 and presented towards the end of the year with the main goal to identify the main drivers of energy consumption and provide a basis to optimise the design of the machines taking into account not only the performance but also the use of energy. For what concerns HVAC, the systems will be very likely similar to what is being operated in the present CERN complex. An optimisation will have to be done to limit the energy consumption. Particular attention will be devoted to the target area and the muon cooling section, where special requirements might be applied, due to the high level of radiation expected and the eventuality of using liquid or gaseous hydrogen as absorbing material or as gas used to inert the secondary emission yield of the RF cavities surfaces. A study will be launched as soon as decisions on the technologies to be used will be assessed. At the same time, at the moment it is not considered a priority to look in details into logistic aspects such as transport of magnets and other equipment, remote handling etc... such studies will be performed once the design of the machines will be stable. A conceptual design of the remote handling in the target area will be part of target studies as during the lifetime of the complex it is to be expected that several targets and surrounding solenoids might have to be exchanged. In this sense the civil infrastructure shall have to foresee a proper ares for the radiation cooling of spent targets and surrounding equipment and hot-cells for the treatment of radioactive waste.

\subsection{General safety considerations
}
\label{sec:Section7_14}
The safety of people and environmental protection are priorities in the concept of a muon collider complex.  Collaboration with CERN's Occupational Health \& Safety and Environmental Protection (HSE) Unit is established to assure the satisfaction of the regulations and applications of best practice in these domains.

Many standard industrial hazards will need to be considered for working in such a facility, such as noise, lighting, air quality, working in confined spaces, which can be satisfied by applying existing CERN safety practices and the Host State's regulations for workplaces.  The ambient temperature in the facility will be of particular interest for access and work.

The creation of a muon collider complex will involve civil works, including tunnels, concrete structures and buildings.  At CERN, all infrastructure shall be designed in accordance with the applicable Eurocodes to withstand the expected loads during construction and operation, but shall also consider accidental actions, such as seismic activity, fire, release of cryogens and the effect of radiation on the concrete matrix and other tunnel construction fabric.  

All buildings, experimental facilities, equipment and experiments installed at CERN shall comply with CERN Safety Code E and other relevant fire safety related instructions.  In view of the special nature of the use of certain areas, in particular underground, and the associated fire risks, CERN's HSE Unit is considered to be the authority for approving and stipulating special provisions.  As the muon collider complex study progresses, detailed fire risk assessments will have to be made, in line with the evolution of the study of the technical infrastructures.  The most efficient protection strategy is one that uses complementary 'safety barriers', with a bottom-up structure, to limit fires at the earliest stages with the lowest consequences, thus considerably limiting the probability and impact of the largest events.  In order to ensure that large adverse events are only possible in very unlikely cases of failure of many barriers, measures at every possible level of functional design need to be implemented:
\begin{itemize}
    \item In the conception of every pieces of equipment (e.g.~materials used in electrical components, circuit breakers, etc.);
    \item In the grouping of equipment in racks or boxes (e.g.~generous cooling of racks, use of fire-retardant cables, and fire detection with power cut-off within each racks, etc.);
    \item In the creation and organisation of internal rooms (e.g.~fire detection, power cut-off and fire suppression inside a room with equipment);
    \item In the definition of fire compartments;
    \item In the definition of firefighting measures, including smoke extraction and fire suppression systems;
    \item Access and egress sufficiently sized in an acceptable range for evacuation and fire service intervention.
\end{itemize}

With the proposed technology solutions for equipment items, personal and process safety aspects, as well as environmental concerns, must be taken into account across the full life cycle from design, fabrication, testing, commissioning, use and dismantling.
For example, extensive use of cryogenics will require considerations of cryogenic and mechanical safety, as well as risk evaluations with regard to potential oxygen deficiency hazards and potential mitigation means compatible with the complex.  
The potential use of gaseous and/or liquid hydrogen will also require particular study, in terms of explosion and flammable gas risk assessments, potential ATEX zoning and particular mitigation measures.  
The electrical hazards present in the complex will need to be assessed and mitigated through sound design practice and execution.  The CERN Electrical Safety Rules, alongside NF C18-510, shall be followed throughout the design process; where exceptions are required, this shall be subject to an appropriate level of risk assessment to evaluate the residual risk, and determine the mitigation strategies required.  For all aspects concerning non-ionising radiation (magnetic fields, electromagnetic compatibility, etc). the relevant CERN Safety Rules shall be followed.

For environmental aspects, efforts will be made to economise of the use of energy and water.  
\begin{itemize}
    \item Atmospheric emissions shall be limited at the source and comply with the regulations in force.  
    \item There will be a rational use of water with the discharge of effluent water into the relevant networks in accordance with the regulations in force.
    \item Use of energy will be made as efficiently as possible, with thermal efficiency for all new buildings
    \item The natural physical and chemical properties of the soil will be preserved.  All the relevant technical provisions related to the usage and/or storage of hazardous substances to the environment shall be fulfilled to avoid any chemical damage to the soil.  Furthermore, the excavated material shall be handled adequately and prevent further site contamination.  All excavated material must be disposed of appropriately in accordance with the associated waste regulations.  
    \item The selection of construction materials, design and fabrication methods shall be such that the generation of waste is both minimised and limited at the source, with the appropriate waste handling and traceability of the waste in place.
\end{itemize}
Noise generated at CERN shall comply with safety requirements of the CERN Safety Rules..  Emissions of environmental noise related to neighbourhoods at CERN shall comply with CERN's Noise footprint reduction policy and implementation strategy.

\begin{flushleft}
\interlinepenalty=10000

\end{flushleft}

\clearpage
\section{Synergies
}
\label{sec:Section08}


The muon collider R\&D will develop a number of important technologies that have application both within particle physics, to other related facilities, and beyond particle physics both in the science community and in the larger society.

\subsection{Technologies
}
\subsubsection{Magnets}

\begin{figure}[!h]
\includegraphics[width=\textwidth]{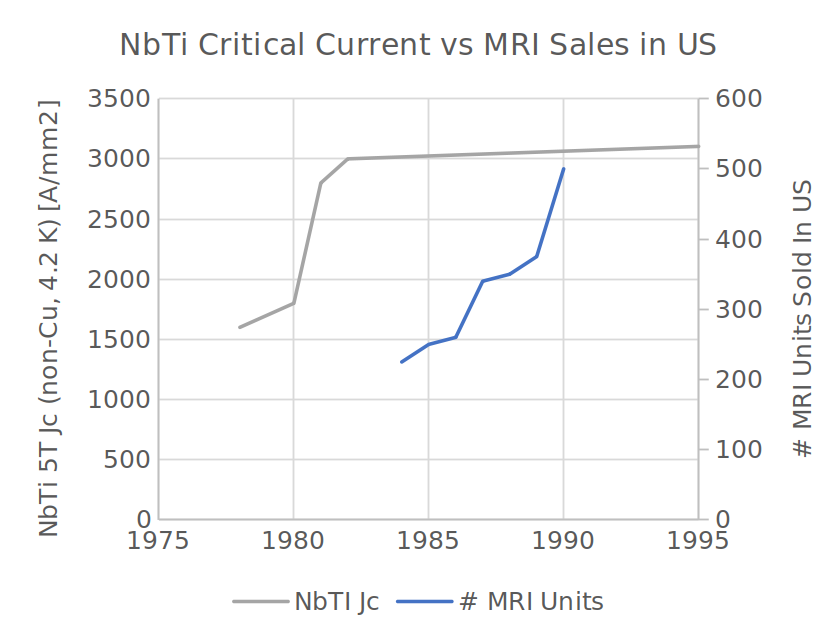}
\caption{Timeline of the performance of industrially produced Nb-Ti, measured as the superconductor critical current density, contrasted to the timeline of MRI units sold in the US. \label{fig:magnet_timeline}}
\end{figure}

Reaching  mature magnet technology required to produce, cool, accelerate and collide muon beams, is a grand challenge for accelerator magnet science. Achieving such advances requires development of new concepts for superconducting and resistive magnets, as well as suitable power converters that manage efficiently the large stored energies and powers required to drive the magnetic circuits. 

This is not new: indeed, the need for fundamental advances in accelerator magnet science has been a characteristic of the development of accelerator magnets since inception. We can recall as good examples the development of Nb-Ti, initiated on large scale by the Tevatron, through HERA, RHIC, the SSC R\&D phase and finally the LHC. A remarkable result of this endeavor is the development of industrial capability for the high performance and high quality Nb-Ti that feeds the MRI market. A demonstration of this achievement is reported in Fig.~\ref{fig:magnet_timeline}, relating the improvement in Nb-Ti critical current density Jc to the growth of the MRI production.

A similar development was necessary for Nb3Sn, to evolve the wire performance beyond the state-of-the-art achieved in the 1980’s and used for the construction of the thermonuclear fusion project ITER. This development was initiated by the DOE US Conductor Development Program in the late 1990’s and beginning of 2000’s, and is now in demonstration phase in the HL-LHC magnets. It is expected that this development may bear fruits of improved performance and production control for future applications of Nb3Sn, e.g.~in future fusion reactors, though a killer application comparable to MRI has not emerged yet.

In the case of the Muon Collider, we plan to profit from the recent advances in HTS magnets and stimulate further developments. This field of magnet science is new, very dynamic, and, most important, the challenges are synergic to several other fields and applications.

Development of the muon collider will deliver important advances in key magnet technologies:
\begin{itemize}
\item Ultra-high field solenoids
\item High field, large bore solenoids
\item High field, low consumption, compact dipoles and quadrupoles
\item Fast ramping dipoles
\end{itemize}

\subsubsection{SRF technology} 
The muon collider proton driver and accelerator chain requires several 1000s of SRF cavities in the range from 300\,MHz to 1300\,MHz. Starting from LEP2, the SRF cavity and associated cryomodule technology has been developed for accelerators over the last few decades for other circular : LHC, RHIC, etc and linear : TESLA, ILC, XFEL, LCLS2, SHINE, etc accelerators for HEP and other application. Muon collider fully rely on this developments and has a lot of synergy with the ongoing SRF R\&D. Furthermore, due to very short life time of the muons, as high as possible accelerating gradient is mandatory to accelerate muons up to 10\,TeV collision energy. This motivation drives the SRF R\&D towards even high gradient not only at 1.3\,GHz but also at lower RF frequencies down to 350\,MHz. In addition to the standard bulk Nb SRF technology which is well developed, new directions potentially open the way to higher gradients including Nb3Sn, thin films, and even HTS coatings. Driven by the large scale HEP accelerator facilities the SRF R\&D has already found and will find even more other applications for light sources, nuclear physics, medical and industrial accelerators. 

\subsubsection{Other technologies}
The R\&D programme to support RF cavities will yield important developments for next-generation RF sources. In particular, significant R\&D will be undertaken on normal conducting RF cavities in the frequency range 300-3000\,MHz with studies focusing on achieving high gradients even in the presence of magnetic fields.

The ability of targets to withstand high beam powers is a fundamental limitation in the performance of many proton accelerators built to create secondary particles. Fixed targets cannot support beam powers about around 2\,MW. Fluid targets can in principle withstand much higher beam powers. The first generation of fluid targets, developed with liquid metals, have been limited by heating effects in the liquid metal. Development of next generation fluid targets would alleviate this fundamental limitation in proton beam current.

The high-field magnet, RF and targetry activities together can have application in the creation of bright muon beams. This technology is, in itself, an important technique that has application in a number of different areas.


\subsection{Technology applications}

\subsubsection{Accelerator magnets}
A first clear synergy is the need for HTS accelerator magnet development for a future collider. The requirements for the 10\,TeV Muon Collider ring are presently beyond the state of the art and will likely need an all-HTS solution to reach peak field values of 16\,T, while operating in the temperature range of 10\,K to 20\,K with a minimal amount of cryogen. These requirements resemble closely those established recently for the hadron stage of the Future Circular Collider, FCC-hh. In fact, the development of high performance and high efficiency HTS accelerator dipoles and quadrupoles will likely profit any future collider at the energy frontier.

Lower field particle accelerators may profit from increased stability and more efficient cooling that can be achieved using HTS. The R\&D on rapid acceleration techniques will yield more efficient rapid cycling magnets that can benefit Rapid Cycling Synchrotrons.

\subsubsection{Fusion}
We have identified several areas where the developments of magnet science and power converters have synergies with the needs of thermonuclear fusion. A good example is the large bore and high field HTS solenoid around the proton target in the muon production and capture channel, resembling similar magnets that produce the field swing in tokamaks (central solenoid), and the pinch solenoids in magnetic mirrors. Another example is the technology to be developed for the high field solenoids in the 6D and final cooling, based on non-insulated HTS windings to ease quench protection. Similar technology has been used in large toroidal field model coils, demonstrating successfully high field, but at the same time indicating that quench management still needs to be improved. A further example is the RF source used for heating of the fusion plasma, which is usually situated in a region having high magnetic field and suffers from the possibility of RF breakdown.

In general, future magnetically confined fusion power plants will need to reduce system complexity to become economically viable, and may require high field to reduce the size of the reactor. This field will hence profit from advances in the HTS magnet technology developed for the Muon Collider, offering simpler cryogenics at improved performance. A last point of contact among the development and studies for the Muon Collider and thermonuclear fusion is the use of power converters delivering high energy in pulses of short duration, high repetition rate and large number of pulses. This is one of the main challenges for the rapid cycling synchrotrons of the muon acceleration stage, whose pulsed power requirements resemble those required for magneto-inertial fusion.

\subsubsection{Material and life science in high magnetic field}
A magnetic field induces new states of matter, and can be used to precisely study them. The ability to measure the properties of these new states, and the measurement resolution tend to grow with field, in most cases with power dependencies. The result is a constant call for higher fields in user facilities. The present state of the art of all-superconducting solenoids for material and life science is the 32\,T LTS/HTS hybrid at the National High Magnetic Field Laboratory of Florida State University (NHMFL). Higher fields can be reached presently only resorting to resistive magnet. The present highest steady state field facility is at the High Magnetic Field Laboratory of the Chinese Academy of Science (CHMFL), with a field of 45.22\,T achieved in 2022. Though functional, facilities of this class are very power-hungry, with continuous electric consumption in the range of tens of MW. The next step in high field user facility has been projected at 40\,T, and is the subject of R\&D and demonstration in laboratories in the US (NHMFL), Europe (Super-EMFL). The work on the final cooling solenoid, a 40\,T-class, 50\,mm bore all-HTS magnet, is clearly highly relevant for the companion developments in worldwide high-field magnet laboratories, promising compact, high-performance, and high-efficiency solutions to future UHF user facilities.

\subsubsection{Nuclear magnetic resonance (NMR) and magnetic resonance imaging (MRI)}
The development of high-field and ultra-high-field HTS solenoids for the 6D and final cooling quoted earlier can also support development of NMR instruments. The present state-of-the-art for industrial NMR machines is the 28.2\,T, 1.2\,GHz system produced by Bruker. This magnet is an hybrid made with an LTS outsert and a HTS insert. Introducing HTS technology has not only boosted field performance, but also significantly reduced the size and consumption of the system. The next step in NMR machines is projected in the range of 30\,to 35\,T, and may profit directly from the technology developed for the HTS final cooling solenoid, exploring a range of field which is comparable, and even higher. This development may also have implications for MRI magnets, though this is presently not as direct as for NMR. The magnet program for a Muon Collider, in particular for the 6D cooling section, will need to demonstrate engineering and reliable operation of HTS solenoids in the medium to high field range (3\,T to 10\,T), and temperatures well in excess of liquid helium (10\,K to 20\,K). Success of this development will raise a definite interest in helium-free technology for MRI systems operating in the present clinical range (i.e.~up to 7\,T) but with simplified cryogenic systems, which will increase penetration of this fundamental tool of modern medicine in remote and less developed areas.

\subsubsection{Low-consumption magnets for nuclear and particle physics, and medical applications}
As is the case of MRI, mentioned earlier, a successful development of HTS accelerator magnet technology for helium-free (or minimal cryogen) operation will offer energy efficient and sustainable options for the beam line magnets used in nuclear and particle physics experiments, as well as for beam production and delivery in radiation therapy. These beam line magnets often tend to require large apertures, implying large power consumption if powered with resistive coils, and may be subjected to significant heat loads from radiation, hence not within the optimal environmental conditions for LTS. Though field range and geometry may not compare directly, energy efficiency and cryogen-free (or minimal cryogen) benefits are a common denominator among the developments for the Muon Collider and beam line magnets for physics and medicine.

\subsubsection{Magnets for neutron spectroscopy}
Magnetic configurations with free access to the magnet bore are useful for neutron spectroscopy, for material and life sciences. Present state-of-the-art of commercial systems is between 15\,T and 17\,T, but there is a clear call for higher fields to  uncover new phenomena. The targets set for next generation instruments range from 30\,T to 40\,T. The development of large bore, large energy and split HTS solenoids is a technology step towards the next generation of all-HTS neutron spectroscopy instruments, which could eventually integrate the improved field performance offered by the development of UHF 40\,T solenoids for the final cooling. 

\subsubsection{Detector magnets for physics search}
Besides their use in colliders, research in particle physics also calls for advances in magnet technology. Such examples are the very high fields that would be needed to improve on the exclusion bounds for the mass of axions as dark mass candidates, or the need for very thin, lightweight, but large dimension solenoids and toroids to be used as detectors in space, operating in helium-free conditions. Such magnets will benefit from HTS technology, opening up options for science that cannot be reached with LTS magnets. The development in the scope of the magnet program of the Muon Collider will contribute to making this technology available.

\subsubsection{Superconducting motors and generators}
The poles of superconducting motors and generators, such as those considered for marine transport and aviation, or wind turbines, are compact and non-planar to limit the dimensions and increase the power density. The challenges coming from compactness (high current density, high stress, and high energy density) and 3D shape (coil winding and joints) are similar to those of accelerator dipole and quadrupole magnets. The technical solutions developed for one field of application could directly applied to the other fields.

\subsubsection{Targets for spallation neutron sources}
Spallation neutron sources have the conflicting requirement of a small beam spot size, to improve the neutron beam quality, and a high power deposition, to improve the neutron beam flux. This is typically achieved using targets comprising materials with high atomic number, but this causes increased beam damage. Flowing liquid metal targets have been used but boiling of the liquid caused cavitation and limited target performance. The fluidised tungsten target that is being studied as part of the muon collider R\&D can alleviate this fundamental limitation; a flowing target can be achieved but one that is resistant to cavitation.

\subsubsection{Microwave RF sources}
Microwave sources and amplifiers use strong magnetic fields to confine the electron beam, with application in communications, RADAR, fusion heating and particle acceleration. In addition some magnetically confined fusion schemes require waveguides to be brought through a region with high magnetic field where breakdown can occur.

\begin{figure}
\includegraphics[width=\textwidth]{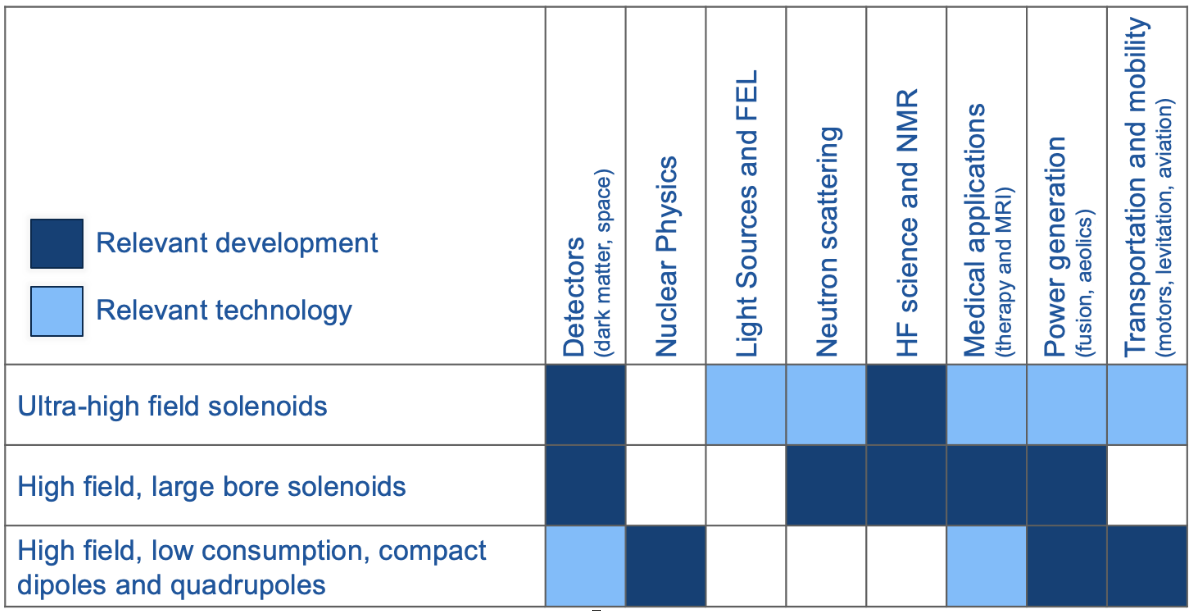}
\caption{Matrix of relevance of the developments required for a Muon Collider (rows) to other fields of science and societal applications (columns). The color shading indicates whether the developments can have direct application to other fields (dark shading) or whether the technology development is relevant, but application to other fields may require further development (light shading). 
\label{fig:magnet_matrix}}
\end{figure}

\subsection{Facilities
}
We have identified a number of facilities that may take advantage of, and benefit, the muon collider R\&D. 

\subsubsection{Colliders}
There are several important R\&D topics that the IMCC is studying that have direct impact on other collider projects. High-Field Magnet and SRF technologies will be required by any future collider. Additionally, the muon collider detector R\&D may be a stepping stone towards achievement of the challenging parameters required by hadron colliders at next-generation energies and luminosities.

\paragraph{Status} We are actively working with the High Field Magnets programme, including leading an EU grant submission. Detector R\&D collaboration is proceeding through the ECFA-led DRD process.

\subsubsection{Charged lepton flavour violation experiments}
Charged lepton flavour violation experiments such as mu2e 
and COMET 
require high-current and high-purity semi-relativistic muon beams. In these experiments pion production and capture is performed using targets immersed in high field solenoids.  Pions decay and the beam is cleaned in systems of long solenoids and dipoles. These pion production systems share many common challenges with the muon collider pion production system. In particular, the target and capture solenoid in mu2e and COMET has strong similarities with the muon collider target, albeit with some less demanding parameters. mu2e phase 2 and the proposed Advanced Muon Facility both represent opportunities to develop novel concepts in muon beam manipulation.

\paragraph{Status} We have had in-depth discussions regarding possible collaboration on the target solenoid with the mu2e team. mu2e expressed significant interest in the HTS concepts that we have developed for the muon collider target. We have had active discussions also with the COMET team and look to further strengthen this area.

\subsubsection{g--2}
The proposed beam that is under study at J-PARC represents an interesting alternative route to low emittance muon beams. In this technique, positively-charged muons are stopped in a target and form muonium. A laser is subsequently directed onto the target and re-ionises the muons, resulting in a very small emittance in the outgoing positively-charged muon beam.

\paragraph{Status} We are seeking further discussions with colleagues from J-PARC and seek to strengthen this area.

\subsubsection{nuSTORM and precision neutrino experiments}
Stored muon beams may be used as a well-characterised neutrino source. nuSTORM has been proposed as a neutrino production facility. Pions are passed into a muon storage ring. Muons arising from backwards decaying pions are captured and circulated, while remnant pions are extracted and pass onto a beam dump. The momentum difference between pion and muon beams enables injection and extraction without the use of pulsed magnets. The energy of the stored muon beam can be changed, yielding precisely characterised neutrino beams at different energies. nuSTORM will be used to study nuclear physics and neutrino scattering cross sections. This will provide a unique insight into nuclear physics. It may significantly reduce systematic uncertainties in the next generation of neutrino oscillation experiments. Additionally such a neutrino source will enable improved capability to reject exotic neutrino physics models.

nuSTORM provides an important opportunity to deliver a physics result using a high-intensity stored muon beam. While the energy and intensity is lower than the muon collider, it would be the highest stored muon beam power and so could be an important stepping stone on the way to a muon collider, were it to be approved.

\paragraph{Status} We are actively collaborating with the nuSTORM team. We have considered the potential to share a target, with O(200\,MeV) pions used for a demonstrator source and O(GeV) pions used for a neutrino source. Consideration is ongoing for the design of the necessary switchyard.

\subsubsection{muSR and low-energy muon beams}
Highly polarised muon beams may be produced using pions stopped in a fixed target. These low energy muon beams may benefit from similar cooling techniques to the ionisation cooling system. The cooling systems are challenging because these muons lose energy much more quickly than the semi-relativistic muons used in the muon collider, so very thin absorbers are required and low-frequency or electrostatic acceleration is preferred. Issues such as breakdown at high voltage and solenoid optics are shared.

\paragraph{Status} We have established a collaboration with experts from the ISIS muon beams instruments. ISIS provides one of the brightest polarised muon beams in the world. The ISIS team are interested to collaborate on advanced instrumentation concepts, for example using vertex detector-like technology as well as muon beam cooling techniques. We are collaborating on the design of a low energy muon cooling system.

\subsubsection{High power proton sources}
The high-power proton source under consideration for use as the muon collider muon source shares many characteristics with proton sources used for neutron spallation and neutrino production. In particular, technical challenges such as heating in the charge exchange foil and high power targetry are shared between these facilities.

\paragraph{Status} The proton driver task is led by ESS and ISIS are active members of the team. Foil heating is an area of active concern and one which we are investigating. We are also considering laser stripping techniques. More generally, the experts from ESS and ISIS are looking at issues such as space charge. The RAL target technology group, who led the construction of the T2K graphite target and are responsible for delivery of the DUNE target, are members of the collaboration.

\subsection{Synergies---summary}
In summary, there are strong motivators for present and future research and development of the underlying technologies that will be required to deliver the Muon Collider, which has direct (i.e.~similar specifications) and indirect (i.e.~similar technology) relevance to other scientific and societal applications. Figure~\ref{fig:magnet_matrix} renders graphically our evaluation of the relevance of the technology developments for a Muon Collider to other fields of science and societal applications.





\clearpage
\section{Development of the R\&D programme
}
\label{sec:Section09}


\noindent

\subsection{Demonstrator
}
\label{sec:Section09_1}

\subsubsection{System overview}
The muon cooling demonstrator is proposed to demonstrate the key components of a full muon cooling channel. It will extend the measurements made by the Muon Ionisation Cooling Experiment (MICE) to include cooling in all six phase-space dimensions, not just four phase-space dimensions, reacceleration of the beam and chaining of multiple cooling cells. The aims of the muon cooling demonstrator are described in Table~\ref{tab:demo_aims}, compared with the achievements of MICE.

\begin{table}[h]
\begin{center}
\caption{Performance for the updated cooling system.}
\label{tab:demo_aims}
\begin{tabularx}{0.9\linewidth}{lXX}
\hline
             & \textbf{MICE}       & \textbf{Cooling Demonstrator} \\
\hline
Cooling type & Transverse & Transverse and Longitudinal \\
Absorber type & Single absorber & Many absorbers \\
Cooling cell & Cooling cell section & Many cooling cells \\
Acceleration & No reacceleration & Reacceleration and stable bucket \\
Beam & Single Particle & Bunched beam \\
Instrumentation & HEP-like & Multiparticle \\
\hline
\end{tabularx}
\end{center}
\end{table}

\subsubsection{Key challenges}
The Demonstrator must meet the following key challenges:
\begin{itemize}
\item Create a muon beam with a sufficiently high flux of particles in the small transverse and longitudinal acceptance of the cooling system.
\item Operate a cooling system, including integrated RF cavities, solenoids and magnets, in a day-to-day manner.
\item Commission and operate the system to transmit beam through the lattice and cool it.
\item Measure the beam before and after cooling to characterise the beam and demonstrate successful emittance reduction.
\end{itemize}

\begin{table}[!hbt]
   \centering
   \caption{Possible hardware parameters for the Demonstrator. Parameters are presented for both a lithium hydride (LiH) and liquid hydrogen (LH2) absorber.}
   \label{tab:c}
   \begin{tabular}{lr}
       \multicolumn{2}{c}{\textbf{Cooling System}} \\
       \textbf{Parameter} & \textbf{Value} \\
       \midrule
       Cell length & 2 m \\
       Peak solenoid field on-axis & 7.2 T \\
       Dipole field & 0.2 T \\
       Dipole length & 0.1 m \\
       RF real estate gradient & 22 MV/m \\
       RF nominal phase & 20$^\circ$ \\
       RF frequency & 704\,MHz \\
       LiH wedge thickness on-axis & 0.0342 m \\
       LiH wedge apex angle & 5$^\circ$ \\
       LH2 wedge thickness on-axis & 0.19\,m \\
       LH2 wedge apex angle & 30$^\circ$ \\
       \bottomrule
   \end{tabular}
   \label{l2ea4-t1}
\end{table}

\subsubsection{Recent achievements}
Two potential sites have been identified on the CERN estate that can host the demonstrator, the TT10 and the TT7 tunnels. The first site would be suitable for a high power (>2MW) target but would require extensive new civil engineering. The second site would be suitable only for a low (10\,kW) power target but would largely reuse existing tunnels. The low power site would have no limitation in terms of maximum instantaneous intensity; the limitation would be on the average current. This site is therefore fully suitable for studies of cooling efficiency, but it would require more time to build large statistical samples. 

The high power site is placed to receive beam from the transfer line from the PS to the SPS and could potentially receive the entire 80\,kW of beam available from the PS, and could, in future be connected to the High Power SPL as designed in Ref.~\cite{Gerigk_HPSPL}. 

We have designed a possible layout for the demonstrator, based on a cooling cell adapted from MAP designs of the cooling channel. The cooling cell has requirements consistent with existing technologies while still yielding low emittances. The parameters of the cooling cell are shown in Table~\ref{tab:c}. The lattice corresponds to a system towards the end of the rectilinear cooling system. The current design is a preliminary one, but the design enables us to set parameters for the cooling cell and other R\&D in order to get started.

Work has begun towards a beam physics design for the target region and transport line that will be used to bring muons into the cooling system. We have a beam physics design for the beam preparation system, establishing the feasibility to create the necessary 100\,ps pulses with a small transverse emittance. We have made rate estimates based on reasonable target horn parameters.

\subsubsection{Planned work}
We will continue with the beam physics design. We hope to have a full start-to-end simulation of the demonstrator from target through to beam stop, including matching sections. Two separate designs are required, for the low-power and high-power siting options which have different constraints on target and beam line. We hope to include studies on the possibility to develop a parasitic, high energy pion beam line for physics experiments such as neutrino production.

We will also proceed with the complete mechanical design of a typical cooling cell in the framework of WP8 of Mucol (see Section~\ref{sec:Section7_8}). 

\subsubsection{Important missing effort}
Currently there is little effort available beyond beam physics designs and preliminary engineering integration studies of the cooling cell. Delivery of a fully engineered target design and beam preparation system is challenging. Detailed consideration of appropriate beam instrumentation, both in the cooling cells for commissioning and in the upstream and downstream matching sections, is not foreseen within the current resource envelope.

\subsection{RF test stand
}
\label{sec:Section09_2}
\subsubsection*{CEA test stand at 704\,MHz}

CEA proposes to use the two 704\,MHz klystrons presently used for ESS cryomodules tests in order to provide up to 2.8\,MW at this frequency. In a classical pillbox cavity at 704\,MHz, this would allow for gradients up to 30\,MV/m. Adding a pulse compressor would allow to reach higher gradients, but would need some extra developments, as there exists no such compressor at this frequency.

For tests under high magnetic field, a magnet is required. The available bore radius of the magnet must be at least 380\,mm to host a pillbox cavity and its waveguides at 704\,MHz. For testing cavities at lower temperature, extra space would be required. A magnet like the AFC magnet used in the frame of the MICE experiment fits with these requirements, at least for tests at room temperature. An alternative to the MICE magnet could be the magnet proposed for the INFN test stand (see below).

To comply with safety regulations, a bunker shielded against radiation and magnetic stray fields will have to be designed and built.

\subsubsection*{\textbf{R}adio \textbf{f}requency cavity in a \textbf{m}agnetic \textbf{f}ield \textbf{t}est \textbf{f}acility (RFMTF) at INFN-LASA }

Cooling beams of muons in flight requires absorbers to reduce the muon longitudinal and transverse momentum, accelerating fields to replace the lost longitudinal momentum inside a static solenoidal magnetic field to focus the muon beams. In order to study the interactions of a static magnetic field with the operation of high gradient accelerating fields, a twofold approach has been conceived.

The first one is related to the study of materials and surface finishing to obtain the highest HV field with a reduced number of breakdown events. This will be done both looking at experimental data available in literature and reported in recent workshops devoted to high gradient structures (MiniMeVArc 14--15 September 2023 CERN and HG2023 16--20 October 2023 Frascati) and designing a devoted test stand. This test stand is going to be installed at LASA within the next 3 months and will allow to test very intense electric fields (up to 100\,MeV/m) embedded in a 1 Tesla parallel static magnetic field. The test is an evolution of one already operational at CERN but it adds the availability of the magnetic field. Moreover this test stand will allow the possibility to reach more than 10\textsuperscript{6} HV pulses for each material tested to work with a suitable statistic. This HV voltage setup represents a compromise between the costs and the time involved in a real RF test bench.

Figure~\ref{fig:LASA_elements} shows the main elements of the setup: the pulsed HV power supply, the magnet to be used, and the geometrical arrangement of the pieces of material involved. Electrical measurements of peak currents, optical analysis of the damaged surfaces along with optical images related to the onset of discharges will be the main diagnostic tools used.

\begin{figure}
\includegraphics[width=\textwidth]{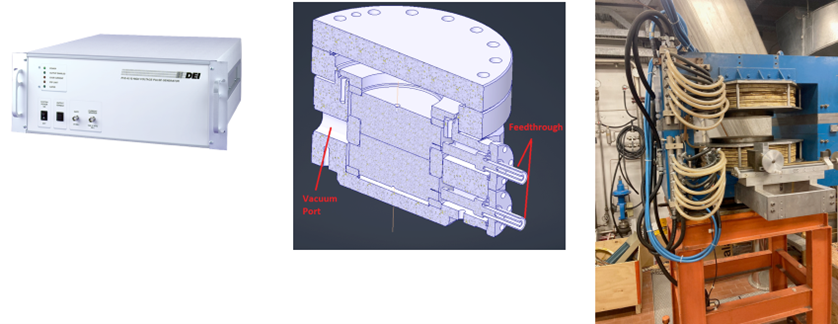}
\caption{main elements of the first experimental set up under preparation at INFN-LASA in Milan to test RF components in 1 T field.\label{fig:LASA_elements}}
\end{figure}

The second approach has been devoted to the study of a cavity suitable to be inserted in the above (below) discussed superconducting solenoid and able to be powered to reach and eventually exceed the limit of 30\,MV/m. This set up will constitute the next relevant step with respect to the activities carried out in the USA in 2005 with electrical fields up to 45\,MV/m and magnetic fields up to 4\,T. 

The first choice that was faced in the work was to carry out an initial evaluation that considered the costs involved in the operation. It immediately became clear that the magnetic field was the most relevant element in this respect. Two perspectives relating to operating in the 800--3000\,MHz range were analyzed. The lower limit constituted a choice more related to what could be the final frequency for the cooling channel cavities, but a cost exactly double compared to the higher frequencies.

For these reasons, a project is currently being finalized which involves a magnet capable of containing a 3~GHz cavity with all the necessary related elements.

Figure~\ref{fig:RFMTF_cavity} shows the electromagnetic design of a cavity of this type. A first constructive mechanical design will be available by spring 2024.

\begin{figure}
\includegraphics[width=\textwidth]{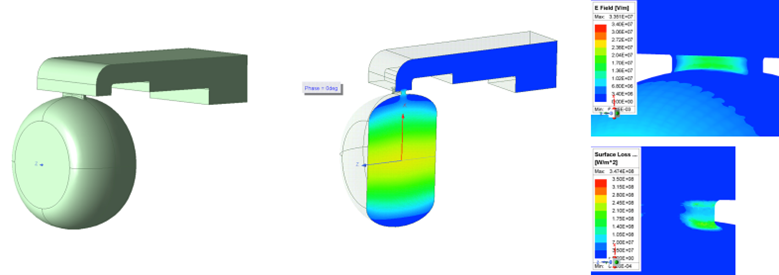}
\caption{Conceptual of the 3 GHz cavity for the test in the RFMFTF capable of 7\,T field. \label{fig:RFMTF_cavity}}
\end{figure}

The high field RFMFTF is conceived in such a way to be primarily a facility for RF studies and investigations. However, it has also the important function to be a first, very significant, integration exercise and a proof-of-principle of the technologies we envisage to use in the cooling cell solenoids. For these reasons, despite the field level makes it in the range possible for LTS, we have decided to pursue a design in HTS at 20\,K. The use of cryogen-free solution also allows the facility to be installed in places where liquid helium is not available.

In a second phase, would the MC studies get momentum and more resources be available, a facilities of 5--7\,T with large bore (700\,mm free aperture) can be built to test the actual 704\,MHz cavities foreseen for the cooling cells.

Following a study of three different configurations, we selected the layout with less use of HTS tapes (by far the single higher cost in the BOM). The parameters of the first RFMTF (300\,mm bore) coils are given in Fig.~\ref{fig:Table_RFMTF}.

\begin{figure}
\includegraphics[width=\textwidth]{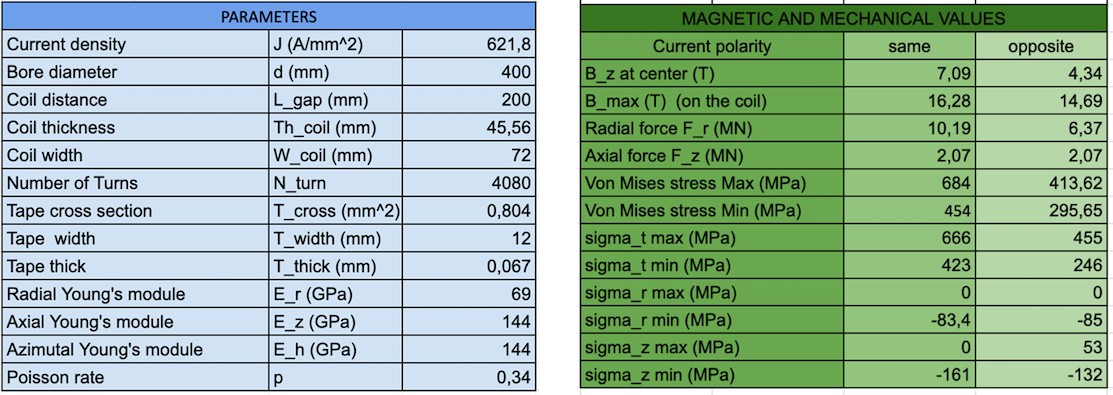}
\caption{Geometrical, mechanical and magnetic parameters of the 300 free bore split coil \label{fig:Table_RFMTF}}
\end{figure}

In Fig.~\ref{fig:RFMTF_assembly} a sketch of the preliminary design of the RFMFTF, with a 3 GHz RF test cavity inserted in the bore, is shown together with some details of the support system and of the coil suspension system are shown.  

\begin{figure}
\includegraphics[width=\textwidth]{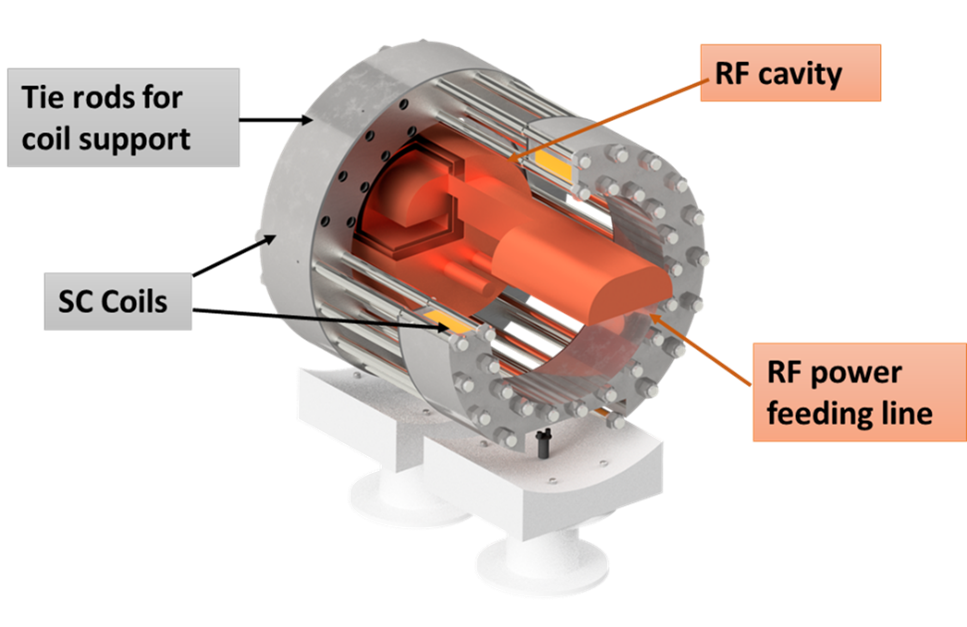}
\caption{Sketch of the RFMFTF with an RF cavity inside the magnet bore. \label{fig:RFMTF_assembly}}
\end{figure}

A first evaluation of the forces generated by misalignment in assembly of the split coil, errors in co-axiality and tilting angle, show that the tolerances are not critical, see table in Fig.~\ref{fig:Split_coils__RFMTF}: few tenths of mm of assembly accuracy should suffice. Various solutions to solve the tolerance and coil force support in an integrated way are under investigations.

\begin{figure}
\includegraphics[width=\textwidth]{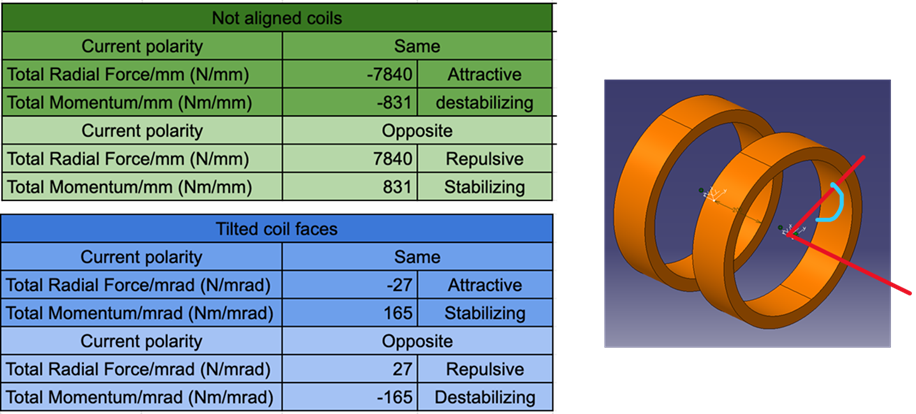}
\caption{Sketch of the RFMFTF with an RF cavity inside the magnet bore. \label{fig:Split_coils__RFMTF}}
\end{figure}

A first rough evaluation of the cost of the magnet system is around 2 M€ in material (of which 7--800 k€ of HTS tape), while the construction and commissioning time from the time t$_{0}$ (design and budget available) is about two years.

\subsection{Magnet test facility
}
\label{sec:Section09_3}
\subsubsection{System overview}
The rectilinear cooling cells need magnetic field ranges from 2\,T up to more than 40\,T, while the bore size ranges from about 800\,mm for the low field initial section to 50--60\,mm for the ultra-high filed solenoids in the final part of the cooling. Then coil length ranges from 2 to 3\,m in the initial party to 500\,mm of the ultra-high field of the final cooling cell. The solenoids may have opposite fields such that the forces maybe attractive or repulsive. While the detailed design of the solenoid of a cooling cell is under way, together with the integration of all elements above mentioned, the design and construction of a magnet test facility is a unique opportunity for an early  exercise of integration,  a study that will certainly be much useful for the detailed integration study to be done for the cooling cell demonstrator, see Section~\ref{sec:Section09_1}.

The magnetic system foreseen for the Radio Frequency in a Magnetic Field Test Facility (RFMFTF) is not only a test bed. As test bed for RF, indeed, the magnet could be in built with more established technology, Nb3Sn or even Nb-Ti superconductor at the prize of requiring liquid helium as coolant.  However, preliminary studies have already singled out the solenoids of the cooling cell as ideal for use of HTS, operating at a temperature higher than that of liquid helium. The main reasons are:
\begin{enumerate}
    \item The solenoidal configuration of the magnets well suites the natural winding topology of the HTS conductor. We think that the multiple stacked layer pancake-wound is the ideal solution for  a conductor that comes in form of a wide and very thin tape (tape cross section: $12 \, \mbox{mm} \, \times 60 \, \mu\mbox{m}$).
    \item The coils operates in steady state mode, with ramping and de-ramping rate that are dictated only by practical  considerations for turn on/off the accelerator complex, and can be as long as 5 to 10\,hours. We think that the steady state and the medium size and stored energy of these magnets (10--30\,MJ) allows for use of some form of non-insulated winding technology. This technology, in principle, makes protection possible even in absence of extra copper for stabilization. The only stabilizing material is the usual 5 to 20\,\textmu m thin copper layer heating the tape. Since the conductor is basically a single or a double tape without any addition, the current density is extremely high, favoring compact windings. If needed, in order to increase conductor-to-conductor resistance a highly resistive metal ribbon can be inserted, at price of making the winding a little more complex. However, other solutions might be explored, like partially conductive resins, use of metal as impregnant etc.
    \item The use of single/double tape winding is a key ingredient of the simple stacked pancake construction, that allows a modular approach to the solenoid construction, with copper inner/outer rings for: i) current transfer, in and out of each single pancake; ii) for mechanical restrain; iii) and for thermal anchoring.
    \item Operation at current densities of 650\,A/mm$^2$ and above (a density that is double of the LHC dipoles, at a field that is double as well, or more) has to go with an operating temperature of 20\,K or so: this for us is a key objective, in order to reduce drastically both the helium inventory (helium is becoming very scarce!) and especially the energy consumption. The use of single/double tape conductor means to stay at current around 1 kA which is sufficiently low current in the feedthroughs to go in cryogen-free cooling mode, or at least without use of liquid helium. For comparison the LHC dipole requires 12\,kA at 2\,K: the heat inlet is ten times higher with an efficiency ten times worse: the cooling cell of the Muon Collider will then be about 100 times less energy intensive than the present LHC dipoles. In principle the HTS cooling cell magnets could be then a hundred times more energy efficient that the LHC dipoles. Not only energy-consumption is reduced , but also the infrastructure  will be greatly reduced in term of electrical power installation and footprint.
\end{enumerate}

\subsubsection{Key challenges}
The Magnetic system of the RFMFTF will be, beside a test bed for RF cavities,  a first test very significant of:
\begin{enumerate}
    \item integration exercise of magnet and RF system in the cooling cells.
    \item a proof-of-principle of the technologies we envisage to use in the cooling cell solenoids. For these reasons, despite the field level makes it in the range possible for LTS, we have decided to pursue {\bf a design in HTS at 20\,K}. The use of cryogen-free solution also allows the facility to be installed in places where liquid helium is not available. 
	\item The test facility will have the shape of solenoid doublet, called also split coils, that could be energized both with same and opposite polarity, exactly like in the Cooling cells. The spit coils will result in very significant net forces of each coil of the order of 200 tons (repulsive when oppositely energized and attractive when in series).
\end{enumerate}

\subsubsection{Recent achievement}
As mentioned in the RF test stand in Section~\ref{sec:Section09_3}, a first preliminary design of a 700\,mm split coil free bore (see Fig.~\ref{fig:RFMTF_assembly} of Section~\ref{sec:Section09_3}), that could host 700\,MHz cavities, ended with a rough evaluation of 5\,MCHF. Therefore a considerable effort has  bene done toward a reduced size system whose cost should be in the more viable 2-2.5\,MCHF range. The inner free bore at room temperature  will be 300\,mm allowing to test easily cavity like the 3 GHz whose test is envisaged at INFN-LASA.

Following a study of three different configurations, we selected for the magnetic system the layout with less use of HTS tapes (by far the single higher cost in the BOM). The parameters of the first RFHMTF (300\,mm bore) coils are in the tables of Fig.~\ref{fig:Table_RFMTF} while in Fig.~\ref{fig:Split_coils_GFR} a sketch of the coil and of the homogeneous region for the same polarity excitation is shown.

\begin{figure}
\includegraphics[width=\textwidth]{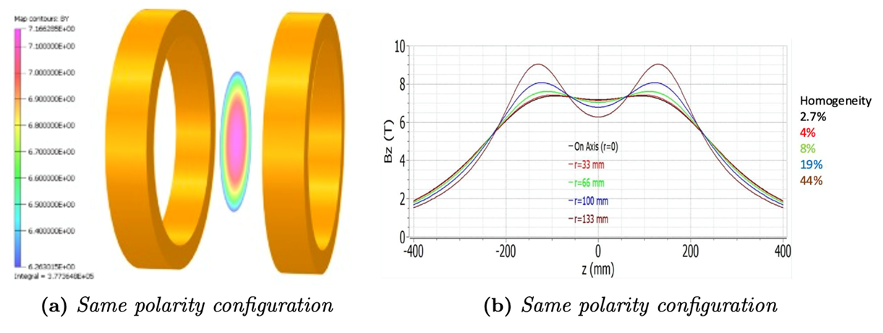}
\caption{Sketch of the spit coil (without cryostat) and good field region ($\pm200$\,mm).} \label{fig:Split_coils_GFR}
\end{figure}

Very recently the forces and torques raising form assembly tolerance have bene evaluated and find manageable since the request assembly tolerance are  in the order of a few hundred $\mu\mbox{m}$, see Fig.~\ref{fig:Split_coils__RFMTF}. Various solutions to solve the tolerance and coil force support in an integrated way are under investigations. A sketch of various components of RFMFTF is shown in Fig.~\ref{fig:Assembly_HTS}.

\begin{figure}
\includegraphics[width=5cm]{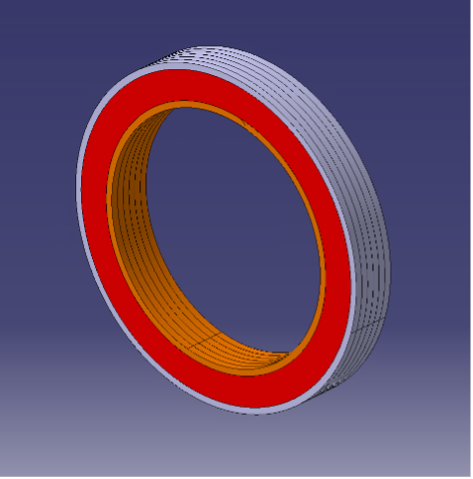} \hfill
\includegraphics[width=5cm]{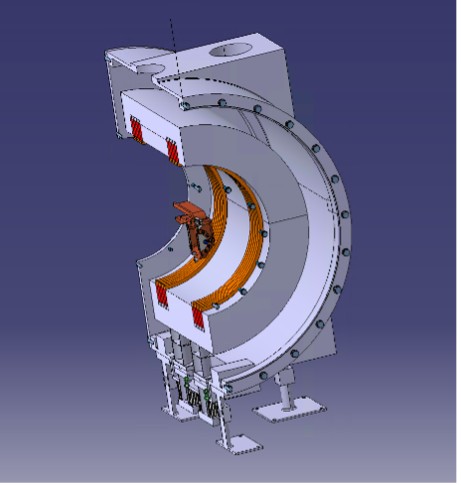} \hfill
\includegraphics[width=5cm]{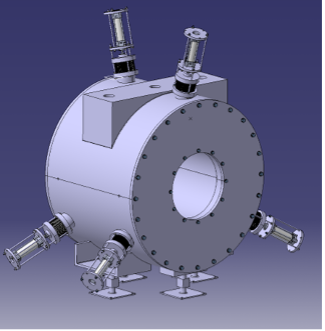}
\caption{Left. Single HTS coil with its restraining ring. Center: view of longitudinal cut of the spit coil system with inside the 3 GHz RF cavity (without ancillary systems). Right: the complete cryomagnetic assembly system  for the 7\,T RFMFTF.} \label{fig:Assembly_HTS}
\end{figure}

A first rough evaluation of the cost of the magnet system is around 2 \mbox{M\texteuro} in material (of which \mbox{700--800 k\texteuro} of HTS tape), while the construction and commissioning time from the time t0 (design and budget available) is about two years.  If at least part of the budget would be allocated by INFN in 2024 the facility can be ready for commissioning before the end of the EU-HE-MuCol Design Study. 

\subsubsection{Planned work}

The design of the magnetic system of the RFMFTF will continue along these lines:
\begin{enumerate}
    \item Thermal design of the magnet, a critical issue for cryogen-free layout. (Jan--May 2024)
    \item Structural analysis (also with construction defects)  and mechanical design of the facility (Feb-June 2024)
    \item Complete cost evaluation (March 2024)
    \item Construction drawings: May--Sept 2024)
    \item Production readiness review (Sept 2024)
\end{enumerate}
A further work that will start in 2024 is the integration of the cooling cell demonstrator. We plan to start after summer and having a first layout for comments and discussion by the IMCC CM2025.

\subsubsection{Important missing effort}

An effort beyond a reasonably detailed design is today not possible because of missing resources, both in personnel and especially in material. As previously stated about 2 MCHF are necessary to build the facility (for the magnetic part: RF budget is not considered here in this section). Mutualization with other projects (INFN/UMIL IRIS project and FCC-hh/HFM project, next EU application on HTS magnets) may help to bridge the resource gap.

\subsection{Other test infrastructure required (HiRadMat, ...)
}
\label{sec:Section09_4}

\noindent
The muon collider proposes a front-end including a production target designed to withstand high-intensity short pulses. While the baseline concept envisions a graphite target, various alternative target concepts are also under examination 
Regardless of the approach, the interaction between the target material and the intense proton pulses results in an immediate adiabatic temperature rise within the core of the production target. This surge is swiftly followed by a dynamic stress-strain response that pushes the material to its limits. Additionally, the high average beam power directed at the target creates challenging high-temperature conditions. These thermal-mechanical conditions, particularly those induced during the pulse, cannot be accurately replicated offline in a laboratory at these specific time-temperature ranges. Beam test facilities like CERN’s HiRadMat facility of SPS are therefore required to study such phenomena. HiRadMat provides multiple test stands capable of hosting experiments and receives LHC-type proton beams at 440\,GeV/c. While the beam parameters differ from those of the muon collider's frontend, adjustments can be made to ``simulate'' optimal resemblance with a muon collider beam and target operational conditions. Utilizing the modular setup of HiRadMat could potentially assist in evaluating and validating the durability of the target design and material selection. Other beam test facilities worldwide may also be considered if suitable. As previously mentioned, the target will operate at high temperatures, potentially reaching up to 2000\textdegree C depending on the chosen concept. Material properties in such extreme conditions are limited yet crucial for accurate modelling and engineering of the target and other beam-intercepting devices for the IMCC. Consequently, it is imperative to characterize thermally and mechanically the materials to be used. Such tests do not necessarily require accelerator-specific facilities but rather laboratories equipped to perform such characterization. 

The absorbers within the muon collider, notably the production target, its beam windows, and the surrounding radiation shielding, will be subjected to mixed fields of protons, neutrons, and electromagnetic showers. The high yearly integrated intensity contributes to the creation of radiation-induced defects that subsequently affect the behaviour of the materials. Quantifying these defects and assessing their impact on material properties are crucial for understanding the long-term survivability of the IMCC components. Facilities like BLIP at the Brookhaven National Laboratory can be exploited to deliberately induce high radiation damage on materials in a controlled environment. Laboratories equipped for post-irradiation examinations can then analyse and characterize these irradiated materials. The RaDIATE collaboration, which is led by Fermilab and strongly collaborates with multiple laboratories including CERN, already has defined a framework for such irradiation and characterization endeavours. Efforts in this direction shall be pursued by the IMCC. 

\begin{flushleft}
\interlinepenalty=10000

\end{flushleft}

\clearpage
\section{Implementation considerations 
}
\label{sec:Section02}

\long\def\nix#1{}

One of the main goals for the IMCC Accelerator R\&D Roadmap is to deliver by 2026 a tentative timeline for the collider project implementation. Considering global uncertainties in planning of the next generation collider facilities, we focus on development of the fastest implementation timeline. This includes identification of possible sites to host the collider and understanding the associated costs and power consumption of the facility. The scope and the timeline of the R\&D program necessary to conduct design studies and to establish feasibility of critical technological components also need to be developed. This chapter provides a snapshot of the current status of the roadmap towards the collider implementation. The timeline being prepared aims to demonstrate all the critical collider and detector technologies by 2040 and to start the collider operation before 2050. The timeline involves a two-stage approach based on the choice of the collider dipoles, for which the technology is expected to have the longest lead-time. 

\subsection{Timeline considerations}
In development of the timeline for the muon collider implementation, we assume that all the required funding and resources will become available. The timeline is thus based primarily on technical considerations, such as the typical amount of time needed to make one iteration on the magnet technology; we refer to it as "technically limited". We also assume that no show-stoppers will be identified during the course of the R\&D program. 

\subsubsection{Primary timeline drivers}
Three R\&D areas have been identified to be on the critical path and therefore define the technically limited timeline:
\begin{itemize}
\item
  {\bf The muon cooling complex and its technology}: development of muon cooling modules is essential in order to integrate superconducting solenoids, normal-conducting RF, robust absorbers and auxiliary structures in a tight space and to verify mechanical robustness of the modules. Integration of several modules into a facility with beam will ensure convincing demonstration of this novel technology.
\item
  {\bf The superconducting magnet technology}: development of HTS-based solenoids for the target/captures systems and the muon cooling channel are essential for production of quality muon beams. The superconducting accelerator magnets in the collider ring define luminosity performance of the machine. 
\item
  {\bf The detector and its technologies}: development of the detector technologies necessary to suppress the beam induced background are essential for ensuring that physics goals of the machine are met.  
\end{itemize}
The timeline for the collider implementation will therefore focus primarily on these three technology developments. Amongst these, we anticipate that the longest time will be needed to develop HTS-based or hybrid collider dipole magnets. However, Nb$_{3}$Sn-based magnets will be available much sooner and on a timescale similar to the other critical technology items. 

\subsubsection{Staging strategy}
The goal of having an operating collider by 2050 demands that all the necessary technologies are mature in approximately fifteen years. The muon cooling technologies and the detector technologies are expected to be available on this timescale. This includes technology to produce HTS solenoids in the muon production and cooling sections as well as magnets in the muon accelerator system. HTS accelerator magnet technology for the collider ring may require longer development. However, Nb$_3$Sn-based magnets are expected to be available within the required time frame and would deliver 11\,T field at 4\, K with an
aperture of about 15\, cm. We therefore consider a staged approach with
two primary staging options under study:

In the {\bf energy-staging} option, the first stage has a lower center-of-mass energy (e.g. 3\,TeV) using LTS magnets, followed by an upgrade to the nominal energy with HTS. A benefit of this option is that the cost of the initial stage is substantially lower than for the full project, which can potentially simplify international negotiations for funding of the collider and thus facilitate faster construction. The 3\,TeV design needs are consistent with 11\,T Nb$_3$Sn magnets. In the second stage, almost the entire complex from the first stage will be reused (with the exception of the collider ring) and a new RCS to accelerate muons to full energy and a new collider ring will be constructed.

In the {\bf luminosity-staging} option, the first stage is already at the full energy but using 11\,T LTS instead of 16\, T HTS dipoles. The resulting increase of collider ring circumference leads to a reduction of luminosity by approximately a factor of one third. In addition, a further reduction arises from weaker focusing of the beams in the interaction regions. A detailed study is necessary to quantify this loss, but the initial estimates indicated a factor of three reduction. In the second stage, the interaction region magnets will be replaced with higher performance HTS ones, similarly to the High Luminosity upgrades of the LHC. In this staging scenario, the majority of the project costs are incurred during the first stage.

The exact choice between the staging options will depend on funding availability as well as potential changes in the physics landscape due to new results from the LHC/HL-LHC and elsewhere. The rate of progress in the high-field magnet development can also contribute to the overall planning, as faster than anticipated progress with HTS collider ring magnets can enable operation at the nominal center-of-mass energy from the start up.

\subsubsection{Muon cooling roadmap}
The muon cooling technology is the most novel of the collider concepts and has to be demonstrated in a facility with beam. The demonstrator will consist of a proton injector, a target to produce muons, a sequence of muon cooling modules and several systems to manipulate and measure the muon beam. Using an existing proton infrastructure will significantly reduce the cost of the demonstrator but limits the number of potential sites. In particular, first considerations have been made to site the demonstrator at FNAL, CERN or ESS; also JPARC could be considered. The demonstrator will have to integrate with the operation of the proton facility. At CERN, two sites that could potentially house the demonstrator have been identified. One could extract the beam from the transfer line between the PS and SPS. The other could use the beam from the booster. The potential timelines for the realization of the demonstrator in the two sites depends on the operation of the proton complex for other purposes, for example the LHC. The developed timeline will take these constraints into account.

The development of the muon cooling modules is an important prerequisite of the demonstrator. The modules integrate normal-conducting RF and superconducting solenoids very tightly and also  include other components. Currently no infrastructure exists that allows to test the RF cavities in a high magnetic field. A previous infrastructure in the US has been dismantled after successful experiments. We are striving to develop a new test infrastructure, different laboratories could potentially host them. These include INFN Milano, STFC, IRFU-CEA, Fermilab, and CERN. 

\subsubsection{Magnet roadmap}
Different magnets are needed in different stages of the muon collider complex. Since HTS has already been successfully used in high-field solenoids, it is anticipated that HTS technology will be used for solenoids in the muon production and cooling sections. These include:
\begin{itemize}
\item
  Large-aperture, high-field solenoids for the muon production target.
\item
  Solenoids with a range of apertures and fields for the 6D cooling.
\item
  Highest-field, small-aperture solenoids for the final cooling.
\end{itemize}

Fixed-field superconducting accelerator magnets are required both for the muon accelerator rings and the collider ring. The technology choice for the magnets in the accelerator ring will be based on the cost optimization. The collider ring is more demanding on the strength of the magnetic field than the accelerator ring because of its impact on the final focusing of the beam and thus machines luminosity performance. 
As explained in the Staging Strategies section, it maybe risky to consider the use of HTS for the collider ring magnets. Therefore, more mature Nb$_3$Sn technology is considered as the baseline for the first stage of the collider operation, while the second stage would rely on HTS or hybrid magnet designs. HTS can still be considered for the first stage if the progress is faster than anticipated.

In addition to superconducting fixed-field magnets, fast-ramping normal-conducting magnets are required for the pulsed synchrotrons in the acceleration complex. The technology is well established, however further development is needed to optimise their cost and power consumption. 

\subsubsection{Detector roadmap}
The first estimates of the detector roadmap are based on experience with constructing the original ATLAS and CMS detectors and their major upgrades for HL-LHC. Almost all the subsystems of a Muon Collider detector require significant R\&D. In particular, the current requirements on the performance of tracking, calorimeter, and muon systems include challenging specifications for granularity, spacial and timing resolutions, and radiation hardness of these systems. Individually these requirements could be satisfied with technologies being developed for the HL-LHC detectors, however achieving all of them in a single detector represents a significant challenge that will require a decade or longer R\&D program. Development of detector technologies should proceed hand in hand with simulation studies aimed to further refine the detector performance specifications with specific physics targets in mind. It should be noted that the detector size and performance requirements are significantly different for the 3\,TeV and 10\,TeV detectors. It is therefore important to continue with detector development also for the 3\,TeV while the staging strategy is being further refined. Certain detector design aspects, such as the forward muon tagging and the potential luminosity detector, are still at the roots of their development and require identification of promising design options and their studies/comparisons. The data acquisition strategies, including novel AI/ML based approaches for on- and off- detector data processing,  and the corresponding data bandwidth, trigger/DAQ and computing requirements need to be evaluated and refined. Construction of large solenoidal magnets for the detectors is currently the largest uncertainty in the schedule due to limited supply of the magnet conductor as well as lack of industry companies capable of producing such magnets. On the bright side, this challenge is not unique to muon colliders, and development of future detector solenoids should be identified as a priority for the field and be pursued in synergistic fashion with other collider, CLFV, Axion Dark Matter and other experiments. Finally, engineering aspects of putting the entire detector together, while preserving capability to access and replace its components, need to be studied. 


\subsubsection{Other R\&D roadmaps}
In addition to the primary R\&D drivers, there are several other R\&D  areas that are important for the overall performance, cost, and power consumption of the muon collider. While these developments are not driving the schedule, it is nevertheless important to ensure that resources are available to ensure progress in the following areas:
\begin{itemize}
\item
The lattice design of the different parts of the collider.
\item
Fast ramping magnets and their power converters.
\item
The muon production target.
\item
The RF system of the muon accelerator and the proton complex.
\item
The neutrino flux mitigation technology.
\item
Cryogenics, instrumentation, vacuum technologies and others.
\end{itemize}

\subsection{Site considerations}
The study is currently focusing on a site-agnostic design. The  implications of specific sites will be investigated at a later time, assuming that additional resources will be available to conduct the work. Specifically, we will investigate the potential to reuse the existing infrastructure and study environment impact of the facility. Initial assessments of the civil engineering costs will also be performed. As of now, potential sites in Europe and the United States have been identified; other regions may propose alternative siting options as the study progresses.

\subsubsection{European sites}
Studies in Europe were limited to identifying at least one site close to CERN. One particularly important concern for this site is the mitigation of the neutrino flux that arises from the collider ring. At CERN, the goal is to ensure that the impact of the neutrino flux on the environment outside of the laboratory site is negligible, similar
to the impact of the LHC. A concept has been developed that can ensure meeting this goal for neutrinos from the arcs. It is based on a magnet mover system that slightly deforms the machine every few hours. Work has started on exploration of the mechanical systems and the implication on the beam dynamics. 

The experimental insertions can lead to a more important neutrino flux where the prolongation of the axis of each detector intercepts the surface of the earth. These locations will see non-negligible neutrino flux and might be suited for neutrino measurements, see Chapter~\ref{sec:Section03}. At CERN one site and orientation of the collider ring has been identified where the neutrino beams would enter the Mediterranean sea on one side and the uninhabited flank of a mountain on the other side. Other possible locations will be explored. 

The use of existing infrastructure will also be considered in the future. In particular, the SPS and LHC tunnels have sizes that make it interesting to consider implementing the accelerating synchrotrons in them. However this will have some impact on the schedule, since the tunnels need to be cleared of existing equipment before. Other relevant facilities that can be reused include the existing proton complex, the cryogenics infrastructure and more.

\subsubsection{US site}
Renewed US interests and contributions in muon collider efforts provide for an ideal opportunity to consider potential US siting. The existing accelerator facilities at Fermilab (e.g.~PIP-II) and the proposed booster upgrade or extension of the PIP-II linac could be developed into proton drivers for a Muon Collider. Furthermore, synergies with the neutrino and intensity frontiers programs in the US, as well as overlaps with nuclear science and industry applications provide additional benefits. This makes Fermilab an attractive siting option.

The generic design presented here needs to be modified for siting of a muon collider at Fermilab, where one is confined within the site bounds of the laboratory. Table~\ref{tab:Table1} shows tentative parameters of a possible Fermilab-based accelerator scenario that produces 5\, TeV beams for a 10\,TeV collider. This is a 4-stage hybrid RCS scenario that is constrained to fit within the Fermilab site, which restricts the circumference of the largest tunnel hosting the final two accelerators to approximately 16.5\, km. The magnetic fields in the highest energy accelerators are set at 15\, T for the superconducting magnets and $\pm$1.8\, T for the rapid-cycling magnets. 

A center of mass energy at the top quark production threshold ($\approx$173\,GeV per beam) could be a convenient choice for a first Collider, particularly if highest intensities or lowest beam emittances may not be initially available, since important physics results could be obtained with lower luminosity. We therefore propose to extend the low energy acceleration chain by adding an additional recirculating linear accelerator from 63\,GeV up to 173\,GeV. This would use 5 passes with 650\,MHz RF and would be similar to the earlier RLA stages.

The first RCS is chosen to include only normal-conducting magnets (up to 1.8\,T) and to fit a 6280\,m circumference so that it could utilize the existing Tevatron tunnel. That limits its peak energy to approximately 450\,GeV. The second RCS must take beam from 450\,GeV to the 1.725\,TeV injection
of the 16.5\,km rings and does not fit well in either the Tevatron or the 16.5\,km tunnel. We thus chose an intermediate circumference of 10.5\,km for RCS2. With the more flexible choice of circumference, we can reduce the high field magnet to 12\,T. Energy range in the 16.5\,km ring is constrained by the peak field available (15T) and by the limited cycling range of -1.8\,T to + 1.8\,T in the RC magnets. The highest energy accelerator cycles from 3.56\,TeV to 5.0\,TeV (RCS4). A second accelerator (RCS3) is included in the 16.5\,km tunnel, and it cycles from 1.725\,TeV to 3.56\,TeV. An example layout of the accelerator complex based on the design considerations presented above is shown in Fig.~\ref{fig:sitefillermap}.
 
\begin{table}
\caption{\label{tab:Table1}Parameters for a Fermilab-based 5\,TeV muon accelerator.}
\centering

\begin{tabular}{| l | l | l | l | l | l | l |}
\hline
\textbf{Parameter} & \textbf{Symbol} & \textbf{Unit} & \textbf{RCSI} & \textbf{RCS2} & \textbf{RCS3} & \textbf{RCS4} \\
\hline
Hybrid RCS &   &   & No & Yes & Yes & Yes \\
\hline
Repetition rate & $f_{\mathrm{rep}}$ & Hz & 5 & 5 & 5 & 5 \\
\hline
Circumference & $C$ & m & 6280 & 10500 & 16500 & 16500 \\
\hline
Injection energy & $E_{\mathrm{inj}}$& GeV & 173 & 450 & 1725 & 3560 \\
\hline
Ejection energy & $E_{\mathrm{ej}}$ & GeV & 450 & 1725 & 3560 & 5000 \\
\hline
Energy ratio & $E_{\mathrm{ej}}/E_{\mathrm{inj}}$ &   & 2.60 & 3.83 & 2.06  & 1.40 \\
\hline
Decay survival rate & $N_{\mathrm{ej}}/N_{\mathrm{in}}$ &   & 0.85 & 0.83 & 0.85 & 0.89 \\
\hline
Acceleration time & $T_{\mathrm{acc}}$ & ms & 0.97 & 3.71 & 8.80 & 9.90 \\
\hline
Revolution period & $T_{\mathrm{rev}}$  &\textmu s & 21 & 35 & 55 & 55 \\
\hline
Number of turns & $N_{\mathrm{turn}}$ &   & 46 & 106 & 160 & 180 \\
\hline
Required energy gain per tum & $\Delta E$ & GeV & 6 & 12 & 11.5 & 8.0 \\
\hline
Average accel.\,gradient & G\textsubscript{avg} & MV/m & 0.96 & 1.15 & 0.70 & 0.48 \\
\hline
Bunch population at injection & N\textsubscript{in} & 10\textsuperscript{12} & 3.3 & 2.83 & 2.35 & 2.0 \\
\hline
Bunch population at ejection & N\textsubscript{ej} & 10\textsuperscript{12} & 2.83 & 2.35 & 2.0 & 1.8 \\
\hline
Vertical norm. emittance & $\epsilon_{v,N} $  & mm-mrad & 25 & 25 & 25 & 25 \\
\hline
Horiz.\,norm.\,emittance &$ \epsilon_{h,N} $ & mm-mrad & 25 & 25 & 25 & 25 \\
\hline
Long, norm.\,emittance  &$ \epsilon_{z,N }$ & eV-s & 0.025 & 0025 & 0.025 & 0.025 \\
\hline
Total straight length & $L_{\mathrm{str}}$ & m & 1068 & 1155 & 2145 & 2145 \\
\hline
Total NC dipole length & $L_{\mathrm{NC}} $ & m & 5233 & 7448 & 10670 & 8383 \\
\hline
Total SC dipole length & $L_{\mathrm{SC}} $ & m &   & 1897 & 3689 & 5972 \\
\hline
Max. NC dipole field & $B_{\mathrm{NC}} $ & T & 1.80 & 1.80 & 1.80 & 1.80 \\
\hline
Max. SC dipole field & $B_{\mathrm{SC}}$ & T &   & 12 & 15 & 15 \\
\hline
Ramp rate & $B'$ & T/s & 1134 & 970 & 440 & 363 \\
\hline
Main RF frequency & $f_{\mathrm{rf}}$ & MHz & 1300 & 1300 & 1300 & 1300 \\
\hline
Total RF voltage  & $V_{\mathrm{rf}}$ & MV & 6930 & 13860 & 13280 & 9238 \\
\hline

\end{tabular}

\end{table}

\begin{figure}[t!]
\centering
\includegraphics[width=0.9\linewidth]{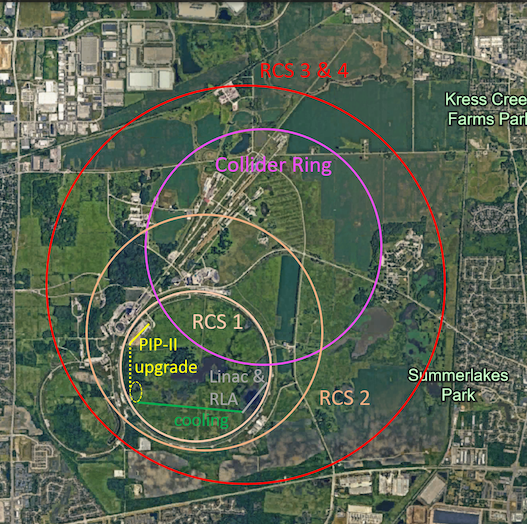}
\caption{Proposed layout of a 10\,TeV center-of-mass energy muon collider complex sited at Fermilab. The proton driver is shown in blue and the cooling channel is shown in green. RCS1 can be located inside the existing Tevatron tunnel. RCS3 and RCS4 are located inside the Accelerator Ring with circumference of 16.5 km, which is the largest we can fit onsite.}
\label{fig:sitefillermap}
\end{figure}

\subsection{Costing and power consumption considerations}
Full costing of the muon collider by 2026 is not possible with the existing resources. The study will therefore focus on the primary cost and power consumption drivers:
\begin{itemize}
\item
  The magnets, the power converters, and the RF systems of the RCSs. In particular, we will study the trade-offs between the power converters and the RF systems, in terms of both cost and power consumption.
\item
  The collider ring magnets and their shielding. In particular, we will study the trade-offs between cost, power consumption and luminosity performance.
\item
  The cost and power consumption of the target solenoid.
\item
  The muon cooling complex.
\item
  The proton complex.
\item
  An approximate estimate of civil engineering and infrastructure costs, based on scaling from existing or planned facilities. 
\end{itemize}

The study will also provide an estimate of the cost range required for the muon collider R\&D program. However, there will be an uncertainty associated with this estimate as current resources do not allow for a fully detailed design of the demonstrator facility. The current resources will also not allow to re-design the initial accelerating linacs; these will rely on past designs by the Muon Accelerator Program in the US.

\nix{
The accelerator R\&D Roadmap foresees two deliverables as input to the strategy
processes: first, a project evaluation report that assesses the muon collider potential;
second, an R\&D plan that describes the path towards the collider. 
The goal of the study is to deliver by 2026 a credible tentative timeline for the project implementation.
Given the uncertainties we will focus on the development of the fastest reasonable implementation timeline.
This requires identifying possible sites to host the collider and understanding the cost and power consumption
of the facility. It also requires understanding the necessary R\&D scope as well
as the time that is required to carry it out.
This chapter provides a snapshot of the current status of the roadmap toward the collider.
We prepared an initial version of a timeline that is based on a two-staged approach. The aim is to have the technologies
available by 2040 and the collider start operation before 2050. The schedule will be based on
a certain level of parallel efforts, starting approval and planning when approaching technical maturity and
starting construction only once maturity is fully achieved.

\subsection{Collider roadmap}
The timeline for the muon collider, as for any future project, depends on
external factors that are beyond our control, in particular the physics
findings at the HL-LHC and the available budget. For the evaluation report we
will therefore assume that the muon
collider will be identified as a priority for particle physics, similar to past
collider projects, and that it will receive the required funding. The timeline
will thus be mostly based on technical considerations, such as the typical time
needed to make one iteration on the magnet technology. If the project has lower
priority and funding the timelines will stretch, if it has higher priority it
might be shortened. We also assume that our work is successfull.

\subsubsection{Key timeline drivers}
The current timeline considerations identified that three technical developments are defining the minimum time to implement a muon collider:
\begin{itemize}
\item
  The muon cooling complex and its technology. The development of muon cooling modules is essential that
  thightly integrate superconducting solenoids, normal-conducting RF, robust absorbers and auxialliary
  technologies in a thight space.
  Integration of several modules into a facility with beam will ensure the full control of this novel technology.
\item
  The superconducting magnet technology. The development of HTS-based solenoids for the target and the muon cooling
  will be the basis for the production of quality muon beams. The superconducting accelerator magnets in the collider
  ring will drive the luminosity performance. 
\item
  The detector and its technologies. The development of the technologies to suppress the
  background and achieve the resolution is the key activity.
\end{itemize}
The timeline for the collider implementation will thus focus on considering these items.
We currently anticipate that the longest time is required to develop HTS-based or hybrid accelerator magnets. However,
Nb$_{3}$Sn-based accelerator magnets will be available on a timescale similar to the other critical technologies.

\subsubsection{Implementation strategies}
We aim for a collider to start physics in the 2040s. This requires the
technologies to be mature for project implementation in about fifteen years.
The muon cooling technologies and the detector technologies are expected to
be available at this timescale. This is also true for the HTS solenoid
technology in the muon production and cooling section and the magnets in the
accelerator system. However, HTS accelerator magnet technology for the
collider is potentially not mature enough in fifteen years but would need more
development time. Nb$_3$Sn-based collider ring magnets are expected to be
mature within fifteen years and could deliver 11\,T field at 4~K with an
aperture of about 15\,cm. We therefore consider a staged approach.
Two main options exist:

In the {\bf energy-staging} approach, the inital stage is at lower energy, for
example 3\,TeV. There is a important physics case already at this energy. In
this strategy, the cost of the initial stage is substantially lower than
for the full project. This could accelerate the decision-making processes. The
3\,TeV design is consistent with Nb$_3$Sn magnets at 11\,T.
In the second stage the whole complex will be reused with the exception of the
collider ring. 
An RCS to accelerate to full energy and a new collider ring will be
added.

In the {\bf luminosity-staging} approach, the inital stage is at the full
energy but using less performant collider ring magnets. This leads to an
important reduction of the luminosity. If 11\,T Nb$_3$Sn dipoles are used instead
of 16\,T HTS dipoles, the resulting increase of collider ring circumference
reduces the luminosity by one third. In addition a further reduction arises from
the interaction regions. A detailed study is required to quantify the loss, but
a factor three reduction might be a good guess.
In the luminosity upgrade, the interaction region magnets will be replaced with
more performant ones, similar to the HL-LHC. However one most likely will not
replace the other collider ring magnets. In this scenario almost the complete
project cost is required in the first stage, which could have important
implications on the timeline.

The choice between the staging options will depend on physics needs and funding
availability. Also the progress in magnet development will play an important
role. Faster progress, in particular of HTS for the collider ring magnets,
combined with strong funding support will make an earlier start of the 10\,TeV
option more attractive.

\subsubsection{Muon cooling roadmap}
The muon cooling technology is the most novel of the collider concept and has to be developed in a facility with beam.
This demonstrator will consist of a proton injector, a target to produce muons, a sequence of muon cooling modules and several systems to manipulate and measure the muon beam. Using an existing proton infrastructure will significantly reduce the cost of the demonstrator but limits the number of potential sites. In particular, first considerations have been made to house the demonstrator at FNAL, CERN or ESS; also JPARC could be considered. The demonstrator will have to integrate with the operation of the proton facility. For example, at CERN two sites that could house the demonstrator have been identified. One could extract the beam from the transfer line between the PS and SPS. The other could use the beam from the booster. The potential timelines for the realisation of the demonstrator in the two sites depends on the operation of the proton complex for other purposes, in particular the LHC. We will develop a timeline that takes these constraints into account.

The development of the muon cooling modules is an important prerequisite of the demonstrator. The modules integrate normal-conducting RF and superconducting solenoids very tightly also including other components. Currently no infrastructure exists to that allows to test the RF cavities in a high magnetic field. A previous infrastructure in the US has been dismantled after successful experiments. We are striving to develop a new test infrastructure, different laboratories could potentially house them. These include INFN Milano, STFC, IRFU-CEA and CERN. However currently the funding is not secured.

\subsubsection{Magnet roadmap}
Several magnet technologies are key. Currently we anticipate the use of
HTS solenoids in the muon production and cooling section:
\begin{itemize}
\item
  Large-aperture, high-field solenoids for the muon production target.
\item
  Solenoids with a range of apertures and fields for the 6D cooling.
\item
  Highest-field, small-aperture solenoids for the final cooling.
\end{itemize}
It is expected that the technology can be mature in 15 years from now since HTS has already been successfully used in high-field solenoids.

In addition high-field accelerator magnets are required both for the muon accelerator rings and the collider ring. The latter are more demanding since their field impacts the luminosity performance. These magnets need not to be ramped in normal operation, which allows to consider different technology choices such as the use of uninsulated coils.
For the collider ring magnets, it maybe premature to consider the use of HTS.
Therefore the currently more mature Nb$_3$Sn technology will be considered as the baseline. HTS is also being explored, both as a potential technology for the first stage, if progress is fast, but more importantly as a technology for the second stage of the project which could use HTS or hybrid magnet designs.
The technology choice for the superconducting magnets in the accelerator ring will be based on an optimisation of the cost.

In addition, fast-ramping normal-conducting magnets are required for the pulsed synchrotrons of the acceleration complex. They represented the longest part of the complex and need development to optimise the cost and power consumption. However the technology is in principle well established and can be developed quickly with enough funding.

\subsubsection{Detector roadmap}
An outline of the detector development timeline is illustrated in Fig.~X, which has been derived based on experience with constructing the original ATLAS and CMS detectors and their major upgrades for HL-LHC. Almost all the subsystems of a Muon Collider detector require significant  R\&D. In particular, the current requirements on the performance of tracking, calorimeter, and muon systems include challenging specifications for granularity, spacial and timing resolutions, and radiation hardness of these systems. Individually these requirements could be satisfied with technologies being developed for the HL-LHC detectors, however achieving all of them in a single detector represents a significant challenge that will require a decade or longer R\&D program. Development of detector technologies should proceed hand in hand with simulation studies aimed to further refine the detector performance specifications with specific physics targets in mind. It should be noted that the detector size and performance requirements are significantly different for the 3\,TeV and 10\,TeV detectors, it is therefore important to proceed with the development of both designs while the staging strategy is being further refined.  Certain detector design aspects, such as the forward muon tagging and the potential luminosity detector, are still at the roots of their development and require identification of promising design options and their studies/comparisons. The data acquisition strategies, including novel AI/ML based approaches for on- and off- detector data processing,  and the corresponding data bandwidth, trigger/DAQ and computing requirements need to be evaluated and refined. Construction of large solenoidal magnets for the detectors is currently the largest uncertainty in the schedule due to limited supply of the magnet conductor as well as lack of industry companies capable of producing such magnets. On the bright side, this challenge is not unique to muon colliders, and development of future detector solenoids should be identified as a priority for the field and be pursued in synergistic fashion with other collider, CLFV, Axion Dark Matter and other experiments. Finally, engineering aspects of putting the entire detector together, while preserving capability to access and replace its components, need to be studied. 

The timeline for the detector development project anticipates about five to six years for generic R\&D to further develop and refine the detector technologies. This will be followed by six to seven years of targeted R\&D to finalize the technology choices and the designs ahead of the TDRs, and a further three to four years for prototyping to develop production grade components and to test them in a detector demonstrator, mimicking the final detector but at a smaller scale. The construction (including pre-production batches) is then planned for five to six years, followed by four years of installation and commissioning in mid-2040s. More detailed schedule will be developed at the time of the CDR which is anticipated in 2030s.

\subsubsection{Other R\&D roadmap}

In addition there are several other items are key for the performance, cost,
power consumption and risk of the muon collider. They can be
sufficiently accelerated not to limit the schedule provided the resources are
available. The most important of these items are:
\begin{itemize}
\item
The background mitigation technology in the detector.
\item
Fast ramping magnets and their power converters.
\item
The lattice design of the different parts of the collider.
\item
The muon production target.
\item
The RF system of the muon accelerator and the proton complex.
\item
The neutrino flux mitigation technology.
\item
Other technologies, such as cryogenics, instrumentation, vacuum and many more.
\end{itemize}
These also need to be addressed to verify that the muon collider can hold its
promises and to develop the R\&D programme for the next phase.

\subsection{Site considerations}
At this moment the study is almost completely site independent. At a later
stage the implications of specific sites will be investigated as much as
resources allow. In particular, the key impact on the environment will be considered
together with potential reuse of existing infrastructure. Also a first-order assessment
of the civil engineering cost and impact on the project will be performed.

\subsubsection{European sites}
Up to now studies in Europe were limited to identifying at least one site close to CERN.
One particularly important concern for this site is the mitigation of the neutrino flux
that arises from the collider ring. At CERN, the goal is to ensure that the impact of the
neutrino flux on the environment outside of the laboratory site is negligible, similar
to the impact of the LHC. A concept
has been developed that can ensure meeting this goal for neutrinos from the arcs. It is
based on a mover system that slightly deforms the machine every few hours. Work
started on exploration of the mechanical systems and the implication on the
beam dynamics. 

The experimental insertions can lead to a more important neutrino flux where
the prolongation of the axis of each detector intercepts the surface of the
earth. These locations will see non-negligible neutrino flux and might be
suited for neutrino measurements, see~\ref{physics-section}. At CERN one site and orientation of the
collider ring has been identified where the neutrino beams would enter the
Mediterranean on side and the uninhabited flank of a mountain on the other side.

Other possible locations will be explored. It should also be noted that a
previous collider ring lattice design for 3\,TeV promised to limit the neutrino
flux to an acceptable level even in direction of the experiments. This design
would compromise the use of the neutrinos for measurements.

The use of existing infrastructure will also be considered in the future. In particular,
the SPS and LHC tunnels have sizes that make it interesting to consider implementing the
accelerating synchrotrons in them. However this will have some impact on the schedule,
since the tunnels need to be cleared of existing equipment before. Other relevant infrastructure
include the existing proton complex, the cryogenics infrastructure and more.

\subsubsection{US site}
Renewed US interests and contributions in muon collider efforts
provide for an ideal opportunity to consider potential US
siting. The existing accelerator facilities at Fermilab (e.g.~PIP-II)
and the proposed booster upgrade or extension of the PIP-II linac
could be developed into proton drivers for a Muon
Collider. Furthermore, synergies with the neutrino and intensity
frontiers programs in the US, as well as overlaps with nuclear science
and industry applications provide additional benefits.
This makes Fermilab an attractive siting option.

A reasonable question to ask is how the generic design presented here can be modified for siting of a muon collider at Fermilab, where one is confined within the site bounds of the laboratory. Table
~\ref{tab:Table1} shows tentative parameters of a possible Fermilab-based accelerator scenario that produces 5\,TeV beams for a 10\,TeV collider. This is a 4-stage hybrid RCS scenario that is constrained to fit within the Fermilab site, which restricts the circumference of the largest tunnel hosting the final two accelerators to approximately 16.5\,km. The magnetic fields in the highest energy accelerators are set at 15\,T for the superconducting magnets and $\pm$1.8\,T for the rapid-cycling magnets. 

A center of mass energy at the top quark production threshold (173\,GeV per beam) could be a convenient choice for a first Collider, particularly if highest intensities or lowest beam emittances may not be initially available, since important physics results could be obtained with lower luminosity. We therefore propose to extend the low energy acceleration chain by adding an additional recirculating linear accelerator from 63\,GeV up to 173\,GeV. This would use 5 passes with 650\,MHz RF and would be similar to the earlier RLA stages.

The first RCS is chosen to include only normal-conducting magnets (up to 1.8\,T) and to fit a 6280\,m circumference so that it could utilize
the existing Tevatron tunnel. That limits its peak energy to
approximately 450\,GeV. Parameters for the 173 to 450\,GeV RCS1 are presented in table ~\ref{tab:Table1}.

The second RCS must take beam from 450\,GeV to the 1.725\,TeV injection
of the 16.5\,km rings and does not fit well in either the Tevatron or
the 16.5\,km tunnel. We thus chose an intermediate circumference of 10.5\,km for RCS2. With the more flexible choice of circumference, we can reduce the high field magnet to 12\,T. See Table ~\ref{tab:Table1} for possible parameters.

Energy range in the 16.5\,km ring is constrained by the peak field available (15T) and by the limited cycling range of -1.8\,T to + 1.8\,T in the RC magnets. The highest energy accelerator cycles from 3.56\,TeV to 5.0\,TeV (RCS4 in Table ~\ref{tab:Table1}). A second accelerator (RCS3) is included in the 16.5\,km tunnel, and it cycles from 1.725\,TeV to 3.56\,TeV.

An example layout of the accelerator complex based on the design considerations presented above is shown in Fig.~\ref{fig:sitefillermap}.
 
\begin{table}
\caption{\label{tab:Table1}Parameters for a Fermilab-based 5\,TeV muon Accelerator.}
\centering

\begin{tabular}{| l | l | l | l | l | l | l |}
\hline
\textbf{Parameter} & \textbf{Symbol} & \textbf{Unit} & \textbf{RCSI} & \textbf{RCS2} & \textbf{RCS3} & \textbf{RCS4} \\
\hline
Hybrid RCS &   &   & No & Yes & Yes & Yes \\
\hline
Repetition rate & $f_{rep}$ & Hz & 5 & 5 & 5 & 5 \\
\hline
Circumference & $C$ & m & 6280 & 10500 & 16500 & 16500 \\
\hline
Injection energy & $E_{inj}$& GeV & 173 & 450 & 1725 & 3560 \\
\hline
Ejection energy & $E_{ej}$ & GeV & 450 & 1725 & 3560 & 5000 \\
\hline
Energy ratio & $E_{ej}$/$E_{inj}$ &   & 2.60 & 3.83 & 2.06 & 1.40 \\
\hline
Decay survival rate & $N_{ej}/N_{in}$ &   & 0.85 & 0.83 & 0.85 & 0.89 \\
\hline
Acceleration time & $T_{acc}$ & ms & 0.97 & 3.71 & 8.80 & 9.90 \\
\hline
Revolution period & $T_{rev}$  &\textmu s & 21 & 35 & 55 & 55 \\
\hline
Number of turns & $N_{turn}$ &   & 46 & 106 & 160 & 180 \\
\hline
Required energy gain per turn & $\Delta E$ & GeV & 6 & 12 & 11.5 & 8.0 \\
\hline
Average accel. Gradient & G\textsubscript{avg} & MV/m & 0.96 & 1.15 & 0.70 & 0.48 \\
\hline
Bunch population at injection & N\textsubscript{in} & 10\textsuperscript{12} & 3.3 & 2.83 & 2.35 & 2.0 \\
\hline
Bunch population at ejection & N\textsubscript{ej} & 10\textsuperscript{12} & 2.83 & 2.35 & 2.0 & 1.8 \\
\hline
Vertical norm. emittance & $\epsilon_{v,N} $  & mm-mrad & 25 & 25 & 25 & 25 \\
\hline
Horiz. norm. emittance &$ \epsilon_{h,N} $ & mm-mrad & 25 & 25 & 25 & 25 \\
\hline
Long, norm. emittance  &$ \epsilon_{z,N }$ & eV-s & 0.025 & 0025 & 0.025 & 0.025 \\
\hline
Total straight length & $L_{str}$ & m & 1068 & 1155 & 2145 & 2145 \\
\hline
Total NC dipole length & $L_{NC} $ & m & 5233 & 7448 & 10670 & 8383 \\
\hline
Total SC dipole length & $L_{sc} $ & m &   & 1897 & 3689 & 5972 \\
\hline
Max. NC dipole field & $B_{NC} $ & T & 1.80 & 1.80 & 1.80 & 1.80 \\
\hline
Max. SC dipole field & $B_{SC}$ & T &   & 12 & 15 & 15 \\
\hline
Ramp rate & $B'$ & T/s & 1134 & 970 & 440 & 363 \\
\hline
Main RF frequency & $f_{rf}$ & MHz & 1300 & 1300 & 1300 & 1300 \\
\hline
Total RF voltage  & $V_{rf}$ & MV & 6930 & 13860 & 13280 & 9238 \\
\hline

\end{tabular}

\end{table}

\begin{figure}[t!]
\centering
\includegraphics[width=0.9\linewidth]{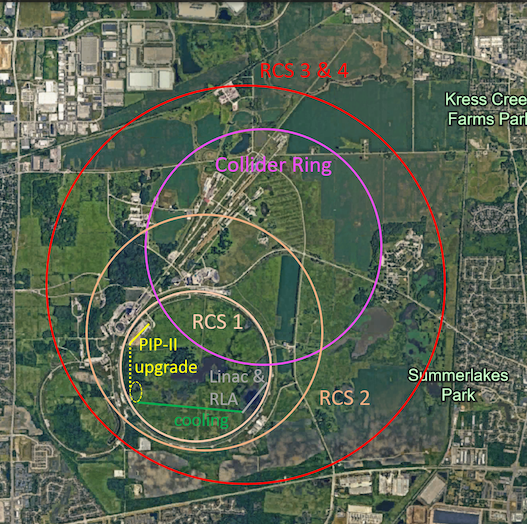}
\caption{Proposed layout of a 10\,TeV center-of-mass energy muon collider complex sited at Fermilab. The proton driver is shown in blue and the cooling channel is shown in green. RCS1 can be located inside the existing Tevatron tunnel. RCS3 and RCS4 are located inside the accelerator ring with circumference of 16.5 km, which is the largest we can fit onsite.}
\label{fig:sitefillermap}
\end{figure}


\subsubsection{Other regions}
At this moment no other sites have been considered. We hope that regional interest will
provide resources to start these studies.

\subsection{Sustainability, environmental impact, cost and power consumption considerations}
By 2026, a full costing of the muon collider is not possible with the existing
resources. The study will focus on the key cost and power consumption drivers:
\begin{itemize}
\item
  The magnets, power converters and RF systems of the RCSs. In particular,
  the trade-off between the power converters and the RF systems, both in cost
  and power consumption.
\item
  The collider ring magnets and their shielding. In particular, the
  trade-off between cost, power consumption and luminosity performance.
\item
  The cost and power consumption of the target solenoid.
\item
  The muon cooling complex.
\item
  The proton complex.
\item
  An approximate estimate of civil engineering and infrastructure cost based on
  scalings.
\end{itemize}
The superconducting magnet cost will have to make assumptions about the future
cost reduction that can be achieved for different technologies.
Current resources will not allow a design of the initial accelerating linacs;
we will rely on past designs by MAP.

The study will also provide a rough overall estimate of the budget required for
the R\&D towards the muon collider. However, this will remain very rough since
the current resources situation will not allow the design of the test facility.

\begin{flushleft}
\interlinepenalty=10000

\end{flushleft}
}
\clearpage




\end{document}